\pdfoutput=1
\documentclass[pdftex]{panda_book}
\svnInfo $Id: P_magn_TDR.tex 710 2009-02-10 12:13:38Z IntiL $ 
\svnKeyword $Revision: 710 $ 
\svnKeyword $Date: 2009-02-10 12:13:38 +0000 (Tue, 10 Feb 2009) $
\svnKeyword $HeadURL: https://einstein.astro.gla.ac.uk/pandatdr/Tex/P_magn_TDR.tex $

\makeatletter

\markpanda{Magnet TDR, Feb.\ 2009}
	  {Technical Design Report -- Magnets}

\makeatother
\usepackage{fix-cm}

\begin{document}

\svnInfo $Id: title.tex 708 2009-02-09 16:52:12Z IntiL $
\svnKeyword $Revision: 708 $ 
\svnKeyword $Date: 2009-02-09 16:52:12 +0000 (Mon, 09 Feb 2009) $

\pagenumbering{roman}
\onecolumn
%
%
\thispagestyle{empty}

\vspace*{-1.7cm}
\hfill \includegraphics[width=5cm]{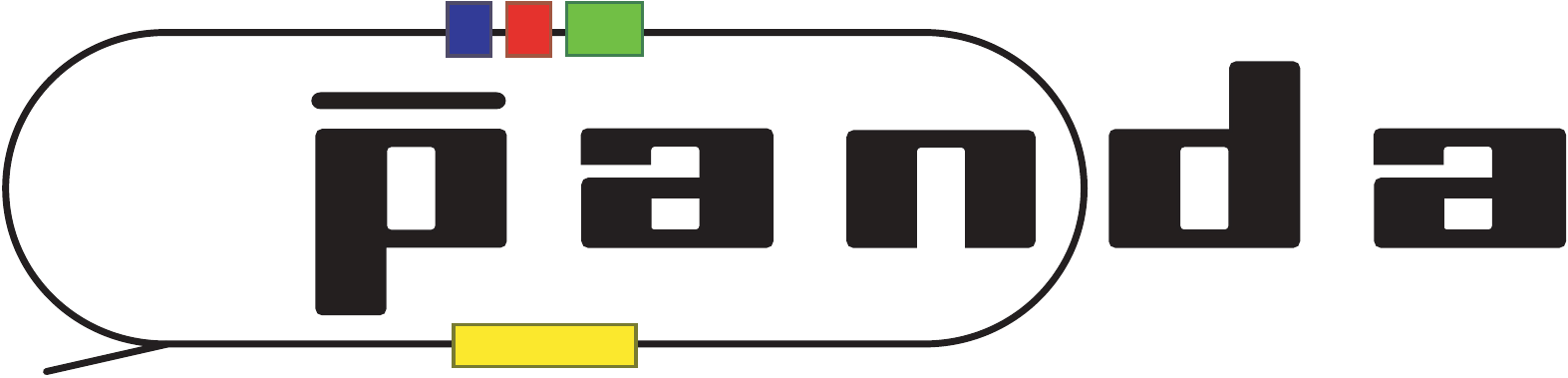}

\vspace*{5mm}
\begin{center}
{\bfseries \sffamily \Huge Technical Design Report for the\\
\vspace*{9mm}
{\fontsize{34}{40}\selectfont  \Panda{}} } \\
\vspace*{2mm}
{\sffamily  (Anti\underline{p}roton \underline{An}nihilations at \underline{Da}rmstadt)\\
Strong Interaction Studies with Antiprotons}\\
\vspace*{7mm}
{\bfseries \sffamily \fontsize{34}{40}\selectfont Solenoid and Dipole\\
\vspace*{3mm}
           Spectrometer Magnets}\\

\vskip 1cm
{\LARGE \sffamily The \Panda{} Collaboration}
%
%

\vskip 1cm
{\LARGE \sffamily February 2009}
\vskip 5mm

\includegraphics[width=0.8\dwidth]{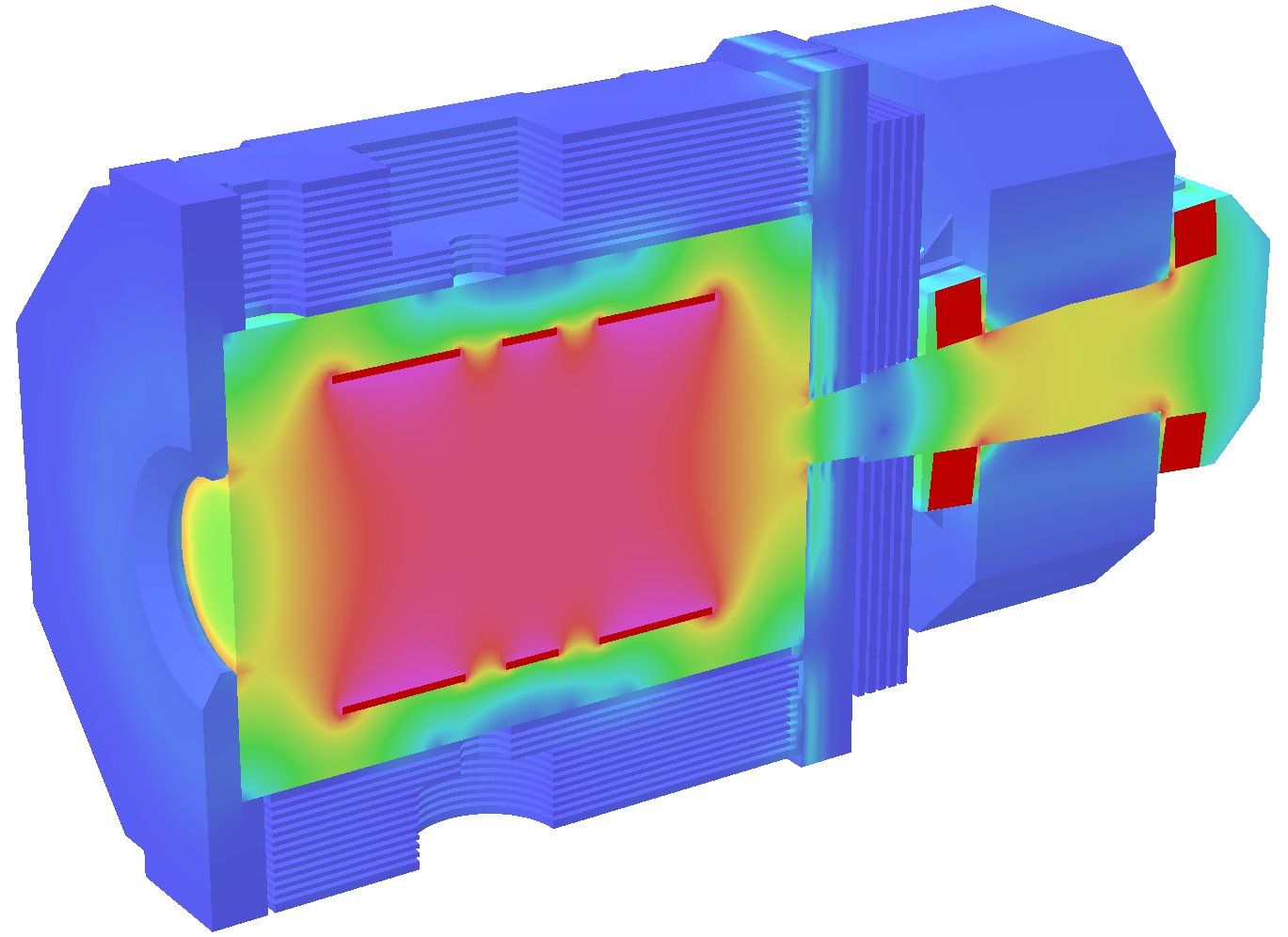}
\end{center}
\vfill
%
%
\cleardoublepage
\begin{center}
\vspace*{3mm }
{\LARGE \bfseries \sffamily The \Panda{} Collaboration}\\
\vskip 7mm
%
%
\svnInfo $Id: authors.tex 707 2009-02-05 17:30:00Z IntiL $

\institem{Universit\"at {\bf Basel}, Switzerland}
\authitem{W.~Erni},
\authitem{I.~Keshelashvili},
\authitem{B.~Krusche},
\authitem{M.~Steinacher}\lastitem
\institem{Institute of High Energy Physics, Chinese Academy of Sciences, {\bf Beijing}, China}
\authitem{Y.~Heng},
\authitem{Z.~Liu},
\authitem{H.~Liu},
\authitem{X.~Shen},
\authitem{O.~Wang},
\authitem{H.~Xu}\lastitem
% ...
%
%

\end{center}
%
%
\vfill
\newpage
\hrulefill\\
{\bfseries \sffamily Editorial Board:}  \\
\begin{tabular}{l l l}
                       & Inti~Lehmann (chair)    & Email: \verb$I.Lehmann@physics.gla.ac.uk$ \\  \ \\
                       & Andrea~Bersani	    & Email: \verb$Andrea.Bersani@ge.infn.it$ \\ \ \\
		       & Yuri~Lobanov	    & Email: \verb$Lobanov@jinr.ru$ \\ \ \\
		       & Jost~L\"uhning	    & Email: \verb$J.Luehning@gsi.de$ \\ \ \\
		       & Jerzy~Smyrski	    & Email: \verb$Jerzy.Smyrski@uj.edu.pl$ \\ \ \\

Technical Coordinator: & Lars Schmitt       & Email: \verb$L.Schmitt@gsi.de$ \\ 
Deputy:                & Bernd Lewandowski  & Email: \verb$B.Lewandowski@gsi.de$ \\ \ \\
Spokesperson:	       &  Ulrich Wiedner    & Email: \verb$Ulrich.Wiedner@ruhr-uni-bochum.de$ \\
Deputy:		       & Paola Gianotti	    & Email: \verb$Paola.Gianotti@lnf.infn.it$ \\
\end{tabular}\\

\hrulefill\\

\vspace*{1cm}

%
\vfill

\cleardoublepage

%
\svnInfo $Id: preamble.tex 550 2009-01-27 10:42:55Z IntiL $

\begin{center}
\vspace*{2cm}
{\Large \bfseries \sffamily Preface}\addcontentsline{toc}{chapter}{Preface}
\vskip 2cm
\begin{minipage}[t]{8cm}
\sloppy\large
This document is the Technical Design Report covering the two large
spectrometer magnets of the \PANDA detector set-up.  It
shows the conceptual design of the magnets and their anticipated
performance.  It precedes the tender and procurement of the magnets
and, hence, is subject to possible modifications arising during this
process.
\end{minipage}
\end{center}
\vspace*{2cm}
\centerline{
}
\vfill
\clearpage
\vspace*{18cm}
\hrulefill\\
\vspace*{2cm}\\
\begin{minipage}[t]{10cm}
\sloppy
The use of registered names, trademarks, \etc in this publication does not
imply, even in the absence of specific statement, that such names are exempt
from the relevant laws and regulations and therefore free for general use.
\end{minipage}
\vfill
%

\onecolumn
\cleardoublepage
\begin{sffamily}
\tableofcontents
\end{sffamily}

\twocolumn

\cleardoublepage
\pagenumbering{arabic}
\setcounter{page}{1}
\svnInfo $Id: exs.tex 708 2009-02-09 16:52:12Z IntiL $

\setcounter{chapter}{-1}
\refstepcounter{chapter}
\chapter*{Executive Summary}
\addcontentsline{toc}{chapter}{Executive Summary}
\label{s:exs}

\AUTHORS{G.~Rosner}

\section*{Physics Case}
\addcontentsline{toc}{section}{Physics Case}

The microscopic structure of dense matter is governed by one of the
four fundamental forces in nature, the strong force. This force
dominates the interaction between nucleons (protons and neutrons) in
atomic nuclei, as it determines the interaction between quarks and
gluons inside nucleons and other hadrons. Achieving a quantitative
understanding of matter at this microscopic level is one of the
exciting challenges of modern physics. The underlying fundamental
theory, Quantum Chromo Dynamics (QCD), is elegant and deceptively
simple. It generates a remarkable richness and complexity of
phenomena, which are far from being completely understood.

The fundamental building blocks of QCD are point-like quarks, which
interact by exchanging gluons, the messenger particles (intermediate
bosons) of QCD.  At very high energies, or distances much smaller than
1\,fm, quarks interact only weakly.  Hence, QCD offers simple
perturbative solutions. At nuclear energy scales, or distances of
about the size of a nucleon, 1\,fm, the interaction between quarks
becomes very strong and a perturbative approach is no longer
applicable. Solving QCD becomes very involved, and is either done by
formulating effective field theories that preserve certain features of
QCD, or solving QCD on the lattice.

At larger distances, the attractive force between quarks becomes so
strong, that it is impossible to separate them - free quarks have
never been observed. Rather they are \emph{confined} to ``their''
hadron.  This very unusual behaviour of the strong force, as compared
to other fundamental forces like gravity or electromagnetism, is
attributed to the self-interaction of gluons. Not only do gluons
mediate the force between quarks, they also interact among themselves,
thus forming ``strings'' or ``flux tubes''. Consequently, the strongly
interacting particles we observe in nature, such as baryons and
mesons, are complex systems of confined quarks and gluons.

Quarkonia, which are states of a quark and an antiquark of the same
flavour, are among the simplest QCD states and therefore well suited
to study confinement. The charmonium system offers a unique
opportunity to study quarkonia, since the low density of states and
their narrowness reduces mixing among them. The best understanding has
so far been achieved for the charmonium states with $J^{PC}=1^{--}$,
because they can be directly formed at electron-positron
colliders.  The big advantage of using antiproton beams is that 
charmonium states of \emph{all} quantum numbers (not only 
$J^{PC} = 1^{--}$ as at $e^{+}e^{-}$ colliders) can be formed directly 
and that the precision of the mass and width measurement only depends 
on the beam quality. (For these so-called
formation processes, the detector resolution is less important. Still,
the detector response needs be optimised to reject background
efficiently for rare events.)  Data on the excited non-$\psi$ states,
the $D$ states of charmonium, will be very instrumental to improve our
theoretical understanding further.  \PANDA's $\bar{p} p$ scans of
charmonium states will be much superior to the experiments performed
at $e^+e^-$ colliders, because of much smaller statistical and
systematic errors. Hence, \PANDA's discovery potential will be
significantly higher.

An important consequence of the gluon self-interaction is the
predicted existence of particles with gluonic degrees of freedom.
These particles would be so-called hybrids consisting of quarks,
antiquarks and gluons, or may even consist of pure ``glue''.  Their
discovery would provide another highly relevant test of QCD in the
non-perturbative regime. The additional degrees of freedom carried by
gluons would allow glueballs and hybrids to have spin-exotic quantum
numbers $J^{PC}$ that are forbidden for normal mesons, by which they
could then be identified uniquely.

The properties of glueballs and hybrids are determined by the
long-distance (low-energy) features of QCD, and their study will yield
fundamental insight into the structure of the QCD vacuum. The most
promising results regarding gluonic hadrons have come from antiproton
annihilation experiments. Two particles, first seen in $\pi$N
scattering with exotic $J^{PC} = 1^{-+}$ quantum numbers, the
$\pi_1(1400)$ and $\pi_1(1600)$, are clearly seen in annihilation at
rest.  The search for glueballs and hybrids has so far been restricted
mainly to the mass region below 2.2\,GeV/c$^2$, where the density of
ordinary quark-antiquark states is high. Experimentally, it will be
very rewarding to go to higher masses, because above 2.5\,GeV/c$^2$,
heavy quark states are few in number and hence can easily be
resolved. (The light quark states form a structureless continuum.)
This is particularly true for the charmonium region. It is expected
that there are a number of exotic charmonia in the 3 to 5.5\,GeV/c$^2$
mass region, which is accessible to \PANDA, and where they could be
resolved and unambiguously identified.

The current quarks inside the nucleon are very light point-like
particles, which contribute only a few percent to the mass of the
nucleon or nucleus/universe.  Nearly all of the mass is thought to be
generated dynamically, the mechanism being related to the spontaneous
breaking of chiral symmetry, one of the fundamental symmetries of QCD
in the limit of massless quarks, or confinement. However, up to now
the detailed structure of hadrons such as protons and neutrons is far
from being understood quantitatively. Antiproton annihilations leading
to electromagnetic final states will provide new information,
complementary to the classical approach of elastic lepton
scattering. There are several ways in which \PANDA will be able to
investigate the structure of the proton.  The most promising
approaches are the measurements of time-like form factors and of
time-like Compton Scattering, crossed channel Deeply Virtual Compton
Scattering (DVCS), and the extraction of the Boer-Mulders structure
function from Drell-Yan data.  The proton time-like form factors have
previously been measured in several experiments in the low
four-momentum-transfer, $Q^2$, region down to threshold. At high
$Q^2$, the only measurements are those performed by E760 and E835 at
Fermilab up to $Q^2$ values of about 15\,GeV$^2/c^2$. However, the
magnetic and electric form factors $| G_M |$ and $| G_E |$ could not be
measured separately, due to limited statistics. This will be possible
at \PANDA.

The phenomenon of confinement, the existence or non-existence of
hybrids and glueballs, the origin of hadron masses and the structure
of the nucleon are long-standing puzzles in contemporary physics. They
will be addressed by the \PANDA experiment at the Facility for
Antiproton and Ion Research, \FAIR.

\section*{The \PANDA Experiment}
\addcontentsline{toc}{section}{The \PANDA Experiment}

The \PANDA collaboration proposes to build a state-of-the-art
universal detector system to study reactions of anti-protons impinging
on a proton or nuclear target internal to the High Energy Storage Ring
(\HESR) at the planned \FAIR facility at \GSI, Darmstadt, Germany. The
detector aims at taking advantage of the extraordinary physics
potential offered by a high-intensity, phase-space cooled anti-proton
beam colliding with a flexible arrangement of targets.

\begin{figure*}[htb]
\begin{center}
\includegraphics[width=\dwidth]{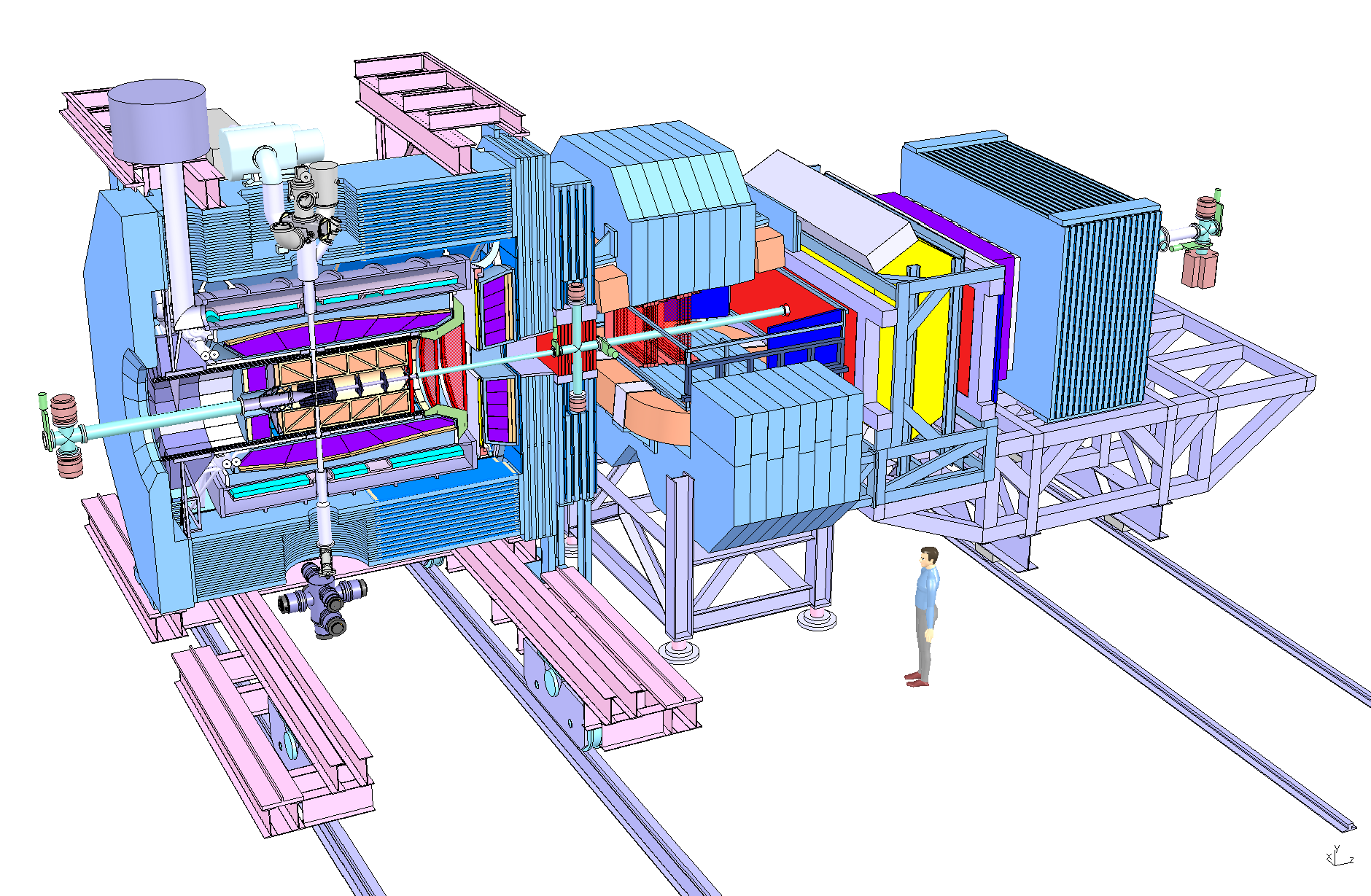}
\caption[Schematic layout of the proposed \PANDA Experiment at \FAIR.]
   {Schematic layout of the proposed \PANDA Experiment at \FAIR.
   The antiproton beam enters the detector from
   the left. The Target Spectrometer is complemented by a Forward
   Spectrometer to ensure full phase space coverage.}
\label{f:exs:panda}
\end{center}
\end{figure*}

The \PANDA detector (see Fig.~\ref{f:exs:panda}) will consist of a
4\,m long, 2\,T superconducting target solenoid spectrometer and a
2\,Tm resistive dipole magnetic spectrometer at forward angles. \PANDA
will incorporate the latest detector technology to achieve excellent
mass, momentum, energy and position resolution, superior particle
identification and large solid angle coverage.

One obtains the maximum acceptance for the physics channels of interest
when placing the detector system around a target internal to the ring
that stores the antiprotons. Experimentally, this is a challenge.
Antiproton beams of superior quality and intensity are
difficult to produce. Although \FAIR will provide the best anti-proton
beams worldwide, its intensity will still be low compared to
conventional particle beams. An appropriate target technology has to
be developed to achieve high luminosity. Studying the decay of charmed
particles requires a precise micro-vertex tracking detector system
close to the interaction region as well as a powerful particle
identification system. The latter should be able to discriminate
hadron species as well as providing an excellent hadron/electron
separation and muon identification. A central tracker will provide
charged particle tracking. The Target Spectrometer detector system will be
completed by a high-resolution electromagnetic calorimeter. A large
solid angle coverage will be achieved by adding a Forward Spectrometer of
similar capabilities. This complex detector arrangement will ensure the
measurement of complete sets of observables, thus enabling \PANDA to
reach its physics goals.

\section*{Large Aperture Magnetic Spectrometers}
\addcontentsline{toc}{section}{Large Aperture Magnetic Spectrometers}

The detection concept for the \PANDA experiment is based on the
reconstruction of charged particle tracks in magnetic fields in
conjunction with calorimetry for neutral particles and muons. Only
with this combination it will be possible to identify the reaction channels
of interest unambiguously. The \PANDA Target Spectro\,meter (TS) will
consist of a superconducting solenoid, which will feature a 1.9\,m free
inner diameter to house a variety of tracking, particle identification
and calorimetric detectors. The large-aperture resistive dipole magnet
plus a set of analogous detectors will constitute the \PANDA Forward
Spectrometer (FS).

The Target Spectrometer will be the central part of the \PANDA
detector, and will enclose the interaction point. The solenoid magnet must
provide a field of 2\,T in the central region, where the trackers will be
located, with a homogeneity of $\pm2$\% and very small radial field
components, so that a 1.5\,m long Time Projection Chamber (TPC) can be
operated reliably. This translates to the requirement that the
integral of the radial component along the $z$ axis $\int B_r(z)/B_0
dz$ from any point inside the tracker to the read-out plane of
the TPC must be less than 2\,mm. At this level of precision the field is
strongly dependent on many parameters, like details of the yoke
geometry and coil arrangement, which in turn influence the design of
several critical detectors inside the solenoid. Above requirements
called for an extensive optimisation process in designing the
solenoid, with many iterations.

For the \PANDA TS magnet we chose to use a superconducting coil.
The TS design is based on an aluminium stabilised, indirectly
cooled superconducting solenoid using internal winding in an aluminium alloy
mandrel.  Aluminium stabilised cables give high stability against
quenches due to the large electrical conductivity of aluminium at low
temperature. The coil cable will withstand large thermal perturbations
(energy releases) before a normal conducting zone starts to grow,
leading to a quench. In addition, when a quench occurs it spreads
more evenly than in other cable types, thus reducing high thermal stresses
which could potentially lead to damages to the magnet.

By using internal winding, the cable will be pressed against the outer
mandrel keeping to a minimum the stress on the epoxy glass insulation.
Internal winding and indirect cooling will greatly reduce the amount of
liquid helium, the cryostat design will be simplified and, since there will be
no liquid helium bath, no pressure vessel will be needed. The helium will be
contained in standard aluminium pipes rated to the maximum pressure
that occurs during a quench or a major failure of the
refrigerator. The use of internal winding and aluminium stabilised
cables has been the technology chosen for many 4$\pi$ spectrometer
solenoids from Cello in the early 80s
to CMS at LHC today.

All detectors of the Target Spectrometer will be accommodated
inside the solenoid, most of them inside the bore of the cryostat.
This is a challenge for the design of the detectors, of their supports
and supplies, and the magnet itself. A particular challenge is
the accommodation of the target, which requires a vertical pipe
traversing the magnet upstream of its centre. Consequently, the coil
will be divided into three sub-coils, which renders the design of the
coil former and of the cryostat much more difficult than those of
other solenoid magnets. The target pipe will pass in between the first
two sub-coils through a warm bore in the cryostat. The iron of the
flux return yoke of the solenoid will act as an active muon range
system.  This is to be achieved by segmenting the yoke into 13 iron
layers in the barrel and 5 iron layers in the downstream end cap
interleaved with Mini Drift Tubes (MDTs).

It will be possible to open the flux return yoke from both upstream 
and downstream
sides by sliding doors to give access to all detectors inside the
solenoid. The whole TS, with all detectors in place, was designed to
be movable from the in-beam position to a maintenance position.

Extensive studies regarding the electrical and mechanical properties
of the solenoid were performed. Stresses and temperature gradients in
the coil were analysed in detail.  A quench protection system was
designed and the behaviour of the coil during a quench
was evaluated. Detailed Finite Element Model (FEM) calculations for the
coil, cryostat, flux return yoke and support structures showed that
all parts can be safely operated. Field studies showed that
all fields fulfil the given requirements; and a powering and emergency
shut-down procedure was developed.

The main purpose of the Forward Spectrometer is to reconstruct
particles emitted from the target at angles below $5^\circ$ vertically
and $10^\circ$ degrees horizontally.  The large aperture resistive dipole
magnet will provide a field integral of 2\,Tm, allowing for a momentum
reconstruction of charged particles to a precision of better than
1\%. The dipole will have an aperture of about $1 \times 3\,$m$^2$ to cover
the aforementioned angular acceptance at about 3.5\,m downstream from
the interaction point. Inside this gap, two large tracking chambers and
scintillation counters will be housed on a dedicated support frame which
will allow their retraction for maintenance. In addition, support
structures were designed, which will allow to mount the drift chambers
at the entrance and exit of the dipole. The remaining detectors of the
FS will be mounted on a platform that can be moved by a drive system
from the maintenance and assembly position in the hall to the in-beam
position. Between the TS solenoid and the FS dipole 5 instrumented
layers of iron will add to the muon detection system and will 
decouple the magnetic
fields of the solenoid and the dipole.

The decision to use resistive race-track coils for the dipole magnet
was taken after evaluating alternative options in great detail. This
choice proved to be the safest and most economic option both with
respect to the investment as well as the running costs. The design was
optimised to satisfy the requirements concerning the bending power,
field homogeneity and acceptance providing sufficient space for the
tracking detectors in the gap. A thorough mechanical analysis of the
coil and frame stability was carried out. Extensive field studies
showed that the bending of tracks traversing the magnet on different
trajectories will vary no more than 10\%. This can easily be handled by
the track reconstruction software of \PANDA. Since the dipole magnet
will be part of the HESR lattice, its current must increase during the
acceleration of antiprotons. Therefore the magnet was designed to
ramp the current from 25\% to 100\% of its maximum value within
60\,s. It was shown that a lamination of 20\,cm is enough to keep the
eddy currents at an acceptable level.

The design of the two large spectrometer magnets plus their support
structures has been performed in a collaborative effort by seven
groups from Germany, Italy, Poland, Russia and the UK. In
Table~\ref{t:orga:inst}, the institutes leading the design of specific
items are listed.

\begin{itemize}
\item Coil and cryostat of the TS solenoid -- INFN, Genoa.
\item Instrumented flux return yoke of the TS solenoid -- JINR, Dubna.
\item Large aperture FS dipole magnet -- University of Glasgow.
\item Support structures for the FS detectors -- CUT and UJ, Krakow.
\item Rail systems and movement of the TS and the FS detector
  platform -- GSI, Darmstadt and CUT, Krakow.
\end{itemize}

The Forschungszentrum J\"ulich (FZJ) takes care of the \PANDA
spectrometers' integration into the HESR, which is particularly
important for the dipole, since it will be part of the
accelerator/storage ring lattice.  A detailed list of institutions and
work packages can be found in Chapter~\ref{s:orga}.

Both spectrometer magnets will act as mounting structures for the
detectors. Therefore the two magnets will need to be in place before any
mounting of the detectors can be started. We have taken this into
account regarding the timelines, which foresee that all magnet
components will be shipped to Darmstadt in 2012.

\cleardoublepage
\svnInfo $Id: intro.tex 712 2009-03-31 16:27:19Z IntiL $

\chapter{Introduction}
\label{s:intro}

\AUTHORS{I.~Lehmann}

\begin{figure}[htb]
 \begin{center} 
 \includegraphics[width=\swidth]{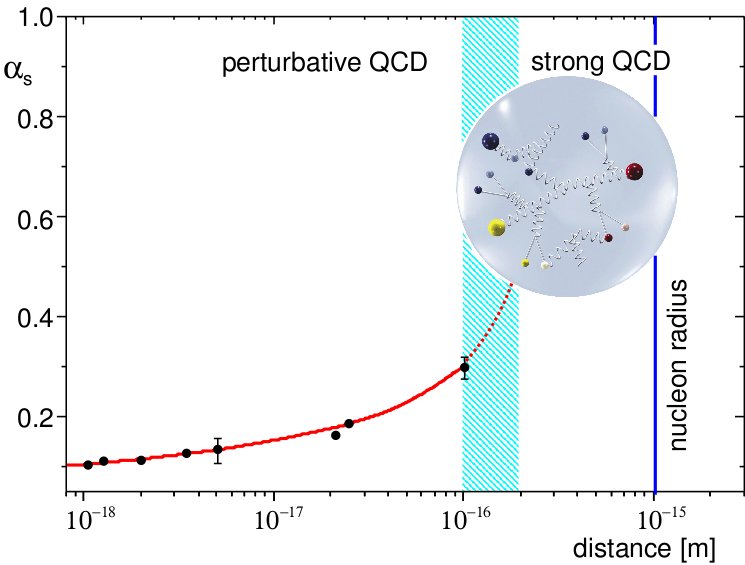}
 \caption[Coupling constant of the strong interaction as a function of
 distance.]{Coupling constant of the strong interaction as a function of
 distance. The data points represent experimental
 values~\cite{PDG2004}. For distances between quarks comparable to the
 nucleon size the interaction becomes so strong that quarks cannot be
 further separated (confinement) and hadrons are formed. \PANDA will
 investigate the properties of the strong interaction in this key
 region for the understanding of matter.} 
 \label{f:intro:alpha_s}
 \end{center}
\end{figure}

The physics of strong interactions is undoubtedly one of the most
challenging areas of modern science.  Quantum Chromo Dynamics (QCD) is
reproducing the physics phenomena only at distances much shorter than
the size of the nucleon, where perturbation theory can be used
yielding results of high precision and predictive power. As the
coupling constant rises steeply at nuclear scales (see
Fig.~\ref{f:intro:alpha_s}) perturbative expansions diverge and a
different theoretical approach is required. However, the strong
interaction keeps providing new experimental observations, which were
not predicted by ``effective'' theories. The latter retain the
fundamental symmetries of QCD, but have problems in describing all the
observed phenomena simultaneously.

The physics of strange and charmed quarks holds the potential
to connect the two different energy domains interpolating between the
limiting scales of QCD. In this regime only scarce experimental data
are available, most of which have been obtained with electromagnetic
probes.

One possible single issue that may greatly advance our understanding
of hadronic structure is the predicted existence of states outside of
the two- and three-quark classifications, which for example could
arise from the excitation of gluonic degrees of freedom. Recent
findings from running experiments at B-factories (see {\it
e.g.} Refs~\cite{Acosta:2003zx,Choi:2003ue}) show that, indeed,
unexpected narrow states unaccounted for in the na\"ive quark models
exist. Experiments focused on the abundant production and systematic
studies of these states are needed. Preferably, these should be
performed using hadronic probes because the cross sections are
expected to be very large in such systems. Results of high 
precision are a decisive element to be able to identify and extract
features of these exotic states. Hadron beams are advantageous also
for the production of hadrons with non-exotic quantum numbers, as
these can be formed directly with high cross sections. Phase space
cooling of the antiproton beam furthermore allows high precision
determination of the mass and width of such states.  Using heavier
nuclei as targets enables us to investigate in-medium properties of
hadrons and to produce hypernuclei, even those containing more than
one strange quark, copiously.

The \PANDA (Antiproton Annihilation at Darmstadt) experiment (see
also Sec.~\ref{s:over} and Refs.~\cite{PANDA:TPR, PANDA:Intro}), which
will be installed at the High Energy Storage Ring for antiprotons of
the upcoming Facility for Antiproton and Ion Research
(FAIR)~\cite{FAIR:BTR}, features a scientific programme devoted to the
following key areas.
\begin{itemize}
 \item Charmonium spectroscopy.
 \item Search for gluonic excitations (hybrids and glueballs).
 \item Multi-quark states.
 \item Open and hidden charm in nuclei.
 \item Open charm spectroscopy.
 \item Hypernuclear physics.
 \item Electromagnetic processes.
\end{itemize}
These and selected further topics will be studied with
unprecedented accuracy.

\section{Topics Addressed at \PANDA}

\begin{figure*}[thb]
 \begin{center} \includegraphics[width=0.83\dwidth]{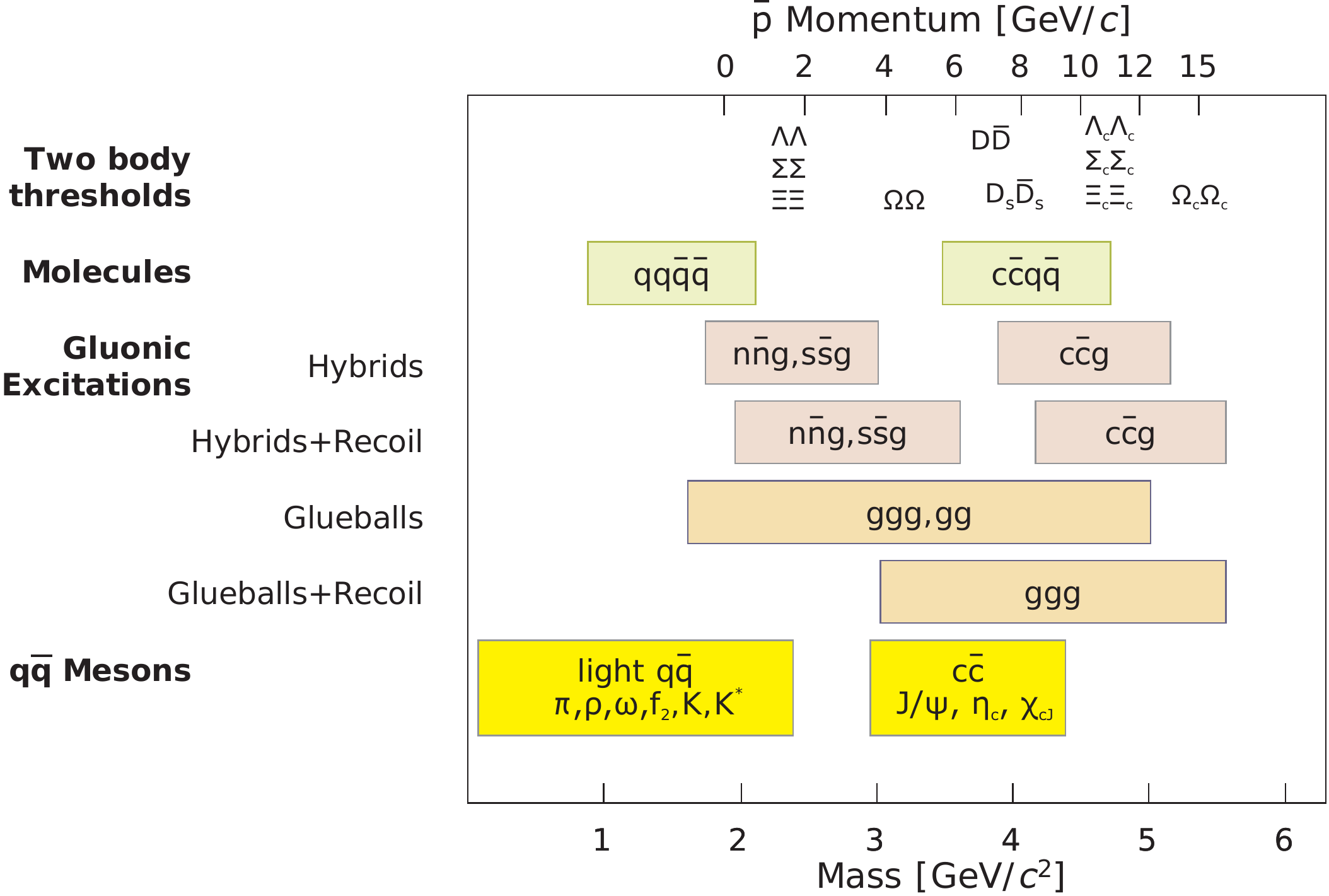}
 \caption[Mass range of hadrons that will be accessible at PANDA.]
 {Mass range of hadrons that will be accessible at PANDA. The
 upper scale indicates the corresponding antiproton momenta required
 in a fixed-target experiment. The HESR will provide 1.5 to 15\,GeV/c
 antiprotons, which will allow charmonium spectroscopy, the search for
 charmed hybrids and glueballs, the production of D meson baryon pairs
 for pairs and the production of hypernuclear studies.} 
 \label{f:intro:panda_range}
 \end{center}
\end{figure*}

In the following major physics topics are briefly introduced. See also
Fig.~\ref{f:intro:panda_range} for an overview. A detailed discussion
and further references can be found in Ref.~\cite{PANDA:PhysBooklet}.

\paragraph*{Charmonium Spectroscopy.}  
The $\ccbar$ spectrum is often referred to as the positronium of QCD,
because the properties of the states can be calculated precisely
within the framework of non-relativistic potential
models.  More recently, results from
quenched Lattice QCD emerged describing the known spectrum rather
well.  Recent findings of states around
4\,$\gevc$ (X(3872), Z(3931), X(3940), Y(3940), Y(4260), Y(4320), to
name only a few)~\cite{RBES1,RBES2,Z3931,X3940} show that the
spectrum, which was believed to be well understood, in fact yields
much more than has been expected.

\PANDA will not only be able to measure those states in a different 
production channel, which may reveal more unexpected states, but also
allow for scans over the width of those states with a precision of
$10^{-5}$ relative to its mass.  This is because one can produce these
states directly in formation and make use of the precisely determined
antiproton momenta in \HESR.  At full luminosity \PANDA will be able
to collect several thousand $\ccbar$ states per day.  Thus properties
and branching ratios will be determined to a unprecedented precision.

\paragraph*{Search for Gluonic Excitations (Hybrids and Glueballs).}
One of the main challenges of hadron physics is the search for gluonic
excitations, {\it i.e.\ }hadrons in which the gluons can act as
principal components. In other words, the state cannot be fully
described in terms of quantum numbers by solely taking its
valence-quark content into account. These gluonic hadrons fall into
two main categories described in the following.  Glueballs are states
where only gluons contribute to the overall quantum numbers while
hybrids consist of a valence quarks as in ordinary hadrons plus one or
more gluons which contribute to the overall quantum numbers.


The additional degrees
of freedom carried by gluons allow these hybrids and glueballs to have
$J^{PC}$ exotic quantum numbers. In this case mixing effects with
nearby $\qqbar$ states are excluded and this makes their experimental
identification easier.  The properties of glueballs and hybrids are
determined by the long-distance features of QCD and their study will
yield fundamental insight into the structure of the QCD vacuum.
Antiproton-proton annihilations provide a very favourable environment
to search for gluonic hadrons. 

\paragraph*{Multi-Quark States.}  These are states which cannot be 
assigned to an arrangement of tree quarks or a quark-antiquark pair as
the classical baryons and mesons.  Similarly to the gluonic
excitations mentioned above they would show up as states outnumbering
the multiplets and their clearest signature would be possible exotic
quantum numbers.  They could be interpreted as hadronic molecules or
octet couplings.  The well known states $a_0 (980)$ and $f_0 (975)$
are suspected to have admixtures of $K \bar{K}$ components. Here,
however, mixing is large and clear statements on their nature are
difficult to draw.  In the charmonium region all states are narrower
and positive identification is much more likely.  This can also be
seen from the current discussion on the nature of the X(3872) and
other states recently found at the B-factories~\cite{RBES1,RBES2}.

\paragraph*{Open and Hidden Charm in Nuclei.}
The study of medium modifications of hadrons embedded in hadronic matter
is aimed at understanding the origin of hadron masses in the context
of spontaneous chiral symmetry breaking in QCD and its partial restoration
in a hadronic environment~\cite{Brown:1994cq}. 
So far experiments have been focused on the light quark sector.
The high-intensity $\pbar$ beam of up to 15\ GeV/c will allow an extension of
this programme to the charm sector both for hadrons with hidden and open charm.
The in-medium masses of these states are expected to be affected primarily
by the gluon condensate.

Another study which can be carried out in PANDA is the measurement of $\jpsi$
and D meson production cross sections in $\pbar$ annihilation
on a series of nuclear targets. The
comparison of the resonant $\jpsi$ yield obtained from $\pbar$ annihilation
on protons and different nuclear targets allows to deduce the
$\jpsi$-nucleus dissociation cross section, a fundamental parameter to 
understand $\jpsi$ suppression in relativistic heavy ion collisions interpreted
as a signal for quark-gluon plasma formation. 

\paragraph*{Open Charm Spectroscopy.} 
The HESR running at full luminosity and at $\pbar$ momenta larger than
6.4\ GeV/c would produce a large number of $D$ meson pairs.  The high
yield ({\it e.g. }up to 100 charm pairs per second around the
$\psi$(4040)) and the well defined production kinematics of $D$ meson
pairs will allow to carry out a significant charmed meson
spectroscopy programme which will include, for example, the rich $D$
and $D_s$ meson spectra.

\paragraph*{Hypernuclear Physics.}
Hypernuclei are systems in which up or down quarks are replaced by
strange quarks. In this way a new quantum number, strangeness, is
introduced into the nucleus.  Although single and double
$\Lambda$-hypernuclei were discovered many decades ago, only 6 events
of double $\Lambda$-hypernuclei were observed up to now. The
availability of $\pbar$ beams at FAIR will allow efficient production
of hypernuclei with more than one strange hadron, making PANDA
competitive with planned dedicated facilities. This will open new
perspectives for nuclear structure spectroscopy and for studying the
hyperon-nucleon and in particular the hyperon-hyperon interaction.

\paragraph*{Electromagnetic Processes.}
In addition to the spectroscopic studies described above, \PANDA will
be able to investigate the structure of the nucleon using
electromagnetic processes, such as Wide Angle Compton Scattering
(WACS) and the process $\pbarp \to \ee$, which will allow the
determination of the electromagnetic form factors of the proton in the
time-like region over an extended $q^2$ region. In addition the
Drell-Yan process will allow to access the transverse nucleon spin
structure.

\section{Experimental Approach}

Conventional as well as exotic hadrons can be produced by a range of
different experimental means.  Among these, hadronic annihilation
processes, and in particular antiproton-nucleon and antiproton-nucleus
annihilations, have proven to possess all the necessary ingredients
for fruitful harvests in the hadron field.
\begin{itemize}
 \item Hadron annihilations produce a gluon-rich environment, a
       fundamental prerequisite to copiously produce gluonic
       excitations.

 \item The use of antiprotons permits to directly form all states
       with non-exotic quantum numbers (formation experiments).
       Ambiguities in the reconstruction are reduced and cross
       sections are considerably higher compared to producing
       additional particles in the final state (production
       experiments). The appearance of states in production but not in
       formation is a clear sign of exotic physics.  

 \item Narrow resonances, such as charmonium states, can be scanned
       with high precision in formation experiments using the small
       energy spread available with antiproton beams (cooled to
       $\Delta p/p = 10^{-5}$).
       
 \item Since exotic systems will appear only in production
       experiments the physics analysis of Dalitz plots becomes
       important. This requires high-statistics data samples. Thus,
       high luminosity is a key requirement. This can be achieved
       using an internal target of high density, large numbers of
       projectiles and a high count-rate capability of the detector.
       The latter is mandatory since the overall cross sections of
       hadronic reactions are large while the cross sections of
       reaction channels of interest may be quite small.  

 \item As reaction products are peaked around angles of $0^\circ$ a
       fixed-target experiment with a magnetic spectrometer is the
       ideal tool.  At the same time a $4 \pi$ coverage is mandatory
       to be able to study exclusive reactions with many decay
       particles.  The physics topics as summarised in
       Fig.~\ref{f:intro:panda_range} confirm that the momentum range of
       the antiproton beam should extend up to 15\,GeV/c with
       luminosities in the order of $10^{32}\,$cm$^{-2}$s$^{-1}$
\end{itemize}

To take full advantage of the HESR beam features, a compact, high
resolution and high angular coverage spectrometer was designed. To
cope with the need of $2\unit{Tm}$ bending power both at a very wide
angular range in the laboratory reference frame, two magnets are
necessary. A solenoid magnet provides the required bending power for
particles exiting at $5 - 140^\circ$ in vertical direction and at $10
- 140^\circ$ in horizontal direction, whereas a dipole magnet provides
bending power for particles exiting at angles smaller than $5^\circ$
in vertical direction and $10^\circ$ in horizontal direction.  The
requirements for both magnets are discussed in Sec.~\ref{s:over:req},
the superconducting solenoid magnet is described in detail in
Chapter~\ref{s:sol} and the dipole magnet is described in
Chapter~\ref{s:dipole}.


\bibliographystyle{tdr_lit}
\bibliography{lit}

\cleardoublepage
\svnInfo $Id: over.tex 634 2009-01-28 18:51:24Z IntiL $
\svnKeyword $Revision: 634 $

\chapter{Spectrometer Overview}
\label{s:over}

\AUTHORS{A.~Bersani, I.~Lehmann, R.~Parodi, A.~Vodopianov, G.~Rosner}

This chapter introduces the facility and the storage ring at which
the \PANDA experiment will be located.  A brief overview over the
detector and its components is given before concluding with the
requirements which are imposed specifically to the two large
spectrometer magnets in Sec.~\ref{s:over:req}.

\svnInfo $Id: fair.tex 708 2009-02-09 16:52:12Z IntiL $

\section{Facility for Antiproton and Ion Research -- \FAIR}
\label{s:over:fair}

The Facility for Antiproton and Ion Research (\FAIR) will be an
accelerator facility leading the European research in nuclear and
hadron physics in the coming decade.  It will address a wide range of
physics topics in the fields of nuclear structure, nuclear matter, studies using
high energy and very slow antiprotons, atomic and plasma physics.
Several topics in applied science and accelerator development will be
addressed as well.  \FAIR builds on the experience and technological
developments from the existing \GSI facility, and incorporates new
technological concepts.  In this document we briefly summarise aspects
relating to the production of antiprotons for the use by the \PANDA
experiment.  Please refer to Refs.~\cite{FAIR:BTR,FAIR:Info} for more
details.

\begin{figure}[htb]
\begin{center}
\includegraphics[width=\swidth]{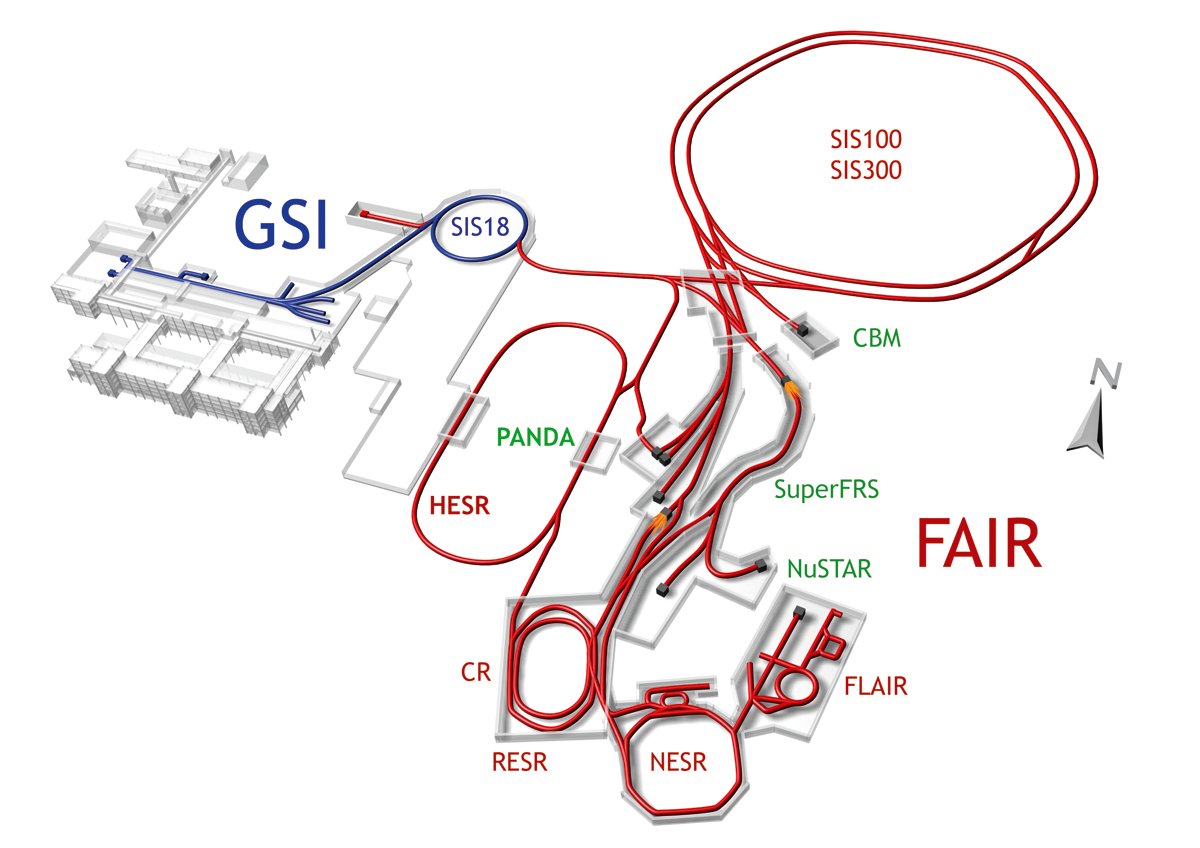}
\caption[Layout of the \FAIR facility.]
{Layout of the \FAIR facility. Shown are: synchrotrons with 
   bending powers up to 18, 100 and 300 Tm the SIS 18, 100 and 300,
   respectively; the Collector Ring CR; the Accumulator Ring RESR; the
   low and high energy experimental storage rings NESR and HESR,
   respectively; and the Fragment Separator FRS.}
\label{f:over:fair-layout}
\end{center}
\end{figure}

The existing \GSI accelerators will be upgraded by the addition of 
a proton-linac and used as injectors for the newly built complex of storage
rings to form the \FAIR facility.  An overview of the layout is given
in Fig.~\ref{f:over:fair-layout}.  To the left, the existing \GSI
facility is shown.  New proposed beam lines and accelerator structures
are shown in red.  This facility will provide intense secondary beams
of antiprotons and rare isotopes which will be used for research at
the main experimental setups labelled in green. 


\PANDA will make use of antiprotons, which will be produced as follows.
Protons will be accelerated to 70\,MeV in a new linear accelerator and
injected in several turns into the existing synchrotron SIS18
and accelerated to 2\,GeV with a repetition rate of 5\,Hz. After four
SIS18 cycles and subsequent transfers to the SIS100 synchrotron, the
protons will be accelerated to 29\,GeV with a ramp rate of 4\,T/s. Up to
$2 \times 10^{13}$ protons will be compressed into a single short
bunch of less than 50\,ns length in order to minimise the heating in
the antiproton production target.  The proton bunch will be directed onto a
nickel target of about 60\,mm length followed by a magnetic horn.  The
cycle of proton acceleration will be repeated every 10\,s. (An upgrade
allowing for a cycle time of 5\,s is foreseen.)  This scheme is
expected to produce a bunch of at least $1 \times 10^8$ antiprotons in
the phase space volume which can be accepted by the magnetic separator
and the Collector Ring (CR). The CR will be used for pre-cooling of
the antiprotons.  Thereafter, the antiprotons will be moderately
compressed to a single bunch and transferred to the RESR storage ring.
The antiprotons will be accumulated to high intensities in the RESR by a
dedicated stochastic cooling system with a rate that matches the speed
of cooling in the CR.  The antiprotons will then be transferred at a
momentum of 3.8\,GeV/c to the High Energy Storage Ring (\HESR), which
hosts \PANDA and will be discussed in detail in the following.

\svnInfo $Id: hesr.tex 708 2009-02-09 16:52:12Z IntiL $

\section{High Energy Storage Ring -- \HESR}
\label{s:over:hesr}

\begin{figure*}[thb]
\begin{center}
\includegraphics[width=\dwidth]{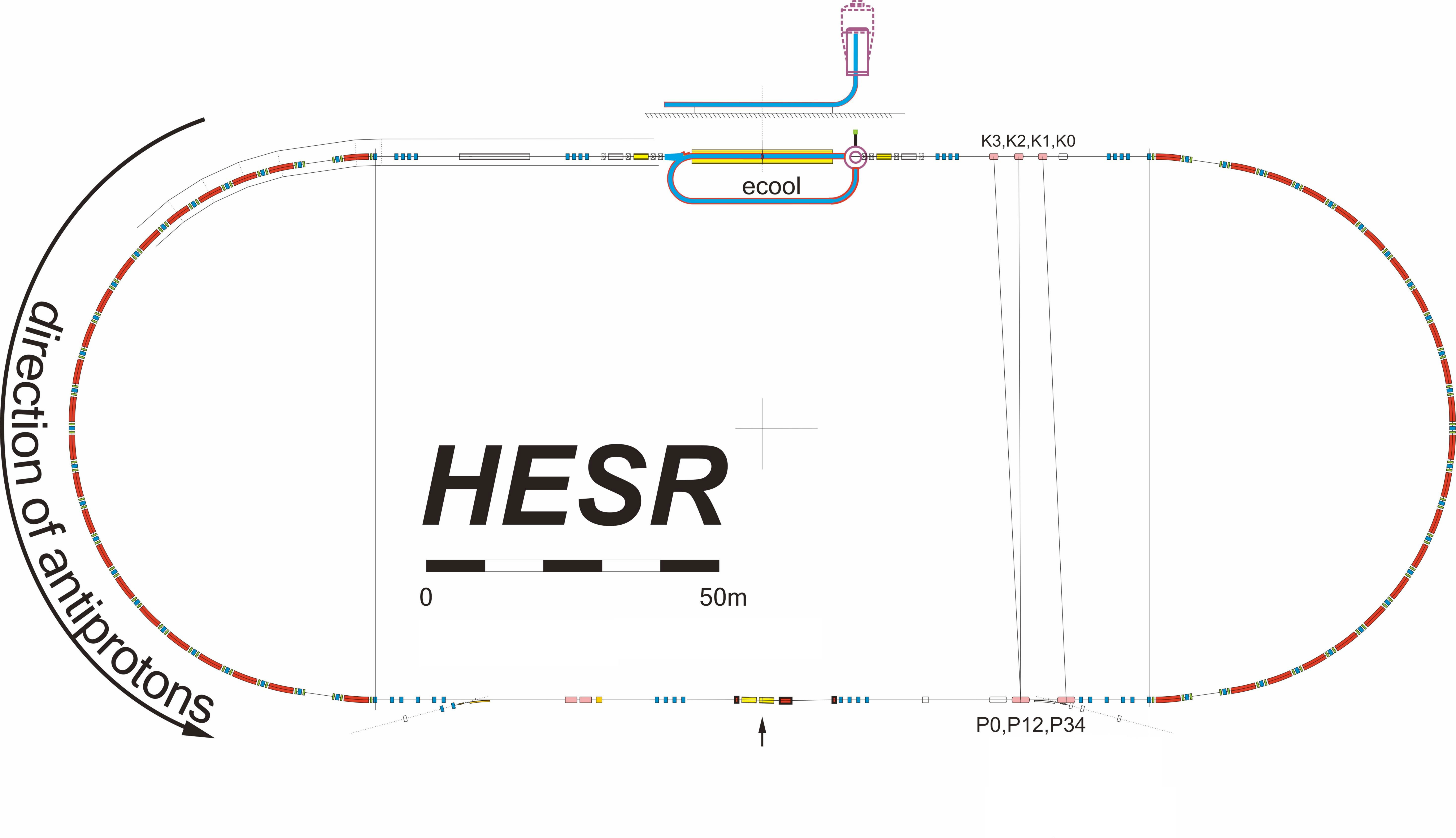}
\caption[Layout of the High Energy Storage Ring \HESR.]
{Layout of the High Energy Storage Ring \HESR.  The following 
  colour code is used for the active components: dipole magnets --
  red, quadrupoles -- blue, solenoids -- yellow, and cavities,
  pickups, kickers and septa -- ros\'e or grey.  The beam is injected
  into the lower straight section. Stochastic cooling ($P0-4$ and
  $K0-3$) and electron cooling (``ecool'') is foreseen.  The location
  of the \PANDA target is indicated with an arrow at the bottom of the
  figure.  A detail of this section is shown in
  Fig.~\ref{f:over:HESR_chicane} later in this document.}
\label{f:over:hesr}
\end{center}
\end{figure*}

The \HESR is dedicated to supply \PANDA with high-quality anti-proton
beams over a broad momentum range from 1.5 to 15\,GeV/c.  In storage
rings the complex interplay of many processes like beam-target
interaction and intra-beam scattering determines the final equilibrium
distribution of the beam particles.  Electron and stochastic cooling
systems are required to ensure that the specified beam quality and
luminosity for experiments at
\HESR~\cite{HESR:FAIR:2006,HESR:FAIR:2008} is achieved.  Two different operation
modes have been worked out to fulfil these experimental
requirements. (Please also refer to Ref.~\cite{lehrach:stori:2008} and
references therein.)

\subsection{Lattice Design and Experimental Requirements}

The \HESR lattice is designed as a racetrack shaped ring with a
maximum beam rigidity of 50\,Tm (see Fig.~\ref{f:over:hesr}). The basic
design consists of FODO cell structures in the arcs. The arc
quadrupole magnets will be grouped into four families, to allow a
flexible adjustment of transition energy, horizontal and vertical
betatron tune, and horizontal dispersion.

One straight section will mainly be occupied by the
electron cooler. The other straight section will host the
experimental installation with internal H$_2$ pellet and cluster jet  target, 
RF cavities, injection kickers and septa. Four stochastic cooling
pickup and kicker tanks will also be located in the straight sections,
opposite to each other.

\begin{figure}[ht]
\begin{center}
\includegraphics[width=\swidth]{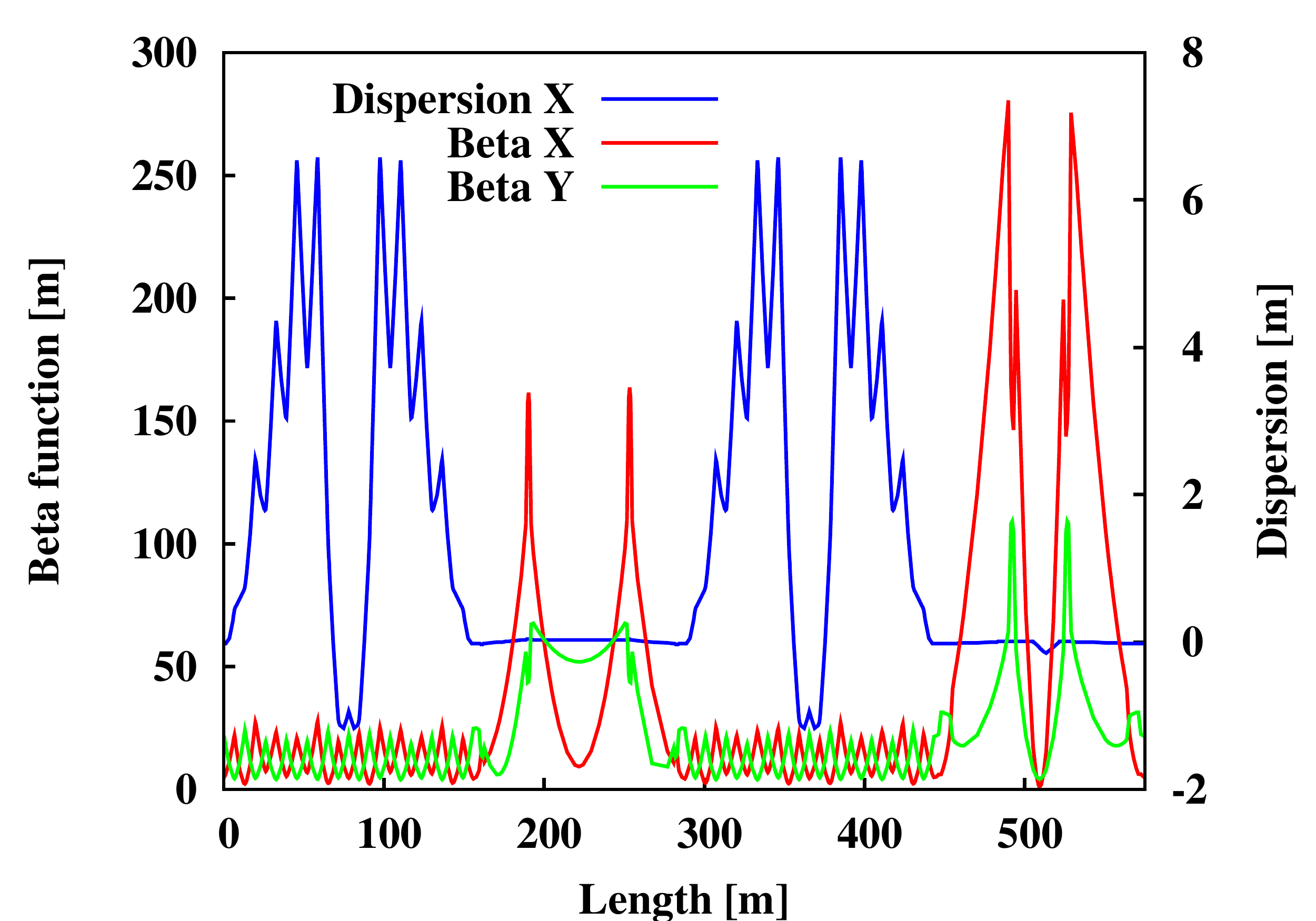}
\caption[Optical functions of \HESR lattice for $\gamma_{tr}$ = 6.2.]
{Optical functions of \HESR lattice for $\gamma_{tr}$ = 6.2
optical setting. Plotted are the horizontal dispersion, horizontal and
vertical betatron function. The electron cooler is located at a
length of $s$= 222\,m, and the target at $s$=509\,m, where a kink
in the horizontal dispersion is generated by the \Panda dipole
chicane.}
\label{f:hesr:optics}
\end{center}
\end{figure}

Special requirements for the lattice are low dispersion in
straight sections and small betatron amplitudes in the range
between 1 and 15\,m at the internal interaction point (IP) of the
\PANDA detector. In addition, the betatron amplitude at the
electron cooler must be adjustable within a large range between 25 and
200\,m. There are by now four defined optical settings: Injection,
$\gamma_{tr} = 6.2,\, \gamma_{tr} = 13.4,\, \gamma_{tr} = 33.2$.
Both betatron tunes will roughly be 7.62 for different optical
settings and natural chromaticities will be ranging in X from -12 to
-17 and in Y from -10 to -13.  Examples of the optical functions of the
$\gamma_{tr} = 6.2$ lattice are shown in
Fig.~\ref{f:hesr:optics}. 

The large aperture spectrometer dipole magnet also deflects the
antiproton beam. To compensate for this two further dipole magnets
surrounding the setup of the \Panda experiment will be used to create a beam
chicane. To provide space for \Panda, the two chicane
dipoles will be placed $4.6\unit{m}$ upstream and $13\unit{m}$ 
downstream the \Panda IP. This 
gives a boundary condition for the placement of the quadrupole
elements closest to the experiment. For symmetry reasons, they have to
be placed at $\pm14\unit{m}$ with respect to the IP. The asymmetric 
placement of the chicane dipoles will result in the experiment axis 
occurring at a small angle with respect
to the axis of the straight section.

Special equipment like multi-harmonic RF cavities, electron and
stochastic cooler will enable a high performance of this antiproton storage
ring to be achieved, and therefore make high precision experiments feasible. Key tasks
for the \HESR design work to fulfil these requirements are:

\begin{table*}[pt]
\begin{tabular}{l l} 
\hline
  \multicolumn{2}{c}{\textbf{Injection Parameters}}\\ 
\hline
  Bunch length &
150\,m long bunches from RESR \\
Transverse emittance & 0.25\,mm $\times$ mrad (RMS) for $3.5\times 10^{10}$ particles, \\
 & scaling with number of accumulated particles:  $\sim N^{4/5}$ \\
Relative momentum spread &
$3.3\times 10^{-4}$ (RMS) for $3.5\times10^{10}$ particles, \\
 & scaling with number of accumulated particles: $\sigma_p/p \sim
N^{2/5}$ \\
Injection momentum &
3.8 {GeV/c} \\
Injection type &
Kicker injection using multi-harmonic RF cavities \\
 & \\
\hline
\hline
  \multicolumn{2}{c}{\textbf{Experimental Requirements}}\\ 
\hline
Ion species & Antiprotons \\
$\bar{p}$ production rate &
$2\times 10^7 ${/s} ($1.2\times 10^{10}$ per 10\,min) \\
Momentum / Kinetic energy range &
1.5 to 15 {GeV/c} / 0.83 to 14.1 {GeV} \\
Number of particles &
$10^{10}$ to $10^{11}$ \\
Target thickness &
$4\times 10^{15} ${atoms/cm$^2$} (H$_2$ pellets) \\
Beam size (radius) at IP & $\sim$ 1\,mm (RMS) \\
Betatron amplitude at IP &
1 -- 15\,m \\
Betatron amplitude E-Cooler &
25 -- 200\,m \\
 & \\ 
\hline
\hline
  \multicolumn{2}{c}{\textbf{Operation Modes}}\\ 
\hline
High resolution (HR) & Peak Luminosity of 2$\times
10^{31}${cm$^{-2}$ s$^{-1}$} for $10^{10}\
\bar{p}$ \\
 & RMS momentum spread $\sigma_p / p \leq 4\times 10^{-5}$, \\
 & 1.5 to 9 {GeV/c} \\
High luminosity (HL) & Peak Luminosity up to 2$\times
10^{32}${cm$^{-2}$ s$^{-1}$} for $10^{11}\
\bar{p}$ \\
 & RMS momentum spread $\sigma_p / p \sim 10^{-4}$, \\
 & 1.5 to {15} {GeV/c},  \\
 \hline
\end{tabular}
\caption{Injection parameters, experimental requirements and operation
  modes.}
\label{t:hesr1}
\end{table*}

\begin{itemize}
 \item Design and testing\footnote{Prototype cavities have been built
   and barrier-bucket operation was performed with stochastic
   cooled beams at COSY. Simulated and measured beam equilibria are in
   good agreement.}  of multi-harmonic RF cavities. One
   cavity is required for barrier-bucket operation to compensate the
   mean energy loss due to beam-target scattering. The second cavity
   is needed for bunch rotation, acceleration and deceleration of the
   beam. In addition, this second cavity will provide for
   bunch manipulation during refill to increase the average
   luminosity.
 \item Technical design study and prototyping of critical elements for
   high-voltage electron cooling system. An electron beam
   with up to 1\,A current, accelerated in special accelerator columns
   to energies in the range of 0.4 to 4.5\,MeV is planned for the
   HESR. The 24\,m long solenoidal field in the cooler section has a
   longitudinal field strength of 0.2\,T with a magnetic field
   straightness in the order of $10^{-5}$. This arrangement allows
   beam cooling between 1.5\,GeV/c and 8.9\,GeV/c. Since its design is
   modular, a future increase of high-voltage to 8\,MV is possible,
   which would make electron cooling feasible in the whole HESR
   momenta range. \item Development and testing of high-sensitivity
   stochastic cooling pickups for the frequency range 2 -- 4\,GHz. The
   main stochastic cooling parameters have been determined for a
   cooling system utilising pickups and kickers with a band-width of
   2 -- 4\,GHz and the option for an extension to 4 --
   6\,GHz. Since stochastic filter-cooling is specified above and
   stochastic time-of-flight cooling below 3.8\,GeV/c, the whole HESR
   momentum range can be covered by the stochastic cooling system.
\end{itemize}

Table~\ref{t:hesr1} summarises the specified injection parameters,
experimental requirements and operation modes.  Demanding requirements
for high intensity and high quality beams are combined in two
operation modes: high luminosity (HL) and high resolution (HR),
respectively. The HR mode is defined in the momentum range from 1.5 to
9 GeV/c. To reach a relative momentum spread down to a few times
$10^{-5}$, only $10^{10}$ circulating particles in the ring are
anticipated. The HL mode requires an order of magnitude higher beam
intensity with reduced momentum resolution to reach a peak luminosity
of 2$\times 10^{32}${cm$^{-2}$ s$^{-1}$} in the full momentum range up
to 15\,GeV/c.

\subsection{Beam Dynamics}

\subsubsection{Closed orbit correction}

The most serious causes of closed orbit distortions are angular
and spatial displacements of magnets. Alignment and measurement
errors of beam position monitors also contribute to closed orbit
distortions. Both types of errors have been included in the
simulations.

The goal of the orbit correction scheme is to reduce
maximum closed orbit deviations to below 5\,mm while not exceeding
1\,mrad of corrector strength. The inverted orbit response matrix
method was utilised to obtain the necessary corrector
strengths. In this simulation the correction scheme consists of 64
beam position monitors and 36 orbit correction dipoles. In order
to verify the possibility to improve the closed orbit, Monte-Carlo
methods have been used. More than 1000 different sets of
displacement and measurement errors have been applied. For all
defined optical settings the effectiveness of the developed closed
orbit correction scheme could be demonstrated. 

Additionally, the influence of the electron cooler
toroids had to be investigated. Toroids are used in beam guiding
systems of the electron cooler to overlap the electron beam with
the antiproton beam. Since antiprotons are much heavier than
electrons, the deflection of the antiprotons by toroids is
much smaller. The deflection angles are different for the two
transverse directions. To compensate this deflection four
additional correction dipoles have to be included in the HESR
lattice around the electron cooler. The inner ones need to be
placed very close to the toroids to keep orbit deviations as small
as possible.

There are a few positions in the straights of the HESR where orbit
bumps will have to be used, e.g. at the target. Therefore, all closed
orbit correction dipoles in the straights are designed to provide
an additional deflection strength of 1 mrad. Investigations have
shown that this is sufficient to set angle and position of the
circulating beam in the desired ranges.

\begin{table*}[ht]
\begin{center}
\begin{tabular}{|l|l|c c c|} 

\hline
\multicolumn{2}{|l|}{Beam momentum} & 1.5\,GeV/c & 9\,GeV/c & 15\,GeV/c \\
\hline 
\hline
\multirow{5}{3cm}{Rel. loss rate $\tau_{loss}^{-1}$ $[$s$^{-1}]$} & Hadronic Interaction & $1.8\times 10^{-4}$ & $1.2\times
10^{-4}$ &
$1.1\times 10^{-4}$ \\
& Single Coulomb & $2.9\times 10^{-4}$ & $6.8\times 10^{-6}$ &
$2.4\times 10^{-6}$ \\
& Energy Straggling & $1.3\times 10^{-4}$ & $4.1\times 10^{-5}$ &
$2.8\times 10^{-5}$ \\
& Touschek Effect & $4.9\times 10^{-5}$ & $2.3\times 10^{-7}$ &
$4.9\times 10^{-8}$ \\ 
\hline 
& Total & $6.5\times
10^{-4}$ & $1.7\times 10^{-4}$ &
$1.4\times 10^{-4}$ \\
\hline 
\hline 
\multicolumn{2}{|l|}{Beam lifetime $t_{pbar} \; [$s$]$ } & $\sim 1540$ & $\sim 6000$ &
$\sim 7100$ \\ 
\multicolumn{2}{|l|}{Max.\ luminosity $L_{max}$ $[10^{32}$cm$^{-2}$ s$^{-1}]$} &
0.82 & 3.22 & 3.93 \\ 
\hline
\end{tabular}
\end{center}
\caption{Upper limits for relative beam loss
rate, beam lifetime ($1/e$) $t_{pbar}$, and maximum average
luminosity $L_{max}$ for a H$_2$ pellet target.}
\label{t:hesr2}
\end{table*}

\subsubsection{Beam equilibria and luminosity estimates}

\textbf{Beam equilibria with electron cooling}

The empirical magnetised cooling force formula by V.V. Parkhomchuk is
generally used for electron cooling~\cite{parkhomchuk:2000}, and an
analytical description for intra-beam scattering~\cite{sorensen:1987}.
Beam heating by beam-target interaction is described by transverse and
longitudinal emittance growth due to Coulomb scattering and energy
straggling~\cite{hinterberger:1989a,hinterberger:1989b}. Beam
equilibria with electron cooled beams in the HESR have been
investigated in detail~\cite{BoineFrankenheim:2006ci}. In the HR mode
an RMS relative momentum spreads are ranging from $7.9\times 10^{-6}$
(1.5\,GeV/c) to $2.7\times 10^{-5}$ (8.9\,GeV/c), and $1.2\times
10^{-4}$ (15\,GeV/c).

\textbf{Beam equilibria with stochastic cooling}

Beam equilibria have been simulated based on a Fokker-Planck
approach. Applying stochastic cooling with a band-width of
2 -- 6\,GHz one can achieve an RMS relative momentum spread of
$5.1\times 10^{-5}$ (3.8\,GeV/c), $5.4\times 10^{-5}$ (8.9\,GeV/c)
and $3.9\times 10^{-5}$ (15\,GeV/c) for the HR mode. In the HL mode
RMS relative momentum spread of roughly $10^{-4}$ can be expected.
Transverse stochastic cooling can be adjusted independently to
ensure sufficient beam-target overlap.

The relative momentum spread can be further improved by combining
electron- and stochastic cooling.

\textbf{Beam losses and luminosity estimates}

Beam losses are the main restriction for high luminosities. Three
dominating contributions of beam-target interaction have been
identified: Hadronic interaction, single Coulomb scattering at large 
angle and
energy straggling of the circulating beam in the target. In
addition, single intra-beam scattering due to the Touschek effect
has also to be considered for beam lifetime estimates. Beam losses
due to residual gas scattering can be neglected compared to
beam-target interaction for a vacuum of $10^{-9}$\,mbar. A
detailed analysis of all beam loss processes in the HESR was
carried out~\cite{lehrach:2006,hinterberger:2006}.

The relative beam loss rate for the total cross section
$\sigma_{tot}$ is given by the expression
\begin{equation}
\tau_{loss}^{-1} = n_t \sigma_{tot} f_0
\end{equation}
where $\tau_{loss}^{-1}$ is the relative beam loss rate, $n_t$ the
target thickness and $f_0$ the revolution frequency of the reference
particle. In Table~\ref{t:hesr2} calculated upper limits for beam losses and
corresponding lifetimes are listed for transverse beam emittances of
1\,mm$\times$mrad, betatron amplitudes of 1\,m at the internal
interaction point, a longitudinal ring acceptance of $\Delta p/p =
\pm 10^{-3}$, and $10^{11}$ circulating antiprotons.

For beam-target interaction, the beam lifetime is calculated to be
independent of the beam intensity, whereas for the Touschek effect
it is found to depend on the beam equilibrium. Beam lifetimes
ranging from 1540\,s to 7100\,s are found. Beam lifetimes at low momenta
strongly depend on the beam cooling scenario and the ring
acceptance. Less than half an hour beam lifetime is too small
compared to the planned antiproton production rate.

The maximum average luminosity depends on the antiproton
production rate $d N_{pbar}$ / dt = 2$\times 10^7$/s and loss rate
\begin{equation}
L_{max} = \frac{d N_{pbar} / dt} {\sigma_{tot}}.
\end{equation}
Estimates of the maximum average luminosities are listed for different beam momenta in
Table~\ref{t:hesr2}. The maximum average luminosity for 1.5\,GeV/c is
below the specified value for the HL mode.

\textbf{Cycle averaged luminosity}

To calculate the cycle averaged luminosity, machine cycles and
beam preparation times have to be specified. After injection, the
beam is pre-cooled to equilibrium (with target off) at 3.8\,GeV/c.
The beam is then accelerated or decelerated to the desired beam
momentum. A maximum ramp rate of 25\,mT/s is specified. After
reaching the final momentum beam steering respectively focusing in
the target and in beam cooler region will take place. Total beam
preparation time $t_{prep}$ will range from 120\,s for 1.5\,GeV/c to
290\,s for 15\,GeV/c.

\begin{figure}[th]
\begin{center}
\subfigure{\includegraphics[width=\swidth]{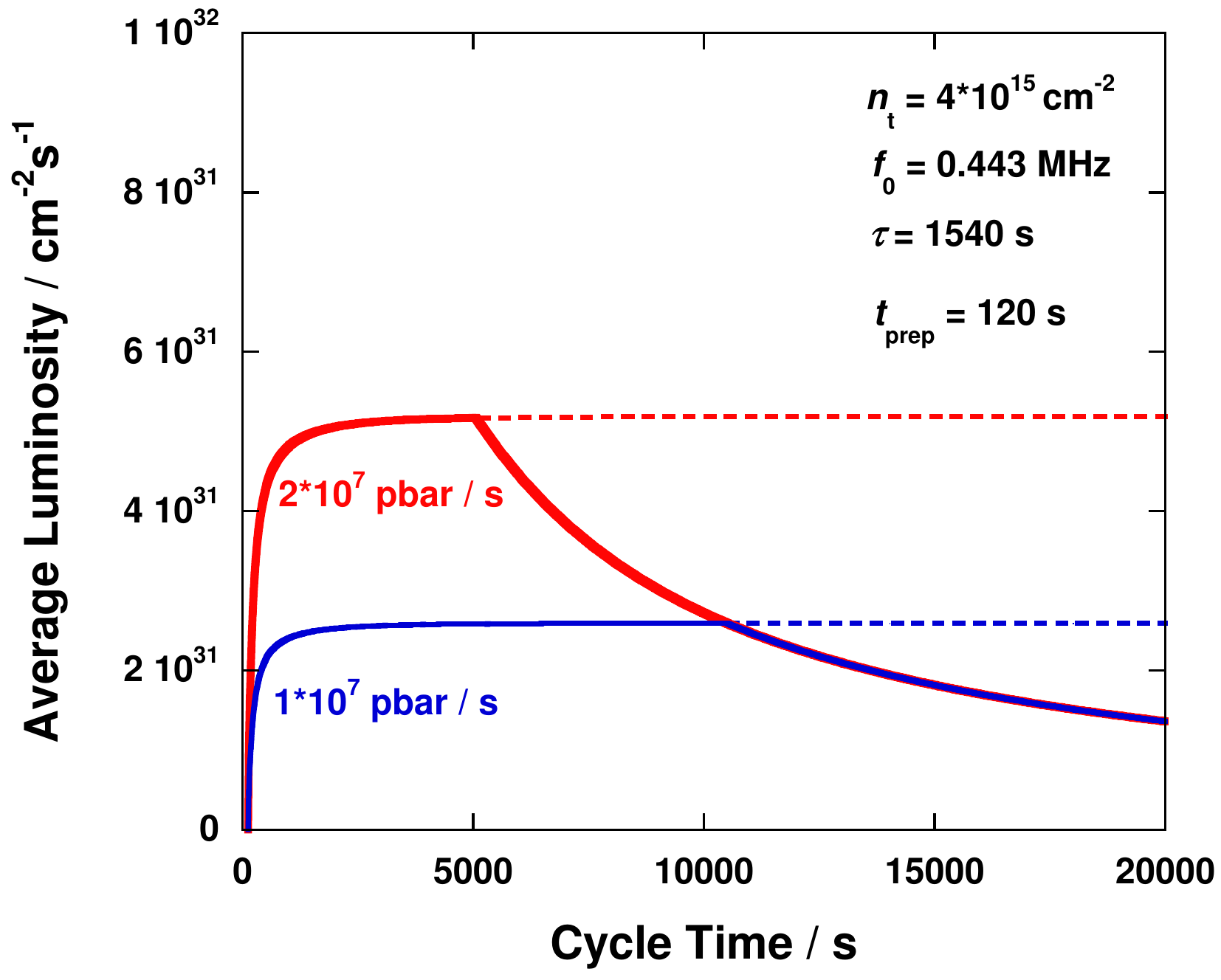}}
\subfigure{\includegraphics[width=\swidth]{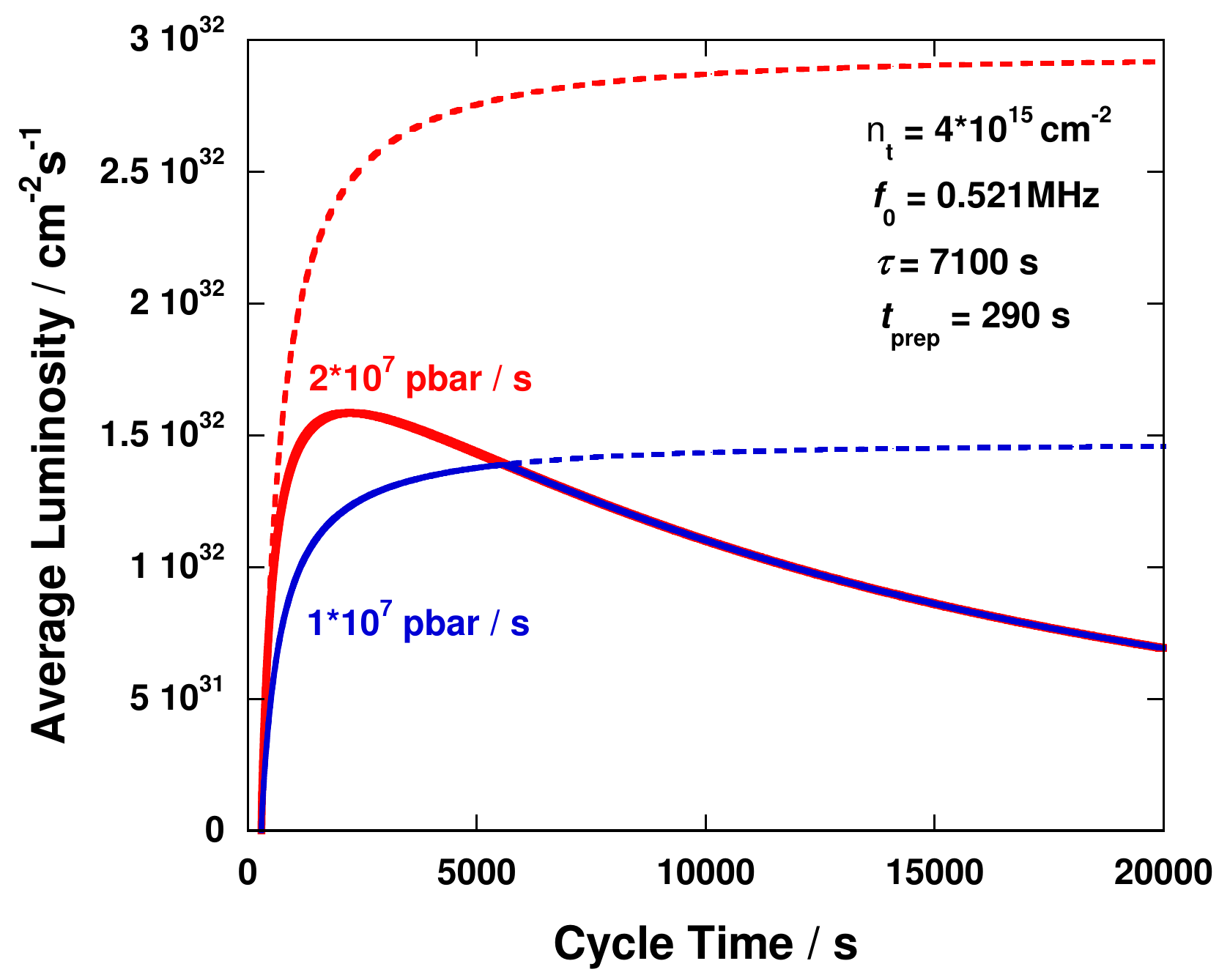}} 
\caption[Cycle averaged luminosity vs. cycle time at 1.5\,GeV/c
  and 15\,GeV/c.]
  {Cycle averaged luminosity vs. cycle time at 1.5\,GeV/c (top)
  and 15\,GeV/c (bottom).  Assuming unlimited maximum number of
  particles inside the \HESR ring one obtains the dashed lines with a
  production rate of 1 or $2 \times 10^7 \bar{p}$/s (blue or red,
  respectively).  In a more realistic scenario the number of particles
  which can be stored is limited to $10^{11}$ (solid
  lines).}  \label{fig:avglum}
\end{center}
\end{figure}

In the HL mode, particles should be re-used in the next
cycle. Therefore the used beam will be converted back to the
injection momentum and merged with the newly injected beam. A
bucket scheme is planned for beam injection and refill procedure,
utilising the second cavities. During acceleration 1\% and
deceleration 5\% beam losses are assumed.  The cycle averaged
luminosity reads
\begin{equation}
\bar{L} = f_0 N_{i,0} n_t
  \frac{\tau \left[ 1 - e^{-t_{exp}/\tau} \right]}{t_{exp} +
  t_{prep}},
\end{equation}
where $\tau$ is the $1/e$ beam lifetime, $t_{exp}$ the experimental
time (beam on target time), and $t_{cycle}$ the total time of the
cycle, with $t_{cycle} = t_{exp} + t_{prep}$. $N_{i,0}$ is the number
of available particles after the target is switched on. The dependence
of the cycle averaged luminosity on the cycle time is shown for
different antiproton production rates in Fig.~\ref{fig:avglum}.

With a limited number of antiprotons restricted to $10^{11}$, as
specified for the HL mode, cycle averaged luminosities of up to
1.6$\times$ 10$^{32}$cm$^{-2}$s$^{-1}$ can be achieved at 15\,GeV/c
for cycle times of less than one beam lifetime. If one does not
restrict the number of injected antiprotons, cycle times should be
chosen longer to reach maximum average luminosities close to
3$\times$10$^{32}$cm$^{-2}$s$^{-1}$. This is a theoretical upper
limit, since the larger momentum spread of the injected beam would
lead to higher beam losses during injection due to limited
longitudinal ring acceptance. Due do short beam lifetimes, more
than $10^{11}$ particles can not be provided at the
lowest momentum. This means that, at low momenta, cycle averaged 
luminosities are expected to be below 10$^{32}$cm$^{-2}$s$^{-1}$.


%
%

\svnInfo $Id: panda.tex 709 2009-02-10 09:03:42Z IntiL $

\section{The \PANDA Detector}
\label{s:over:panda}

The main objectives of the design of the \PANDA experiment 
are to achieve $4\pi$ acceptance, high resolution
for tracking, particle identification and calorimetry, high rate
capabilities and a versatile readout and event selection. To obtain a
good momentum resolution the detector will be composed of two magnetic 
spectrometers: the {\em Target
  Spectrometer (TS)}, based on a superconducting solenoid magnet surrounding
the interaction point, which will be used to measure at large angles and the {\em Forward
  Spectrometer (FS)}, based on a dipole magnet, for small angle tracks. A
silicon vertex detector will surround the interaction point. In both
spectrometer parts, tracking, charged particle identification,
electromagnetic calorimetry and muon identification will be available to
allow to detect the complete spectrum of final states relevant for the
\PANDA physics objectives.

\begin{figure*}[htb]
\begin{center}
\includegraphics[width=0.9\dwidth]{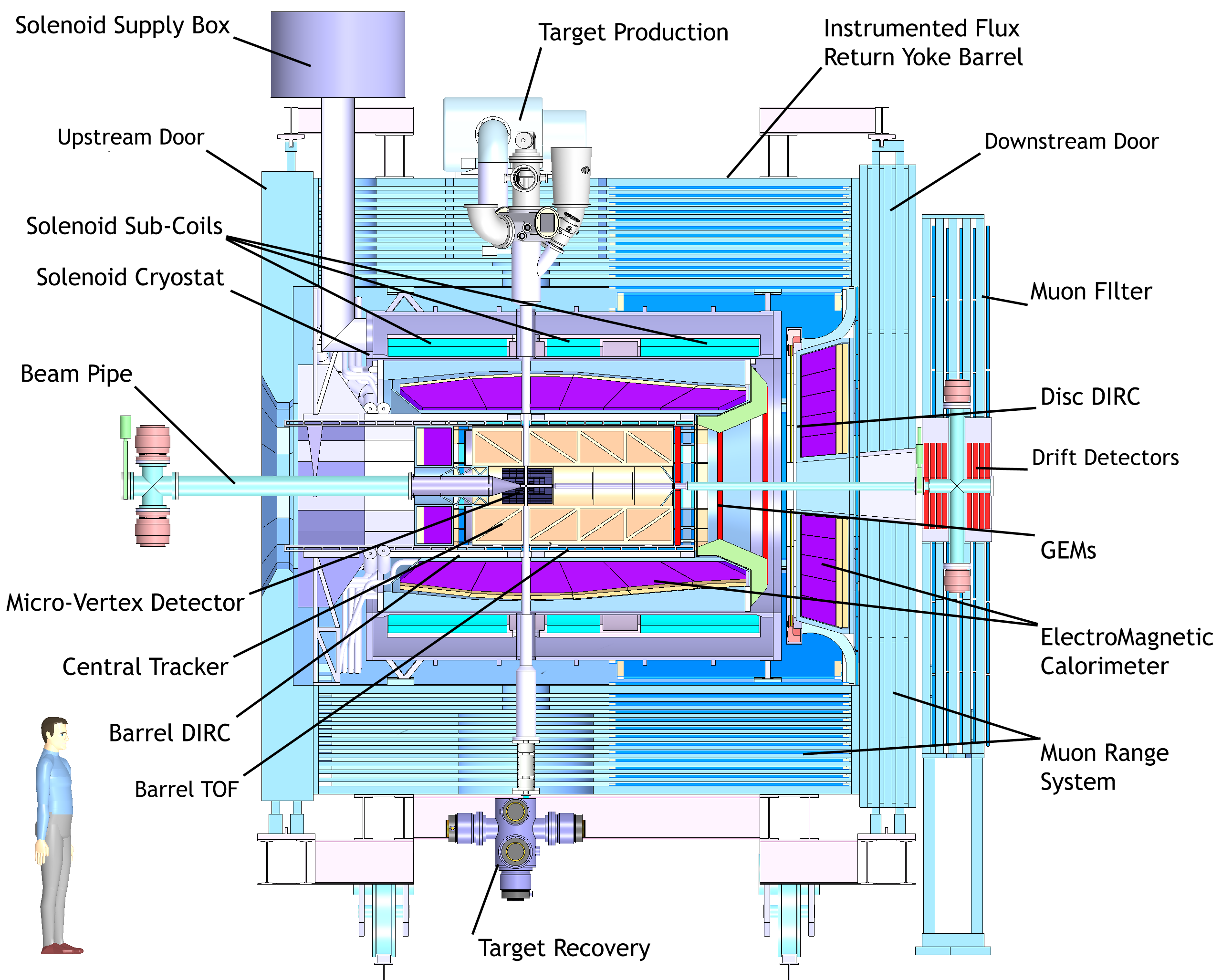}
\caption[Artistic side view of the Target Spectrometer (TS) of \PANDA.]
{Artistic side view of the Target Spectrometer (TS) of \PANDA. 
  To the right of this the Forward Spectrometer (FS) follows, which is
illustrated in Fig.~\ref{f:over:fs_view}.}
\label{f:over:ts_view}
\end{center}
\end{figure*}

\subsection{Target Spectrometer}
The Target Spectrometer will surround the interaction point and measure
charged tracks in a highly homogeneous (better than $\pm$2\%) solenoidal 
field of $2\unit{T}$. In the manner of a collider detector it will contain
detectors in an onion shell like configuration. Pipes for the
injection of target material will have to cross the spectrometer
perpendicular to the beam pipe.

The Target Spectrometer will be arranged in three parts: the barrel
covering angles between 22$\degrees$ and 140$\degrees$, the forward
end cap extending the angles down to 5$\degrees$ and 10$\degrees$ in
the vertical and horizontal planes, respectively, and the backward end
cap covering the region between about 145$\degrees$ and 170$\degrees$.
Please refer to Fig.~\ref{f:over:ts_view} for an overview.

\subsubsection{Target}
\label{sec:det:ts:tgt}

The compact design of the detector layers nested inside the
solenoidal magnetic field, combined with the request of minimal
distance from the interaction point to the vertex tracker, leaves only a very
restricted space for the target installations.  In order to reach the
design luminosity of $2\times 10^{32}$ s$^{-1}$cm$^{-2}$ a target thickness
of about $4\times 10^{15}$ hydrogen atoms per cm$^2$ is required assuming
$10^{11}$ stored anti-protons in the \HESR ring.

These conditions pose a real challenge for an internal target
inside a storage ring. At present, two different, complementary
techniques for the internal target are being developed further: the
cluster-jet target and the pellet target. Both techniques are capable
of providing sufficient densities for hydrogen at the interaction
point, but exhibit different properties concerning their effect on the
beam quality and the definition of the interaction point. In addition,
internal targets also of heavier gases, like helium, deuterium,
nitrogen or argon can be made available.

For non-gaseous nuclear targets the situation is different, in
particular in the case of the planned hyper-nuclear experiment. In these
studies, the whole upstream end cap and part of the inner detector
geometry will be modified.

\paragraph*{Cluster-Jet Target}
The expansion of pressurised cold hydrogen gas into vacuum through a
Laval-type nozzle leads to a condensation of hydrogen molecules
forming a narrow supersonic jet of hydrogen clusters. The cluster size varies
from $10^{3}$ to $10^{6}$ hydrogen molecules tending to become larger at higher
inlet pressure and lower nozzle temperatures.  Such a cluster-jet
with density of $10^{15}$ atoms/cm$^2$ acts as a very diluted target since it
may be seen as a localised and homogeneous monolayer of hydrogen atoms
being passed by the antiprotons once per revolution.

Fulfilling the luminosity demand for \PANDA still requires a density
increase compared to current applications. Additionally, due to
detector constraints, the distance between the cluster-jet nozzle and
the target will be larger than usual. The great advantage of cluster 
targets is the
homogeneous density profile and the possibility to focus the
antiproton beam at highest phase space density. Hence, the interaction
point is defined transversely but has to be reconstructed
longitudinally in beam direction. In addition the low $\beta$-function
of the antiproton beam keeps the transverse beam target heating effects
at the minimum. The possibility of adjusting the target density along
with the gradual consumption of antiprotons for running at constant
luminosity will be an important feature.

\paragraph*{Pellet Target} 
The pellet target features a stream of frozen hydrogen micro-spheres,
called pellets, traversing the antiproton beam perpendicularly. 
Typical parameters for pellets at the interaction point are the rate
of $1.0 -1.5\times 10^{4}\unit{s^{-1}}$, the pellet size of 
$25 - 40\unit{\mu m}$, and
the velocity of about 60 m/s. At the interaction point the pellet train
has a lateral spread of $\sigma\approx1\unit{mm}$ and an interspacing of
pellets that varies between $0.5$ to $5\unit{mm.}$ With proper adjustment of
the $\beta$-function of the coasting antiproton beam at the target
position, the design luminosity for \PANDA can be reached in time
average. The present R\&D is concentrating on minimising the
luminosity variations such that the instantaneous interaction rate
does not exceed the rate capability of the detector systems. Due to the large 
number of interactions expected in every pellet, and thanks to the
foreseen pellet tracking system, a resolution in the vertex position 
of $50\unit{\mu m}$ will be possible with this target.

\paragraph*{Other Targets} are under consideration for the
hyper-nuclear studies where a separate target station upstream will
comprise primary and secondary target and detectors. Moreover, current
R\&D is being undertaken for the development of a liquid helium target and a
polarised $^3$He target. A wire target may be employed to study
antiproton-nucleus interactions.

\subsubsection{Solenoid Magnet}
The magnetic field in the Target Spectrometer will be provided by a
superconducting solenoid coil with an inner radius of $105\unit{cm}$ and a
length of $2.8\unit{m}$. The maximum magnetic field needs to be $2\unit{T}$. The field
homogeneity is foreseen to be better than $2\%$ over the volume of the
vertex detector and central tracker.  In addition the transverse
component of the solenoid field should be as small as possible, in
order to allow a uniform drift of charges in the time projection
chamber. This is expressed by a limit of $\int B_r/B_z dz < 2\unit{mm}$ for
the normalised integral of the radial field component.

In order to minimise the amount of material in front of the
electromagnetic calorimeter, the latter will be placed inside the magnetic
coil. The tracking devices in the solenoid will cover angles down to
$5\degrees{}/10\degrees{}$ where momentum resolution is still
acceptable. The dipole magnet with a gap height of $1.4\unit{m}$ provides a
continuation of the angular coverage to smaller polar angles.

The cryostat for the solenoid coils is required to have two warm bores of $100\unit{mm}$
diameter, one above and one below the target position, to
allow for insertion of internal targets.

The proposed \Panda Target Spectrometer solenoid is comparable by dimensions and
field to the solenoid built in the late eighties for the ZEUS
experiment at HERA, the proton electron collider of the DESY
laboratory at Hamburg.

The winding construction proposed for the solenoid is based on the
well proven technique used for the superconducting coils used since
the beginning of the eighties in the High Energy Physics and nuclear
physics experiments.  We propose to use the same technology used for
superconducting solenoids like CELLO and ZEUS (DESY), ALEPH, DELPHI,
ATLAS, CMS (CERN), BABAR (SLAC), CDF (FERMILAB), BELLE (KEK), FINUDA,
KLOE (LNF\_INFN).

\subsubsection{Micro-Vertex Detector}
\begin{figure}[hbt]
\begin{center}
\includegraphics[width=\swidth]{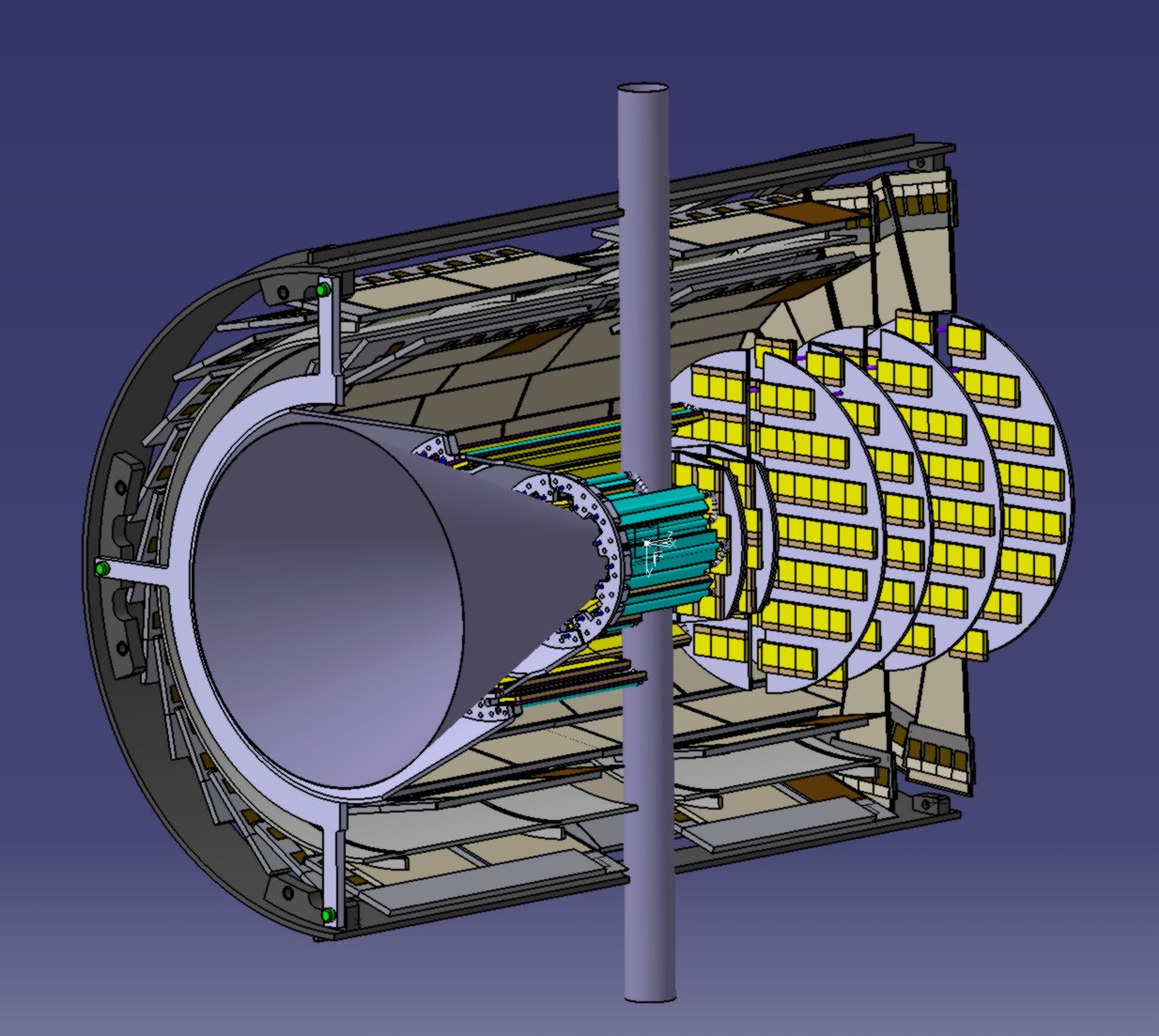}
\caption[The Micro-Vertex Detector (MVD) of the Target Spectrometer]
  {The Micro-Vertex Detector (MVD) of the Target Spectrometer
  surrounding the beam and target pipes seen from upstreams.  The
  outer barrel layers are cut for visibility.}
\label{fig:det:mvd}
\end{center}
\end{figure}

The design of the micro-vertex detector (\Mvd) for the Target
Spectrometer is optimised for the detection of secondary vertices from
\D and hyperon decays and maximum acceptance close to the interaction
point. It will also strongly improve the transverse momentum
resolution. The setup is depicted in Fig.~\ref{fig:det:mvd}.

The concept of the \Mvd is based on radiation hard silicon pixel
detectors with fast individual pixel readout circuits and silicon
strip detectors. The layout foresees a four layer barrel detector with
an inner radius of $2.5\unit{cm}$ and an outer radius of $13\unit{cm}$. The two
innermost layers will consist of pixel detectors and the outer two
layers will consist of double sided silicon strip
detectors.

Eight detector wheels arranged perpendicular to the beam will achieve
the best acceptance for the forward part of the particle spectrum.
Here again, the inner four layers will be made entirely of pixel detectors,
the following two will be a combination of strip detectors on the outer
radius and pixel detectors closer to the beam pipe. Finally the last 
two wheels, made entirely of silicon strip detectors, will be placed
further downstream to achieve a better acceptance of hyperon cascades.

\subsubsection{Central Tracker}
The charged particle tracking devices must handle the high particle
fluxes that are anticipated for a luminosity of up to
several $10^{32}\,$cm$^{-2}$s$^{-1}$.
The momentum resolution $\delta p/p$ has to be on the percent level.
The detectors should have good detection efficiency for secondary
vertices which can occur outside the inner vertex detector
(e.g.\  $\Ks$ or $\Lambda$).
This will be achieved by the combination of the silicon vertex detectors
close to the interaction point (\Mvd) with two outer systems. One
system will cover a large area and is designed as a barrel around the
\Mvd. This will be either a stack of straw tubes (\Stt) or a
time-projection chamber (\Tpc). The forward angles will be covered using
three sets of GEM trackers similar to those developed for the
{\INST{COMPASS}} experiment at \INST{CERN}. The two options for
the central tracker are explained briefly in the following.

\paragraph*{Straw Tube Tracker (\Stt)}
\label{sec:det:ts:stt}

\begin{figure}[htb]
\begin{center}
\includegraphics[width=\swidth]{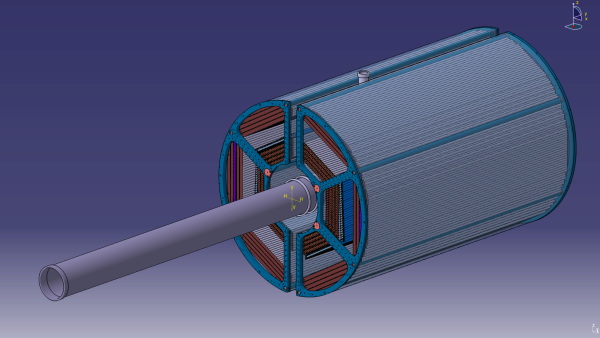} 
\caption[Straw Tube Tracker (STT) of the Target Spectrometer]
  {Straw Tube Tracker (STT) of the Target Spectrometer with beam and
  target pipes seen from upstreams.}
\label{fig:exp:ts:stt}
\end{center}
\end{figure}

This detector will consist of aluminised Mylar tubes called {\em
straws}, which will be self supporting by the operation at
$1\unit{bar}$ overpressure. The straws are to be arranged in planar
layers which are mounted in a hexagonal shape around the \Mvd as shown
in Fig.~\ref{fig:exp:ts:stt}. In total there are 24 layers of which
the 8 central ones are tilted to achieve an acceptable resolution of
$3\unit{mm}$ also in z (parallel to the beam). The gap to the
surrounding detectors will be filled with further individual
straws. In total there will be 4200 straws around the beam pipe at
radial distances between 15$\,$cm and 42$\,$cm with an overall length
of 150$\,$cm. All straws have a diameter of $10\unit{mm}$. A thin and
light space frame will hold the straws in place, the force of the wire
however is kept solely by the straw itself. The Mylar foil is 30
$\mu$m thick, the wire is made of 20$\,\mu$m thick gold plated
tungsten. This design results in a material budget of 1.3 \% of a
radiation length.

The gas mixture used will be Argon based with CO$_2$ as quencher. It
is foreseen to have a gas gain no greater than 10$^5$ in order to
warrant long term operation. With these parameters, a resolution in
$x$ and $y$ coordinates of less than 150$\,\mu$m is expected.

\paragraph*{Time Projection Chamber (\Tpc)}
\label{sec:det:ts:tpc}
A challenging but advantageous alternative to the \Stt is a \Tpc, which
would combine superior track resolution with a low material budget and
additional particle identification capabilities through energy loss
measurements. 

\begin{figure}[htb]
\begin{center}
\includegraphics[width=\swidth]{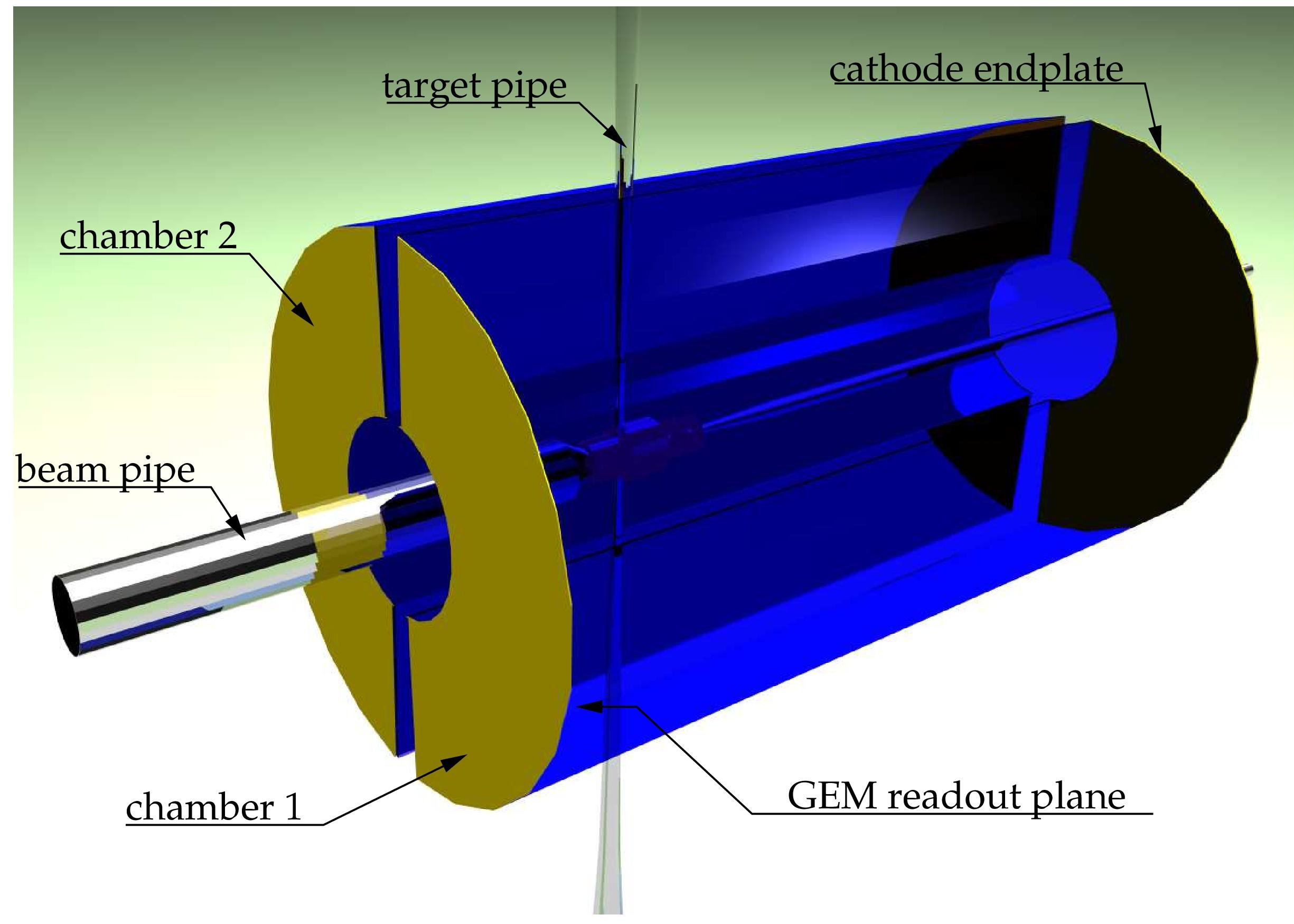}
\caption[GEM Time Projection Chamber (TPC) of the Target Spectrometer]
  {GEM Time Projection Chamber (TPC) of the Target Spectrometer with
  beam and target pipes seen from upstreams.}
\label{fig:exp:ts:tpc}
\end{center}
\end{figure}

The planned \Tpc depicted in a schematic view in 
Fig.~\ref{fig:exp:ts:tpc} will consist
of two large gas-filled half-cylinders enclosing the target and beam
pipe and surrounding the \Mvd.  An electric field along the cylinder
axis separates positive gas ions from electrons created by ionising
particles traversing the gas volume. The electrons drift with constant
velocity towards the anode at the upstream end face and create an
avalanche detected by a pad readout plane yielding information on two
coordinates.  The third coordinate of the track comes from the
measurement of the drift time of each primary electron cluster.  In
common TPCs the amplification stage typically occurs as in multi-wire
proportional chambers. These are gated by an external trigger to avoid
a continuous back flow of ions in the drift volume which would distort
the electric drift field and jeopardise the principle of operation.
In \Panda the interaction rate will be too high and there is no fast
external trigger to allow such an operation. Therefore a novel readout
scheme will be employed which is based on Gas Electron Multiplier (GEM)
foils as amplification stage.

From the viewpoint of the \Panda solenoid magnet, the compatibility with
the \Tpc requires a very good homogeneity of the solenoid field
with a low radial component. The solenoid magnet was designed, anyhow,
to comply with the most stringent requirements coming from both solutions.

\paragraph*{Forward GEM Detectors}
Particles emitted at angles below 22\degrees{} which are not covered
fully by the Straw Tube Tracker or \Tpc will be tracked by three
stations of Gas Electron Multiplier (GEM) detectors placed 1.1\,m, 
1.4\,m and 1.9\,m downstream of
the target.  The chambers have to sustain a high counting rate of
particles peaked at the most forward angles due to the relativistic
boost of the reaction products as well as due to the small angle
$\pbar$p elastic scattering. With the envisaged luminosity, the
expected particle flux in the first chamber in the vicinity of the 
$5\unit{cm}$ diameter beam pipe will be about 
$3 \times 10^{4}\,$cm$^{-2}$s$^{-1}$. 
Gaseous micro-pattern detectors based on
GEM foils as amplification stages are chosen. These detectors have
rate capabilities three orders of magnitude higher than drift
chambers. In the current layout there will be three double planes with two
projections per plane. 

\subsubsection{Cherenkov Detectors and Time-of-Flight}
Charged particle identification of hadrons and leptons over a large
range of angles and momenta is an essential requirement for
meeting the physics objectives of \Panda. There will be several
dedicated systems which, complementary to the other detectors, will
provide means to identify particles. The main part of the momentum
spectrum above 1 GeV/$c$ will be covered by Cherenkov detectors. 
Below the Cherenkov threshold of kaons several other processes have 
to be employed for particle identification.
In addition a time-of-flight barrel will identify slow particles.

\paragraph*{Barrel Time-of-Flight}
For slow particles at large polar angles particle identification will
be provided by a time-of-flight detector. In the Target Spectrometer
the flight path is only in the order of 50 -- 100 cm. Therefore the
detector must have a very good time resolution between 50 and 100 ps.

As detector candidates scintillator bars and strips or pads of
multi-gap resistive plate chambers are considered. In both cases a
compromise between time resolution and material budget has to be
found. The detectors will cover angles between 22\degrees{} and
140\degrees{} using a barrel arrangement around the \Stt/\Tpc at 
42 - 45$\,$cm radial distance.

\paragraph*{Barrel DIRC}
At polar angles between 22\degrees{} and 140\degrees{}, particle
identification will be performed by the detection of internally
reflected Cherenkov (\Dirc) light as realised in the {\INST{BaBar}}
detector~\cite{Staengle:1997xp}.  It will consist of 1.7$\,$cm thick
fused silica (artificial quartz) slabs surrounding the beam line
 at a radial distance of
45 - 54 $\,$cm. At {\INST{BaBar}} the light was imaged across a large
stand-off volume filled with water onto 11\,000 photomultiplier
tubes. At \Panda{}, it is intended to focus the images by lenses onto
micro-channel plate photomultiplier tubes (MCP PMTs) which are
insensitive to magnet fields. This fast light detector type allows a
more compact design and the readout of two spatial coordinates. 

The DIRC design with its compact radiator mounted close to the
\Emc will minimise the conversions. Part of these conversions will be
recovered with information from the DIRC detector, as was shown by 
{\INST{BaBar}}~\cite{Adametz05}.

\begin{figure*}[htb]
\begin{center}
\includegraphics[width=0.75\dwidth]{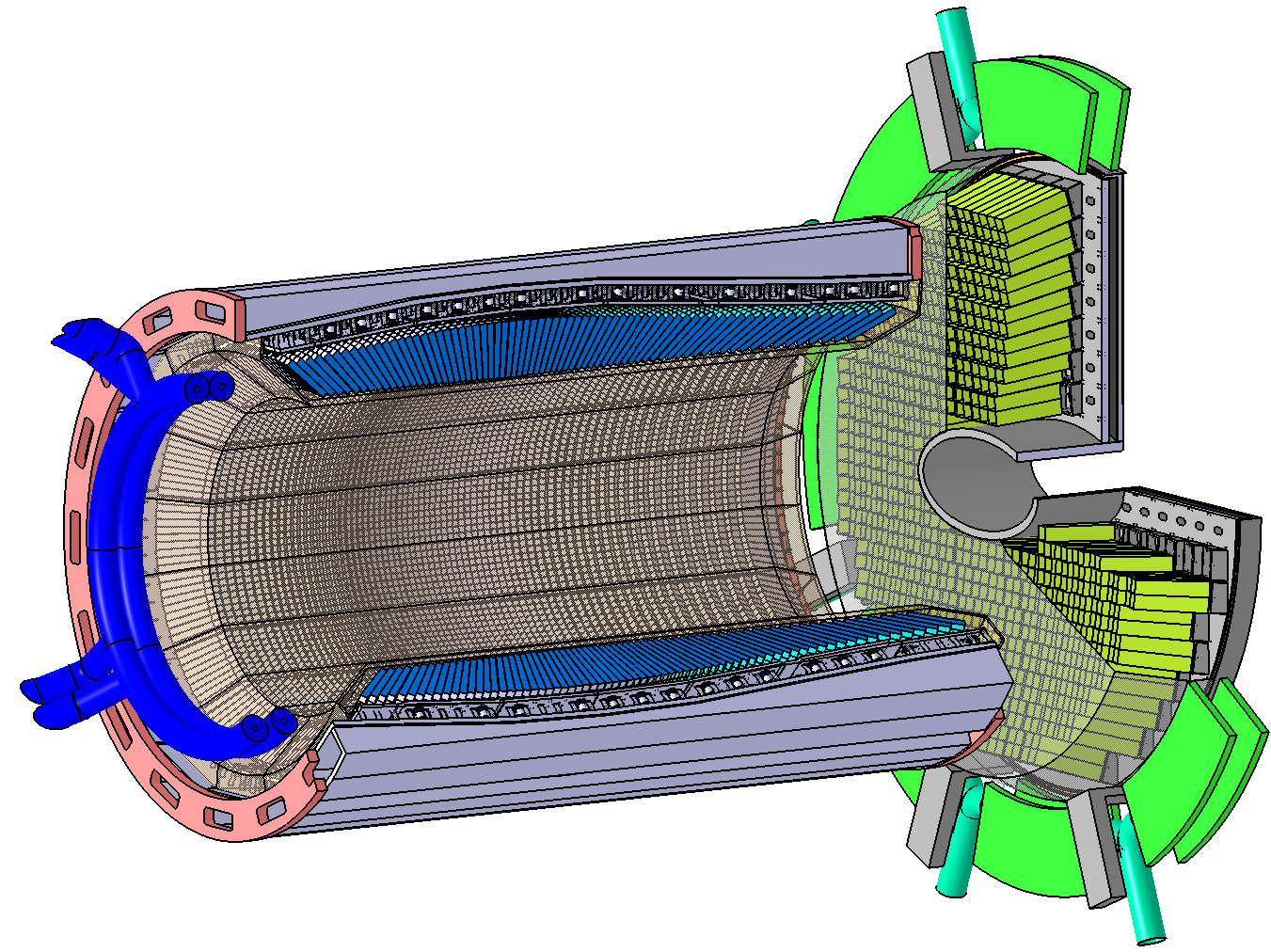}
\caption[Barrel and forward end cap of the Electro-Magnetic Calorimeter.]
{Barrel and forward end cap of the Electro-Magnetic Calorimeter 
  (EMC) with its mounting structures and cooling pipes.  These
  structures will be mounted directly inside the cryostat and forward
  end of the flux return yoke.}
\label{fig:emc}
\end{center}
\end{figure*}

\paragraph*{Forward End-Cap DIRC} 
A similar concept can be employed in the forward direction for
particles between 5\degrees{} and 22\degrees{}. The same radiator,
fused silica, is to be employed however in shape of a disk. At the rim
around the disk the Cherenkov angle will be measured either directly
by focusing elements or indirectly using time-of-propagation (ToP).
In the first option focusing will be done by mirroring quartz elements
reflecting onto MCP PMTs.  Once again two spatial coordinates plus the
propagation time for corrections will be read.  In the ToP design this
angle is reconstructed by using the known path of the photon and its
time of propagation.  This will be achieved by the use of alternating
dichroic mirrors transmitting and reflecting different parts of the
light spectrum .  The disk will be 2\,cm thick and will have a radius
of 110\,cm. It will be placed directly upstream of the forward end cap
calorimeter.

\subsubsection{Electromagnetic Calorimeters}
Expected high count rates and a geometrically compact design of the
Target Spectrometer require a fast scintillator material with a short
radiation length and Moli\`ere radius for the construction of the
electromagnetic calorimeter (\Emc). Lead tungsten (PbWO$_4$) is a
high density inorganic scintillator with sufficient energy and time
resolution for photon, electron, and hadron detection even at
intermediate energies~\cite{Mengel:1998si,Novotny:2000zg,Hoek:2002ss}.

The crystals will be 20 cm long, i.e. approximately 22$\,X_0$, in
order to achieve an energy resolution below 2\percent{} at
1$\,\gev$~\cite{Mengel:1998si,Novotny:2000zg,Hoek:2002ss} at a
tolerable energy loss due to longitudinal leakage of the shower.
Tapered crystals with a front size of $2.1\times 2.1\,\cm^2$ will be
mounted in the barrel EMC with an inner radius of 57$\,$cm.  This
implies 11360 crystals for the barrel part of the calorimeter.  The
forward end cap EMC will be a planar arrangement of 3600 tapered
crystals with roughly the same dimensions as in the barrel part, and
the backward end cap EMC comprises of 592 crystals.  The readout of
the crystals will be accomplished by large area avalanche photo diodes
in the barrel and vacuum photo-triodes in the forward and backward end
caps. The light yield can be increased by a factor of about 4 compared
to room temperature by cooling the crystals down to -25$\degC$. The
arrangement of the barrel and forward end cap calorimeters is shown in
Fig.~\ref{fig:emc}.

The EMC will allow to achieve an e/$\pi$ ratio of 10$^3$ for momenta above
0.5$\,\gevc$.  Therefore, e-$\pi$-separation will not require an
additional gas Cherenkov detector in favour of a very compact geometry
of the EMC. For further details please refer to Ref.~\cite{PANDA:TDR:EMC}.

\subsubsection{Muon Detectors}
In the barrel region the yoke is segmented and will consist 
of 13 layers in total: the
innermost layer will have a thickness of 6\,cm equal to the outermost one, 
in between 11 layers of 3\,cm thickness will be located. The gaps for the
detectors will be 3\,cm wide. This is enough material for the absorption of
pions in the momentum range in \PANDA at these angles. In the forward
end cap more material is needed. Since the downstream door of the
return yoke has to fulfil constraints for space and accessibility,
the muon system will split in several layers.  Six detection layers will be
placed around five iron layers of 6\,cm each within the door, and a
removable muon filter with additional five layers of 6\,cm iron is
located in the space between the solenoid and the dipole. This filter
has to provide cut-outs for forward detectors and pump lines and has
to be built in a way that it can be removed with few crane operations
to allow easy access to these parts. The integration with the muon system 
was the most challenging requirement for the instrumented flux return yoke
design.

As detector within the absorber layers rectangular aluminum drift
tubes will be used as they were constructed for the COMPASS muon detection
system.  They are essentially drift tubes with additional capacitive
coupled strips read out on both ends to obtain the longitudinal
coordinate.

\begin{figure*}[htb]
\begin{center}
\includegraphics[width=0.95\dwidth]{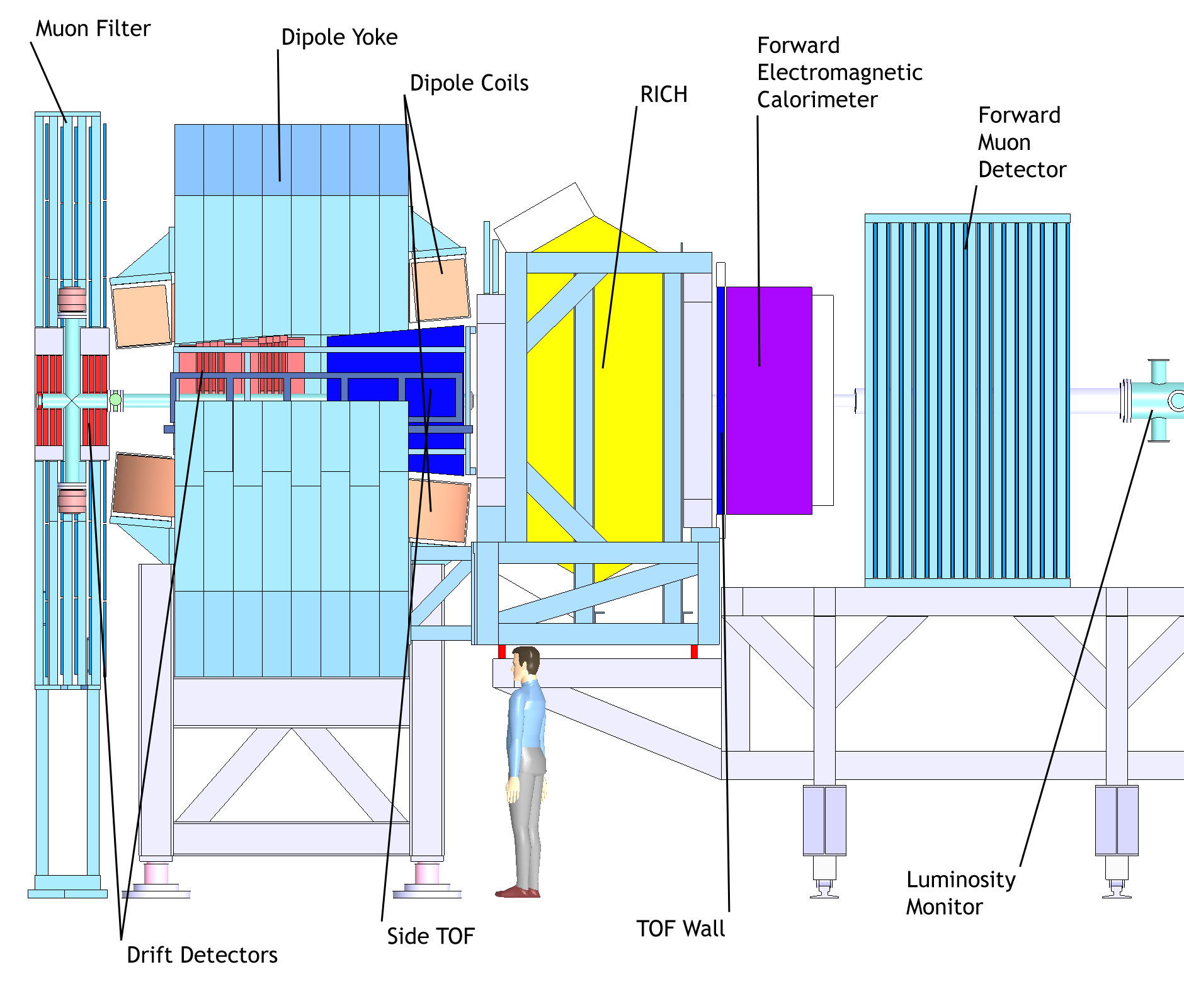}
\caption[Artistic side view of the Forward Spectrometer (FS) of \PANDA.]
{Artistic side view of the Forward Spectrometer (FS) of \PANDA. 
  It is preceded on the left by the Target Spectrometer (TS),
which is illustrated in Fig.~\ref{f:over:ts_view}.}
\label{f:over:fs_view}
\end{center}
\end{figure*}

\subsubsection{Hypernuclear Detector}
The hypernuclei study will make use of the modular structure of
\Panda{}. Removing the backward end cap calorimeter will allow to
add a dedicated nuclear target station and the required additional
detectors for $\gamma$ spectroscopy close to the entrance of
\Panda{}. While the detection of anti-hyperons and low momentum $K^+$
can be ensured by the universal detector and its PID system, a
specific target system and a $\gamma$-detector are additional
components required for the hypernuclear studies.

\paragraph*{Active Secondary Target}
The production of hypernuclei proceeds as a two-stage process. First
hyperons, in particular $\Xi\bar{\Xi}$, are produced on a nuclear
target. In some cases the $\Xi$ will be slow enough to be captured in
a secondary target, where it reacts in a nucleus to form a double
hypernucleus.

The geometry of this secondary target is determined by the short mean
life of the $\Xi^-$ of only 0.164$\,$ns. This limits the required
thickness of the active secondary target to about 25--30$\,$mm. It
will consist of a compact sandwich structure of silicon micro
strip detectors and absorbing material. In this way the weak decay
cascade of the hypernucleus can be detected in the sandwich structure.

\paragraph*{Germanium Array} 
An existing germanium-array with refurbished readout will be used for
the $\gamma$-spectroscopy of the nuclear decay cascades of
hypernuclei. The main limitation will be the load due to neutral or
charged particles traversing the germanium detectors. Therefore,
readout schemes and tracking algorithms are presently being developed
which will enable high resolution $\gamma$-spectroscopy in an
environment of high particle flux.


\subsection{Forward Spectrometer}

The Forward Spectrometer (FS) will cover all particles emitted in
vertical and horizontal angles below $\pm 5^\circ$ and $\pm 10^\circ$,
respectively.  Charged particles will be deflected by an integral
dipole field of 2\,Tm.  Cherenkov detectors, calorimeters and
muon counters ensure the detection of all particle types.  Please
refer to Fig.~\ref{f:over:fs_view} for an overview.

\subsubsection{Dipole Magnet}
A 2$\,$Tm dipole magnet with a window frame, a 1$\,$m gap, and more
than 2$\,$m aperture will be used for the momentum analysis of charged
particles in the Forward Spectrometer.  In the current planning, the
magnet yoke will occupy about 1.6$\,$m in beam direction starting from
3.9$\,$m downstream of the target.  Thus, it covers the entire angular
acceptance of the Target Spectrometer of $\pm$10\degrees{} and
$\pm$5\degrees{} in the horizontal and in the vertical direction,
respectively.  The bending power of the dipole on the beam line causes
a deflection of the antiproton beam at the maximum momentum of
15$\,\gevc$ of 2.2\degrees{}.  The designed acceptance for charged
particles covers a dynamic range of a factor 15 with the detectors
downstream of the magnet.  For particles with lower momenta, detectors
will be placed inside the yoke opening. The beam deflection will be
compensated by two correcting dipole magnets, placed around
the \Panda{} detection system.

\subsubsection{Forward Trackers}
\label{sec:det:fs:trk}

\begin{figure}[htb]
\begin{center}
\includegraphics[width=\swidth]{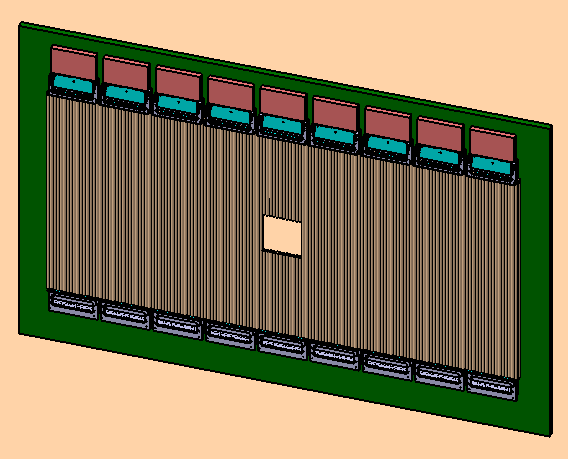} 
\caption[Forward Spectrometer Time-Of-Flight Detector.]
{Double layer of straw tubes with preamplifier cards and gas manifolds
mounted on rectangular support frame.
The opening in the middle of the detector is foreseen for the beam pipe.}
\label{fig:det:fs:trk:dc1}
\end{center}
\end{figure}

The deflection of particle trajectories in the field of the dipole
magnet will be measured with three pairs of tracking drift detectors.
The first pair will be placed in front, the second within and the
third behind the dipole magnet.  Each pair will contain two autonomous
detectors as described below.  Thus, in total, 6 independent detectors
will be mounted.  Each tracking detector will consist of four
double-layers of straw tubes (see Fig.~\ref{fig:det:fs:trk:dc1}), two
with vertical wires and two with wires inclined by a few degrees.  The
optimal angle of inclination with respect to vertical direction will
be chosen on the basis of ongoing simulations.  The planned
configuration of double-layers of straws will allow to reconstruct
tracks in each pair of tracking detectors separately, also in case of
multi-track events.
 
\subsubsection{Forward Particle Identification}
\paragraph*{RICH Detector}
To enable the $\pi$/$K$ and $K$/p separation also at the highest
momenta a \Rich detector is proposed. The favoured design is a dual
radiator \Rich detector similar to the one used at
\INST{HERMES}~\cite{Akopov:2000qi}. Using two radiators, silica
aerogel and C$_4$F$_{10}$ gas, provides $\pi$/$K$/p separation in a
broad momentum range from 2 to 15$\,\gevc$.  The two different indices
of refraction are 1.0304 and 1.00137, respectively.  The total
thickness of the detector is reduced to the freon gas radiator
\mbox{(5\%$\,X_0$),} the aerogel radiator (2.8\%$\,X_0$), and the
aluminum window (3\%$\,X_0$) by using a lightweight mirror focusing
the Cherenkov light on an array of photo-tubes placed outside the
active volume.

\paragraph*{Time-of-Flight Wall}
A wall of slabs made of plastic scintillator and read out on both ends
by fast photo-tubes will serve as time-of-flight stop counter placed at
about 7$\,$m from the target. In addition, similar detectors will be
placed inside the dipole magnet opening, to detect low momentum
particles which do not exit the dipole magnet.
 With the expected time resolution of $\sigma\,=\,50\,$ps
$\pi$/$K$ and $K$/p separation on a 3$\,\sigma$ level will be possible
up to momenta of 2.8$\,\gevc$ and 4.7$\,\gevc$, respectively.

\subsubsection{Forward Electromagnetic Calorimeter}
\begin{figure}[htb]
\begin{center}
\includegraphics[width=\swidth]{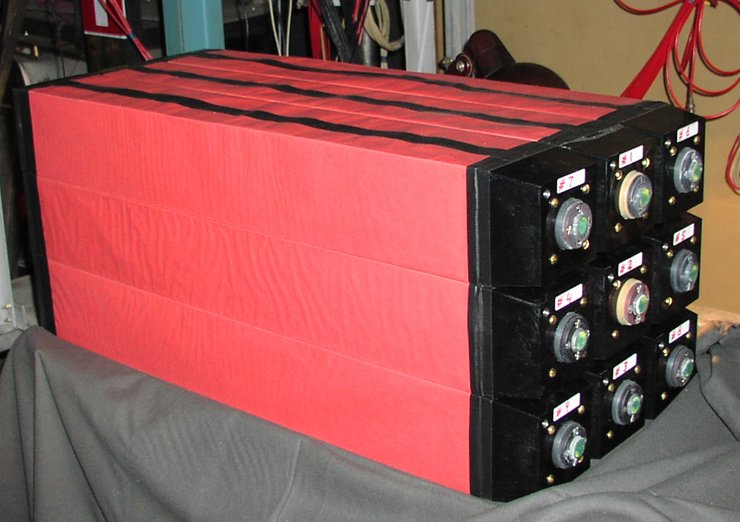} 
\caption[Shashlyk module for the FS calorimeter]
{$3 \times 3$ matrix of prototype Shashlyk modules as they should 
  be employed for the \PANDA Forward Electromagnetic Calorimeter.}
\label{fig:det:fs:shashlyk}
\end{center}
\end{figure}

For the detection of photons and electrons a {Shashlyk}-type
calorimeter with high resolution and efficiency will be employed. The
detection is based on lead-scintillator sandwiches read out with
wave-length shifting fibres passing through the block and coupled to
photo-multipliers. The technique has already been successfully used in
the \INST{E865} experiment~\cite{bib:emc:E865} and it has been adopted
for various other experiments like PHENIX and LHCb.  An energy
resolution of $4\%/\sqrt{E}$~\cite{bib:emc:KOP99} has been achieved.
A view of a 3x3 matrix of Shashlyk modules with lateral size of
110\,mm~x~110\,mm and a length of 680\,mm ($=\,20X_0$) is shown in
Fig.~\ref{fig:det:fs:shashlyk}.  To cover the forward acceptance, 351
such modules arranged in 13 rows and 27 columns at a distance of
7.5$\,$m from the target are required.

\subsubsection{Forward Muon Detectors}
For the very forward part of the muon spectrum, a further range
tracking system consisting of interleaved absorber layers and
rectangular aluminium drift-tubes is being designed, similar to the
muon system of the Target Spectrometer, but laid out for higher
momenta. The system allows discrimination of pions from muons,
detection of pion decays and, with moderate resolution, also the
energy determination of neutrons and anti-neutrons.
The forward muon system will be placed at about 9\,m from the target.

\subsection{Luminosity monitor}
In order to determine the cross section for physical processes, it is
essential to determine the time integrated luminosity $L$ for
reactions at the \PANDA interaction point that was available while
collecting a given data sample. In the following we concentrate on elastic
antiproton-proton scattering as the reference channel.  For most other
hadronic processes that will be measured concurrently in \PANDA the
precision is rather poor with which the cross sections are known.

The basic concept of the luminosity monitor is to reconstruct the
angle of the scattered antiprotons in the polar angle
range of 3-8\,mrad with respect to the beam axis.  Due to the large
transverse dimensions of the interaction region when using the pellet
target, there is only a weak correlation of the position of the
antiproton at e.g.\ $z=+10.0$\,m to the recoil angle. Therefore, it is
necessary to reconstruct the angle of the antiproton at the luminosity
monitor. As a result the luminosity monitor will consist of a sequence
of four planes of double-sided silicon strip detectors located as far
downstream and as close to the beam axis as possible. The planes are
separated by 20 cm along the beam direction. Each plane consists of 4
wafers (e.g. 2\,cm$\times$5\,cm $\times\ 200\,\mu$m, with 50\,$\mu$m
pitch) arranged radially to the beam axis. Four planes are required
for sufficient redundancy and background suppression. The use of 4
wafers (up, down, right, left) in each plane allows systematic errors
to be strongly suppressed. 

The silicon wafers will be located inside a vacuum chamber to minimize
scattering of the antiprotons before traversing the 4 tracking
planes. The luminosity monitor can be located in the space between the
downstream side of the forward muon system and the HESR dipole
needed to redirect the antiproton beam out of the \PANDA chicane back
into the direction of the HESR straight stretch 
(i.e.\  between $z=+11$\,m and $z=+13$\,m downstream of the target). 

As pilot simulations show, at a beam momentum of 6.2\,GeV/$c$ the
proposed detector measures antiprotons elastically scattered in the
range $0.0006 $(GeV)$^2< -t < 0.0035$ (GeV)$^2$, which spans the
Coulomb-nuclear interference region.  Based upon the granularity of
the readout the resolution of $t$ could reach $\sigma_t \approx
0.0001$ (GeV)$^2$. In reality this value is expected to degrade to
$\sigma_t \approx 0.0005$ (GeV)$^2$ when taking small-angle scattering
into account. At the nominal \PANDA interaction rate of $2\times 10^{7}$/s
there will be an average of 10\,kHz/cm$^2$ in the sensors. In
comparison with other experiments an absolute precision of about 3\%
is considered feasible for this detector concept at \PANDA, which will
be verified by more detailed simulations.

\subsection{Data Acquisition}
In \PANDA, a data acquisition concept is being developed to be
as much as possible matched to the high data rates, to the complexity of the
experiment and the diversity of physics objectives and the rate
capability of at least $2\times 10^{7}$ events/s.

In our approach, every sub-detector system is a self-triggering entity.
Signals are detected autonomously by the sub-systems and are preprocessed.
Only the physically relevant information is extracted and transmitted.
This requires hit-detection, noise-suppression and clusterisation at
the readout level. 
The data related to a particle hit, with a substantially reduced rate
in the preprocessing step, is marked by a precise time stamp and
buffered for further processing.
The trigger selection finally occurs in computing nodes which
access the buffers via a high-bandwidth network fabric. The new
concept provides a high degree of flexibility in the choice of trigger
algorithms. It makes trigger conditions available which are
outside the capabilities of the standard approach.

\subsection{Infrastructure}
\label{s:over:infra}
The target for antiproton physics will be located in the straight section
at the east side of the \HESR.  At this location an experimental hall
with 43\,m $\times$ 29\,m floor space is planned.  To allow for access
during \HESR operation the beam line is shielded by a concrete
radiation shield of 2$\,$m thickness on both sides and is covered on
top by concrete bars of 1$\,$m thickness.  Within the elongated
concrete cave the \Panda{} detector together with auxiliary equipment,
beam steering, and focusing elements will be housed.  It is planned that 
it will be possible to open the roof of the
cave as well as its wall, so that heavy components can be hoisted in
by crane.

\begin{figure*}[ht]
  \begin{centering}
    \includegraphics[width=0.95\dwidth]
            {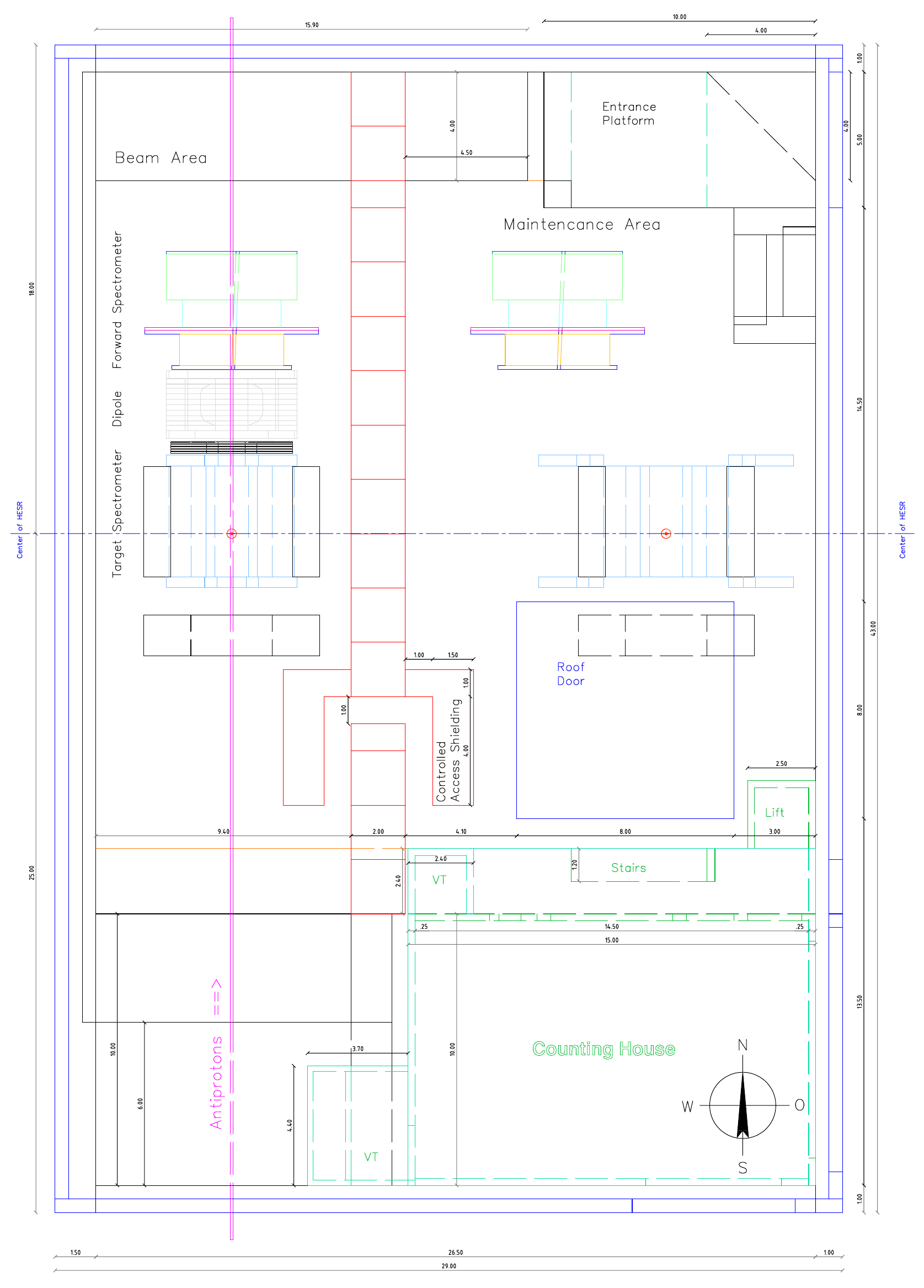}
\caption[Floor plan of the \PANDA hall]{Floor plan of the \PANDA hall 
    showing the location of the solenoid and dipole magnets. The
    solenoid with its detectors and the platform with the FS detectors
    can be moved from the in-beam position inside the \HESR ring
    (western area) to a parking position depicted in the eastern
    area.  In the maintenance position the solenoid is
    shown with open doors, which is required for service access to the
    inner detector components.  Please also refer to
    Sec.~\ref{s:over:infra}.}
\label{f:over:Panda_Hall}
  \end{centering}
\end{figure*}

The shielded beam line area for the \Panda experiment including
dipoles and focusing elements is foreseen to have 37\,m $\times$
9.4\,m floor space and a height of 8.5\,m with the beam line at a height
of 3.5\,m.  The general floor level of the \HESR is planned to be 2\,m higher.
This level will be kept for a length of 4\,m in the north as well
as the south of the hall (right part in~\Reffig{f:over:Panda_Hall}),
to facilitate transport of heavy equipment into the \HESR tunnel.

The Target Spectrometer with electronics and supplies will be mounted
on rails which makes it retractable to parking positions outside of
the \HESR{} beam line area (i.e.\ into the eastern part of the hall 
in~\Reffig{f:over:Panda_Hall}).  The experimental hall will provide
additional space for delivery of components and assembly of the
detector parts.  With the concrete blocks in place, this area will be
sufficiently shielded from radiation to allow access during
commissioning and running of \HESR.  In the south corner of the hall,
a counting house complex with five floors is foreseen. The lowest
floor will contain various supplies for power, high voltage, cooling
water, gases etc. The next higher level is planned for readout electronics
and data processing.  The third level will house the online
computing farm.  The fourth floor will be at level with the surrounding
ground and will house the control room, a meeting room and social
rooms for the shift crew.  Above this floor, hall electricity supplies
and ventilation is placed.  A crane spans the whole area
with a hook at a height of about 10$\,$m.  Sufficient (300$\,$kW)
electric power will be available.

Liquid helium coolant may come from the main cryogenic liquefier for
the \INST{SIS} rings.  Alternatively, a separate small liquefier
(50$\,$W cooling power at 4$\,$K) would be installed. The supply point
will be at the north-east area of the building.  All other cabling,
which will be routed starting at the counting house, will join these
supply lines at the end of the rails system of the Target Spectrometer
at the eastern wall.  The temperature of the building will be
moderately controlled.  More stringent requirements with respect to
temperature and humidity for the detectors have to be maintained
locally.  To facilitate cooling and avoid condensation, the Target
Spectrometer will be kept in a tent with dry air at a controlled
temperature.

\svnInfo $Id: req.tex 709 2009-02-10 09:03:42Z IntiL $

\begin{figure*}[ht]
  \begin{centering}
    \subfigure[Northern cross section]{
      \includegraphics[width=.7\dwidth, angle=0]
            {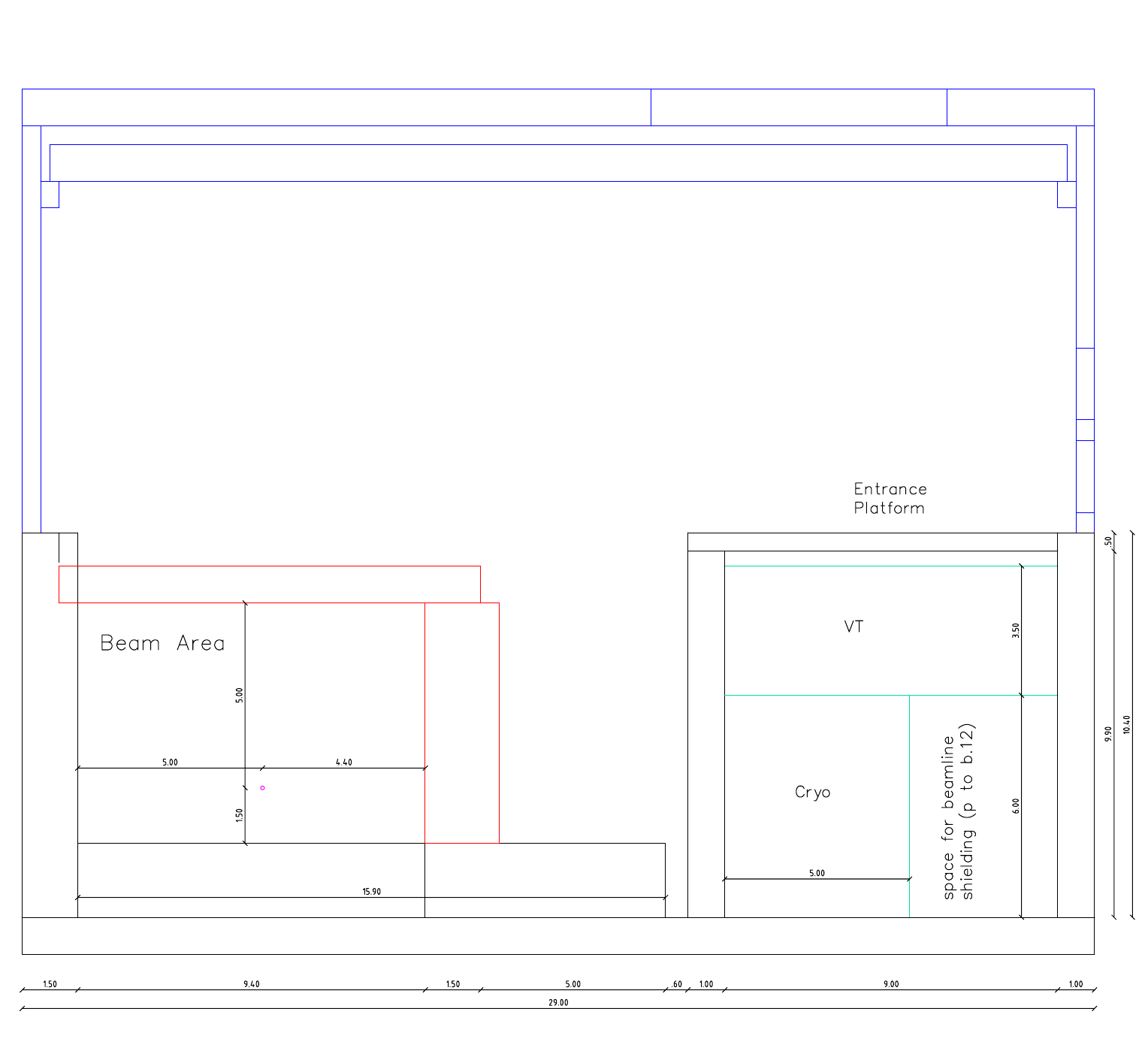}}
    \subfigure[Southern cross section]{
       \includegraphics[width=.7\dwidth, angle=0]
            {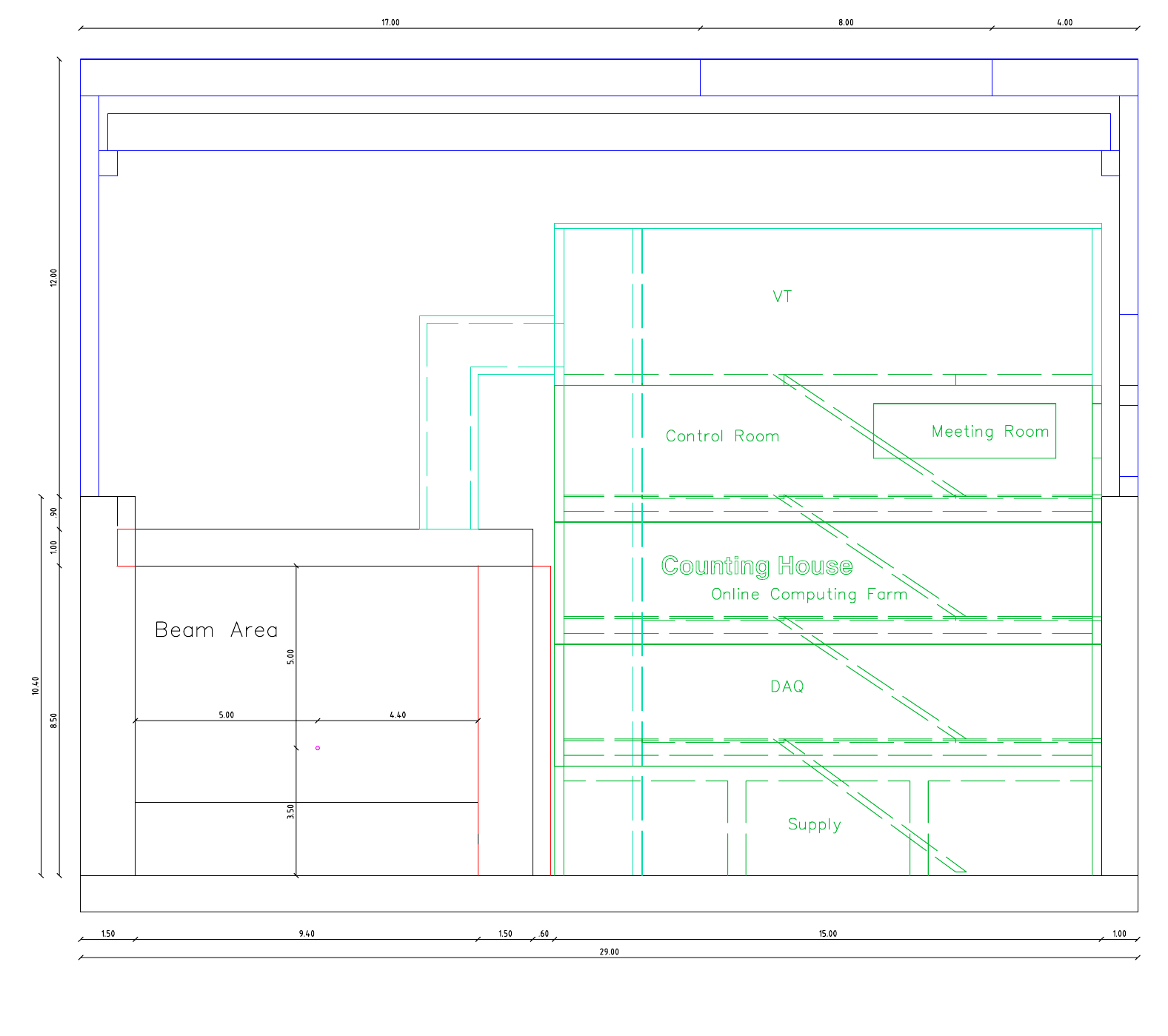}}
\caption[Cross sections of the Panda hall in the east-west plane.]
   {Cross sections of the Panda hall in the east-west plane.  In the
   northern view the shielding of the beam area and the entrance
   platform with the usage of the lower area for service technology
   (VT) and the cryogenic interface is shown.  The southern view shows
   the levels of the counting house together with the interface block
   to the beam area.}
\label{f:over:Panda_Hall_cut}
  \end{centering}
\end{figure*}

\section{Requirements for the Spectrometer Magnets}
\label{s:over:req}

In this section we summarise the most relevant requirements to the two
spectrometer magnets.  How these requirements will be met is discussed in
the Performance and Detector Integration sections of the respective
chapter.

\subsection{Overall Requirements}

The major overall requirements are that the spectrometer magnets need
to provide magnetic fields such that the identification and momentum
reconstruction of charged tracks in \PANDA becomes feasible to the
required level.  Simulations have shown that a 2\,T field along the
axis of the beam is required for the Target Spectrometer, while an
integral of 2\,Tm bending power is needed at highest beam momenta to
obtain the required momentum resolution of 1\% in the Forward
Spectrometer.  Both magnets need to leave enough free space to host all
detectors required for particle identification.

Space constraints both at the in-beam position inside the HESR tunnel
as well as inside the hall constrain the maximum extent of the magnets
(see Fig.~\ref{f:over:Panda_Hall}).  The installation of detectors and
their supplies within the magnets and the sequence of their
installation is an important constraint to the design of the
spectrometer magnets.

Concerning the \PANDA solenoid, the main difference from the
previously built solenoids of same type is the need to accommodate the
pipe for the molecular hydrogen target (cluster or pellet).  This pipe
needs to go through all the iron yoke, the superconducting coil and
all the detectors, not to mention the cryostat housing the
superconducting coil and the thermal shields.  To allow the passage of
the target pipe the coil needs to be wound leaving a gap in the coil
itself. The coil geometry needs also to be chosen in a way to
guarantee an easy track reconstruction and eventually the use of a TPC
as central tracker.

The last two requirements call for good field uniformity ($\pm$2\%)
and an integrated radial field component $B_r$ (normalised to the
normal operation field of 2\,T) lower than 2\,mm for any path parallel to
the solenoid axis inside the central tracker volume.

In the \PANDA Target Spectrometer we plan to meet the requirements on field
quality by opening a second gap in the winding at a position fairly
symmetric to the ``target pipe gap'' and carefully adjusting the
sub-coil lengths.  This solution will also allow for the use of a single
current density in all the coil winding, with noteworthy
simplification compared to the two-current density design of earlier
magnets as in BABAR, or the use of additional compensating coils at
the main coil end, as in ALEPH and DELPHI.
                                   
The spectrometer resolution in the barrel part of the detector (down
to 22 degrees) will be fairly constant, the bending power of the solenoid
remaining about 1\,Tm for any track passing the central tracker.
For the end cap detector (from 22 to 5 degrees in the vertical plane) the
bending power will decrease with smaller angles due to the
finite length of the coil.

Another requirement on the Target Spectrometer magnet is to be
magnetically self screening, to allow the use of turbo-pumps close to
the iron return yoke (for the hydrogen molecular beam and the \HESR
pumping).  The magnet design meets the goal reducing the value of the
stray fields at the pump location down to less than 5\,mT.

The requirements for the Forward Spectrometer can be summarised by
providing a momentum resolution of 1\% or better for all charged
particles emerging below $5^\circ$ and $10^\circ$ in the vertical and
horizontal planes, respectively.  This can be achieved by providing a
rigidity of about 2\,Tm and space for tracking detectors within the
aperture of the magnet.  In the horizontal plane the opening must be
such that particles with a factor of 15 lower momenta than the beam
can still traverse the full magnet and be detected in the subsequent
detectors.  Neutral particles and muons will be either identified by
the forward calorimeters and range systems downstream of the dipole or
in the intermediate muon filter.  More details on the requirements are
given in Sec.~\ref{s:req:fs}.

\subsection{Accelerator Interface}

\begin{figure}[ht]
  \begin{centering}
    \includegraphics[width=\swidth]
            {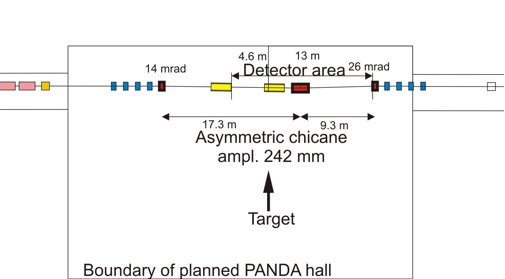}
\caption[Detail of the \HESR layout showing the \PANDA experiment chicane.]
  {Detail of the \HESR layout showing the straight section
  where the \PANDA experiment is located.  The \PANDA detectors will
  be within the ``Detector area''.  The \PANDA solenoid and
  compensating solenoid (to the left of it) are indicated in yellow,
  the \PANDA dipole and the 2 dipoles of the chicane in red, and the
  quadrupoles in blue.}
\label{f:over:HESR_chicane}
  \end{centering}
\end{figure}

The \Panda detector magnets will interact with the antiprotons stored in
High Energy Storage Ring (\HESR) through the magnetic field seen by
the beam circulating in the storage ring. For a smooth operation of
the storage ring the integral of the magnetic field, apart from the
machine optics, seen by the circulating beam in a complete round must
be zero.  The asymmetric chicane which will accomplish this is sketched
in Fig.~\ref{f:over:HESR_chicane}.

In the Target Spectrometer (TS) the longitudinal component of the
magnetic field $B_z$ of the solenoid, oriented along the machine axis,
is 2\,T at maximum over a length of several metres.  Thus, the
integral of the field along the beam of the \PANDA Target Spectrometer
is $\sim7\unit{Tm}$ and influences the divergence of the beam
significantly.  To avoid coupling effects between the beam motions in
the transverse planes a compensating coil of similar strength will be
installed as close as possible to the \Panda solenoid, {\it i.e.\
}after the focusing quadrupoles and upstream of the \PANDA solenoid.
Due to the magnetic field integral of $7\unit{Tm}$ a superconducting
solenoid with a magnetic length of less than 2.5\,m is used, sharing
the refrigeration system with the \Panda Target Spectrometer.  Both
solenoids are operated such that a full compensation of the $B_z$
integral is reached.  Normally, they will be set to the nominal
current at the beginning of a measurement period and remain unchanged
for the full period.  A lower field setting is only required for
measurements at some beam energies below injection energy.

In the Forward Spectrometer (FS) the dipole magnet deflects the whole
of the beam.  Thus the transverse component $B_y$ of the dipole field
must be compensated.  This is realised using two dipole magnets
amounting to the same integral bending power of the \PANDA dipole.
They are arranged asymmetrically around the dipole in order to keep
the necessary space for the solenoidal magnets and detectors, {\it
i.e.\ }at 17.3\,m and 9.3\,m before and after the centre of the
dipole, respectively.  Thus they form a chicane in the straight
section of the \HESR with bending angles of 14\,mrad and 26\,mrad,
respectively.  Thus the beam reenters the straight section after the
whole chicane to the nominal orbit of the \HESR, and the total field
integral in vertical direction $B_y$ is balanced to zero.

The \HESR ring will provide \Panda with antiprotons with momenta
between $1.5$ and $15\unit{GeV/c}$.  These will be injected at the
fixed energy of $3.8\unit{GeV/c}$ and \HESR will operate as a slow
synchrotron, ramping from the injection energy to the final energy
required for the experiment.

To keep the antiproton beam on the nominal orbit, the chicane dipoles,
including the Forward Spectrometer dipole, must ramp
simultaneously with the \HESR magnets, hence the chicane being an
integral part of the \HESR accelerator.  The ramp time of the \HESR of
about 60\,s would introduce too large time delays of the magnetic
field due to eddy currents if a solid iron core had been chosen for
the dipole.  Therefore, a moderate lamination is mandatory.

\subsection{Target Spectrometer}
\label{s:sol:req}

\AUTHOR{I.~Lehmann, A.~Bersani, Y.~Lobanov}

The requirements on the solenoid magnet are manifold, the primary one
being to deliver a 2\,T field parallel to the beam line. This field
should be uniform to a high degree over the region of the \PANDA
tracking detectors.  In Sec.~\ref{s:sol:int} the details of the
integration of the target and detectors are discussed.

\subsubsection{Magnetic Field Requirements}
\label{s:sol:field}

In addition to the main requirement that the magnetic field has to be 2\,T
in the solenoid centre, there are several requirements for field
homogeneity and stray field, which are summarised in the following.

\paragraph{Tracker Region}

\begin{table*}[ht]
  \centering
  \begin{tabular}{|c|c|c|}
\hline
Criteria & Absolute Value & Definition\tabularnewline
\hline
\multicolumn{3}{c}{\bf Micro Vertex Detector}\tabularnewline
\hline
\multicolumn{3}{|c|}{Dimensions, longitudinal: $-0.25$\,m$ < z < 0.2$\,m,
                                \quad radial: $ 0.0$\,m$ < r < 0.12$\,m }\tabularnewline
\hline
\multirow{2}{2cm}{$\frac{\Delta B}{B_0}<2\%$}
  & \multirow{2}{4cm}{$\Delta B :=  | B(r,z) - B_0 | $}
  & $B_0 := $ nominal field, {\it i.e.} 2\,T\tabularnewline
 &  & $B(r,z) := $ field at any point in given region \tabularnewline
\hline
\multicolumn{3}{c}{\bf Central Tracker}\tabularnewline
\hline
\multicolumn{3}{|c|}{Dimensions, longitudinal: $-0.4$\,m$ < z < 1.1$\,m,
                                \quad radial: $0.15$\,m$ < r < 0.42$\,m}\tabularnewline
\hline
\multirow{2}{2cm}{$\frac{\Delta B}{B_0}<2\%$}
  & \multirow{2}{4cm}{$\Delta B := | B(r,z) - B_0 | $}
  & $B_0 := $ nominal field, {\it i.e.} 2\,T\tabularnewline
 &  & $B(r,z) := $ field at any point in given region \tabularnewline
\hline
\multirow{2}{2.8cm}{$I_B(r,z_0) < 2$\,mm}
  & \multirow{2}{4.5cm}{$I_B(r,z_0)
    := \int\limits_{z_0}^{-400}{\frac{B_r(r,z)}{B_z(r,z)}\ dz} $}
  & $z_0 :=$ any $z$ value in region \tabularnewline
&  & $B_r , B_z :=$ radial, long.\ field components, resp.  \tabularnewline
\hline
  \end{tabular}

  \caption[Criteria for the homogeneity of the solenoidal magnetic
  field.]{Criteria for the homogeneity of the solenoidal magnetic
  field in the region of the Micro Vertex Detector (MVD) and central
  trackers of \PANDA. The first respective rows for the MVD and
  central tracker show the longitudinal ({\it i.e.} along the $z$
  axis) and radial dimensions of the trackers, and hence the validity
  region of the give criteria.  The criteria which must be kept are
  listed in the first column, where the definitions of the variables
  are given in columns 2 and 3.  }
\label{t:sol:field:req_tr}
\end{table*}

Three tracking systems will be used to reconstruct charged particle tracks
in \PANDA. (Please also refer to Fig.~\ref{f:over:ts_view}.)  The
inner tracker will be an array of Micro Vertex Detectors (MVD) surrounding
the beam pipe closely and extending up to 12\,cm radially.  The
central tracker will either be a Straw Tube Tracker (STT) or
Time Projection Chamber (TPC).  Any of the two tracking devices will 
occupy a length of 1.5\,m
extending to a radius of 42\,cm. In addition, three layers of GEM
detectors will be used to track particles exiting at small angles.  In
the region occupied by the MVD and the central tracker there are very
stringent requirements on the magnetic field homogeneity.  The
absolute magnitude of the field must not vary by more than 2\% from
the nominal field of 2\,T over the whole tracker region.
Furthermore, the radial component of the magnetic field $B_r$ must
remain as low that any integral along $z$ to the central tracker
read-out plane, located at $z = -400\,$mm, results in a value below
2\,mm if started from any given point inside the central tracker.
This ensures that charges which are produced inside the TPC at any $z=
z_0$ do not experience too large an offset to be assigned to the right
track.  These requirements on the magnetic field are summarised in
Table~\ref{t:sol:field:req_tr}.

\paragraph{Field Limits}
\label{s:req:stray-field}

As much as fields are required in certain areas, some components, in
particular certain read-out electronics and vacuum pumps, restrict the
maximum tolerable field strength.  Often this is directional, {\it
e.g.} a photo-multiplier tube is able to withstand different maximum
fields along its axis than perpendicular to it.  This includes fields
inside and outside the flux return yoke.  Clearly, the former fields
will be generally strong while the latter will be shielded largely by the
return yoke.  Stray fields become important at the following places.

\begin{enumerate}
\item At the location of the readout of both DIRC detectors inside
      the flux return. \label{i:req:stray:DIRC}
\item At the location of the turbo-molecular pumps for the target
      generator and dump. The latter ones will be further from the
      solenoid centre and hence experience smaller
      fields.\label{i:req:stray:target}
\item The pumping stations for the beam line before and behind the
      solenoid are denoted by ``s1'' and ``s2'' in
      Table~\ref{t:req:stray-field},
      respectively.\label{i:req:stray:pumps}
\end{enumerate}

The maximum allowable fields and the locations in the \PANDA
coordinate system where those become applicable are listed in
Table~\ref{t:req:stray-field}.

\begin{table}[ht]
 \begin{center}
 \begin{tabular}{|l|c|c|c|c|}
  \hline
  Detector  & Item & Radius   & $z$      & $B_\mathrm{max}$ \\
            &      & $[$m$]$ & $[$m$]$ & $[$mT$]$ \\
  \hline
  \hline
  Disc DIRC   & \ref{i:req:stray:DIRC}   & 1.4 &  2   & 1200 \\
  Barrel DIRC & \ref{i:req:stray:DIRC}   & 0.7 & -1.6 & 1200 \\
  Target gen  & \ref{i:req:stray:target} & 2   & 0    & 5 \\
  Target dump & \ref{i:req:stray:target} & 2.3 & 0    & 5 \\
  Pump s1     & \ref{i:req:stray:pumps}  & 0   & -2.3 & 5 \\
  Pump s2     & \ref{i:req:stray:pumps}  & 1   & 3    & 5 \\
  \hline
  \end{tabular}

 \caption[Maximum tolerable magnetic fields for sensitive
    detector components or pumps.]
    {Summary of maximum tolerable magnetic fields for sensitive
    detector components or pumps.  The item number from the list in
    Sec.~\ref{s:req:stray-field} is given.  The radial and
    longitudinal extent $z$ from the interaction point are given as
    rough indication for their locations.  Most of the components
    actually cover extended non-trivial volumes. (See also
    Sec.~\ref{s:sol:int})}

 \label{t:req:stray-field}
 \end{center}
\end{table}

\subsubsection{Engineering Requirements}
\label{s:sol:req:mech}

The overall dimensions are driven by the space planned for the central
detectors of \PANDA and the field requirements discussed above.
Radially a free diameter of 1.9\,m is
required for the trackers, time-of-flight counters, Barrel DIRC, and
Electro-Magnetic Calorimeter (EMC) surrounding the target pipe.
Downstream of the interaction point, {\it i.e.\ } in positive $z$
direction, the length of 2485\,mm is required to allow for good
tracking capabilities, the accommodation of a Disc DIRC, the end cap
of the EMC and a double layer of muon counters.  This takes into
account the space requirements of the coil cryostat and DIRC detector
and is the result of an iterative optimisation process.  In upstream
direction, {\it i.e.\ }in negative $z$ direction, 1585\,mm are
required to allow for the installation of the read-out of the Barrel
DIRC. The central dimensions of the magnet are driven by the need to
provide a good central field, to shield outside installations from fringe
fields, and to form a range system for the detection of
muons in most of the angular range.  Simultaneously, the target and
beam pipe require pumping as close to the interaction point as
possible.  The fall-back solution for placing the Barrel DIRC readout
outside the yoke has been taken into consideration.  Thus a suitable
compromise taking into account performance and cost has been derived,
which is shown in the following.  Table~\ref{t:sol:sum-dimen}
summarises the overall boundaries of the solenoid.

\begin{table}[ht]
 \begin{center}
 \begin{tabular}{|c|c|c|c|c|}
  \hline
  Bounds & radius & axial length & $z_\mathrm{min}$ & $z_\mathrm{max}$\tabularnewline
  of & / mm & / mm & / mm & / mm\tabularnewline
  \hline
  \hline
  $C_i$  & 950 & 3090 & -1190 & 1900\tabularnewline
  \hline
  $C_o$ & 1340 & 3090 & -1190 & 1900\tabularnewline
  \hline
  $Y_i$ & 1490 & 4070 & -1585 & 2485\tabularnewline
  \hline
  $Y_o$ & 2300 & 4875 & -1970 & 2905\tabularnewline
  \hline
  \end{tabular}

 \caption[Summary of magnet and cryostat maximal dimensions.]
 {Summary of magnet and cryostat maximal dimensions hereafter
 known as the Magnet Volumes.  Given are the radially inner and central
 bounds of the cryostat and yoke, denoted by $C_i, C_o, Y_i$ and
 $Y_o$, respectively.  All major components of the magnet will be
 accommodated within the given boundaries, whereas the surrounding
 volumes are free for detector installations with the exception of
 where mounting rods and supply lines for the magnet are placed.
 These are specified in detailed drawings in the following sections.
 Within these volumes space is foreseen for the placement of muon
 detectors and the routing of cables and supplies.}
 \label{t:sol:sum-dimen}
 \end{center}
\end{table}

In order to allow both installation but also routine maintenance of
the detectors the solenoid has to be removed from the \HESR ring, such
that the machine development with all the remaining components
of \HESR can continue.  That means that the whole solenoid with all
detectors placed inside needs to be moved by more than 10\,m and the
procedure should not take more than one week in total.  It should be
guaranteed that the whole magnet can be aligned at the operation
position with respect to some reference points to a precision of about
1\,mm.  At the same time the beam axis with respect to the floor level
is fixed at 3.5\,m.  The available space below the iron yoke is
planned to be about 1.2\,m but it will be permissible to introduce grooves in
the \Panda Hall floor of less than 0.5\,m depth, to place rails.
Another requirement concerns the access to the target dump.  Here an
access route with a free headroom of more than 85\,cm and at least
1.5\,m width is required.  The movement should aim at minimising
vibrations and deformations in order to protect sensitive equipment.

\subsubsection{Yoke \& Cryostat Interaction}

The solenoid yoke must support the weight of the cryostat and the
attached detectors within.  The unbalanced axial forces generated by
the energised coils should be transmitted to the yoke. This unbalanced
axial force should be minimised as far as practicable and should not
exceed 20\,tonnes.  All cryogenic equipment mounted on top of the
cryostat should be located upstream of the target to maximise
available space for detectors and wiring in the forward region. The
cryostat chimney will be located as far upstream as possible in order
to simplify cryostat insertion into the yoke.

\begin{table*}[ht]
\begin{center}
\begin{tabular}{|l|c|c|c|c|c|}
\hline
Detector      & Supported     & Magnet          & \multicolumn{3}{|c|}{Mass} \\
              & via           & attachment      & Det.   & Cable & Total     \\
              &               & point           & [kg]   & [kg]  & [kg]      \\
\hline
\hline
MVD             & Central Tracker & Cryostat    & 15     & 100  & 115 \\

Central Tracker & direct      & Cryostat        & 45     & 100         & 145 \\

Barrel TOF      & direct      & Cryostat        & 380    &  50  & 430\\

Barrel DIRC slabs & direct    & Cryostat        & 360    & --          & 360\\

Barrel DIRC read out  & direct & Yoke (upstr.)  & 800    & 200  & 1000 \\

Barrel EMC      & direct      & Cryostat        & 20,000 & 1500 & 21,500\\

Barrel Muon   & direct        & Yoke (uniform)  & 4,000  & 200  & 4,200\\

Forward GEMs  & direct        & Cryostat        & 110    &  40         & 150 \\

Forward DIRC  & Forward EMC   & Yoke (downstr.) & 1,000  & 200  & 1,200\\

Forward EMC   & direct        & Yoke (downstr.) & 5,190  &  810        & 6,000\\

Forward Muon   & direct       & Downstr.\ door  & 1,000  &  50  & 1,050 \\

Backward EMC plug & direct    & Yoke (upstr.) & 1,000 &  500   & 1,500 \\
\hline
\hline
\multirow{2}{2cm}{Totals} &
      \multicolumn{4}{|l|}{Supported by cryostat}            & 22,700 \\
    & \multicolumn{4}{|l|}{Supported by upstream yoke}       & 2,500 \\
    & \multicolumn{4}{|l|}{Supported by downstream yoke}     & 7,200 \\
    & \multicolumn{4}{|l|}{Supported uniformly in yoke or door}  & 5,250 \\
\hline
\end{tabular}
\caption[Detectors located inside the
  solenoid, their total masses and main support point on the solenoid.]
  {\label{t:over:det-masses}Detectors located inside the
  solenoid, their total masses and main support point on the solenoid.
  The detectors are attached either to both ends of the cryostat
  simultaneously or the upstream or downstream ends of the yoke.  It
  should be noted that, where supplied with a possible mass range,
  only the upper limits are reported.}
\end{center}
\end{table*}

\subsubsection{Assembly and General Detector Access}

The solenoid should be mounted on a movable rail-guided carriage to
be transported from the assembly area to its operational position.
The downstream end cap of the solenoid should open up. The two semi-segments
should slide apart on skids. The upstream end cap of the solenoid
should also open in order to allow access for detector installation,
wiring and maintenance and to facilitate cryostat installation.

\subsubsection{Target Integration}

A warm bore of 100\,mm in diameter should be foreseen between the two
parts of the magnet coils, and a hole of 350\,mm through the barrel
yoke is required to accommodate the internal target system.  A
rectangular recess is required from a radial distance of 2\,m to the
interaction point with dimensions of 1\,m $\times$ 1.2\,m in $x$ and
$z$, respectively.  The recess may have rounded corners.  Furthermore,
in the remaining part of the yoke the hole needs to be opened further
than 350\,mm to accommodate the pumping cross.  In the region of the
Turbo Molecular Pumps, the magnetic field must not exceed 5\,mT. The
closest pumps will be installed in the generation part 2\,m vertically
from the interaction point.  The fore pumps will be located such that
they neither interfere with the yoke iron nor with the foreseen
support construction.

The target generation system will be accessible during normal
maintenance to allow to switch between pellet and cluster jet
target. Only a partial modification of the target dump will be
required when the target system will be changed: anyhow, to make the
first installation of the target dump easier, a minimum clearance of
$85\unit{cm}$ between the floor and the support structure of the yoke
barrel is foreseen.  This clearance is foreseen over the whole width
(in $x$ direction) of the support structure, such that any equipment
can be brought in from this direction.  The magnet weight is
transferred to the floor with two symmetric structures at the end of
the barrel and under the end doors.

\subsubsection{Accommodation of Muon Detectors}

In order to allow the detection of muons, to facilitate muon and pion
separation, both the forward doors and the main barrel of the solenoid
yoke will be laminated.  Simulations have shown the need for a range
system of staggered muon detectors and iron covering forward angles up
to at least $70^\circ$ azimuthal angle.  Such a system becomes more
effective the more layers are provided.  In an iterative process taking
into account space and cost considerations the following optimum
solution was found to provide a good overall muon reconstruction
efficiency.  In the barrel part in total 13 layers of muon detectors
will be placed, while 5 layers will be accommodated in the downstream
doors.  The innermost layer will be a double layer allowing
to reconstruct two coordinates, while the intermediate layers will
consist of single layers.  Each layer of muon detectors requires a
free space of 25\,mm.

To allow for the fact that the iron layers have a typical flatness
tolerance of the order of 3\,mm per metre, thus a 30\,mm gap is
foreseen for the muon detectors between the layers of iron.  Between
the cryostat and flux return a minimal space of 10\,cm will be left such
that in addition to the double layer of muon counters further space
for supplies, tolerances and alignment are available.  In the forward
region the 6 layers of detectors will be augmented by a range system
installed between the solenoid and dipole magnets.  The chosen
solution is detailed in the sections concerning the design of the flux
return yoke.

\subsubsection{Detector Support}

There are three distinct areas where mountings for the detector
support will be attached.
\begin{itemize}
\item At the upstream end of the yoke barrel, support structures
      will be mounted.  These hold cables and supply lines, and serve
      as mounting points for the Barrel DIRC and the upstream side of
      a frame to support the inner detectors and the beam pipe.
\item At the downstream end of the yoke barrel similar support
      structures will be installed to hold the forward end-cap
      detectors and their housings and supplies.  They will also hold
      the cable and supply lines from parts of the EMC.
\item Special ribs on the cryostat will serve as mounting points
      for the Electromagnetic Calorimeter (EMC).  The
      downstream side of the frame holding the inner detectors will be
      mounted such that part of the load is effectively resting on
      the downstream rib of the cryostat.
\end{itemize}

The Electromagnetic Calorimeter (EMC) will be attached, as shown in
Figure \ref{fig:EMC-Yoke-Attachment}, to the inner surface of the
cryostat shell via special ribs to which the support structure of the
EMC will be attached by screws. The EMC and all of the other detectors
inside the cryostat must be supported by attachment to the solenoid
yoke via the cryostat. Attachment points for all other detectors will
come from the eight corners of the octagonal yoke. The currently known
detector weights and attachment points are shown in
Table \ref{t:over:det-masses}.

\subsubsection{Cable Routing}

Signal cables and supply lines from all detectors except the MVD and
central tracker will require to be fed to the outside of the yoke
through cut-outs at the upstream and downstream ends of the barrel.
The inner and central trackers will route their supplies trough the
upstream opening of the door and hence do not affect the following
considerations.  Apart from an integral cross section for cables some
supply line require a minimum clearance.  This is the case for the
cooling lines of the EMC which, including their insulation, have a
diameter of 12\,cm.  In Table~\ref{t:req:routing} all the requirements
concerning the space for cable and supply routing are listed.  These
space should preferably be distributed along the circumference of the
flux return yoke.

\begin{table}[ht]
\begin{center}
\begin{tabular}{|l|c|}
\hline
Detector & Cross section \\
& $[$mm$^2]$ \\
\hline
\hline
\multicolumn{2}{|c|}{Downstream routing}\\
\hline
GEM stations & 40,000    \\
Forward DIRC & 40,000    \\
Forward EMC  & 160,000   \\
Muon counters & 160,000       \\
\hline
Total        & 400,000   \\
\hline
\hline
\multicolumn{2}{|c|}{Upstream routing}\\
\hline
Barrel EMC   & 203,800   \\
Barrel DIRC  & 20,000    \\
Muon counters & 200,000       \\
\hline
Total        & 423,800   \\
\hline
\end{tabular}
\caption[Minimum required space for the routing of cables and supply
  lines.]{Minimum required space for the routing of cables and supply
  lines at the upstream and downstream ends of the flux return yoke.
  Cross sections are given by the area of packed cables and
  supplies. A minimum dimension of 120\,mm is given by the diameter of
  the cooling lines of the EMC.  This limits the geometrical design of the
  cut-outs.}
\label{t:req:routing}
\end{center}
\end{table}

\subsection{Forward Spectrometer}
\label{s:req:fs}

\AUTHOR{I.~Lehmann}
 
The Forward Spectrometer at \PANDA is required to cover particles
emitted with angles below 5 and 10 degrees in vertical and horizontal
direction, respectively.  A field integral of 2\,Tm is necessary in
order to achieve the required resolution of 0.5 to 1\% in $\Delta p/p$
for protons, pions and kaons with momenta up to 12\,GeV/c.  This
resolution is essential to identify and study several benchmark
channels, e.g.\ to study conventional and exotic charmonium states
decaying into $D \bar{D}$ and $\Lambda \bar{\Lambda}$ production.  The
dipole magnet of the Forward Spectrometer needs to have a large
aperture as it is located at 3.5\,m downstream of the interaction
point.  At the same time the magnet needs to reach the field integral
over a length of less than 2.5\,m.

\begin{table}[ht]
\begin{center}
\begin{tabular}{|l|c|}
\hline
Description & Value \\
\hline
\hline
Field integral      & 2\,Tm \\
Bending variation   & $\leq \pm 15$\%  \\
Vert.\ Acceptance   & $\pm 5^\circ$ \\
Horiz.\ Acceptance  & $\pm 10^\circ$ \\
Ramp speed          & 1.25\%/s \\
Total length in $z$ & $\leq 2.5$\,m \\
\hline
\end{tabular}
\caption[Main requirements for the Forward Spectrometer.]{Main 
  requirements for the Forward Spectrometer.  The field integral
  maximum variation of the bending angle is valid for charged
  particles with momenta of no lower than one fifth of the beam
  momentum originating at the target within the given acceptance.}
\label{t:req:fs}
\end{center}
\end{table}

The dipole magnet will form part of the accelerator lattice and, hence,
will need to be ramped synchronously with the ring magnets.  To avoid orbit
changes and beam losses the synchronisation of the field needs to be
accurate during the whole ramping procedure.  The maximum ramp speed
will be at 1.25\% change of current per second relative to the maximum
current.  The ramp will generally not be started from zero current but
rather at 25\% of the maximum current.  Both ramp-up to full current
and ramp-down to 10\% current will be required as standard procedure.
The main requirements are summarised in Table~\ref{t:req:fs}.

\bibliographystyle{tdr_lit}
\bibliography{lit}

\cleardoublepage
\svnInfo $Id: solenoid.tex 692 2009-01-30 10:43:35Z IntiL $ 

\chapter{Target Spectrometer}
\label{s:sol} 

\AUTHORS{A.~Bersani, Y.~Lobanov, I.~Lehmann}

The Target Spectrometer will form the central part of the \PANDA detector.
The solenoid is designed such that it will leave a warm bore of 1.9\,m
diameter around the interaction point with more than 4\,m free length.
The asymmetric location of the interaction point with respect to the
magnet's centre will guarantee optimal track reconstruction capabilities
in the solenoidal field.  As the target requires vertical feed pipes
the superconducting coil of the solenoid will be split at this point and
the cryostat will exhibit a warm bore of 100\,mm diameter.  To balance the
forces and guarantee the required field homogeneity the coil will,
hence, be split in 3 interconnected parts.  The flux return yoke is
designed to simultaneously act as a range system for the muon
detection for all angles below $70^\circ$.

A particular challenge for the design of the magnet has been the
asymmetric design of the detection systems inside the magnet.  The
interplay of the coil arrangement and flux return yoke geometry has
been optimised in a detailed iterative process taking into account
various limitations due to detectors and supply lines.  The most
stringent constraints were imposed by the DIRC detectors, as their
read out systems feature bulky optical systems.  These considerations
led to an asymmetric design of the yoke with respect to the coil.  It
has been a challenge to design the coil, cryostat and flux return yoke
such that all these requirements were met without compromising on the
field homogeneity in the region of the central tracker.

An overview of the design of the magnet is presented in the following
section, describing the main topics guiding the conceptual design.  In
the following Secs.~\ref{s:coil} and \ref{s:yoke} the designs of the
coil and cryostat and the instrumented flux return are detailed,
respectively.  This is followed by a section on the integration of the
targets and detectors into the spectrometer.  The chapter concludes
with details on the performance of the presented design.

\svnInfo $Id: concept.tex 694 2009-01-30 11:32:42Z IntiL $

\section{Conceptual Design}
\label{s:conc}

\AUTHORS{I.~Lehmann, A.~Bersani, Y.~Lobanov}

\subsection{General Concepts}

\begin{figure*}[ht]
 \begin{center}
  \includegraphics[width=0.9\dwidth]{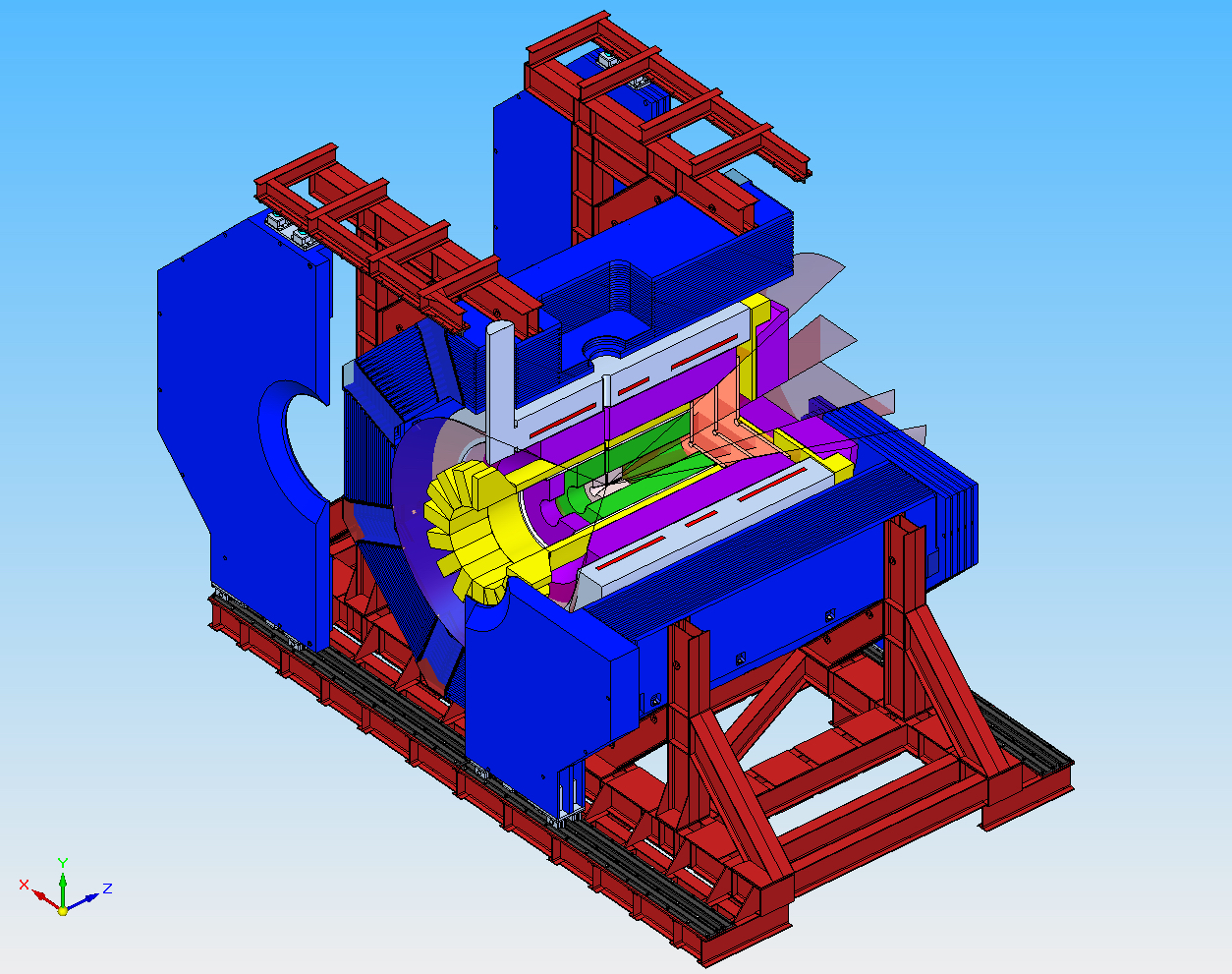}

  \caption[View of the Target Spectrometer]{View of the Target Spectrometer
     with the away doors open.
     For illustration one quarter is cut on this view revealing the
     cryostat, coil and inner detectors.  The antiproton beam would
     enter from the left along positive $z$.  The cones originating at
     the interaction point indicate the acceptance regions of the
     different detection systems.}
\label{yoke:overview}
\end{center}
\end{figure*}

The Target Spectrometer magnet has been designed taking into account
the requirements given in Sec.~\ref{s:sol:req}.  A superconducting
solenoid with external iron return yoke has been chosen, which allows
to achieve a longitudinal field of 2\,T and keep enough space for the
detectors surrounding the interaction point.  All barrel detectors, except
the muon chambers, will be hosted inside the warm bore of the solenoid
cryostat.  The laminated flux return yoke will operate as a range system
for the detection of muons and their discrimination against pions.  An
overview of the solenoid with all the detectors installed can be found
in Fig.~\ref{yoke:overview}.

The design was optimised in an iterative procedure taking into account
all the detector specific issues, while simultaneously optimising the
magnet design in functionality, reliability and cost effectiveness.
The design has undergone a dedicated review by independent
international experts.  This detailed procedure and advice by
companies building in the past similar magnets led to the present design.  The
main features are listed in Table~\ref{t:guidelines}.

\begin{table}[ht]
\begin{center}
\begin{tabular}{|l|c|}
\hline
Parameter & Value\\
\hline\hline
SC Material & NbTi\\
Cable Type & Rutherford\\
Stabilisation & Pure Al\\
Operating Current & $\le5000\unit{A}$\\
Current Density & $<60\unit{A/mm^2}$\\
Winding Layers & 2\\
Residual Axial Force & $\le30\unit{t}$\\
Field in Iron & $\le2\unit{T}$\\
\hline
\end{tabular}
\caption{Key features of the \PANDA solenoid.}
\label{t:guidelines}
\end{center}
\end{table}

Though the magnet design is elaborated from the requirements at \PANDA
it has many similarities with existing and operating magnets at other
research institutions.  A brief compilation of a few magnets is given in
Table~\ref{t:sol:comp}.  This shows that it is technically possible to
build such a magnet.  Moreover, the
experience gained in those projects helped in designing the \PANDA
solenoid.  Though these solenoids are quite similar, they all lack the
special feature of the \PANDA solenoid that the vertical target pipe
is transversing the cold mass of the cryostat at about 2/3 of its
length.

\begin{table*}[ht]
\begin{center}
\begin{tabular}{|l|c c c c|c|}
\hline
                  & ZEUS & ALEPH & BABAR & BESIII & \PANDA\\
\hline\hline
Location          & DESY & CERN & SLAC & IHEP & FAIR\\
Year completed    & 1988 & 1986 & 1997 & 2008 & 2012 \\
Central field (T) & 1.8  & 1.5  & 1.5  & 1.0  & 2.0\\
Inner bore (m)    & 1.85 & 4.96 & 2.8  & 2.38 & 1.9\\
\hline\hline
\multicolumn{6}{|c|}{Cold mass parameters} \\
\hline
Length (m)        & 2.5  & 7    & 3.46 & 3.52 & 2.7\\
Energy (MJ)       & 12.5 & 137  & 25   & 9.5  & 20\\
Current (A)       & 5000 & 5000 & 4600 & 3250 & 5000\\
Weight (t)        & 2.5  & 60   & 6.5  & 4.0  & 4.5 \\
Cable cross section (mm) & $4.3\times15$ & $3.6\times35$
& $4.9\times20$ & $3.7 \times 20$ & $3.4\times25$\\
Current density ($\unit{\frac{A}{mm^2}}$) & 78 & 40 & 47 & 41 & 59\\
\hline\hline
\multicolumn{6}{|c|}{Yoke parameters} \\
\hline
Length (m)        & 15   & 10.6 & 6.0  & 5.8  & 4.9 \\
Outer radius (m)  & 4.20 & 4.68 & 2.92 & $\sim 3$ & 2.30 \\
Iron layers       & 10   & 23   & 20   & 9    & 13   \\
Total weight (t)  & 1962 & 2580 & 580  &      & 300 \\
\hline
\end{tabular}
\caption{Comparison of key parameters of some solenoid magnets
   and the proposed \PANDA solenoid.}
\label{t:sol:comp}
\end{center}
\end{table*}

\begin{figure*}[ht]
\begin{center}
\includegraphics[width=0.9\dwidth]{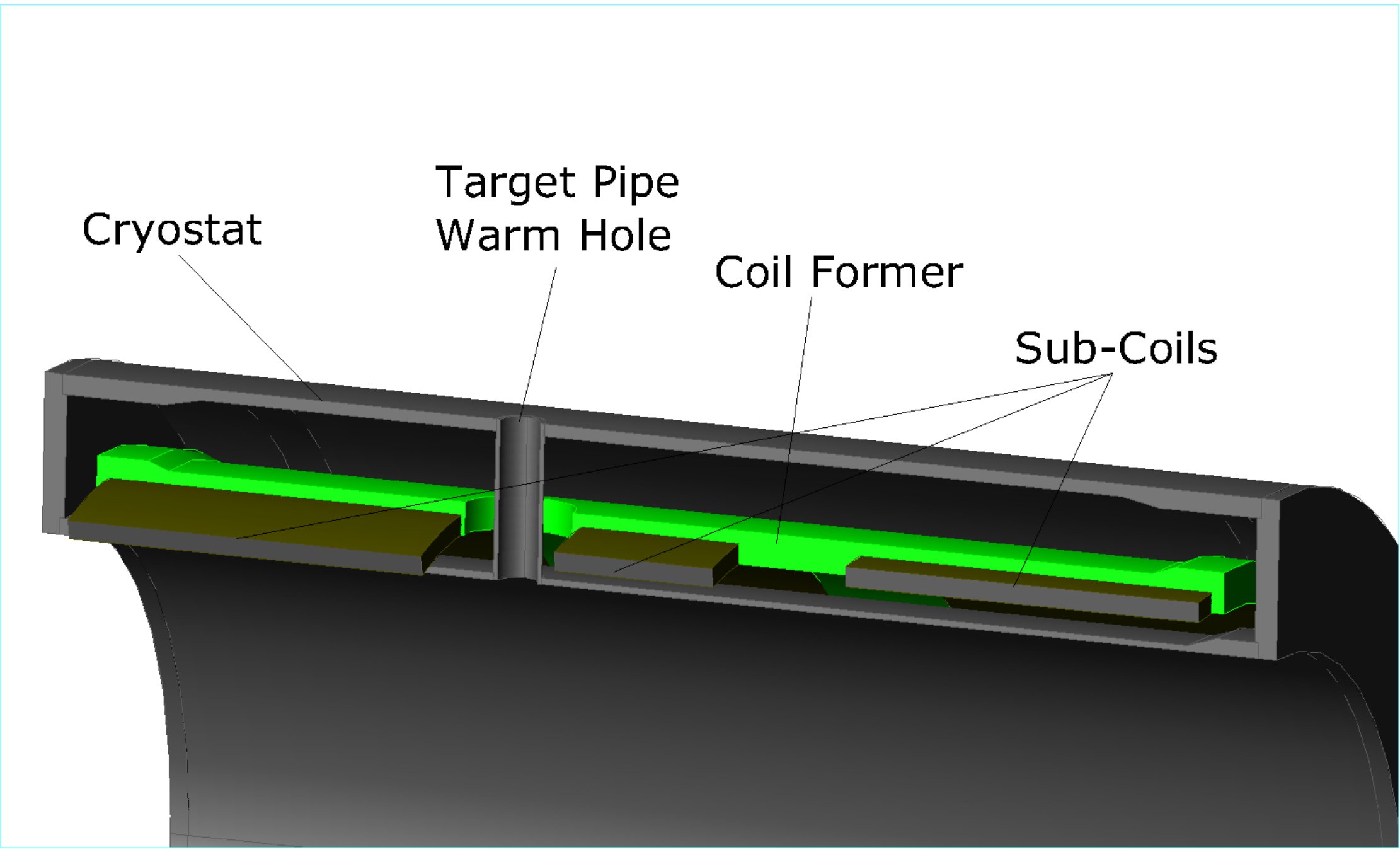}
\caption[Schematic view of part of the coil.]{Schematic view of part of 
the coil.  The cryostat, coil former
  and 3 sub coils are shown, while the thermal screens and cooling
  pipes are omitted.}
\label{f:coil}
\end{center}
\end{figure*}

\subsection{Cable \& Coil Design}

The basic feature of the design is the use of a Rutherford-type cable
made of NbTi superconductor encased in an aluminum stabiliser that
allows for adequate quench protection. A schematic cross section of
the coil and cryostat section is shown in Fig.~\ref{f:coil}.  A
aluminium-stabilised conductor is used which is wound to the inside of
the aluminum alloy coil former.  The energy to cold mass ratio is
$\sim5.2\unit{kJ/kg}$: this value is similar to many previously built
magnets, such as TOPAZ~\cite{Yamamoto:1986zd}, CDF~\cite{Minemura:1985yv},
ALEPH~\cite{Baze:1987ct}, ZEUS~\cite{Acerbi:1985yd}, H1~\cite{:1986ge},
DELPHI~\cite{:1983ac}.  Since we do not
require high transparency for particles, there has been no need to
push this ration to even larger values, as it was done for some other
magnets (e.g.~\cite{Yamamoto:2008zze}).  This allows for a design which 
can safely cope with a
residual axial force of 30\,t.  The coil former cross section is optimised
to minimise the stresses on the cable, while allowing
the pre-stress of the coil with the coil former end flanges.  Cooling
will be achieved by liquid helium which runs in pipes welded on the outside
of the coil former.


Our concept has benefited from experience gained over the past 25
years with thin superconducting solenoids.  The technique was first
developed for CELLO~\cite{coil:desportes}, the first thin solenoid,
and has been improved in subsequent designs.  Although specifically
tailored to meet the requirements of \PANDA, this design is similar to
many operating detector magnets.  Table~\ref{t:sol:comp} shows the
main features of some of these solenoids compared to the \PANDA
design.

\begin{figure*}[ht]
\includegraphics[width=\dwidth]{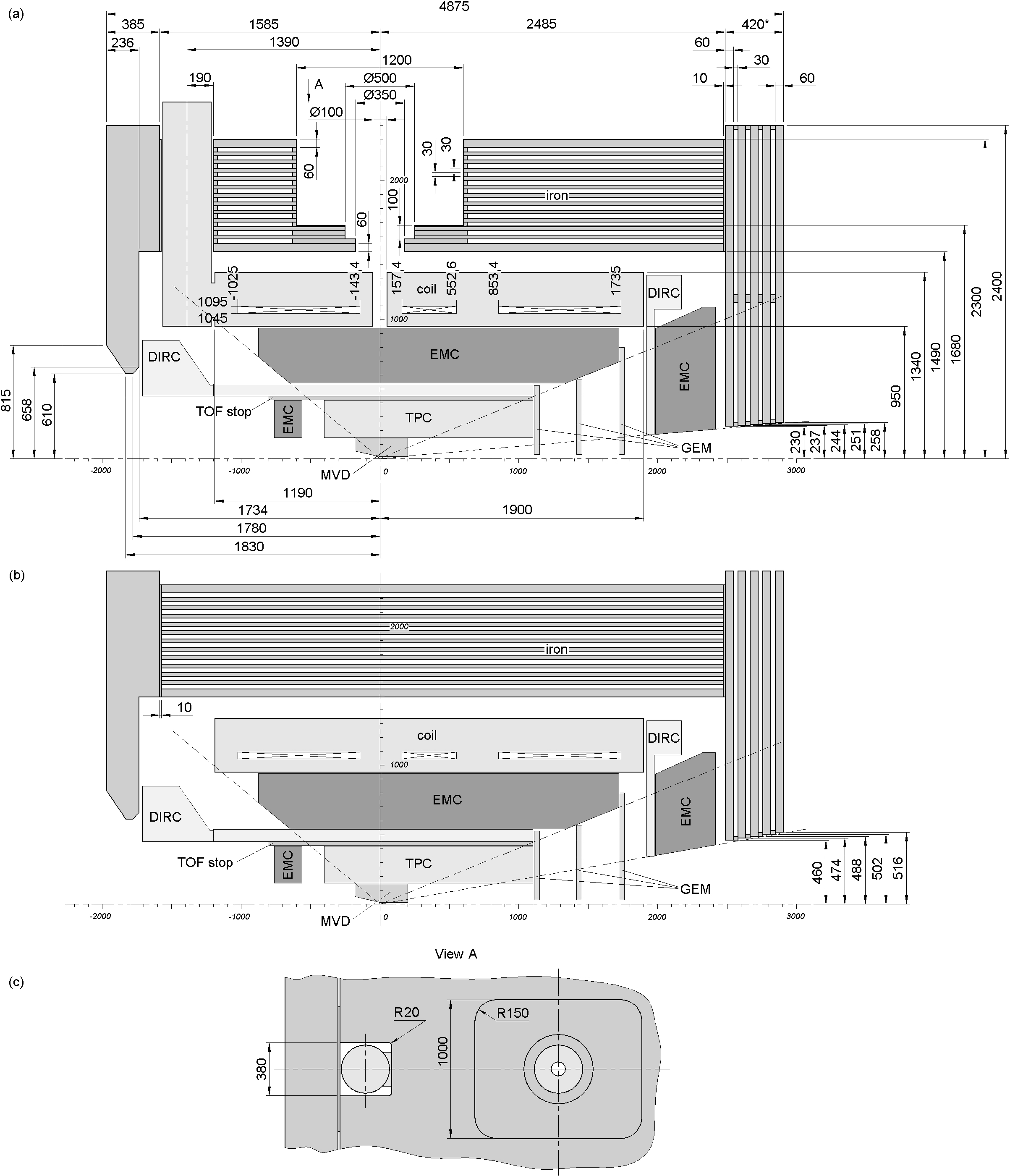}
\caption[Cross sections of the Target Spectrometer showing the
yoke layout]{Cross sections of the Target Spectrometer showing the
yoke layout: a) upper half cross section in the $z-y$ plane
showing the recess for the target generator at the top, b) half
cross section in the $z-x$ plane c) top view of the upstream top
surface showing the cut out for the cryogenic chimney and the
target recess.  The detectors are indicated schematically for
illustration only.} \label{yoke:x-sec}
\end{figure*}

\subsection{Flux Return Yoke Design}

\subsubsection{Lamination}

A cross section of the planned solenoid is shown in Fig.~\ref{yoke:x-sec}.
According to the muon range system requirements, the barrel part will
consist of 13 steel plates interleaved by 30\,mm gaps
for allocation of muon detector planes and signal amplifiers. The
thickness of the inner and outer steel plates will be 60\,mm and of all
intermediate plates 30\,mm. The downstream end cap will consist of 5
steel plates of 60\,mm thickness with 30\,mm gaps between plates
for accommodation of muon detector panels. The upstream end cap will
not be laminated. In order to ensure the unhampered insertion of muon
counters, special templates will be used to control the gap
thickness between the steel plates during the yoke manufacturing.

\subsubsection{End Caps and Passages}

\begin{figure}[ht]
\begin{center}
\includegraphics[width=\swidth]{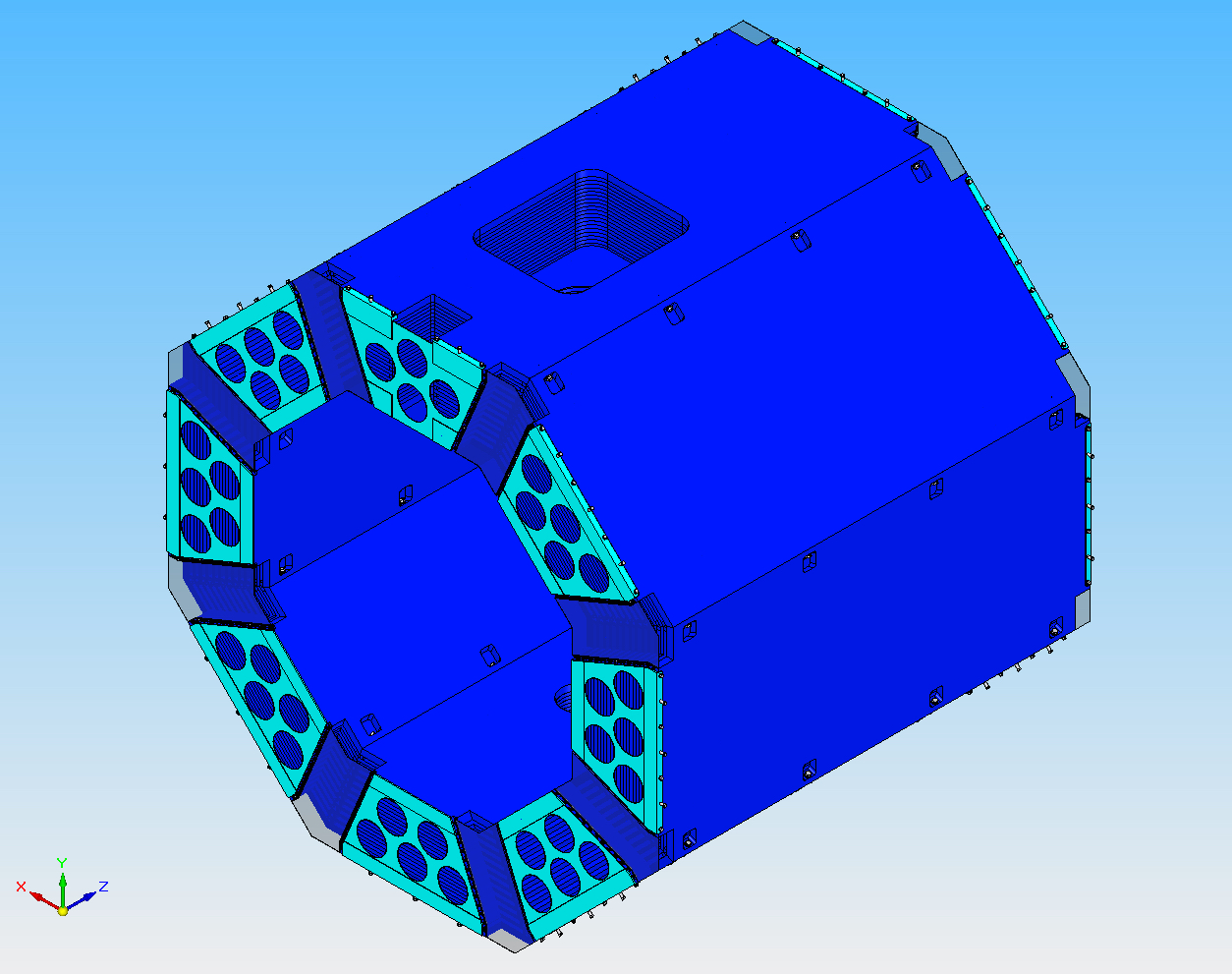}
\caption[Barrel part of the yoke seen from the upstream
end]{Barrel part of the yoke seen from the upstream end.  On top
 the target recess is visible and on the front the face spacers are
 shown in light blue.}
\label{yoke:Barrel}
\end{center}
\end{figure}

The passages for cables and tubes across the yoke are foreseen at the
upstream and downstream ends of the yoke barrel. The radial passages
and steel spacers between the end caps and the barrel are shown in
Fig.~\ref{yoke:Barrel}. These spacers are intended to compensate
possible gaps between some barrel beams and the end caps. The spacers
are designed such that they can be taken out before moving the doors
to ensure an undisturbed opening of the doors.  Additionally a 1
degree deviation of the door sliding direction from the line
perpendicular to the beam axis is foreseen.  The gaps for cables will
be closed by lids to protect their contents from damage when the doors
are moving.  The upper beam of the yoke barrel will contain recesses for
the target and for the cryostat chimney.  The first recess hosts
turbo-molecular pumps which are sensitive to magnetic fields, and the
latter contains the cooling pipes, current leads and wiring for
gauges.  The lower barrel beam will have a smaller recess for the target
dump.

\subsubsection{Support Frames}

The barrel of the iron yoke is to be surrounded by two outer support
frames. Their positions and dimensions are shown in
Sec.~\ref{s:yoke}. The frames will ease the mounting of the 8 segments of
the yoke and ensure the overall rigidity of the yoke, such that both
mechanical and magnetic loads result in deformations of the yoke by
only a few millimetres (see Sec.~\ref{s:yoke:deform}).
Simultaneously, these frames will take the load of the whole Target
Spectrometer and transfer it to the 4 wheel carriages or 6 fixed
support points which suspend the movable system on two rails. The
doors which can be opened independently will be attached and supported by
the same system.  An overview is shown in Fig.~\ref{yoke:overview}
while the details are discussed in Sec.~\ref{s:yoke}.


\subsection{Model for Magnetic Analysis}

\begin{figure}[ht]
\begin{center}
\includegraphics[width=\swidth]{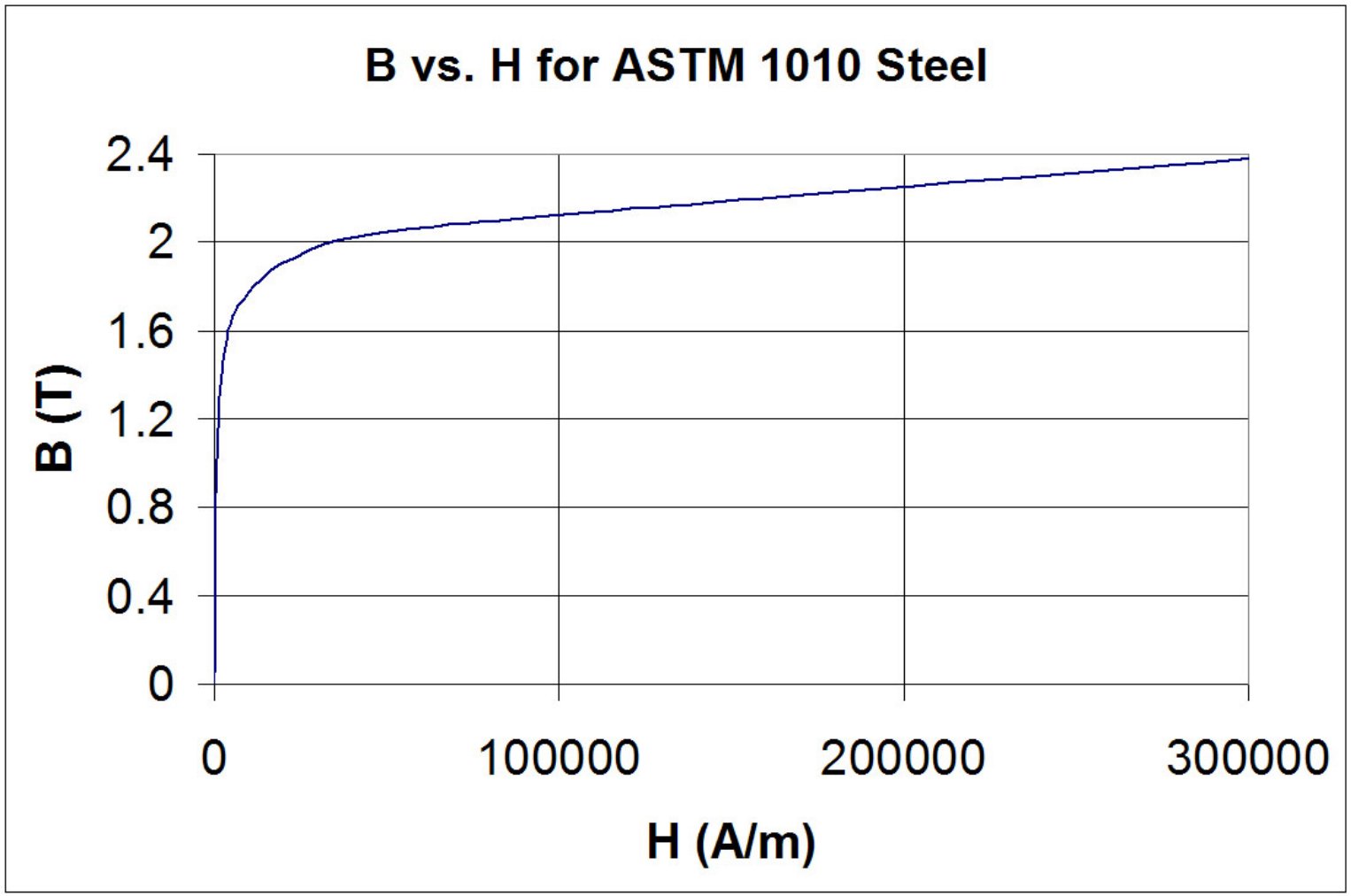}
\caption{Magnetic induction $B$ {\it versus} field $H$ for AISI 1010
low carbon steel used for the magnetic calculations.}
\label{BH}
\end{center}
\end{figure}

During the project optimisation, several different Finite Element
Model (FEM) codes were used: Ansoft Maxwell~\cite{coil:maxwell},
ANSYS~\cite{coil:ansys} and TOSCA~\cite{mag:TOSCA_2003}.  A comparison
of their results gave us an estimate of the accuracy of the
calculations.  The iron properties used for computation are those of
standard AISI 1010 low carbon steel.  We included the material's
non-linear behaviour when saturating.  The dependence used for the
magnetic induction $B$ on the field $H$ is shown in Fig.~\ref{BH}.

\begin{figure}
\begin{center}
\includegraphics[width=\swidth]{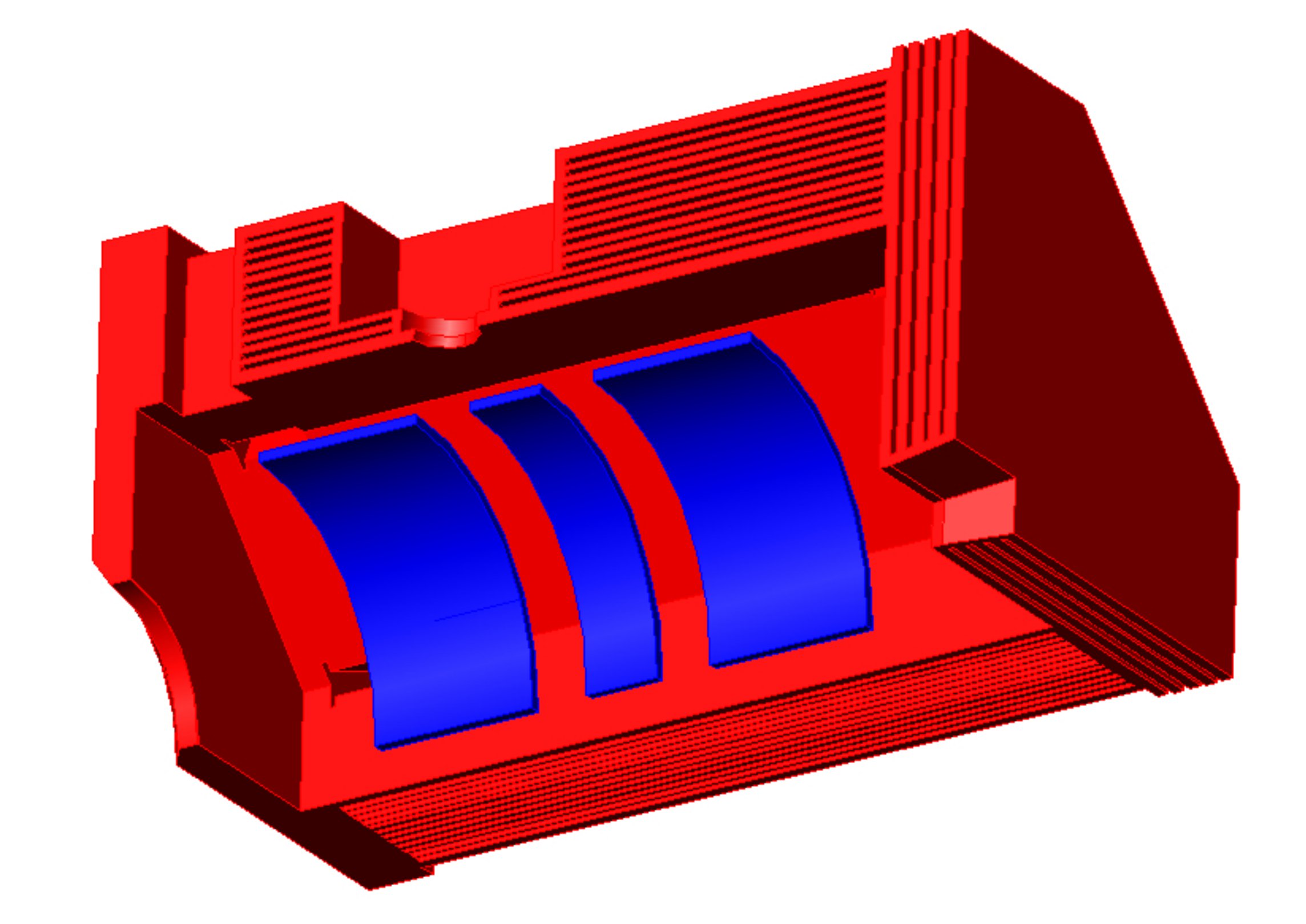}
\caption{Partial model used for the 3D FEM calculations.}
\label{quarto}
\end{center}
\end{figure}

The magnetic analysis is based on a two--dimensional axially symmetric
model and on a complete three--dimensional model. These models include
the solenoid, flux return yoke and the doors.  The three--dimensional
calculations were performed with an accurate 3D model of a quarter of
the system and adequate symmetry conditions (see Fig.~\ref{quarto}).

All the different calculations gave results which showed deviations
smaller than $0.1\%$ in the tracker region.  This indicates that we,
indeed, can trust the results to such a level of accuracy.

\subsection{Summary}

Following the guidelines from the detector requirements, as laid out in
Sec.~\ref{s:over:req}, a detailed conceptual design was developed for the
\PANDA solenoid.  The main parameters which were achieved during the
iterative optimisation procedure are listed in the following.
\begin{itemize}
\item{Induction field in outer tracker region: $2\unit{T}\pm1.6\%$.}
\item{Radial field integral
  $\int{B_r(r,z)/B_z(r,z)\, dz}$ in
  the tracker region: $0 - 2\unit{mm}$.}
\item{Residual axial force on the coil: $20\unit{t}$.}
\item{Design complies with all requirements from the detectors, supplies,
  space and infrastructure restrictions.}
\end{itemize}
The two following sections detail the proposed design of the coil,
cryostat, flux return yoke and support structures.  A detailed
analysis of the performances of the whole solenoid magnet will be
laid out in Sec.~\ref{s:sol:perf}. Sec.~\ref{s:sol:int} describes
the target and detector integration in the system.

\svnInfo $Id: coil.tex 714 2009-03-31 17:34:18Z IntiL $ 

\section{Coil and Cryostat}
\label{s:coil}

\AUTHOR{A.~Bersani}

\subsection{Introduction}

The magnet design provides a magnetic field of $2\unit{T}$ with a
uniformity of $\pm1.6\%$ in the tracking region. This will be obtained
with a single current density (i.e. a single cable cross section) in
three sub-coils connected in series. The coil will be split in three
connected sub-coils to leave a gap at the place where the target pipe
intersects the magnet. The second sub-coil is needed for symmetry
reasons to obtain a good field homogeneity. The central coil will be
$395.2\unit{mm}$ in length with 104 turns. Two end coils will be
$882.6\unit{mm}$ in length with 232 turns. A better field uniformity
may be obtained by reducing the axial length of the two end regions
and increasing the current to generate the same field, but this would
cause a reduction in stability against thermal disturbance.

For the initial design, the maximum allowed current density in the
conductor has been limited to the maximum currently attainable for
magnets of this kind, i.e., $80\unit{A/mm^2}$ (ZEUS magnet). In these
conditions, a cross section of $\sim80\unit{mm^2}$ for the conductor
corresponds to a maximum current of $\sim6400\unit{A}$: our choice of
$5000\unit{A}$ will give a good margin for operations. The main
parameters of the \PANDA~solenoid coil are listed in Table~\ref{para}.

\begin{table}[ht]
\begin{center}
\begin{tabular}{|l|c|}
\hline
Parameter & Value\\
\hline\hline
Central Induction & $2\unit{T}$\\ 
Conductor Peak Field & $2.8\unit{T}$\\
Uniformity in the Tracking Region & $\pm2\%$\\
Winding Length & $2.8\unit{m}$\\
Winding Mean Radius & $1075\unit{mm}$\\
Amp Turns & $5.68\cdot10^6$\\
Operating Current & $5000\unit{A}$\\
Inductance & $1.7\unit{H}$\\
Stored Energy & $21\unit{MJ}$\\
Total Length of Conductor & $8000\unit{m}$\\
\hline
\end{tabular}
\caption{The main parameters of the \PANDA~solenoid winding.}
\label{para}
\end{center}
\end{table}

The coil will be symmetric w.r.t.\ its centre: the resulting magnetic
field will essentially be symmetric inside the cryostat warm bore. The
asymmetries in the magnetic field will arise from the asymmetries of
the instrumented flux return yoke, which cannot be avoided due to the
space requirements from the detectors.
 
\subsection{Coil and Cryostat Design} 

\subsubsection{Aluminium Stabilised Conductor} 
The conductor will be composed of a superconducting Rutherford cable
embedded in a very pure aluminium matrix by a co-extrusion process that
ensures a good bond between aluminium and superconductor.
Table~\ref{scavo} shows the main parameters of the conductor.

\begin{table}[ht]
\begin{center}
\begin{tabular}{|l|c|}
\hline
Parameter & Value\\
\hline\hline
Conductor Type & NbTi\\
& Pure Al--stabilised\\ 
& Co--extruded\\
Aluminum RRR & $>500$\\ 
Conductor Unit Length & $1.6\unit{km}$\\ 
Number of Lengths & $5$\\ 
Bare dimensions & $3.4\times24.6\unit{mm^2}$\\ 
Insulated dimensions&  $3.8\times 25\unit{mm^2}$\\ 
Superconducting Cable & Rutherford type\\
Dimensions & $8\times1.15\unit{mm^2}$\\ 
Strands Diameter & $0.8\unit{mm}$\\ 
Number of Strands & $20$ \\
Cu:Sc & $1.5:1$ \\
Filament Diameter & $20\unit{\mu m}$ \\
$I_c(B=2.5\unit{T},\, T=4.5\unit{K})$ & $> 10 \unit{kA}$ \\
Insulation Type & Fibreglass Tape \\
Insulation Thickness & $0.4 \unit{mm}$ \\
\hline
\end{tabular}
\caption{Conductor parameters.}
\label{scavo}
\end{center}
\end{table}

\begin{figure}
\begin{center}
\includegraphics[width=\swidth]{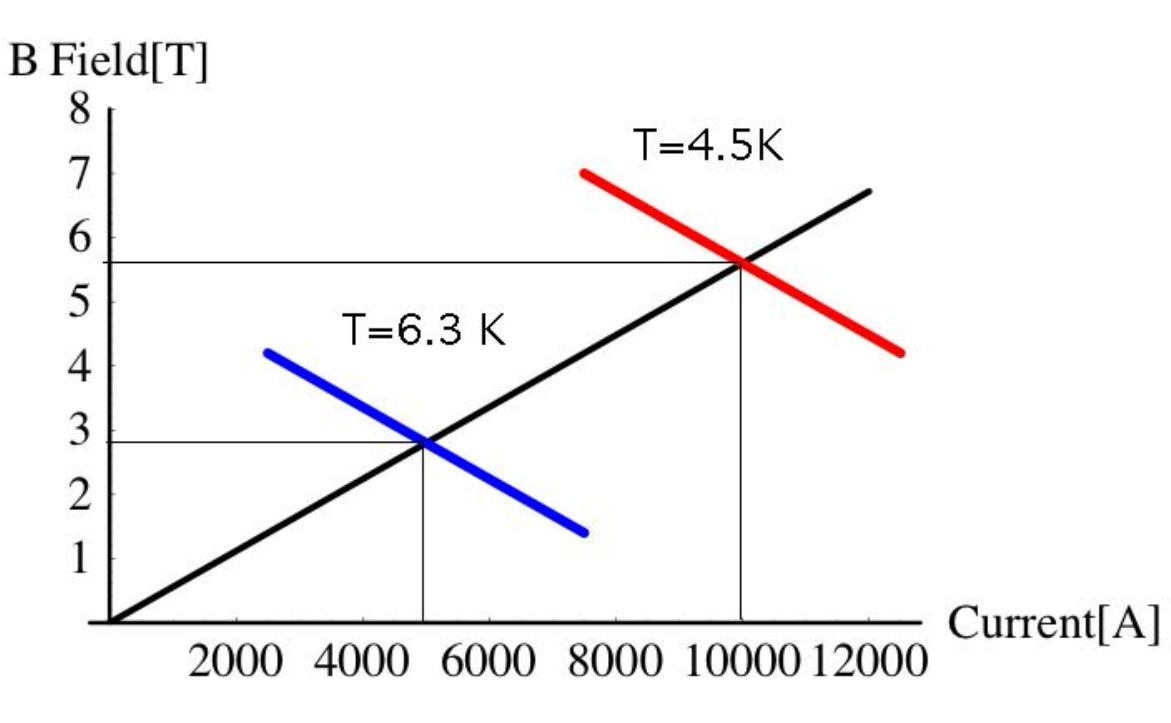}
\caption[Superconducting cable work point calculation.]
{Superconducting cable work point calculation. The red line
represents the current sharing vs. magnetic field curve at
$4.5\unit{K}$ calculated for our cable: the cable was chosen to be
critical at twice the current and twice the magnetic field w.r.t.\ the
work ones. The blue line represents the sharing current vs. magnetic
field at the work condition ($I=5000\unit{A}$, $B=2.8\unit{T}$): the
sharing current temperature is here $6.3\unit{K}$, giving us a safety
margin $\Delta T=1.8\unit{K}$.}
\label{critica}
\end{center}
\end{figure}

The operating current for this conductor will be $50\%$ of the critical
current at twice the peak field, giving a large safety margin. In the
case of local heating up to $5.2\unit{K}$, there will still be a
significant margin on the critical current ($I=0.6I_c$). At $
2.8\unit{T}$, the conductor critical temperature will be $T_c= 8.15
\unit{K}$, and the current sharing temperature will be $6.3 \unit{K}$. This
values can be calculated using the following scaling functions,
so-called Lubell functions~\cite{coil:lubell}.
\begin{equation}
T_c(B)=9.25\unit{K}\left(1-\frac{B[\unit{T}]}{14.5\unit{T}}\right)^{0.59}
\end{equation}
describes the critical temperature as a function of the magnetic
field on the conductor, and
\begin{equation}
J_c(B)=J_0\left(1-\frac{T}{T_c(B)}\right)
\end{equation}
describes the critical current density as a function of the
magnetic field on the conductor and of the temperature.  The working 
point curves are shown in Fig.~\ref{critica}.

A simple method to evaluate the stability of the winding consists of
considering the enthalpy margin per unit length between the operating
and the sharing temperature. This stability parameter will be
$~0.5\unit{J/m}$ for the \PANDA solenoid, which is identical to the
value obtained for the ALEPH and BaBar magnets.

The required superconductor cross section is then calculated to be
$4\unit{mm^2}$, which means 18 strands are needed, assuming
$\alpha=1.2$ and a strand diameter $0.8\unit{mm}$.  Together the
strands will form a Rutherford cable, 80\% compacted, with a cross
section of $1.3\times8\unit{mm^2}$

The cross section of the bare cable will then be $ 3.4 \times
24.6 \unit{mm^2}$.  The coil winding can be made using six lengths of
conductor, four $1600\unit{m}$ long and two $800\unit{m}$ long,
requiring five electrical joints in total.  Each joint must have a
resistance of less than $5 \cdot 10^{-10}\,\Omega$.  The length of
each joint should be optimised limiting the power dissipation to a few
milliwatts.  It can be manufactured either by tungsten-inert-gas (TIG)
welding (as in the Atlas barrel toroid and CMS), or soft soldering
(after electro--deposition of copper).
 
\subsubsection{Electrical Insulation}
Electrical insulation is an important aspect of solenoid design and
manufacture. Two categories of insulation are required: ground plane
insulation between the coil and support cylinder, and turn--to--turn
insulation.

\begin{itemize}
\item{The ground plane insulation must operate at relatively high
voltages during quench conditions and will be subjected to strict quality assurance
controls. The materials will be fully characterised, electrically and
mechanically, before the coil manufacturing, and insulation tests
between the cable and the coil former will be performed at
$2000\unit{V}$ during the whole winding process. The design of the
quench protection systems is based on a maximum voltage to ground of
$500\unit{V}$. The ground plane insulation will be made by a
$0.8\unit{mm}$ layer of fibreglass fabric inserted between the support
cylinder and the solenoid outer layer during the winding. Summing up
this insulation layer to the cable insulation we deduce the need for a
ground insulation of $\sim1\unit{mm}$.  We are confident from the
previous experience on similar solenoids (BaBar, Zeus, CMS) that this
ground insulation thickness will be adequate to withstands the quench
voltage of $\sim500\unit{V}$. The insulation will be fully tested at
$2\unit{kV}$ along the winding process.}
 
\item{The conductor will be insulated with a double wrap of $0.125
\unit{mm}$ fibreglass tape during winding to give an insulation thickness
of $0.2 \unit{mm}$ ($80\%$ compacted). The resulting turn--to--turn
insulation thickness will be $ 0.4 \unit{mm}$ and will be fully
impregnated in the bonding process. Electrical tests will be carried
out during winding to detect any failure of insulation. The tests will
include regular and continuous testing for turn--to--turn and
turn--to--ground insulation. }
\end{itemize}

At the end of the winding process, the fibreglass insulation will be 
fully vacuum impregnated with high strength epoxy resin. The coil and
coil former compound, after the vacuum impregnation, will behave like a 
monolithic cylinder, making turn--to--turn movements very unlikely.

\subsubsection{Winding Support} 
\begin{figure}
\begin{center}
\includegraphics[width=\swidth]{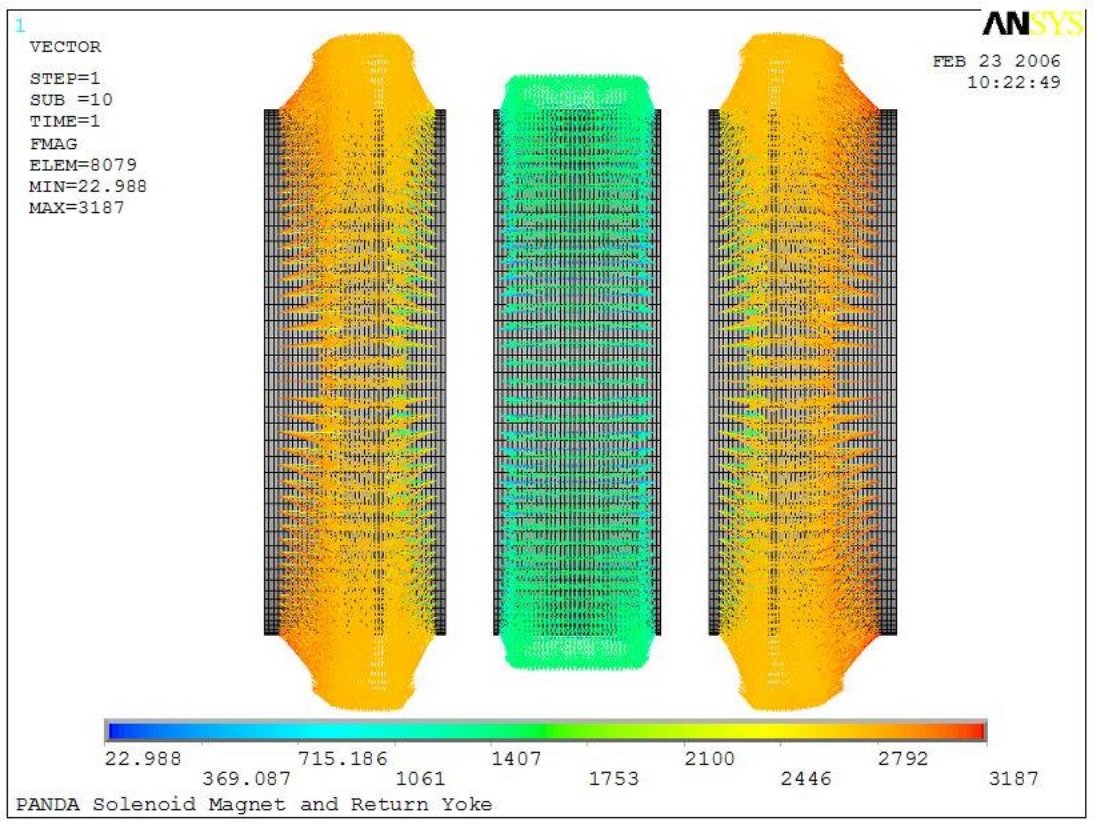}
\caption[The force distribution on the coils.]{The force distribution 
 on the coils (beam direction is left--to--right).}
\label{forza}
\end{center}
\end{figure}

The winding will be supported by an external aluminum alloy cylinder
similar to other existing detector magnets. The winding support is
designed for all aspects of force containment, i.e., its weight and
the radial and axial magnetic forces. Fig.~\ref{forza} shows these
magnetic forces on the solenoid.

The coil former will be built from a single rolled plate of aluminum
alloy.  The plate will be made of EN-AW 5083 aluminum alloy, heat
treated to specification H111 or superior. The CMS experience has
demonstrated very good results by a combination of EN-AW 5083, H321
alloy and metal-inert-gas (MIG) welding controlled procedure, being
able to avoid stress relieving.  Further investigations are under
process on this subject.
 
The maximum radial pressure, $\sim2.90\unit{MPa}$, will be generated
in the first and last sub-coils. The aluminum alloy support cylinder
surrounds the coiled conductor to counteract these radial pressures
and prevent coil movement. An extended stress analysis of the solenoid
coil-support cylinder assembly has been developed to investigate the
behaviour of the high-ductility pure aluminium stabiliser and epoxy
resin under the high radial pressures generated by magnetic
forces. The cable was thus simulated including the material non-linear
stress-strain curve. As a result, plastic deformations are expected to
occur in the coils during the first charge. The amount of these
deformations will be small and will ensure that the pure aluminum
stabiliser will not be stressed beyond the elastic limit in the
subsequent charges. This will help prevent premature quenching during
coil energising. Nevertheless, the support cylinder will be capable to
contain the deformations of the coils while remaining in the elastic
field.

An integrated compressive axial force of $\sim8\unit{MN}$ will be
induced in the winding. According to our calculations, the
distribution of the axial force within the coil will be complex. The
central part will be slightly axially stressed by a force of less than
$1\unit{MN}$.  For the calculations of the axial stress inside the
winding material we considered the maximum total force we for a
sub-coil, i.e., $4.3\unit{MN}$.  This would lead to an axial stress of
$13\unit{MPa}$ on the pure aluminum, with only the winding supporting
the axial forces.  However, since the axial force will be transmitted
to the outer cylinder, the stress will be considerably lowered.  Due
to that the shear stress between the winding and outer supporting
cylinder will be less than $3\unit{MPa}$.  This low value of shear
stress will allow the winding and support cylinder to be mechanically
coupled through an epoxy impregnation without applying any axial
pre-stress to the winding (as it was done for the ZEUS and BaBar
magnets).  Epoxy impregnation can support a shear stress higher than
$30\unit{MPa}$, providing a high safety margin.  This will lead to
significant simplifications and cost savings in the winding
fabrication.

The current design causes axial de-centring forces on the coil due to
the iron asymmetry and a residual force of $0.2\unit{MN}$ is applied
to the winding.  This force will have to be supported by specifically
designed and calculated structures.  For this purpose, 16 axial bars,
made of high-resistance Titanium alloy, have been foreseen, together
with 16 radial bars, which account for the weight of the barrel and
possible forces due to a misalignment of the assembly with respect to
the central axis.

\begin{figure}
\begin{center}
\includegraphics[width=1.0\swidth]{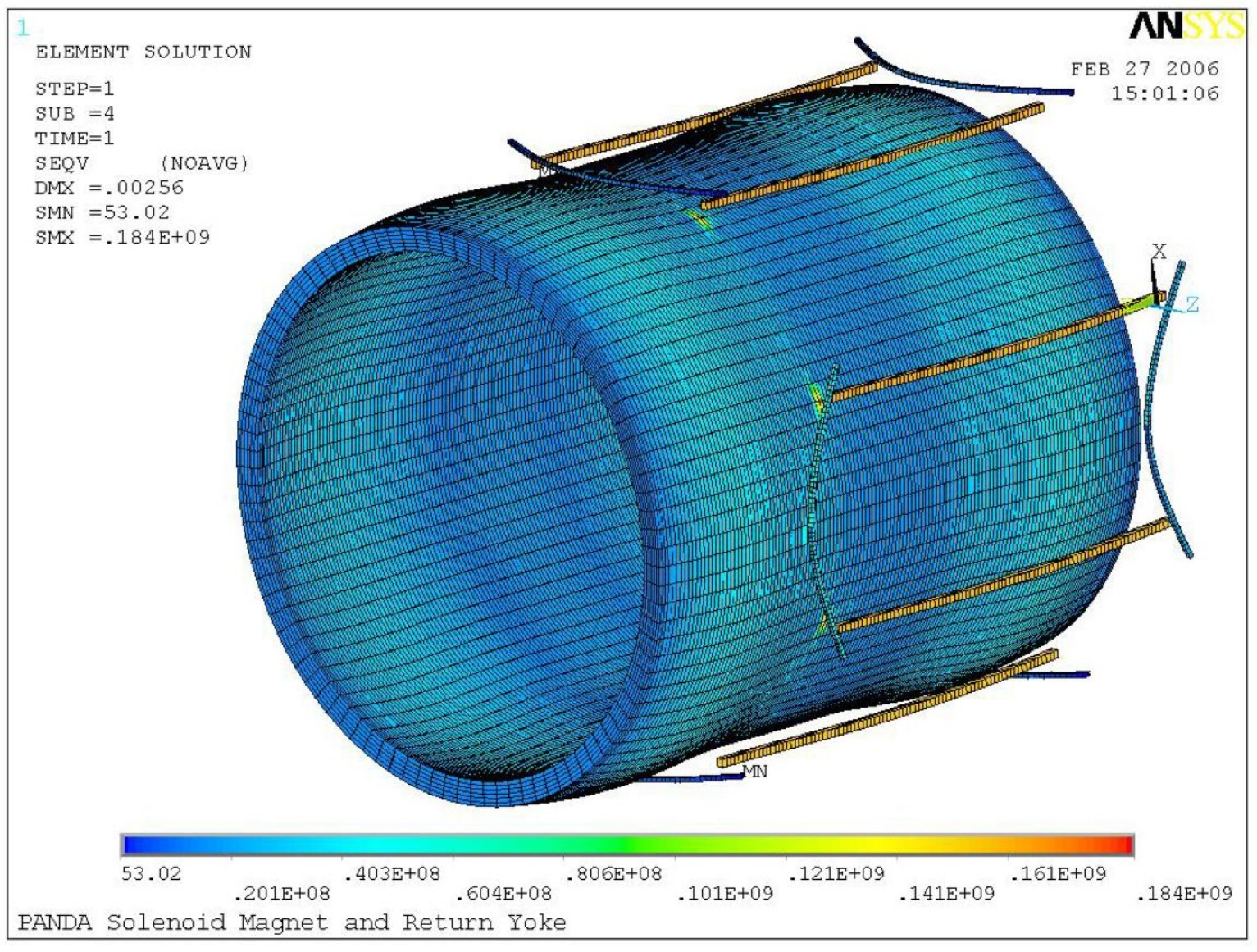}
\caption[Stress on solenoid coil, barrel and supports.]{Calculated Von 
Mises stress on the solenoid coil, coil former and supports during
nominal operations.}
\label{stress}
\end{center}
\end{figure}

While the preliminary analyses described above decoupled the effects
of radial and axial magnetic forces on the coils and support cylinder,
a comprehensive 3D FEM analysis has been made simulating the coil and
cylinder assembly under the effect of the magnetic field during
nominal operations. The results of this calculation (see
Fig.~\ref{stress}), confirming previous analyses, are pointed out
below:

\begin{itemize}
\item{ The radial pressure generated in the windings will cause the pure
aluminium stabiliser to exceed its elastic limit, showing permanent
deformations after the first charge. Nevertheless, the amount of the
plastic deformations will stay negligible and stresses will remain
within the elastic range during the subsequent charges.}
 
\item{ The shear stress transmitted through the epoxy resin to the
aluminium support cylinder will be fairly low if compared to the resin
capabilities.}

\item{ The stability of the whole assembly is ensured by the aluminium
alloy barrel.  The analyses show that the cylinder is capable to
contain the radial and axial deformations of the windings without
showing permanent deformations.}

\item{ Axial and radial supports have also been included in the
model. The decentering forces caused by the asymmetry of the return
flux and the weight of the assembly will be well supported by the
suspension system, the stresses calculated for the bars being well
below the elastic limit for titanium alloys (Ti 6Al 4V ELI - grade 23
or Ti 5Al 2.5Sn ELI - grade 6).}

\end{itemize}

Table~\ref{coldo} shows the main features of the cold mass. The values
are given at a temperature of $4.2\unit{K}$. The dimensions at room
temperature are higher by a factor of approximately $1.004$.

\begin{table}[ht]
\begin{center}
\begin{tabular}{|l|c|}
\hline
Parameter & Value\\
\hline\hline
Winding: & \\
ID & $2100\unit{mm}$ \\
OD & $2200\unit{mm}$ \\
Length & $2780\unit{mm}$\\ 
Weight & $2.2\unit{t}$ \\
&\\
Supporting Cylinder: & \\
Material & Al alloy 5083 \\
ID & $2200\unit{mm}$ \\
OD & $2260\unit{mm}$ \\
Length & $2860\unit{mm}$\\ 
Weight & $1.8\unit{t}$ \\
&\\
Ground Insulation: & \\
Material & Fibreglass epoxy \\
Thickness & $0.8\unit{mm}$ \\
&\\
Total Solenoid Weight: & $4\unit{t}$ \\
&\\
Nuclear Interaction Length: & \\
(Assuming Aluminum) & \\
Maximum & $0.2\unit{\lambda_{int}}$ \\
Minimum & $0.15\unit{\lambda_{int}}$ \\
\hline
\end{tabular}
\caption{Cold mass ($4.5\unit{K}$) parameters.}
\label{coldo}
\end{center}
\end{table}

\subsection{Quench Protection and Stability} 

\subsubsection{Protection Concept}
\begin{figure*}[th]
\begin{center}
\includegraphics[width=5in]{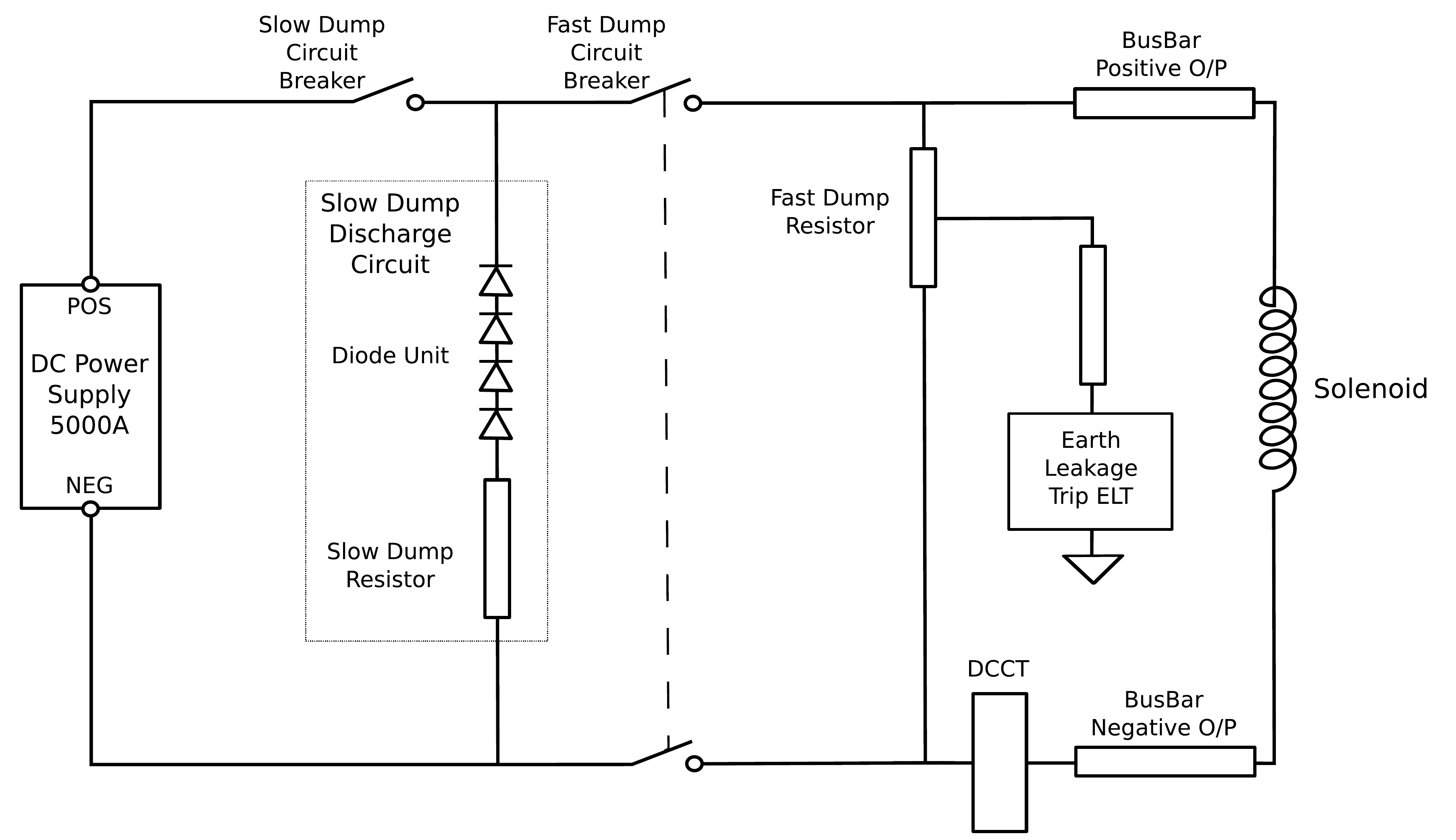}
\caption{Solenoid power and and quench protection concept.}
\label{proto}
\end{center}
\end{figure*}

The solenoid will be protected by an external dump resistor which will
determine the current decay under quench conditions and allow
extraction of $75\%$ of the stored magnetic energy. The quench
protection concept is shown in Fig.~\ref{proto}, and quench parameters
are given in Table~\ref{quench}. The protection concept is based on
two main criteria.

\begin{table}[ht]
\begin{center}
\begin{tabular}{|l|c|}
\hline
Parameter & Value\\
\hline\hline
Operating Current & $5000\unit{A}$ \\
Stored Energy & $21\unit{MJ}$ \\
Inductance & $1.7\unit{H}$ \\
Quench Voltage & $500\unit{V}$ \\ 
Protection Resistor & $0.1\unit{\Omega}$\\ 
Time Constant & $\sim17\unit{s}$ \\
Adiabatic Peak Temperature & $100\unit{K}$ \\
Overall Current Density & $53\unit{A/mm^2}$ \\
Aluminum Stabiliser RRR & $1000$ \\
\hline
\end{tabular}
\caption{Quench parameters.}
\label{quench}
\end{center}
\end{table}
 
Since the calculated stored energy for the magnet is $\sim21\unit{MJ}$
at an operating current of $5000\unit{A}$, an inductance of
$1.7\unit{H}$ is expected. To have a sufficiently fast and effective
quench spreading and energy extraction, a $0.1\unit{\Omega}$ resistor 
was chosen. The time constant of this circuit during a fast discharge 
will therefore be $17\unit{s}$.

\begin{itemize}
\item{A voltage limit of $500\unit{V}$ across the solenoid will apply
during fast discharge. Centre­tapping of the fast dump resistor to
ground will limit the voltage to ground to $250 \unit{V}$. The
centre--tapped resistor will also allow the measurement of ground
leakage currents as a safety and diagnostic tool.}
\item{An upper temperature limit of $100\unit{K}$ will apply during
quench conditions. This limit will give very good safety margins
against the peak temperature and thermally induced stresses at
quench.}
\end{itemize} 

During a quench, it is foreseen that a large fraction of the liquid helium
present in the cooling circuit will boil off. The circuit is dimensioned 
to face a possible failure of the supply system, which will make the pressure
in the pipes as high as at the outlet of the liquefier, i.e. $20\unit{bar}$.
If the pressure should get higher, a proper system of safety valves will open
to avoid damages.

The quench detection-protection circuit shown in Fig.~\ref{proto} will
act also as emergency discharge system, to be used any time a fast
shutdown of the solenoid would be needed and we would like to avoid an
excessive heating of the cold mass, which can lead to a quench of the
winding.  This rapid discharge will be used to protect the solenoid
(or other sub-detectors) against failure of any external
infrastructure, such as power supply, main power and refrigeration
failures.

When a failure in the infrastructure operation was detected, 
the ``slow dump breaker'' (shown in Fig.~\ref{proto}) would open, and the 
magnet current would be dumped by the slow dump resistor through the diode 
unit.  The value of this resistor is chosen to prevent the transition of 
the superconducting coil produced by the power dissipated by the magnetic 
flux variation and the eddy currents in the coil former.  Simulations of 
the transient behaviour of the winding showed that a discharge time 
constant of $600\unit{s}$ will prevent any unwanted heating effect. Given 
the solenoid inductance of $1.7\unit{H}$, the discharge time is fixed 
by the value of the slow dump resistor: in our set-up, the resistor value 
is $3\unit{m\Ohm}$, rated for a peak power dissipation of $15\unit{kW}$. 

\subsubsection{Quench Analysis}
A preliminary quench analysis of the \PANDA~solenoid has been made using
a code developed for indirectly cooled solenoid design. The code
models the thermal and inductive behaviour of the solenoid in order to
account for quench--back effects and heat transfer to the support
cylinder. This analysis shows that quench--back is predicted about two
seconds after opening the protection circuit breakers.

\begin{figure}
\begin{center}
\includegraphics[width=\swidth]{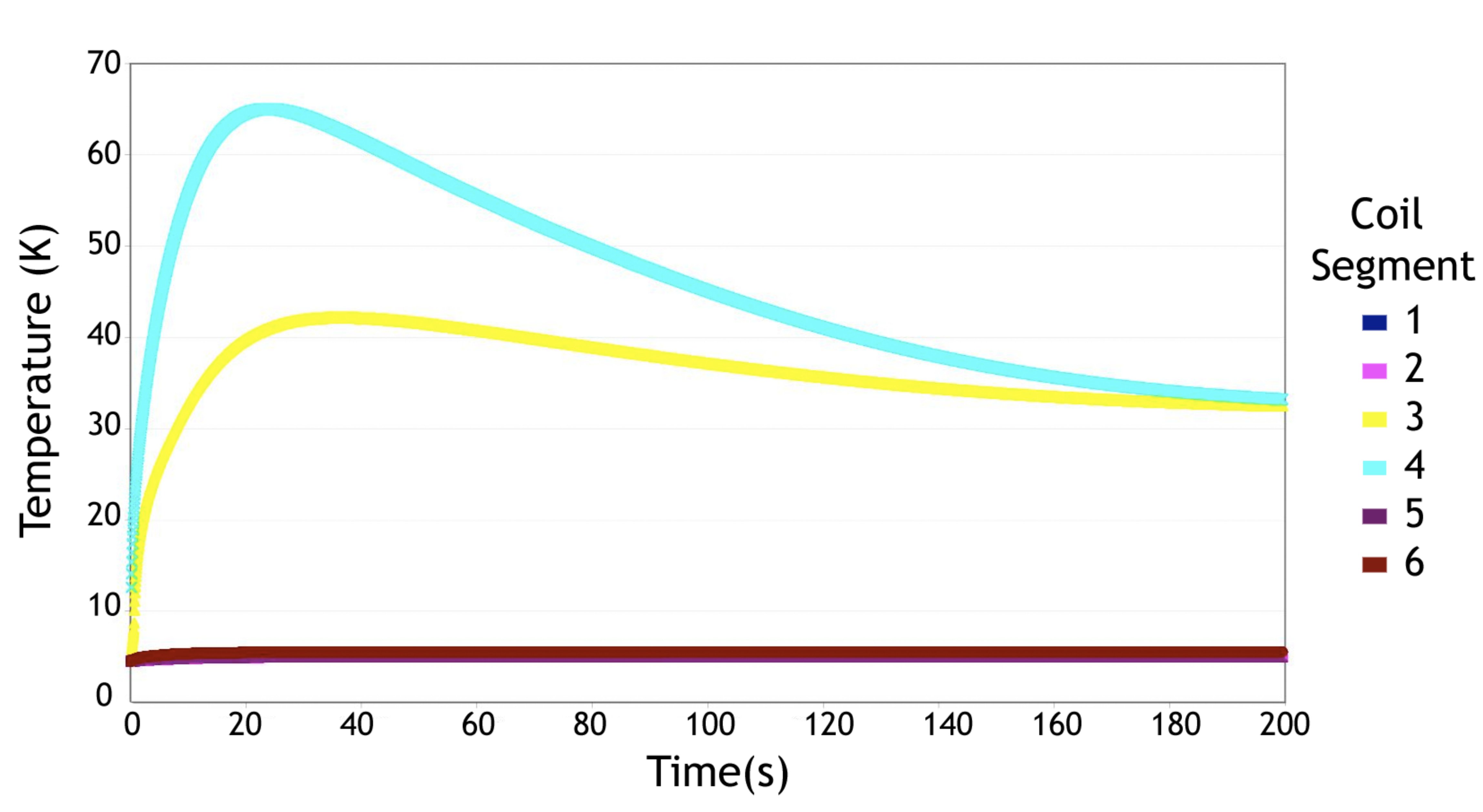}
\caption[Temperature evolution in the coil after a quench.]{Temperature 
  evolution in the coil after a quench, as a function of time. The
  coil was divided into 6 sectors, the 3 sub-coils divided into the 2
  layers. The calculation refers to the worst scenario, i.e.\ the
  quench is generated in the downstream sub-coil, internal layer,
  labelled as number 4. The coil sectors are numbered this way:
  1. external upstream; 2. external central; 3. external downstream;
  4. internal downstream; 5. internal central; 6. internal upstream
  (courtesy AS-G Superconductors).}
\label{qt}
\end{center}
\end{figure}

Several different calculations were performed (courtesy AS-G
Superconductors, formerly known as Ansaldo superconductors),
simulating a local temperature rise in different locations on the
coil.  It was seen that the worst situation would be encountered when
the quench was generated in the downstream, internal coil. The
simulated temperatures of the various windings are plotted in
Fig.~\ref{qt}. In no scenario the temperature on the superconducting
cable exceeded $65\unit{K}$, and in every situation the temperature,
after a short period, will drop to $\sim40\unit{K}$.

\vspace{0.2in}

\begin{figure}
\begin{center}
\includegraphics[width=\swidth]{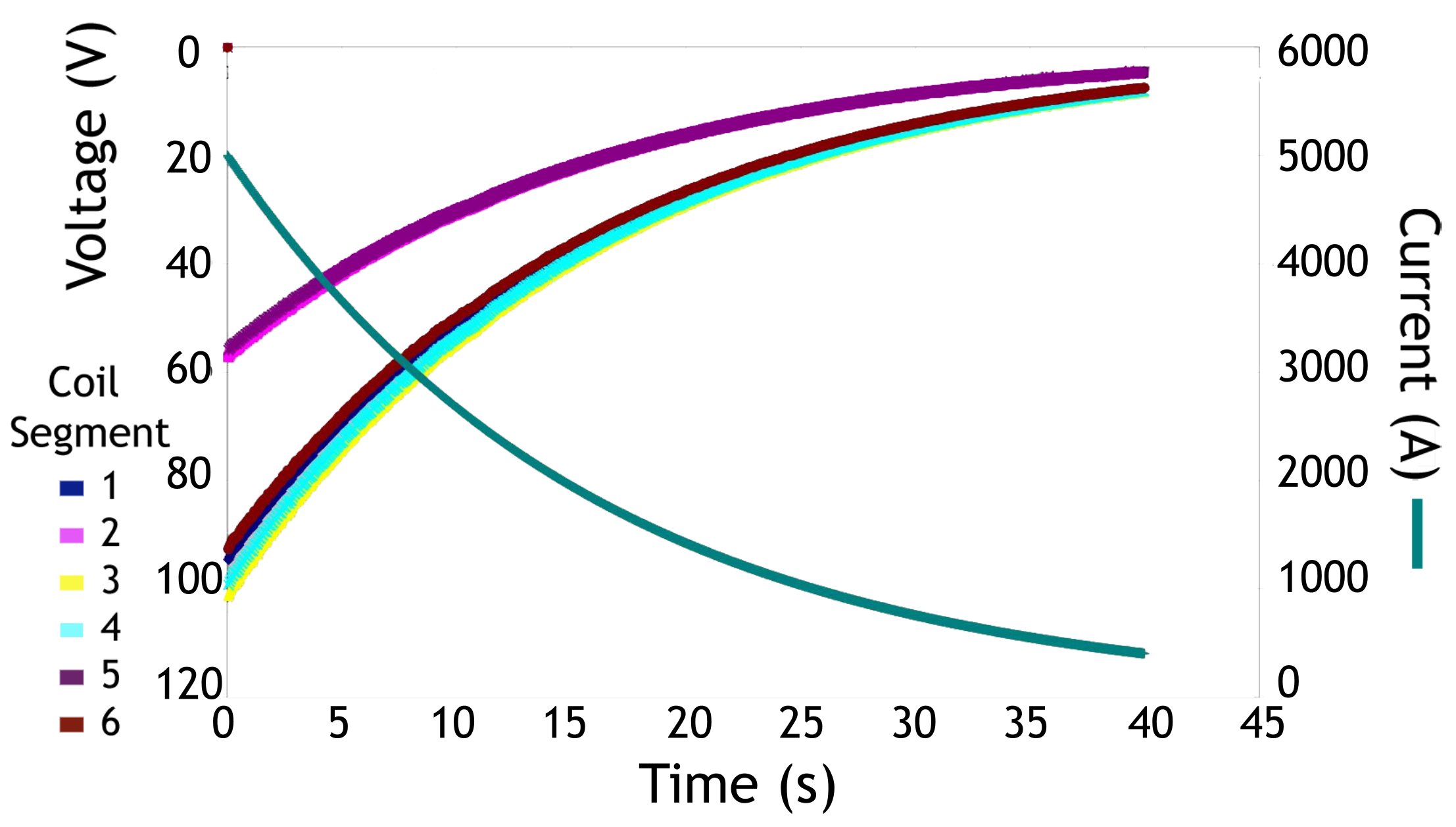}
\caption[Voltage and current  evolution in the coil after a quench.]
  {Voltage and current evolution in the coil after a quench as a
  function of time. The coil was divided into 6 sectors, the 3
  sub-coils divided into the 2 layers. The calculation refers to the
  same scenario as \ref{qt} (courtesy AS-G Superconductors).}
\label{qv}
\end{center}
\end{figure}

The role of the coil former as a quench spreader has been studied
comparing the temperature rise with and without the coil former
itself: this case is similar to the one of a coil former made of a
poor heat and current conductor, such as stainless steel, carbon fibre
or similar. Even in this configuration, the quench would remain limited to
the coil in which it started and in the neighbouring one, and the
temperature would never exceed $80\unit{K}$.

Fig.~\ref{qv} shows the voltage at the ends of each winding and the
current in the coil after a quench: a time constant of
$\sim17\unit{s}$ is obtained, confirming the value of inductance of
$\sim 1.7\unit{H}$ obtained from pure magnetic calculations.

\subsubsection{Stability}
 
The \PANDA solenoid coil will be indirectly cooled using the technology
established for existing detector magnets such as DELPHI and
ALEPH.  Reliable operation of those magnets has demonstrated that
safe stability margins can be achieved using high--purity,
aluminum--clad superconductors in a fully bonded, indirectly cooled
coil structure.
 
The conductor stability has been estimated using an analysis code in
order to establish the minimum quench energy (MQE) for transient
heat pulses.  The computed minimum quench energy is $1.4\unit{J}$.
The computed minimum quench length (MQZ) is $0.6\unit{m}$.  These
margins are considered to be safe for the \PANDA solenoid due to its
low--stress design.

\subsection{Cold Mass Cooling}
 
\subsubsection{Cooldown} 
The cold mass cool down will be accomplished by circulating cold
helium gas either directly from the refrigerator or from a storage
Dewar with gas mixing.

\subsubsection{Operating Conditions}. 

\begin{figure}[th]
\begin{center}
\includegraphics[width=\swidth]{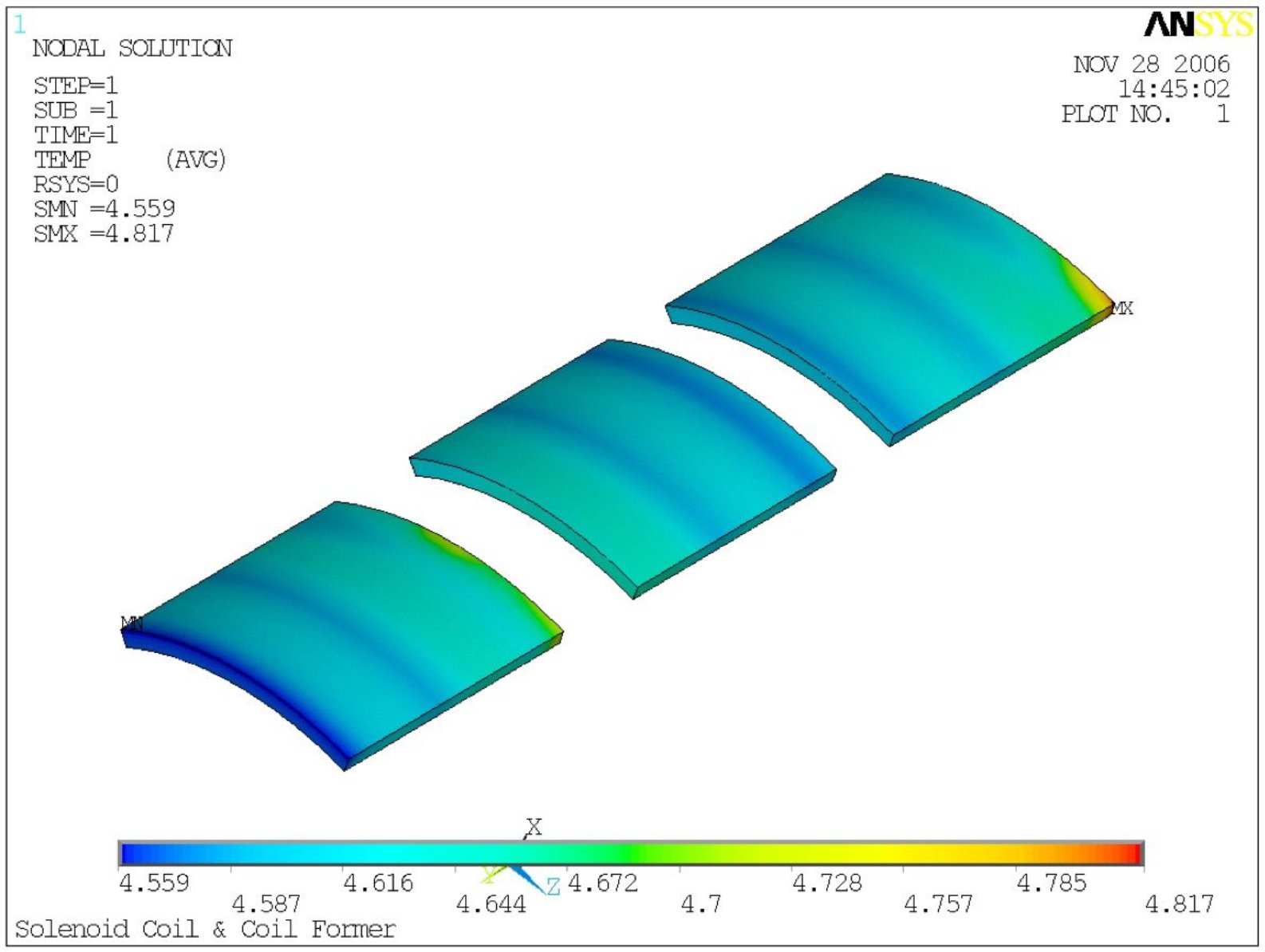}
\caption[Temperature distribution in the coil.]
  {Temperature distribution in the coil during normal operation. The
  calculation was performed on a section of the assembly taking into
  account the coil former assembly, all the heat leaks and the
  symmetry of the system.}
\label{temp}
\end{center}
\end{figure}

\begin{figure}[th]
\begin{center}
\includegraphics[width=\swidth]{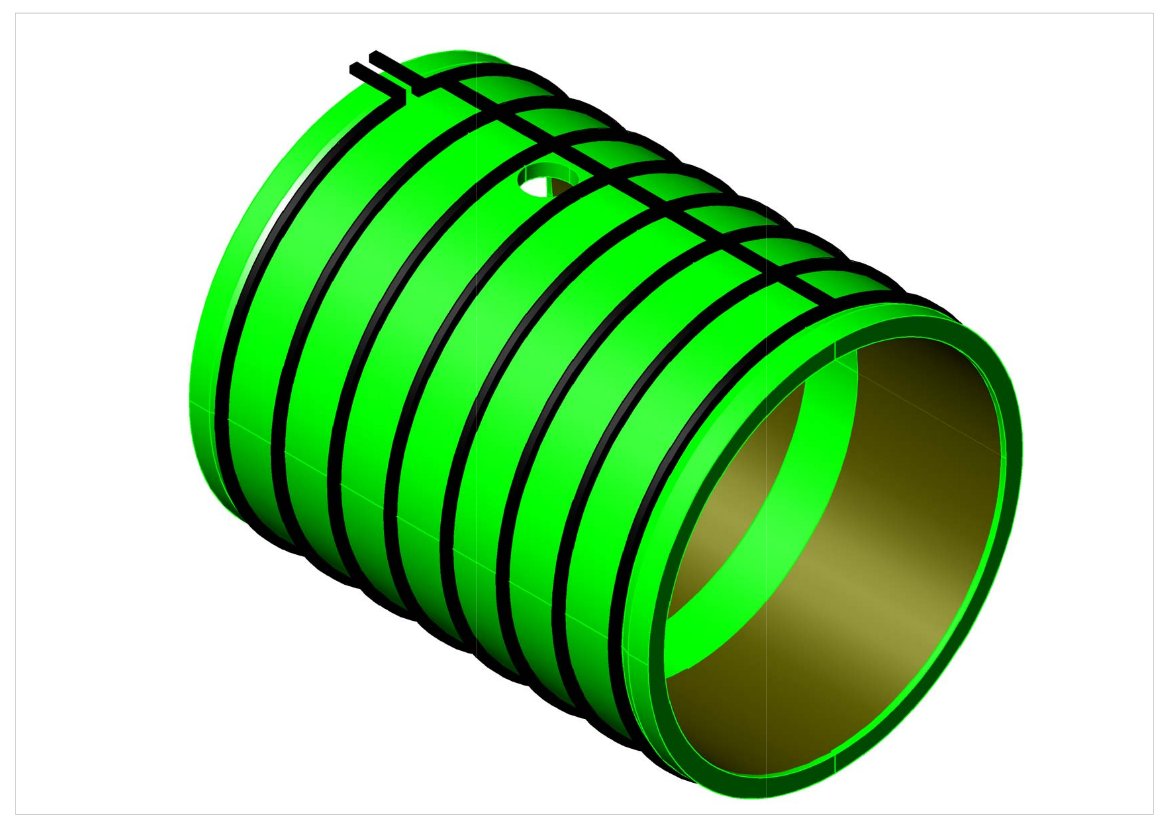}
\caption[Cold mass cooling circuit.]
{Cold mass cooling circuit. The cryogenic supply chimney
passes through a cut--out in the barrel flux
return (shown top left).}
\label{elio}
\end{center}
\end{figure}

Under operating conditions, the cold mass will be cooled by
circulating two--phase helium in pipelines welded on the coil support
cylinder. The cooling circuit will be driven either by forced or
natural (thermo--syphon) convection. This technology is established
and yields the simplest operational mode. The thermo--syphon cooling
circuit is designed for high flow rates to ensure the correct quality
factor for the helium.  The circuit will be fed through a manifold at
the bottom of the support cylinder. The cooling circuits will be
welded to the support cylinder surface with a spacing of
$\sim0.3\unit{m}$ to limit the temperature rise. The cooling pipes
will terminate in an upper manifold. The circuit will be designed to
provide operation during quench conditions. In order to confirm these
assumptions, a finite element model of 1/8 (thanks to the axial
symmetry) of the coil--barrel has been developed.  The maximum
temperature rise will be equal to $0.31\unit{K}$ compared to the
design temperature of $4.5\unit{K}$ (see Fig.~\ref{temp}).  This
assumes the estimated static heat loads (reported in the next
paragraph), a pipe diameter equal to $20\unit{mm}$ and a liquid helium
mass flow of $28\unit{g/s}$.  The conceptual layout of the cold mass
cooling circuit (natural convection solution) is shown in
Fig.~\ref{elio}.

\subsubsection{Heat Loads}
The estimated static heat loads for the solenoid are given in
Table~\ref{cryo}. Eddy current heating in the support cylinder will
cause additional heat loads during charging of the solenoid. However,
for a solenoid charging time of $30\unit{min}$, the estimated
transient power will be $8\unit{W}$, which is small compared to static
heat loads.
 
\subsection{Cryostat Design}\label{sec:coil:cryodesign}
 
\subsubsection{Vacuum Vessel}

The cryostat consists of an annular vacuum vessel equipped with
radiation shields and super-insulation. The vacuum vessel is designed
to satisfy a number of basic criteria listed in the following.

\begin{enumerate}
\item{To support vacuum loads in accordance with recognised pressure
vessel codes.}
\item{To carry the cold mass and radiation shield weight through the
insulating supports.}
\item{To support magnetic loads (20 metric tons of axial forces with a safety
factor of 4, as foreseen for suspended loads) during nominal operations.}
\item{To operate with deflections of less than 1\,mm under all loads when
mounted in the flux return barrel.}
\item{To carry the weight of the inner detectors. (The Barrel EMC
weight is already 20 metric tons.) The calculation has been made
assuming a total weight attached to the cryostat of 50 metric tons.}
\end{enumerate}

\begin{figure*}[th]
\begin{center}
\includegraphics[width=0.7\dwidth]{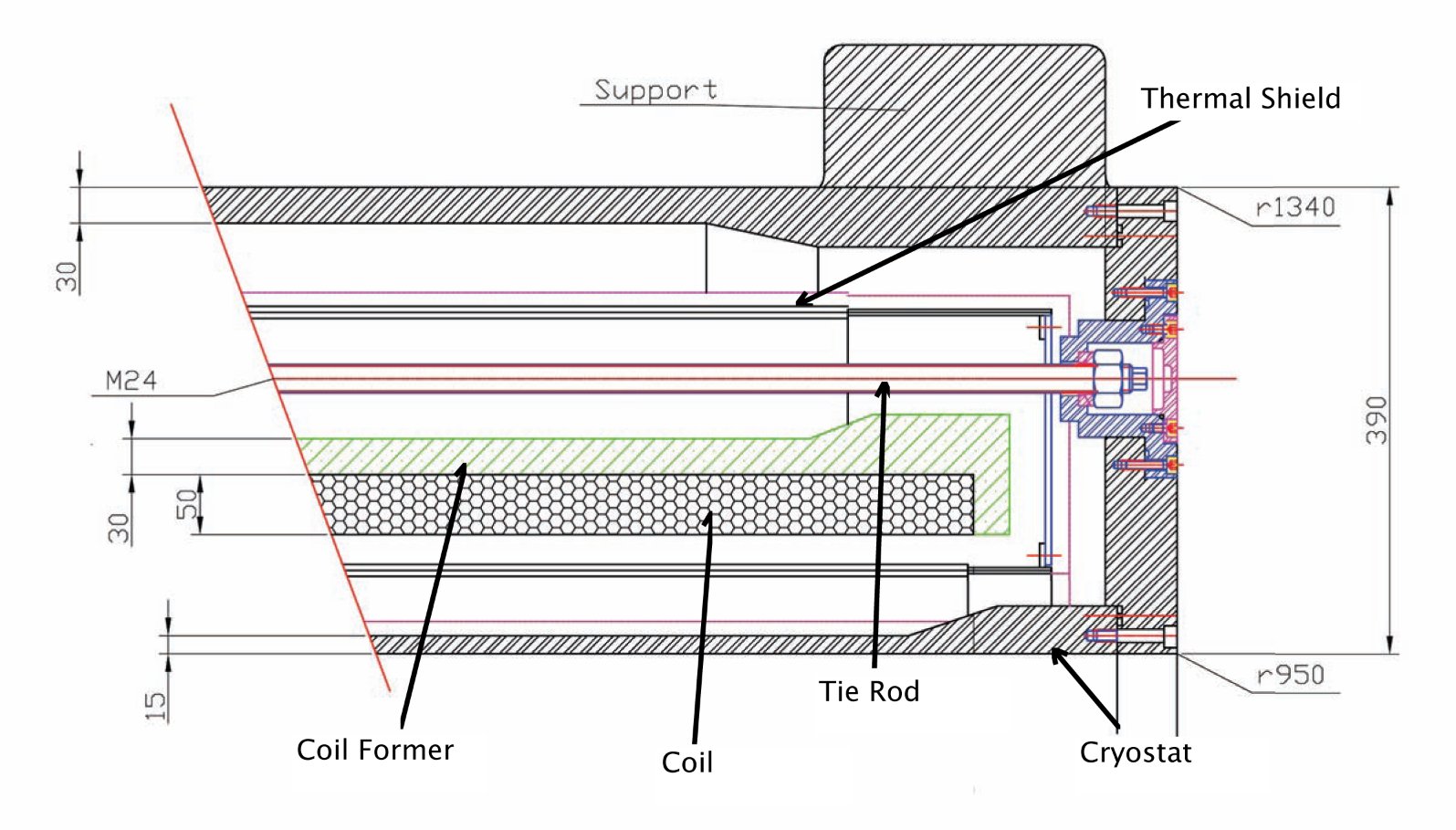}
\caption[Detail of the cold mass suspension.]
{Cross section of the vacuum vessel detailing the downstream 
  end where the rods which suspend the cold mass inside the vessel 
  are attached.  These take the axial load during operation.}
\label{cryo_sec}
\end{center}
\end{figure*}

The vacuum vessel is designed as two concentric cylinders with thick
annular end plates, all of aluminum alloy 5083; its basic parameters
are given in Table~\ref{vessel}.  A cross section of the downstream
end is shown in Fig.~\ref{cryo_sec}.  The minimum thickness of the
different parts composing the cryostat has been designed, will be
fabricated and will be inspected according to the intent of the EC
Euronorm codes for Unfired Pressure Vessels, EN
13445 \cite{EuroCode13445}, but will not be code-stamped.  For steel
structures, the allowable design stresses follow the standard
guidelines of European standards
\cite{EuroCode3}.  Bolted connections and fasteners will 
conform to their recommended torques and allowable stresses depending 
on the connection. 

\begin{table}[ht]
\begin{center}
\begin{tabular}{|l|c|}
\hline
\multicolumn{2}{|c|}{Envelope Dimensions}\\
\hline
\hline
Inner Radius & $950\unit{mm}$ \\
Outer Radius & $1340\unit{mm}$ \\
Length & $3200\unit{mm}$ \\
Inner Cylinder Thickness & $10\unit{mm}$\\
Outer Cylinder Thickness & $30\unit{mm}$\\
End Plates Thickness & $50\unit{mm}$\\
Materials & AL5083, AISI 304 \\
\hline
\multicolumn{2}{c}{}\\
\hline
\multicolumn{2}{|c|}{Design Loads}\\
\hline
\hline
Vessel Weight & $6\unit{t}$ \\
Cold Mass & $4\unit{t}$ \\
Calorimeter & $18\unit{t}$ \\
Other Detectors & $3\unit{t}$ \\
\hline
\end{tabular}
\caption{Vacuum vessel parameters.}
\label{vessel}
\end{center}
\end{table}

\begin{figure}[th]
\begin{center}
\includegraphics[width=\swidth]{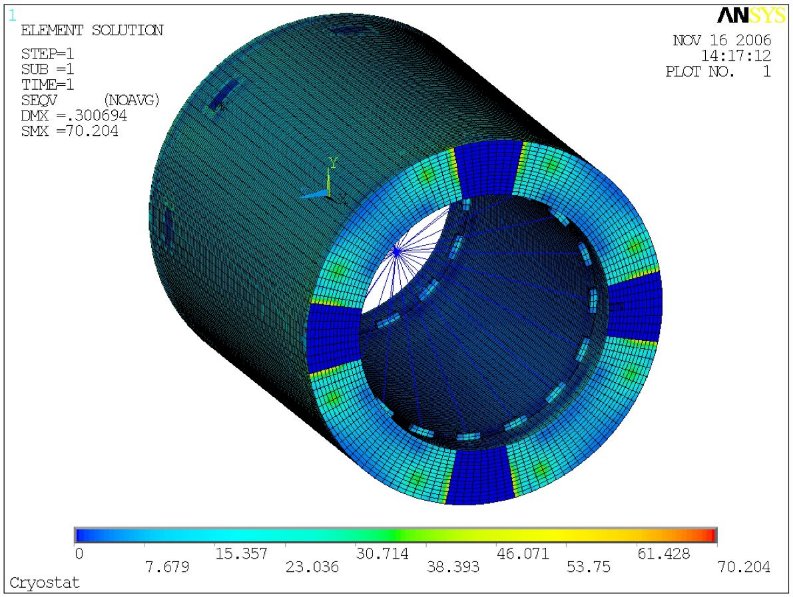}
\caption{Cryostat FEM model, equivalent stress with all loads applied.}
\label{cryostat}
\end{center}
\end{figure}

A preliminary finite element structural analysis of the vessel has
confirmed that design criteria (1)$\div$(4) can be met with reasonable
safety factors. In our model maximum vessel deflections are less than
$1\unit{mm}$, and stress levels peak up to $70\unit{MPa}$ with all
loads applied (see Fig.~\ref{cryostat}).  The cryostat could also be
made of AISI 304 Stainless Steel; in this case, a increased stiffness
and increased stress levels are expected.

\subsubsection{Thermal Shielding}  The cryostat will be equipped 
with radiation shields, which will operate at 40--80\,K, and
super-insulation.  The shields are cooled by helium gas supplied
directly from the refrigerator.  About 30 layers of super-insulation
separate the vacuum vessel walls from the radiation shields. Another
five layers will be installed between the shields.

\subsubsection{Services}  Cryogenic supplies and current supplies 
will be connected from a service turret to the cryostat through the
service chimney in a suitable slit at the backward end of the yoke
barrel. Current leads and local control valves will be mounted in the
services turret.

We foresee to use standard copper counter-flow cryogenic current leads
rated to the running current of the magnet: this corresponds to a
liquid helium consumption of $17\unit{l/h}$.  Cryogenic relief valves
will also be mounted in the service turret for quench and
refrigeration failure conditions.

\subsection{Coil Assembly and Transport}
 
The coil will be assembled inside the cryostat at the manufacturer's
plant. Electrical and cryogenic connections will be made inside the chimney
so that the coil can be tested before shipping.

A complete cooldown will be carried out from room temperature to the
operating temperature of $4.5\unit{K}$.  The cooldown will allow
checking of cooling time, temperature control, heat loads, and full
operation of sensors.  A magnetic test will also be performed at low
field ($30\%$ of the operating current) to check superconductor
operation, the joint resistance, and the additional losses due to the
eddy currents in the outer structural cylinder at the coil ramp-up.
 
Before delivering the magnet, but after the tests at the factory, the
end flanges will be dismounted to allow a hard connection of the cold
mass to the cryostat walls. Depending on the transport facilities, the
cryogenic chimney may also be dismounted. In this case, the electrical
and cryogenic connections also must be dismounted and protected against
breakage.

The setup of the yoke, cryostat and solenoid is compulsory for the
installation of all the detectors of the Target Spectrometer. This is
due to the fact that the cryostat itself provides mechanical support
for the various detectors. Therefore, before the detector assembly,
tests and commissioning of the magnet have to be completed.


\subsection{Cryogenic Supply System and Instrumentation}

The operation of the superconducting solenoid requires both liquid
helium (4.5\,K) and cold helium gas ($20\unit{K} - 100\unit{K}$) for
cooldown and refrigeration of the thermal shields.  Similar systems
have been used successfully throughout the HEP community.  A summary
of the cryogenic loads is given in Table~\ref{cryo}.

\begin{table}[ht]
\begin{center}
\begin{tabular}{|l|c|}
\hline
\multicolumn{2}{|c|}{Magnet Heat Loads at 4.2K}\\
\hline\hline
Item Parameter & Load \\
\hline
Cold Mass & $4000\unit{kg}$\\ 
Total Surface Area & $44\unit{m^2}$ \\
Radiation Heat Flux (Design) & $0.07 \unit{W/m^2}$\\
Radiation Heat Load (Design) & $3.1\unit{W}$ \\
Conduction Heat Load & $2.2\unit{W}$ \\
Cable Joints & max $2\unit{W}$\\
Cryogenic Chimney & $10\unit{W}$\\
Gas Load & $1\unit{W}$\\
Eddy Currents ($1500\unit{s}$ ramp time)& $10\unit{W}$ \\
Total $4.5\unit{K}$ (SF of $2$) & $57\unit{W}$\\
\hline
\multicolumn{2}{c}{}\\
\hline
\multicolumn{2}{|c|}{Magnet Shield Heat Loads at 60K} \\
\hline\hline
Item Parameter & Load\\
\hline
Shield Mass & $1000\unit{kg}$\\
Total Surface Area & $44\unit{m^2}$\\
Radiation Heat Flux (Design) & $1.3\unit{W/m^2}$ \\
Radiation Heat Load (Design) & $57\unit{W}$ \\
Conduction Through Supports & $17\unit{W}$ \\
Conduction To Shields\footnote{This is a rough, conservative estimate:
actual load due to the shields depends on the shield geometry and
technology.} & $150\unit{W}$\\
Diagnostic Wires & $1\unit{W}$\\
Gas Load & $2\unit{W}$\\
Eddy Currents ($1500\unit{s}$ ramp time)& $10\unit{W}$\\
Total $60\unit{K}$ & $237\unit{W}$ \\
\hline
\end{tabular}
\caption{Cryogenic heat loads.}
\label{cryo}
\end{center}
\end{table}

The solenoid will be equipped with a full set of instrumentation
sensors for monitoring, control, and diagnostic
purposes. Instrumentation includes temperature sensors for the cold
mass, shield cryogenic flow monitoring, and strain gauges in the coil
support cylinder. Voltage taps will monitor the electrical resistance
of the conductor joints and will provide quench detection. The quench
detection systems will be hardwired to interlocks. The solenoid
instrumentation and controls will be integrated with the \PANDA
experiment and refrigeration controls.

\begin{figure*}[th]
\begin{center}
\includegraphics[width=\dwidth]{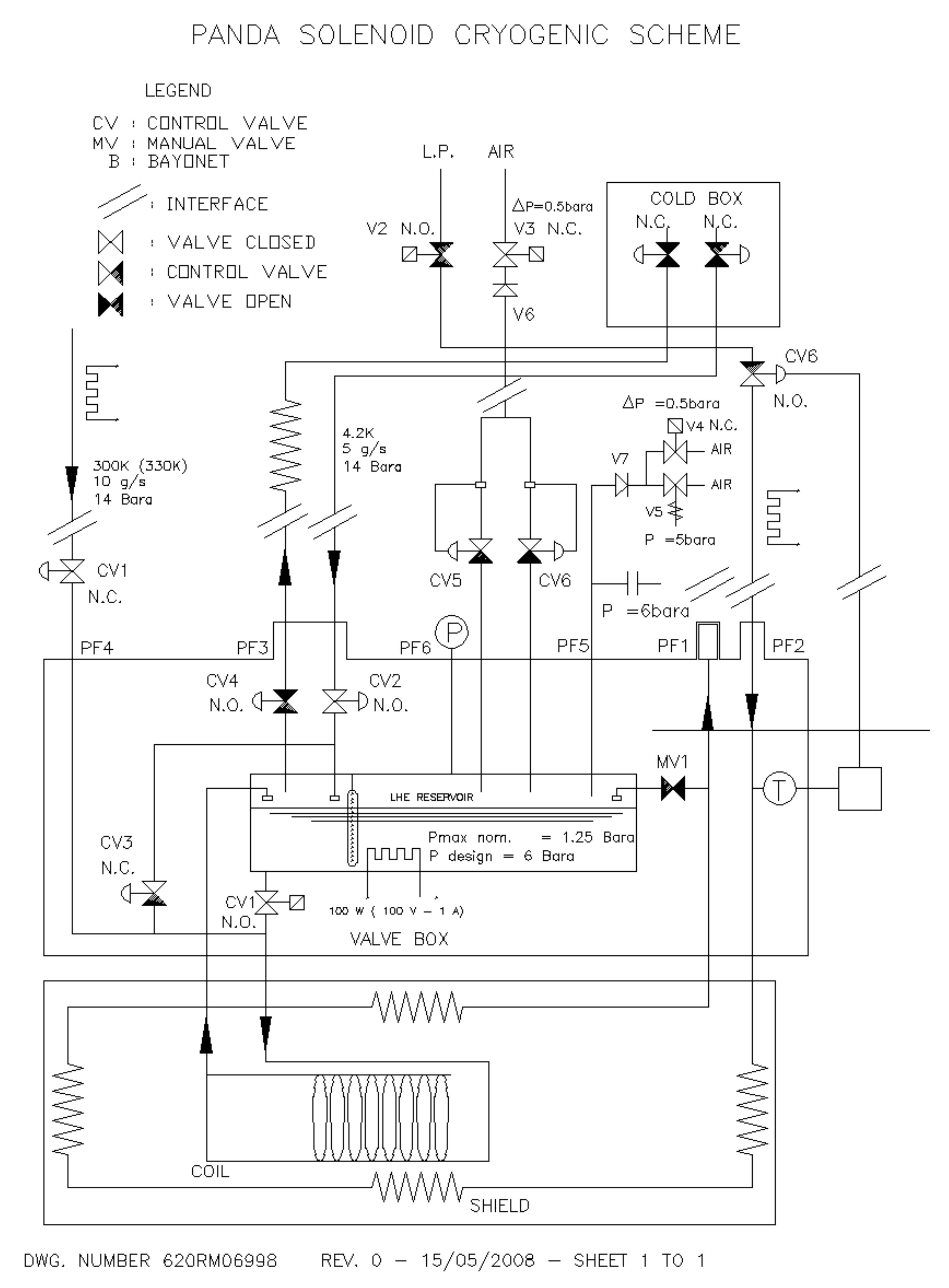}
\caption{Proximity cryogenics scheme.}
\label{prox}
\end{center}
\end{figure*}

The fully automatic liquid helium plant will be equipped with a
process control system and all required logic and software necessary
for all operational modes.  Control and monitoring of the cryogenic
plant and the magnet coil, together with remote control and monitoring
of the compressor room, will be carried out from a control room
adjacent to the plant room. Main operating parameters will be
delivered to the \PANDA data acquisition and monitoring systems. A
scheme of the proximity cryogenics is shown in Fig.~\ref{prox}.

\subsection{Cryostat and Cryogenic System Safety}

\subsubsection{Vacuum System Safety}

The cryostat and cryogenic turret of the \Panda solenoid, as reported
in Section~\ref{sec:coil:cryodesign}, will be built according to the
EC specifications for the Boiler and Pressure-Vessel
Code \cite{EuroCode13445}.  The worst case scenario concerning the
safety of the cryostat would arise, if a leak evolved in the part of
the liquid helium (LHe) cooling circuit that runs through the vacuum
vessel.  Only in this case the pressures balance could actually be
reversed, and the pressure inside the cryostat could exceed the
atmospheric pressure.

The cryostat vacuum vessel, turret and all connections was verified to
be resistant resist up to a pressure of $150\unit{kPa}$ from the
inside.  This provides a good safety margin, as the the coolant
circuit will be working at an absolute pressure of $125\unit{kPa}$ and
we foresee safety valves (see below).  For the outer shell the
stability against a pressure reverse is easily guaranteed, since it is
designed to work with $100\unit{kPa}$ from the outside.  This is
naturally different for the inner wall of the cryostat, which is
designed to keep, under normal operation, the atmospheric pressure
from inside.  However, the inner wall of the cryostat is designed to
keep safely the full load of the detector, mainly the weight of the
EMC with a minimum deformation.  The extra thickness needed to fulfil
this requirement is more than enough to guarantee the cryostat against
buckling of the inner wall in case of a pressure reversal due to a
leak in the helium circuit.

A system of safety valves has been foreseen to prevent an excessive
pressure rise in the cryostat vessel and chimney. A relief valve
opening at $130\unit{kPa}$ (absolute pressure) is foreseen.  This
corresponds to an over-pressure of $30\unit{kPa}$ relative to
atmospheric pressure to which the vessel is subject from the outside.
In case of LHe vaporisation the relief valve would open, venting the
helium gas to the atmosphere preventing damages to the cryostat
vessel.  As a redundant safety measure, a burst disk opening at
$150\unit{kPa}$ is foreseen to cover the unlikely event that the
relief valve malfunctions.

\subsubsection{Proximity Cryogenics Safety}

The choice of indirect cooling of the solenoid will greatly reduce the
amount of liquid helium in the cryogenic system.  The pipes,
guaranteeing the coil refrigeration, will contain $\sim60\unit{l}$ of
refrigerant.  This includes the two manifold placed on the top and on
the bottom of the winding for the thermo syphon circulation. Another
$\sim50\unit{l}$ of liquid helium will be stored in the current leads
bath housed in the reservoir contained in the cryogenic turret of the
cryostat.

The operating pressure of the \Panda coil cryogenic system will be
$125\unit{kPa}$, corresponding to a coolant temperature of
$4.5\unit{K}$.  The two phases liquid helium used by the cooling
system will be fed to the control Dewar trough a transfer line from a
$4000\unit{l}$ buffer-storage Dewar, connected to the helium liquefier
during normal operation.  This buffer tank will give the coil cryogenic
system the capability to operate for more than a day in case of
refrigerator shutdown.  The buffer storage will decouple also the
refrigerator from the solenoid cryogenics in case of refrigerator
failure.

In the worst case, the gas would be transferred into the
buffer-storage at the full pressure of the run compressor of the
liquefier ($20\unit{bar}$) rather than into the cooling circuit of the
solenoid.  In this case it would be vented to the air trough the
relief valve or burst disks.  Furthermore, it will help that the
buffer-storage and the control Dewar will be connected trough a long
liquid helium transfer line. This line will hydraulically decouple the
two reservoirs in the scenario of a liquefier failure with the
$20\unit{bar}$ helium gas at the temperature of the last heat
exchanger ($\sim20\unit{K}$) transferred into the system.

As an additional safety measure, we decided to design all the cooling
loops and distribution lines inside the cryostat (including the
intermediate radiation shields) for a nominal operating pressure of
$20\unit{bar}$.

The control Dewar in the cryo-chimney, operating at $125\unit{kPa}$,
is designed for a pressure of $600\unit{kPa}$ (6 absolute bar) and
will be equipped with a relief valve venting the vessel to the
atmosphere for an overpressure $\Delta p=50\unit{kPa}$ ($p =1.5$\,bar)
in case of helium boiling off due to a quench or vacuum failure.  A
burst disc opening at $600\unit{kPa}$ protects the inner vessel of the
control Dewar in the extreme case of a total malfunction of the relief
valves.

\subsubsection{Seismic Safety Considerations}

The coil and cryostat of the \Panda solenoid are designed to safely
keep in operation static forces exceeding ten times the weight of the
object. For this reason, from a static point of view, these objects
are safe for earthquakes not exceeding magnitude of 3, corresponding
to the strongest event registered in the Darmstadt surroundings in the
last decades.

A static analysis gives only part on the information needed to
guarantee the seismic safety of a complex and composite device as
the \PANDA detector. To ascertain the seismic safety of the system, a
transient analysis of the system using an excitation force reproducing
the frequency spectra of the Darmstadt seismic activity is needed. The
results of seismic analyses are strongly dependent from the
component geometry, mechanical features and suspensions. They will be
performed as the most important check once the detector layout fully
defined and the different options on the sub detector choice
and mounting procedure are well defined and frozen.


\svnInfo $Id: yoke.tex 713 2009-03-31 17:14:58Z IntiL $

\section{Instrumented Flux Return}
\label{s:yoke}

\AUTHOR{Y.~Lobanov}

\subsection{Introduction}

The main purpose of the iron yoke of the Target Spectrometer will be
to serve as flux return for the magnetic flux produced by the
superconducting solenoid.  The details of design of the return yoke
will contribute significantly to the shape and the value of the field
in the central region, where it is needed for momentum
determination. The effect on the absolute value of the magnetic
induction field is due to the strong reduction of the reluctance of
the magnetic circuit, that enhances the field strength, allowing for a
smaller coil at equal fields. In the region of the central tracker the
contribution of the yoke to the total magnetic field will be about
12\%.

With a properly studied shape, in addition, it is possible to adjust the
field flux lines orientation, reducing the radial components of the
field itself. Only with a dedicated and optimised design of the flux
return yoke, in close interplay with the coil design, the challenging
requirements concerning the radial field component in the central
region (see Sec.~\ref{s:sol:req}) can be reached.

Furthermore, the presence of this great amount of iron provides the
shielding of the outside area from the strong fields produced by the
coil.  In this region magnetic fields are unwanted and there are
strict requirements for the maximum allowable stray fields, ranging 3
orders below the required field in the central region.

The flux return yoke must serve as range system for the identification
of muons, too. A system of interleaved detectors and absorbing
material is ideally suited for the detection of muons and their
discrimination versus pions. Thus, the whole barrel and downstream
door will be fully laminated, to allow maximal instrumentation for muon
detection. Besides that, the yoke will provide a solid frame for
mounting the cryostat and the detectors.

The yoke will consist of a barrel part, with octagonal shape, which
will be laminated in 13 layers.  A hole of 350\,mm in diameter through
the barrel part of the yoke is foreseen to accommodate the internal
target system, and another one to host the cryostat chimney.  Both
ends will be closed with end doors.  The downstream door will be
laminated in 5 layers, while the upstream door does not require
lamination.  The upstream door will have a central round hole while
the downstream door will feature a rectangular one which opens
$5^\circ$ and $10^\circ$ in the vertical and horizontal planes,
respectively.  Both end doors will be separated into two halves.  The
end doors will be sliding on skids. Thus they can be opened
independently to allow access for maintenance of the inner detector
systems.

At both ends of the barrel part, in the octagon corners, proper slits
are foreseen, to allow the routing of cables and services for the
detector hosted inside the yoke itself.

\begin{figure}[ht]
\begin{center}
\includegraphics[width=\swidth]{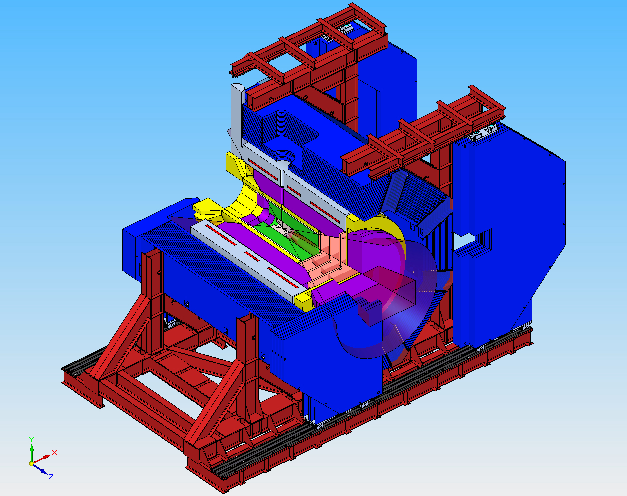}
\caption[View of the Target Spectrometer from downstream]{View of
the Target Spectrometer from downstream showing the
  yoke design and the opening doors.}
\label{yoke:3Dview:ds}
\end{center}
\end{figure}

The Target Spectrometer, hence the instrumented flux return and
everything attached to it, will be mounted as a whole on a movable
rail-guided carriage to be transported from the assembly area to its
operation position.  An overview of the full Target Spectrometer
from the downstream side is shown in Fig.~\ref{yoke:3Dview:ds}.

\begin{figure*}[ht]
\begin{center}
\includegraphics[width=\dwidth]{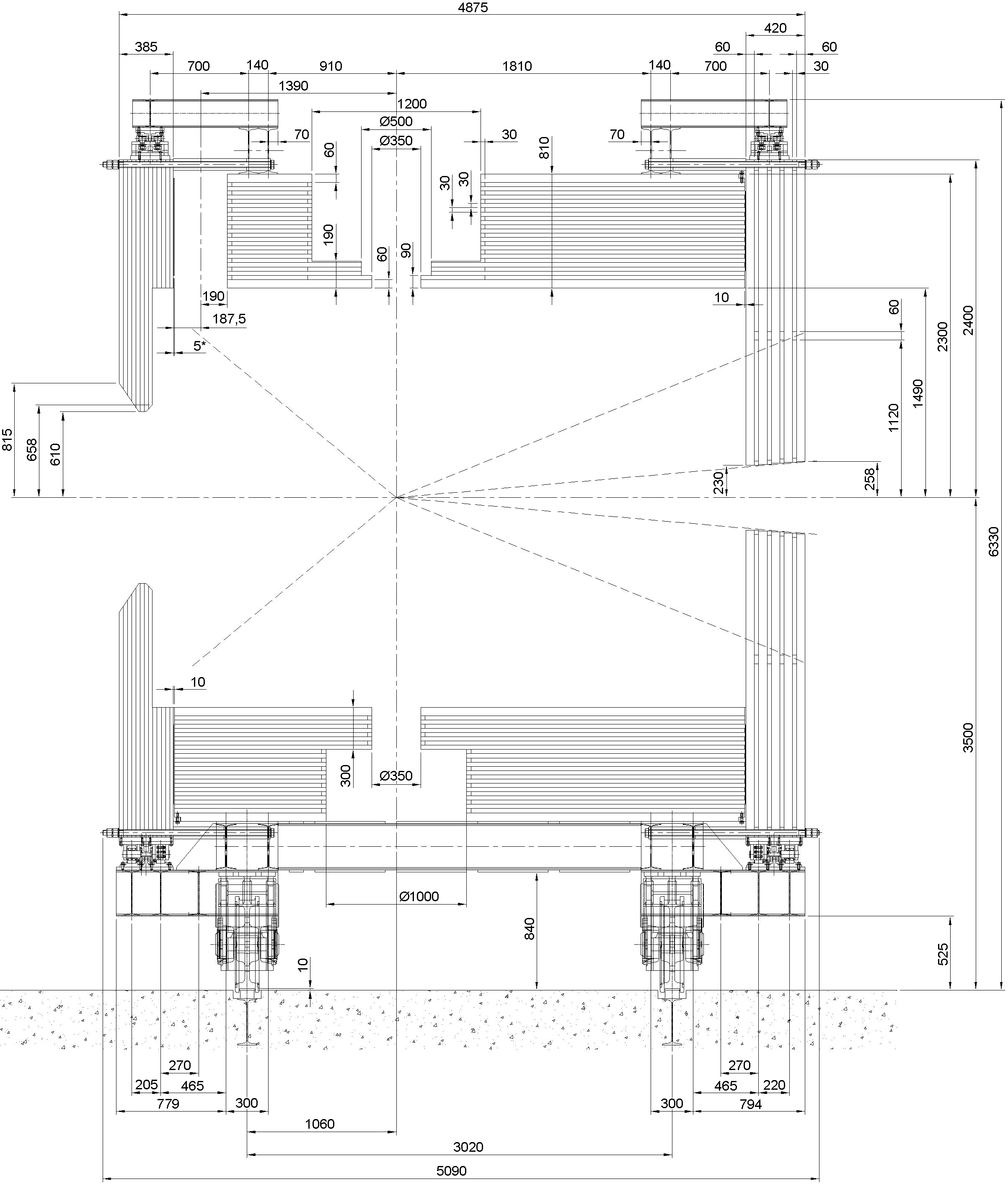}
\caption[Cross-section of the flux return yoke and the support
  structure in the $z-y$ plane]{Cross-section of the flux return yoke and the support
  structure in the $z-y$ plane.  The beam would come in from the left
  and interact with the target at the line crossing in the middle of
  the drawing.  The recesses for the target and the cut out for the
  cryogenic chimney are clearly visible.  The view shows the door
  opening mechanism to the left and right and the carriage on a rail
  system at the bottom.}
\label{yoke:Yoke+frames_section}
\end{center}
\end{figure*}

\subsection{Structural Design}
\label{s:sol:yoke:struct}

The flux return yoke will be divided in a barrel part and end caps on
each end which will be realised in two halves each, such that they can
be opened as sliding doors. The barrel octagon with its support frame
will form an independent self supporting unit which also carries the
weight of the cryostat and all the nested detectors. During operations,
all parts of the yoke, including the doors, will be
rigidly fixed to each other to provide the best mechanical
stability.

\subsubsection{Barrel Part}

In our design the barrel in realised by joining eight beams of 4.05\,m
length and a trapezoidal shape in an octagon with an outer diameter of
4.6\,m.  The structural strength will be obtained with an external
support structure, consisting of two rings surrounding the barrel on
both ends.  Two lower beams under the rings and several perpendicular
beams will form a carriage, which suspends the whole system.  The
cross section of the iron yoke is shown in
Fig.~\ref{yoke:Yoke+frames_section}.

\begin{figure}[ht]
\begin{center}
\includegraphics[width=\swidth]{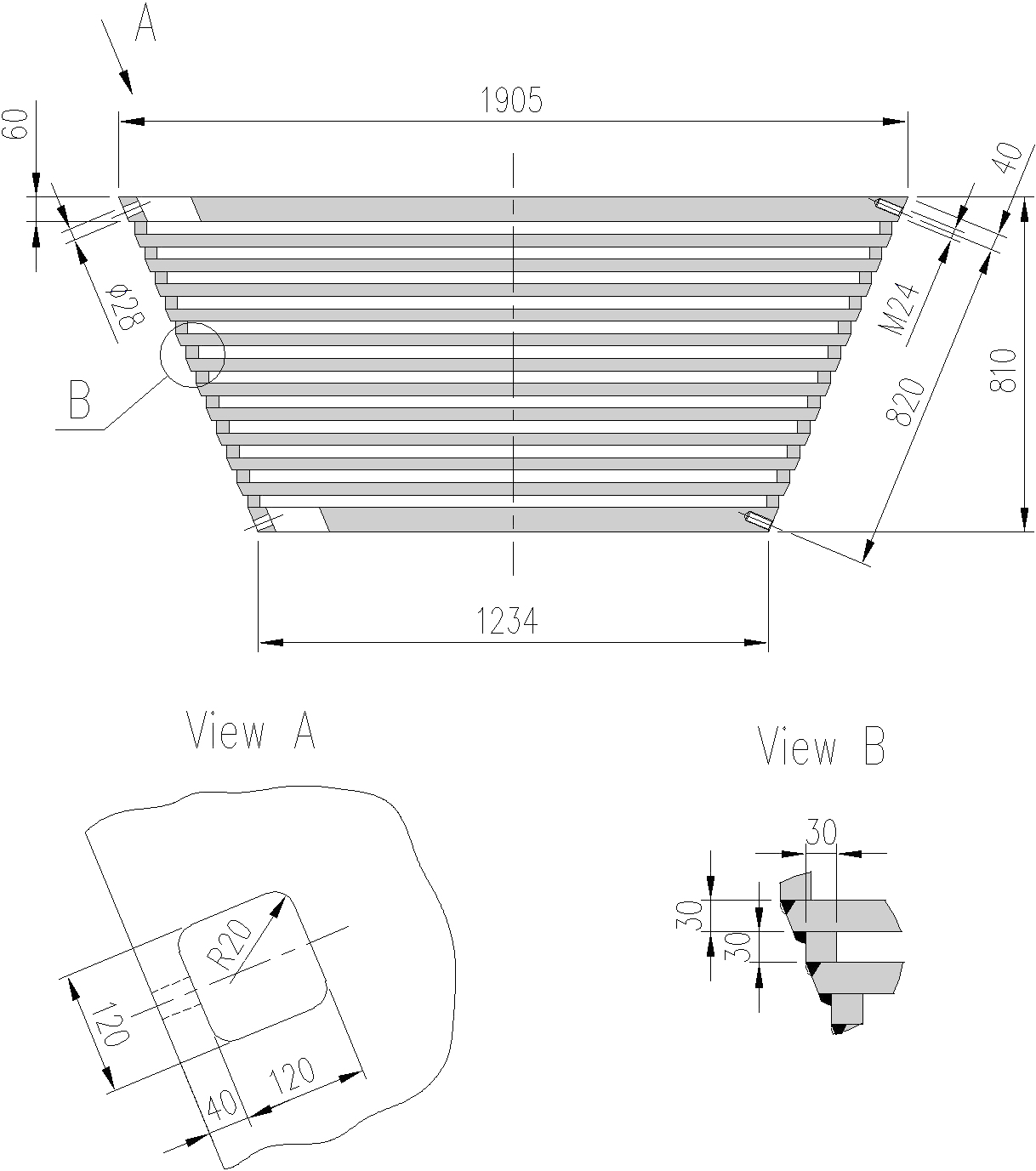}
\caption[Cross-section and 2 details of one beam of the octagonal
  barrel]{Cross-section and 2 details of one beam of the octagonal
  barrel. View A shows a top view of the recess for the mounting of
  the beam to the adjacent neighbour.  View B details the connection
  of the plates.}
\label{yoke:Beam10e}
\end{center}
\end{figure}

The beams will be constructed as stack of plates which will be welded
using spacers of 30\,mm height at both longitudinal edges ({\it i.e.\
}along the $z$ axis).  This will ensure that the beams can be equipped
with detectors in all their section, except a small area around the
corners.  As all connections will be realised along the edge of the
beams, the installation and service of the detectors can be done
independently of the mounting of the yoke.  Please refer to
Fig.~\ref{yoke:Beam10e} for details.  Each beam will be fixed to the
adjacent beam by means of eight bolts at every butt of the inner and
outer plates (Fig.~\ref{yoke:Beam10e}).  Additionally, interlock
connections will be used at the barrel beam interfaces.

\begin{figure}[ht]
\begin{center}
\includegraphics[width=\swidth]{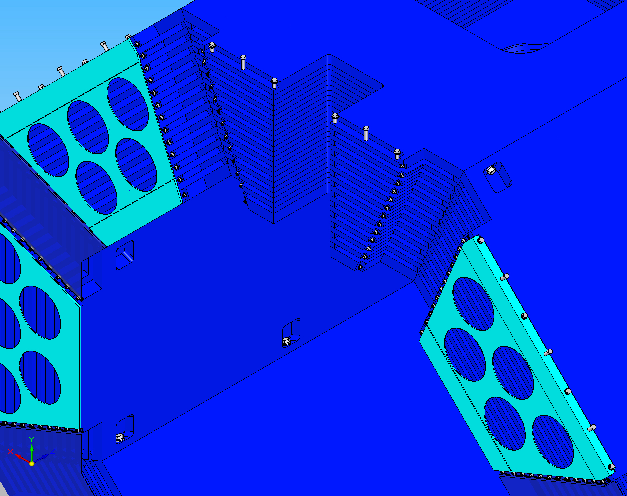}
\caption{Upstream part of the yoke barrel showing the cut-outs for
cables and pipes}
\label{yoke:Cut-outs}
\end{center}
\end{figure}

According to the requirements on the minimum dimensions of the cables
and pipes crossing the barrel part of the flux return (see
Table~\ref{t:req:routing}), proper recesses in the barrel are
foreseen. These recesses will be provided at the corners of the
octagon, both at the upstream and downstream ends of the barrel.  The
size of each recess will be $140 \times 420\unit{mm^2}$. The shape and
positions of the recesses were chosen taking into account the required
space for cables and pipes, the maximum mechanical stability of the
flux return and the minimisation of space inside the yoke that could
not be occupied by muon counters.  At the same time, the influence of
the recesses on the magnetic field in the central tracker region will
be insignificant, as shown in Sec.~\ref{s:sol:perf}.  The recesses
will be closed by lids to protect their contents from during doors
closing and opening. The layout of the upstream part of the yoke
barrel with the cut-outs for cables and pipes is shown in
Fig.~\ref{yoke:Cut-outs}

\begin{table}[ht]
\begin{center}
\begin{tabular}{|c|c|c|c|}
\hline
Layer & Thickness & $r_{min.}$ & $r_{max.}$\tabularnewline
      &  / mm &  / mm &  / mm\tabularnewline
\hline
\hline
1 & 60 & 1490 & 1550\tabularnewline
\hline
2 & 30 & 1580 & 1610\tabularnewline
\hline
3 & 30 & 1640 & 1670\tabularnewline
\hline
4 & 30 & 1700 & 1730\tabularnewline
\hline
5 & 30 & 1760 & 1790\tabularnewline
\hline
6 & 30 & 1820 & 1850\tabularnewline
\hline
7 & 30 & 1880 & 1910\tabularnewline
\hline
8 & 30 & 1940 & 1970\tabularnewline
\hline
9 & 30 & 2000 & 2030\tabularnewline
\hline
10 & 30 & 2060 & 2090\tabularnewline
\hline
11 & 30 & 2120 & 2150\tabularnewline
\hline
12 & 30 & 2180 & 2210\tabularnewline
\hline
13 & 60 & 2240 & 2300\tabularnewline
\hline
TOTAL  & 450 (iron) & 1490 & 2300\tabularnewline
\hline
\end{tabular}
\caption[Table detailing the radial iron lamination
  of the solenoid yoke]{Table detailing the radial iron lamination of
  the solenoid yoke.  All dimensions are given in the radial
  direction, normal to the octagonal face of the yoke, beginning from
  the innermost layer to the outermost.  The layers are counted
  accordingly outward from the centre.}

\label{t:yoke:barrel-lamin}
\end{center}
\end{table}

\begin{figure*}[ht]
\begin{center}
\includegraphics[width=0.8\dwidth]{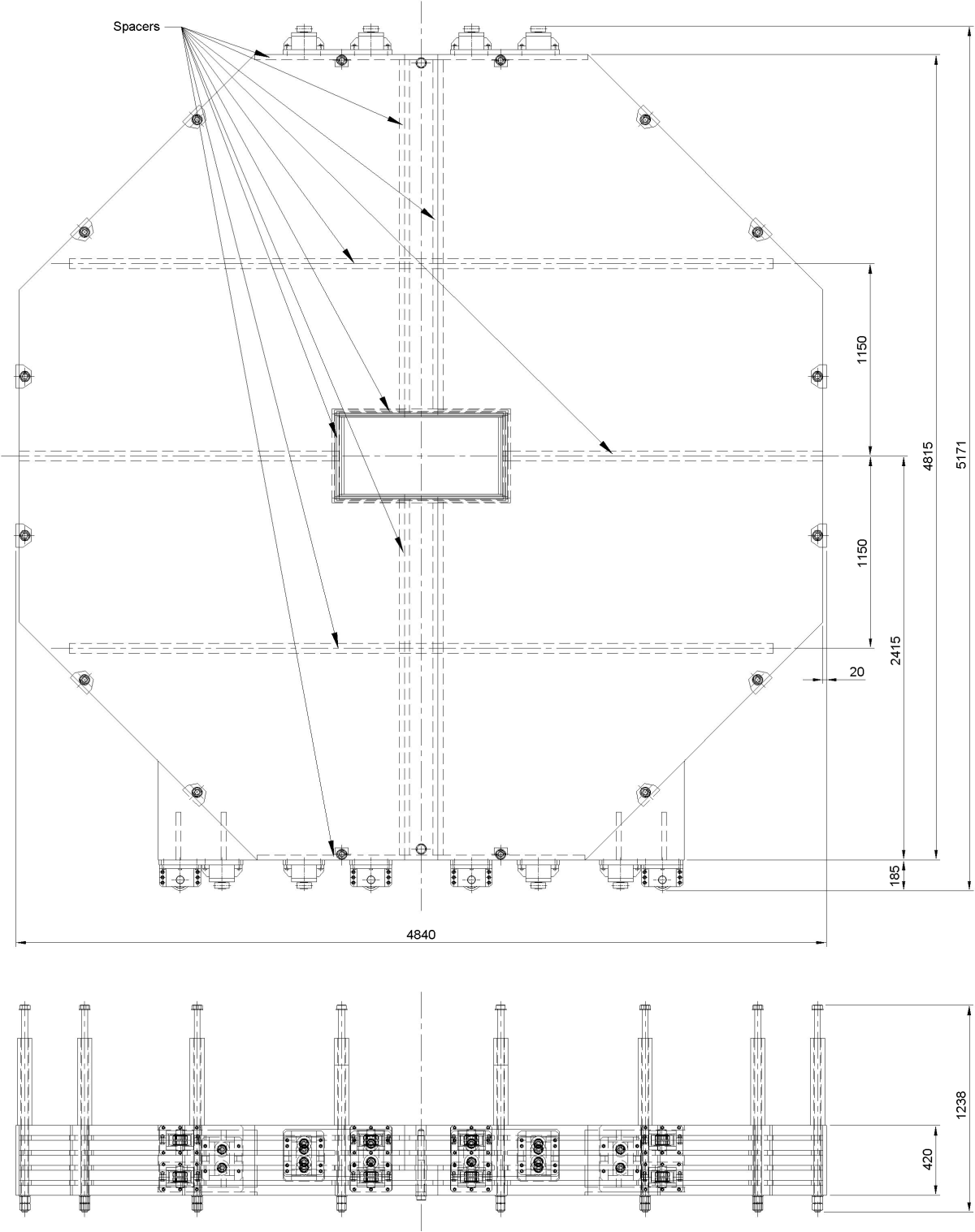}
\caption[Downstream end cap of the flux return yoke]{Downstream
end cap of the flux return yoke. A view in the
  $x-y$ plane is shown in the upper panel, while the lower panel shows
  a top view of the doors and the mounting rods.  The rails for the
  door movement are omitted in this figure for better visibility.}
\label{yoke:DownstreamEndCap10e}
\end{center}
\end{figure*}

\subsubsection{Upstream and Downstream End Doors}

The upstream end cap will consist of 13 steel plates of 30\,mm
thickness consolidated in a package by means of welding trough holes
under pressure.  The beam consolidation of the solid yoke of the ALICE
magnet has been done in the same way.  It will be constructed in two
parts and can be opened to the sides.

An innovative element of the \Panda design is the position of the readout
plane of the DIRC: the photon detectors will be placed inside the iron yoke,
allowing shorter silica slabs and making the design of the readout
volume easier. On the other hand, particular care must be put on the
optimisation of the magnetic field in the volume occupied by the
photon detectors. This constraint and the need to shape the magnetic field
in the central region, led to a very accurate design of the upstream
doors, which where optimised not only to achieve a tolerable field in
the DIRC readout region and to achieve a sufficient uniformity of the
magnetic field in the central tracker volume, but to minimise the
residual axial magnetic force too.

The chosen shape, position and thickness of the upstream end doors
originate from an optimisation process where the main focus was
laid on the following topics:
\begin{itemize}
\item to optimise the field homogeneity in the central tracker
      region, in particular to minimise the integral over the radial
      component in the region of the outer tracker;
\item to minimise the net axial magnetic force on the coil;
\item to reduce the level of induction in the Barrel DIRC readout area
      to a level where the electronic systems can reliably operate.
\end{itemize}

\begin{figure}[th]
\begin{center}
\subfigure[General view]{
\includegraphics[width=\swidth]{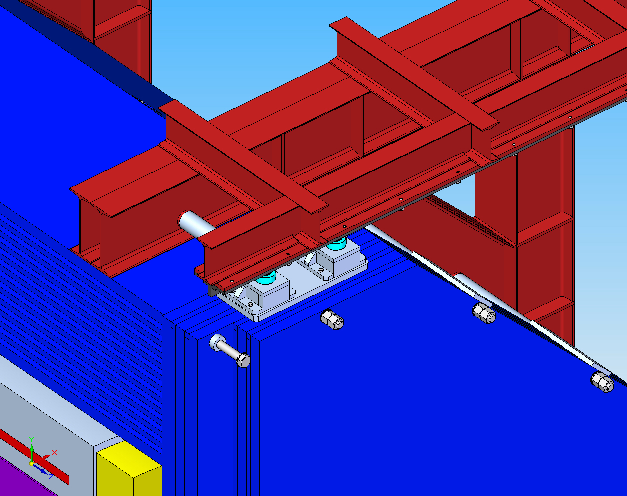}}
\subfigure[Side view]{
\includegraphics[width=\swidth]{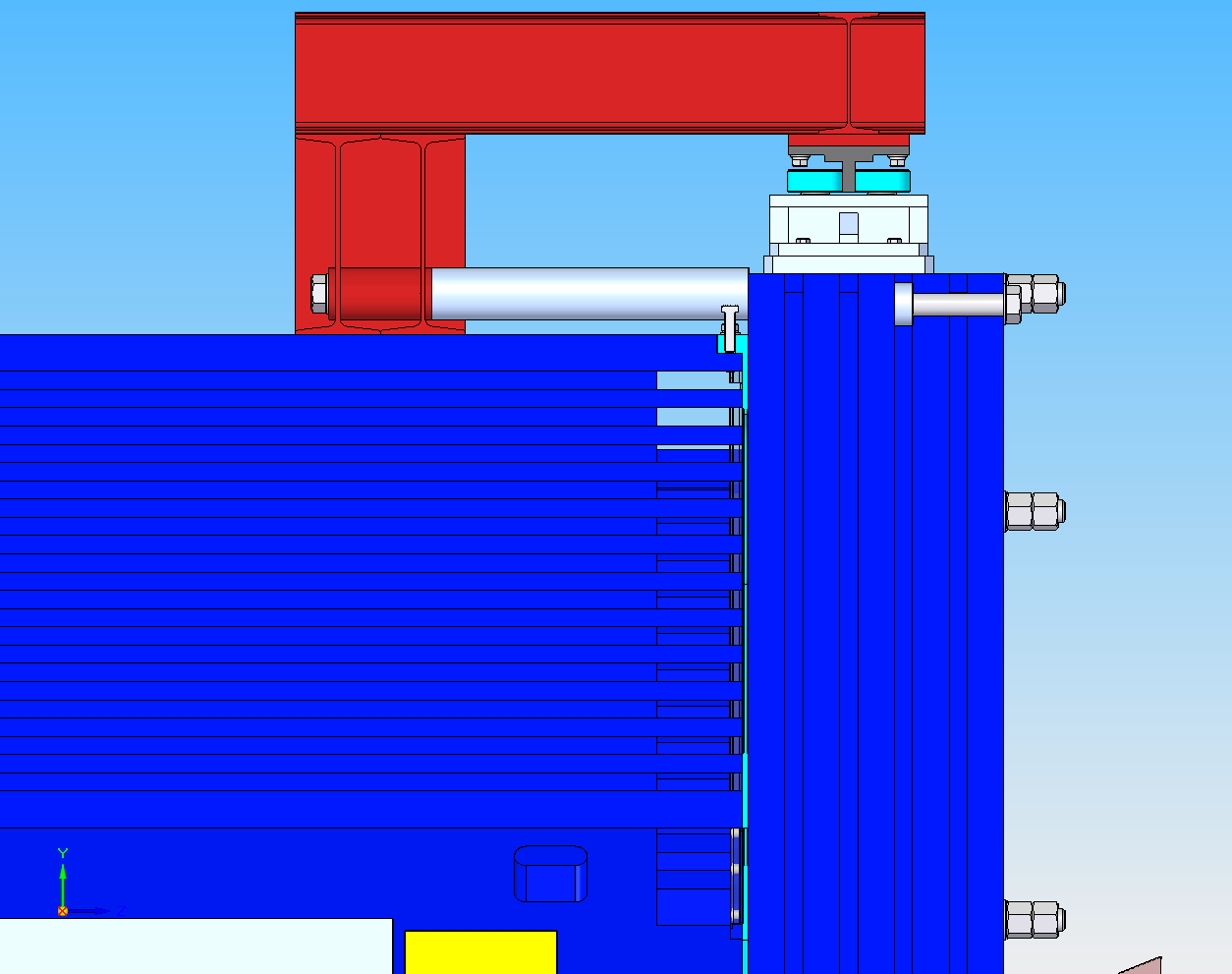}}
\caption[Upper part of the downstream end cap]{Upper part of the
downstream end cap showing the guiding rollers for the door opening}
\label{yoke:Rollers_up}
\end{center}
\end{figure}

\begin{figure}[th]
\begin{center}
\subfigure[General view]{
\includegraphics[width=\swidth]{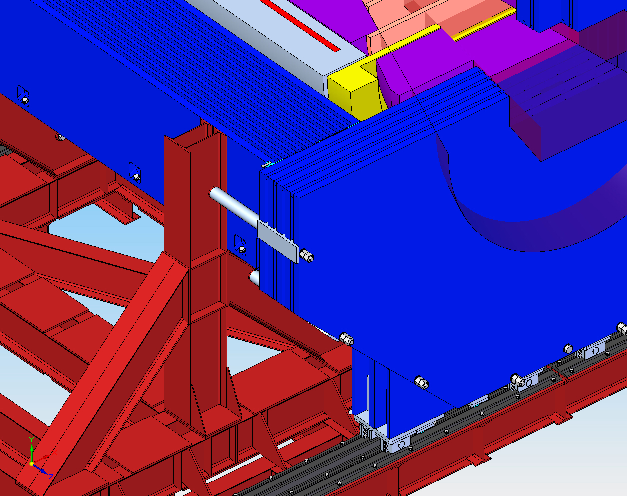}}
\subfigure[Side view]{
\includegraphics[width=\swidth]{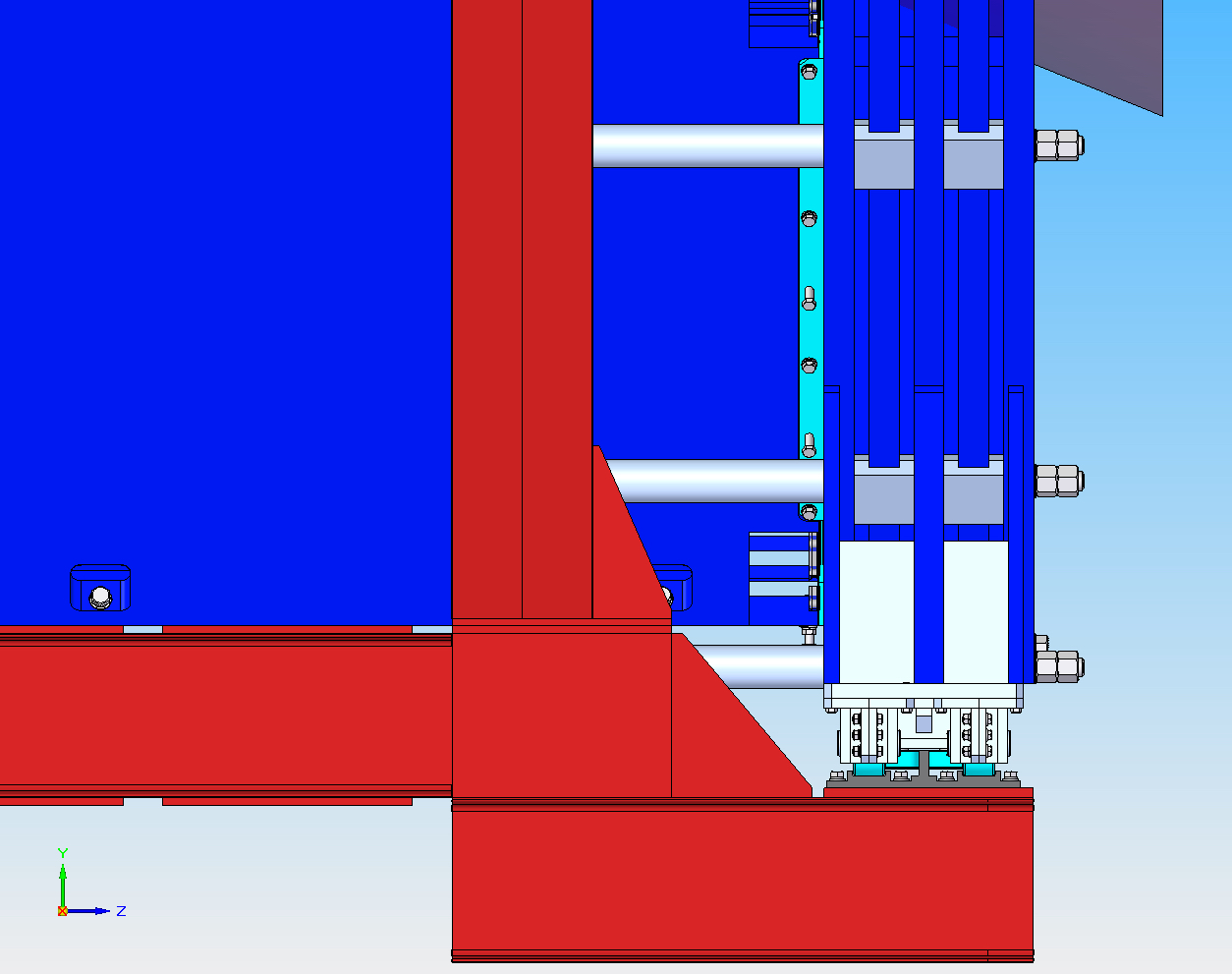}}
\caption[Lower part of the downstream end cap]{Lower part of the
downstream end cap showing the load-bearing wheels for the door
opening} \label{yoke:Rollers_down}
\end{center}
\end{figure}

\begin{table}[th]
\begin{center}
\begin{tabular}{|c|c|c|c|}
\hline
Layer & \multicolumn{3}{c|}{}\tabularnewline
\cline{2-4}
Axially Outward & Thickness & $z_{min.}$ & $z_{max.}$\tabularnewline
from Centre &  / mm &  / mm &  / mm\tabularnewline
\hline
\hline
1 & 60 & 2485 & 2545\tabularnewline
\hline
2 & 60 & 2575 & 2635\tabularnewline
\hline
3 & 60 & 2665 & 2725\tabularnewline
\hline
4 & 60 & 2755 & 2815\tabularnewline
\hline
5 & 60 & 2845 & 2905\tabularnewline
\hline
TOTAL  & 300 & 2485 & 2905\tabularnewline
\hline
\end{tabular}
\caption[The axial iron lamination of the solenoid yoke forward
doors]{The axial iron lamination of the solenoid yoke forward
doors, beginning from the innermost layer to the outermost. }
\label{t:yoke:door-lamin}
\end{center}
\end{table}

This process led to the current design which complies with all
requirements (see also Sec.~\ref{s:sol:req}) and enables to foresee a
fall back solution for the Barrel DIRC, in which the slabs would be
prolonged through the door and the readout box would be placed outside
the flux return yoke.

The downstream doors will provide gaps to accommodate detectors, to be
used as range system for the muons, similar to the barrel.  Each doors
consist of 5 steel plates, each of them being 60\,mm thick. The plates
are separated by means of 30\,mm thick spacers and welded to each
other. The positions of the spacers are shown in
Fig.~\ref{yoke:DownstreamEndCap10e}.  The thickness and position of
the downstream door will conform to the space requirements for the
detectors placed inside it and close to it and to the magnetic field
requirements in the tracker region.  The downstream end cap will also
effectively shield the stray fields of the solenoid into the
Forward Spectrometer; and minimise fields at the location of the
sensitive turbo-molecular pumps between the two spectrometers.  The
axial lamination is summarised in Table~\ref{t:yoke:door-lamin}.

\begin{figure*}[ht]
\begin{center}
\includegraphics[width=\dwidth]{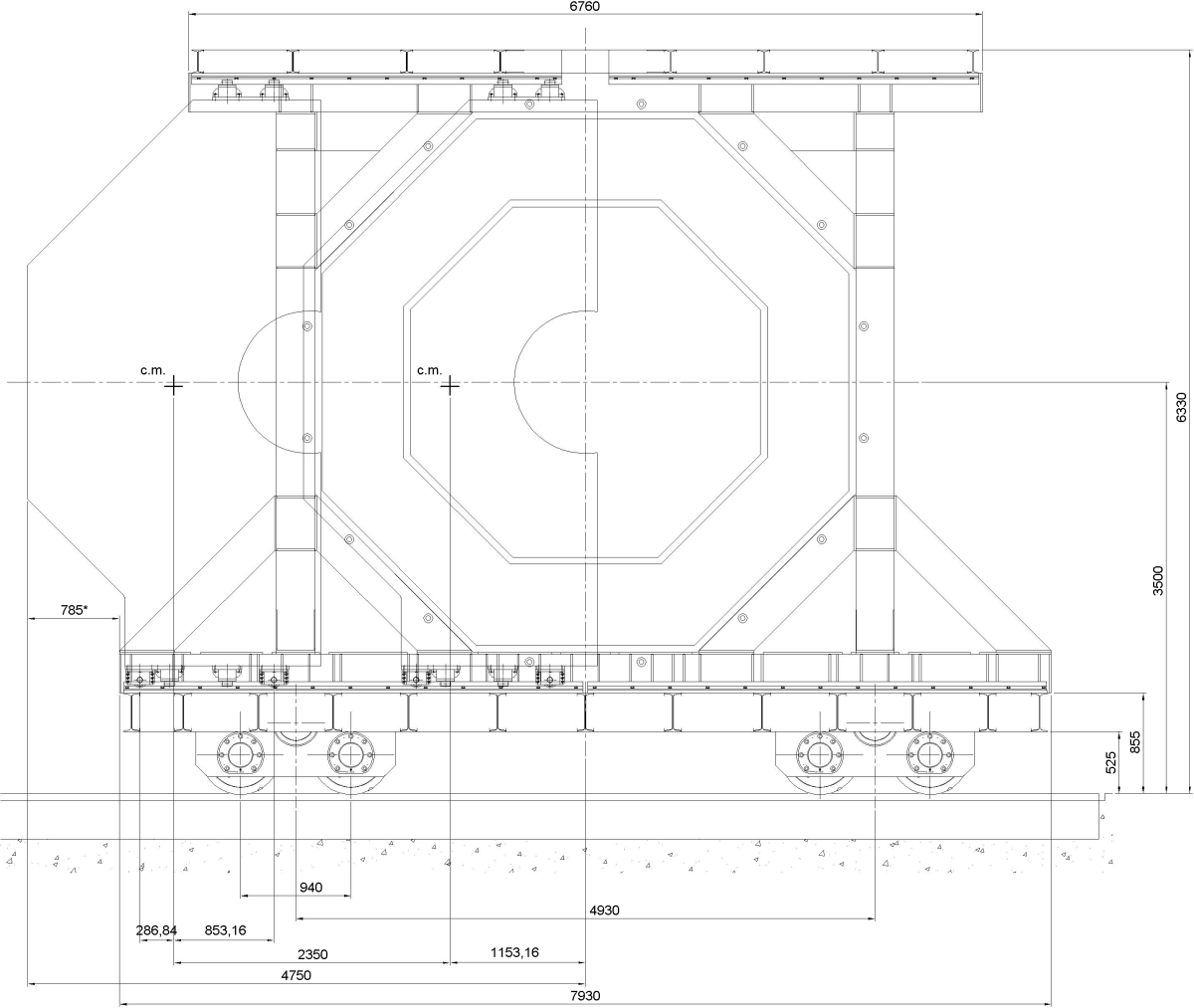}
\caption[Front view of the solenoid yoke]{Front view of the
  solenoid yoke showing the maximum dimensions for the system with
  fully opened door and its centre of mass in both the closed and open
  position.}
\label{yoke:Front_view}
\end{center}
\end{figure*}

\subsubsection{Door Opening System}

Both the upstream and downstream doors will be mounted on rails, so
that they can be opened reliably on a regular basis.  A system of
load-bearing wheels and horizontal guiding rollers will be mounted at
the top and bottom of each door to provide the moving capabilities
(see Figs.~\ref{yoke:Rollers_up} and~\ref{yoke:Rollers_down},
respectively).  The rollers systems for the upstream and downstream
doors will be identical.  In order to ensure a friction-less door opening
spacers will be mounted on the faces of the barrel which can be
(dis)mounted when the doors are closed.  To further ease the opening
procedure a 1 degree opening angle is foreseen with respect to the $x$
axis, {\it i.e.\ }$89^\circ$ w.r.t.~the beam axis.

For safety reasons, the sliding doors will have to be fixed to the
barrel by bolts during the magnet operation as well as during the
transport of the whole solenoid from the assembly area to the in-beam
position or {\it vice versa}.  In the process of assembly and in the
operation position, when the doors may be opened, there will be
additional skid supports to bear the weight of the doors.  The doors
can be opened after removing the attachment bolts.

A front view of the solenoid yoke with dimensions for closed and fully
open doors is shown in Fig.~\ref{yoke:Front_view}.  The suspension has
been designed in a way that stability is guaranteed both with closed and
open doors, while keeping the space occupied by the support structure
minimal for access at the in-beam position.

\subsubsection{Weights of Yoke Parts}

The total weight of the solenoid laminated flux return yoke will be
272\,t. The weight of the carriage and of the space frames will be about
20\,t. The weights of the yoke parts are given in the
Table~\ref{yoke:Weights}. The total weight of the iron for the \PANDA
solenoid will hence be 292\,t.

\begin{figure*}[ht]
\begin{center}
\includegraphics[width=0.7\dwidth]{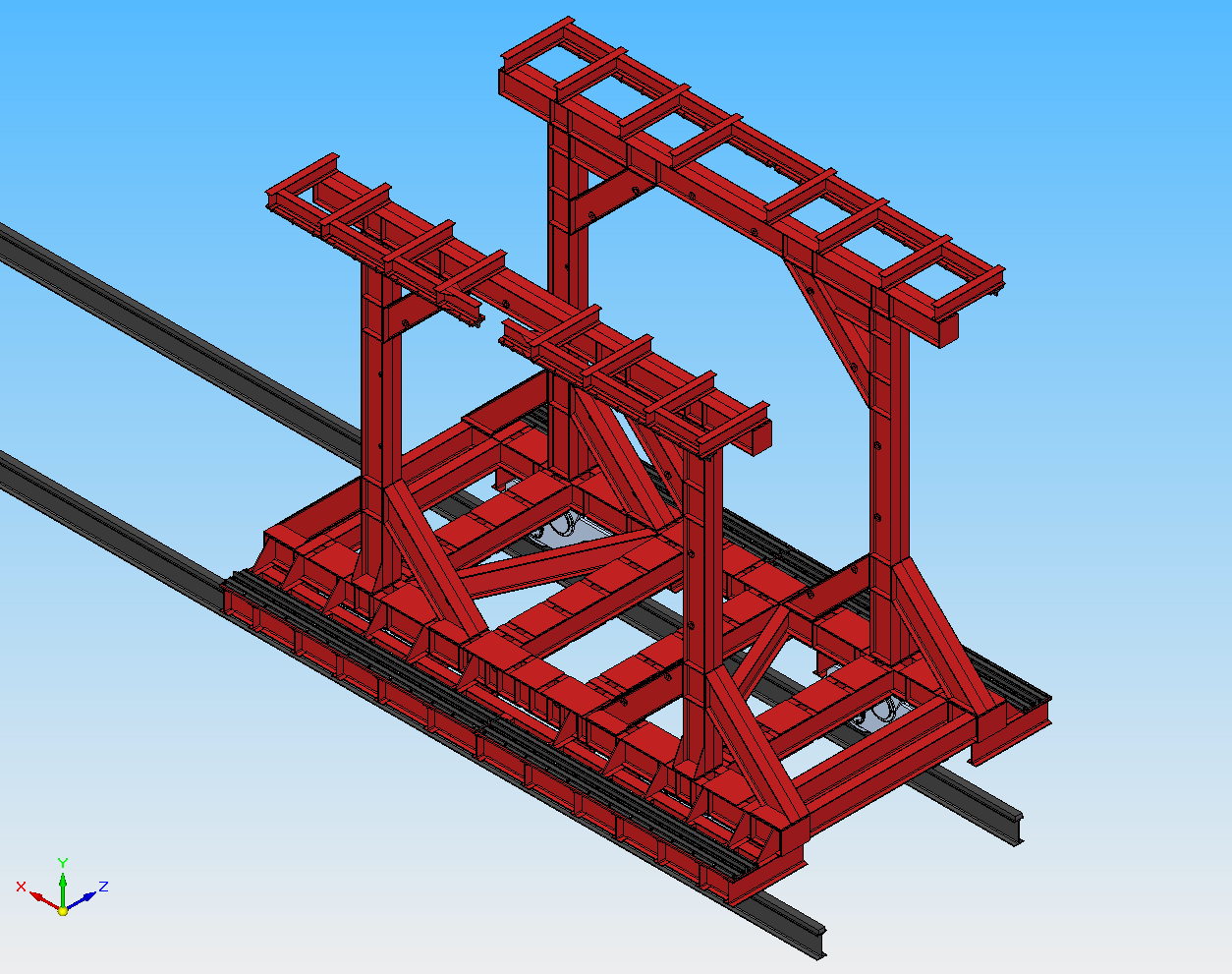}
\caption{Solenoid support frames} 
\label{yoke:Traveling_platform}
\end{center}
\end{figure*}

\begin{table}[ht]
\begin{center}
\begin{tabular}{|l|c|}
\hline
  \hspace{2cm}Item  & Weight  \\
                    & $[$t$]$ \\
\hline
\hline
\multicolumn{2}{|c|}{\bf Upstream door with }  \\
\multicolumn{2}{|c|}{\bf supports, wheels, rollers etc.}  \\
\hline
 Left part   &  23.879  \\
 Right part  &  23.670  \\
 {\bf Total:}  &  {\bf 47.549}   \\
\hline
\hline
\multicolumn{2}{|c|}{\bf Octagonal barrel} \\
\hline
 Upper beam & 20.117 \\
 Inclined side beams (4 pcs.)  &  22.298 \\
 Vertical side beams (2 pcs.)  &  22.298 \\
 Bottom beam  & 20.509 \\
 End face patch plates (16 pcs.) & 0.243 \\
 {\bf Total:} &  {\bf 178.302} \\
\hline
\hline
\multicolumn{2}{|c|}{\bf Downstream door with } \\
\multicolumn{2}{|c|}{\bf supports, wheels, rollers etc.}  \\
\hline
 Left part   &  23.475  \\
 Right part  &  23.061  \\
 {\bf Total:}  &  {\bf 46.536}   \\
\hline
\hline
{\bf Support frames and carriage}
  &  {\bf 19.685} \\
\hline
{\bf Total weight of iron pieces:}
  &  {\bf 292.072} \\
\hline
\end{tabular}
\end{center}
\caption{Summary of weights of the individual yoke parts in tons.}
\label{yoke:Weights}
\end{table}

\subsection{Support and Transport}

\subsubsection{Support Frame}

To ensure the stability of the yoke and to ensure its geometry to
remain precisely under control after long operation and many power
cycles, an outside space frame is necessary.  This will also ensure
that sizable seismic loads can be withstood and hence the
operational safety will be guaranteed.  At the same time, this frame
will allow the attachment of the opening mechanism and suspension of
the doors.  The frame will be welded from standard I-shaped steel beams
and will be fixed by bolts to the outer beam plates.  In addition it will
improve the reproducibility of the solenoid yoke properties
between the tests at the company and the final assembly on site.

The support frame will be integrated with the horizontal beams of the
platform that will carry the total weight of the magnet. The vertical
beams of the frame and the platform are constructed with IPE 330
beams 
(EURONORM 19-57, 330~mm in height, 140~mm in width) connected by arc
welding and by bolts M20. The platform will be placed on 4 railing
wheel carriages for transportation of the solenoid from the assembly
position to the operation position in the experimental hall which is
described below.  A general view of the support frames is shown in
Fig.~\ref{yoke:Traveling_platform}.

\subsubsection{Movement and Positioning}

\AUTHOR{E.~Lisowski, I.~Lehmann}

\paragraph{Conceptual Design}

The design of the movement of the \PANDA solenoid is based on the
considerations in Sec.~\ref{s:sol:req:mech}.  As the whole system will
weigh about 300\,t these requirements pose a serious challenge to the
design of both the support frame as well as the drive system.

A range of different options was considered. The favoured solution is
based on the use of rails and wheels which can be driven
directly in conjunction with a jacking system to align and suspend the
system in both the operation position and the maintenance position.
This solution seems advantageous compared to the use of Hilman rollers
or air cushions.  Both latter options require an additional system
guaranteeing the precise and semi-automatic movement.  In addition,
the rail system is a reliable and conservative solution.  The
evaluation of the details and optimisation studies are still under
way.  Thus, the solution shown here has to be considered as conceptual
only.

\begin{figure}[ht]
\begin{center}
\includegraphics[width=\swidth]{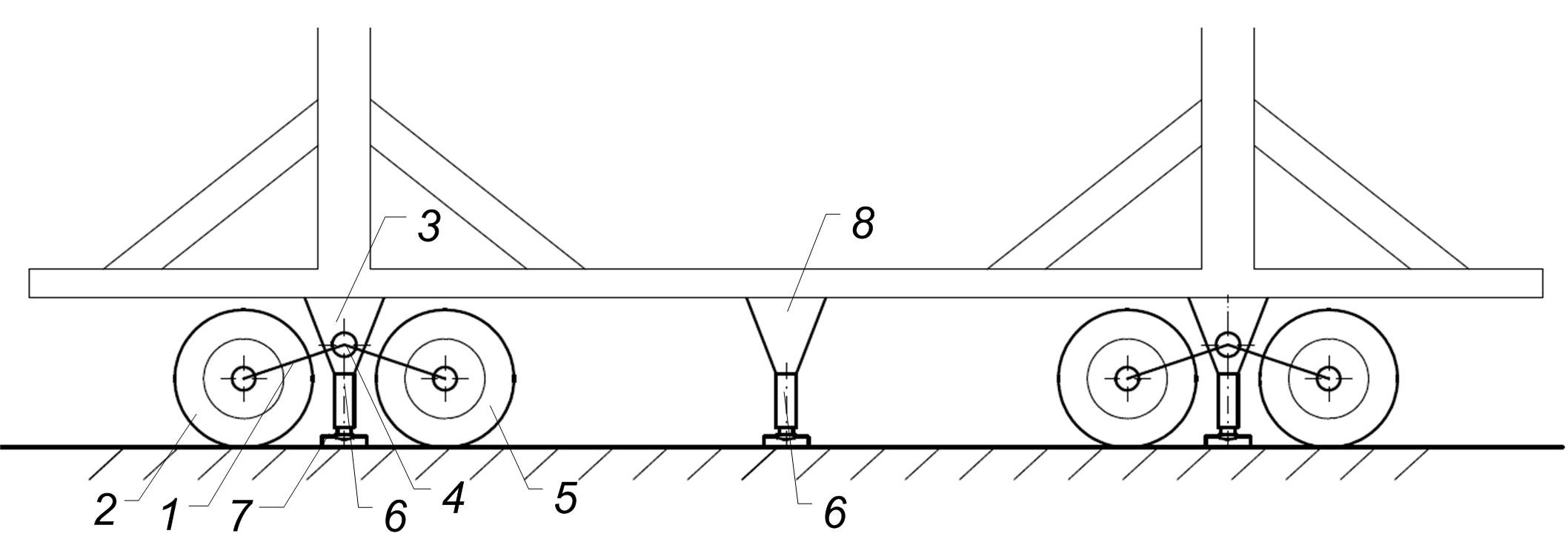}
\caption[Sketch of the concept for the drive and positioning
  system of the Target Spectrometer]{Sketch of the concept for the
  drive and positioning system of the Target Spectrometer seen from
  upstreams.  1~--~carriage, 2~--~wheel, 3~--~carriage support,
  4~--~carriage underframe, 5~--~wheel, 6~--~precise adjustable height
  support, 7~--~lifting structure, 8~--~central support.}
\label{fig:pos_ts:bd}
\end{center}
\end{figure}

Studies using a 2-dimensional Finite Element Model (FEM) have been
used to optimise the arrangement of support points.  A solution with 6
support points at any operational position,{\it i.e.\ }the in-beam
position and the maintenance position, and 4 support points during
movement seemed favourable.  A sketch of the conceptual design is
shown in Fig.~\ref{fig:pos_ts:bd}.  The results of the FEM
calculations for this solution are discussed in the following
(Sec.~\ref{s:yoke:deform}).

The 4 identical carriages will have 2 wheels each and will be located
directly under the vertical beams.  The solution using levers in each
carriage allows for an uniform load on each wheel during the movement.
Once the desired position will have been reached the load will be
taken by 6 precisely adjustable supports, 4 of which will be attached
to the carriages and 2 additional ones at the centres of both main
horizontal beams.  Such, a support will be provided which imposes
minimal deformations to the solenoid at the operational and
maintenance positions. The deformation will stay within tolerable
limits during the movement (see Sec.~\ref{s:yoke:deform}).  The wheels
will be subjected to the load only during the time of the movement.
This will avoid wheel deformations which may occur when subjected to
the load for long periods.  In addition, this system will allow for a
precise positioning of the solenoid with respect to an external
reference frame.  The lifting can be achieved using various solutions.
The favoured solution is the use of computerised hydraulic jacks which
can be mechanically locked as soon as the desired positioning will be
achieved.

\begin{figure}[ht]
\begin{center}
\includegraphics[width=\swidth]{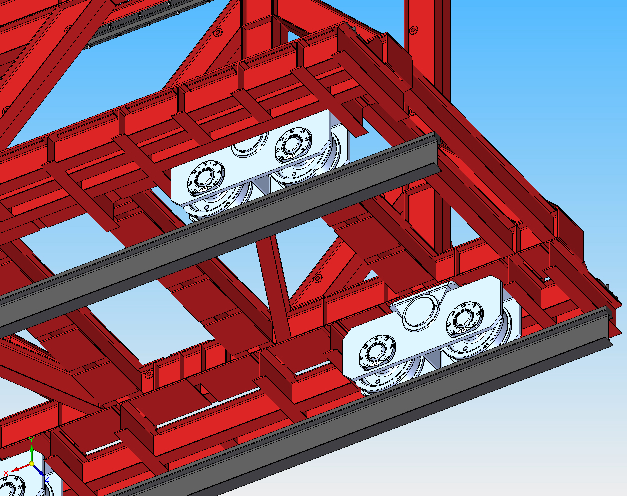}
\caption{Details of solenoid movable platform and transportation carriage} 
\label{yoke:Traveling_platform_detail}
\end{center}
\end{figure}

An artistic view from below the floor which illustrates the placement
of the carriages on the support structure of the solenoid is shown in
Fig.~\ref{yoke:Traveling_platform_detail}.

\begin{figure}[ht]
\begin{center}
\includegraphics[width=\swidth]{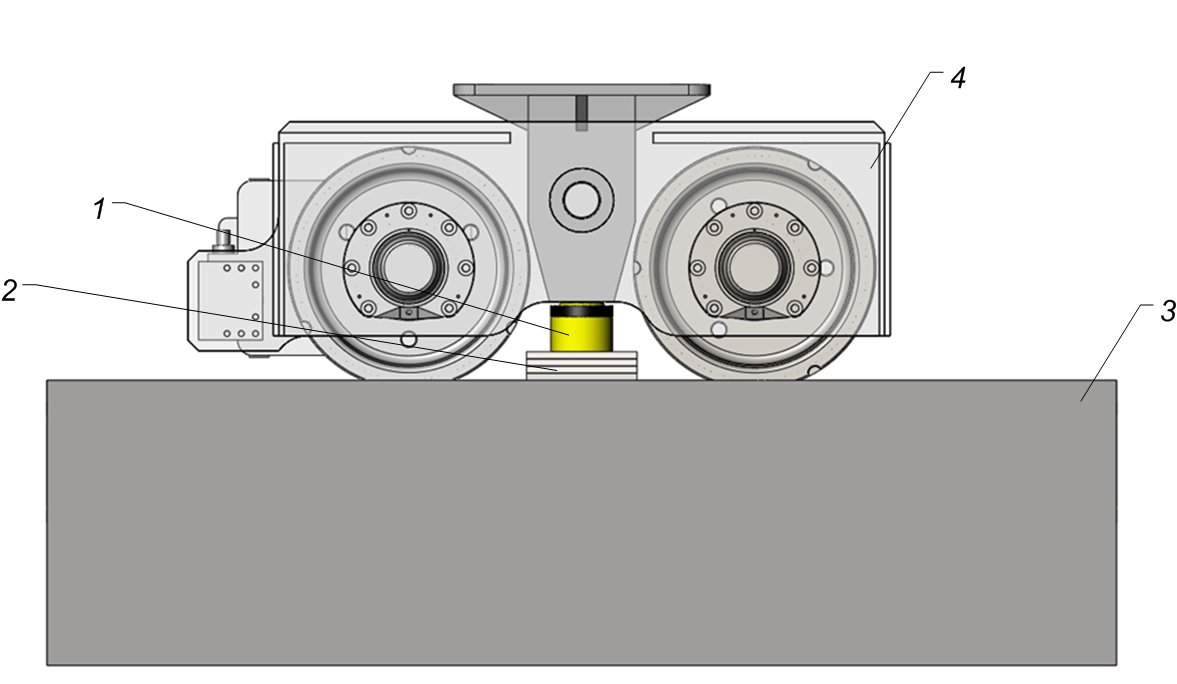}
\caption[View of one carriage perpendicular to the rails at the
  operational position]{View of one carriage perpendicular to the rails at the
  operational position, where the weight will be taken by the jacking
  system: 1~--~hydraulic cylinder, 2~--~steel plates and mounting
  clamp resting on the support profile (invisible), 3~--~foundation,
  4~--~carriage. The front steel plate of the carriage is shown as
  translucent to reveal the wheels and support inside the carriage.
  Gear box, brake and motor are hidden almost completely behind the
  carriage on the left side in this view.}
\label{fig:pos_ts:b_02}
\end{center}
\end{figure}

\paragraph{Carriage Design}
A model of a carriage with hydraulic jack in place is presented in
Fig.~\ref{fig:pos_ts:b_02}.  Each carriage will consist of two wheels
with rims on both sides, a structure with joint (which will be fixed
to the spectrometer frame), a cylinder (which will form the axis joint
for the carriage and the structure), a gearbox, a brake and an
electric motor. This structure will allow to get a uniform load on
every wheel.  Thus it is permissible to use a single motor and brake
for a carriage.  In our design we use 4 identical carriages, {\it
i.e.\ }four electric motors and four gearboxes.  The overall dimension
of each carriage are planned to be $1.97 \times 1.17 \times 0.83$\,m
($L \times W \times H$) including the gearbox and motor with a wheel
diameter of 630\,mm.

\begin{figure}[ht]
\begin{center}
\includegraphics[width=\swidth]{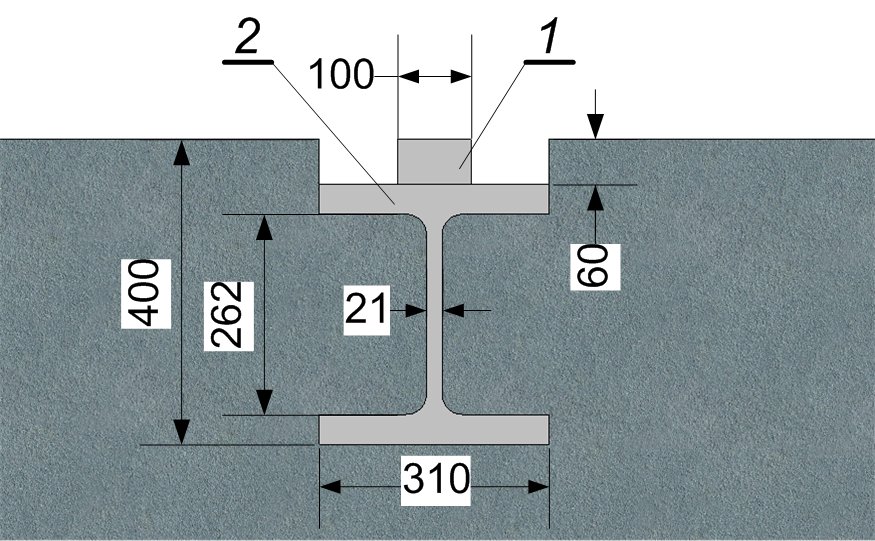}
\caption[Cross section of the support profile and rail]{Cross
section of the support profile (2) and rail (1) inside
  the foundation with dimensions given in millimetres.}
\label{fig:pos_ts:Rail}
\end{center}
\end{figure}

\paragraph{Rail Tracks}
Rails will be placed inside the floor in a way that their upper
surfaces are level to the floor.  The proposed rails are acc.\ DIN
1017, realised in St 70-2/E steel, welded on I profile HE300M acc.\
EN~10034:1994.  Using special clamps covering the rails the whole load
will be transferred to the I-profiles rather than the rails when the
system will be lifted and during any period without frequent movement
(see Fig.~\ref{fig:pos_ts:Rail}).

\subsubsection{Operating Data of Drive System}
The drive system will consist of four electric motors, which motion is
controlled in an open system by an electronic controller. The drive
velocity can be smoothly changed.

The drive system data are anticipated as follows:
\begin{tabbing}
\hspace*{0.25cm}\=\hspace*{4.5cm}\=\hspace*{3cm}\=\hspace*{2.7cm}\= \kill
    \> Wheel numbers   \> 8\\
    \> Driven wheel    \> 4\\
    \> Carriages               \> 4\\
    \> Electric motor  \> 4\\
    \> Spectrometer support \> 6\\
    \> Lifting points \> 4\\
    \> Controllers    \> 1\\
  \> Sliding friction coefficient  \>0.15\\
    \> Adhesion friction coefficient \>0.2\\
    \> Wheel load   \> 375000 N \\
    \> Acceleration \> 0.1 m/s$^2$ \\
    \> Deceleration \> 0.1 m/s$^2$ \\
    \> Speed \> 1.5 m/min \\
    \> Open loop control \> 1 \\
\end{tabbing}

\subsection{Deformations and Stresses}
\label{s:yoke:deform}

\subsubsection{Mechanical Analysis}

An enhanced Finite Element mechanical analysis of the return yoke has been
performed.  The design criteria for the yoke and support frames
correspond to the building norms and codes for steel constructions:
Eurocodes~3~\cite{EuroCode3}.

\subsubsection{Material Properties}

The material to be used for all modules will be low carbon steel AISI
1010 (DIN 1.1121).  The chemical composition of this steel is given in
Table~\ref{yoke:chem_prop}, and the mechanical properties in
Table~\ref{yoke:mech_prop}.  The dependence of the magnetic induction
$B$ on the magnetisation field $H$ in the material is shown in
Fig.~\ref{BH}.

\begin{table*}[ht]
\begin{center}
\begin{tabular}{|c|c|c|c|c|c|c|c|c|}
\hline
  C  &  Si  &  Mn  &  S  &  P   &  Cr  &  Ni  &  Cu  & N$_2$ \\
\hline
 0.07-0.14  &  0.17-0.37  &  0.35-0.65  &  0.04  &  0.035  &
   0.15  &  0.25  &  0.25  &  0.08 \\
\hline
\end{tabular}
\end{center}
\caption[Chemical composition of AISI 1010 steel.]
{Composition of elements alloyed with iron in AISI 1010 steel, in \%.}
\label{yoke:chem_prop}
\end{table*}

\begin{table}[ht]
\begin{center}
\begin{tabular}{|c|c|}
\hline
  Item  &  Value  \\
\hline
 Coefficient of thermal   &   \\
 expansion at 18$^\circ$C, degree$^{-1}$
     &  $1.3\cdot 10^{-5}$  \\
\hline
 Young's modulus, GPa &  198  \\
\hline
 Tensile strength  &   \\
 (annealed sample) $\sigma_u$, MPa  &  340 \\
\hline
 Yield stress of   &   \\
 annealed sample $\sigma_y$, MPa  &  210  \\
\hline
 Elongation, \%  &   29  \\
\hline
\end{tabular}
\end{center}
\caption{Mechanical properties of AISI 1010 steel.}
\label{yoke:mech_prop}
\end{table}

\subsubsection{Magnetic Forces on the Doors}

The axial magnetic force acting on the upstream door will be
$1.85\,$MN in total. The downstream end cap will be subjected to a
total axial magnetic force of $-1.40\,$MN.

The forces acting on the downstream end cap plates are given in
Table~\ref{yoke:down_endcap_forces}.  The strength of the force
imposes additional constraints on reinforcement of the end cap
plates. This reinforcement must not prevent insertion of the muon
chambers.

\begin{table}[ht]
\begin{center}
\begin{tabular}{|c|c|}
\hline
 Plate No. & $F_z\,[\unit{kN}]$ \\
\hline
 1  &  -816.7  \\
 2  &  -379.9  \\
 3  &  -178.1  \\
 4  &  -27.1   \\
 5  &  -0.3    \\
\hline
\end{tabular}
\end{center}
\caption{Axial forces expected to be acting on the iron plates of the downstream
end door.}
\label{yoke:down_endcap_forces}
\end{table}

\subsubsection{Deformation of the Yoke}

Deformation of the yoke in its cross section is calculated for three
load cases:
\begin{enumerate}
\item
gravity load G = G$_{1}$ + G$_{2}$ + G$_{3}$ = 3460~kN, where
G$_{1}$ = 2050 kN -- the weight of the yoke barrel with support
frames, G$_{2}$ = 410~kN -- the weight of the cryostat with
attached detectors and G$_{3}$ = 1000~kN -- the weight of the end caps;
\item
gravity load and magnetic forces;
\item
gravity load, magnetic forces and additional vertical and horizontal
  seismic loads $P_y = 0.11\unit{G}$, $P_x = 0.15\unit{G}$.
\end{enumerate}

\begin{figure}[ht]
\begin{center}
\includegraphics[width=\swidth]{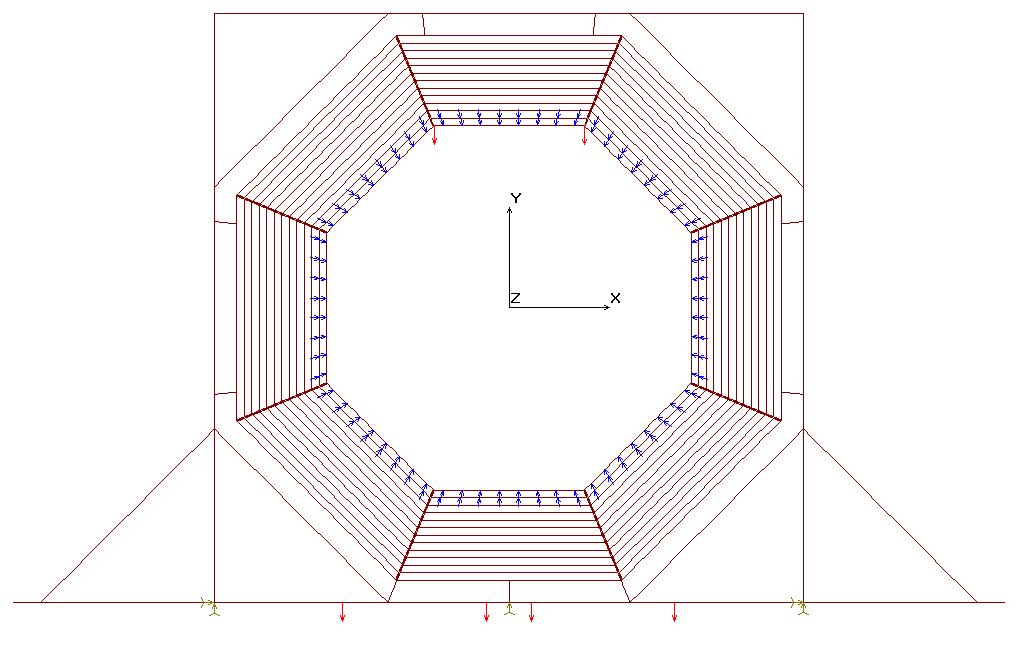}
\caption[2D FE yoke model]{2D FE yoke model showing application of
the doors, cryostat and detectors weight (red arrows), magnetic
pressure (blue arrows), outer support frame and fixation points at
the basement} \label{yoke:Yoke_model}
\end{center}
\end{figure}

\begin{figure}[ht]
\begin{center}
\includegraphics[width=\swidth]{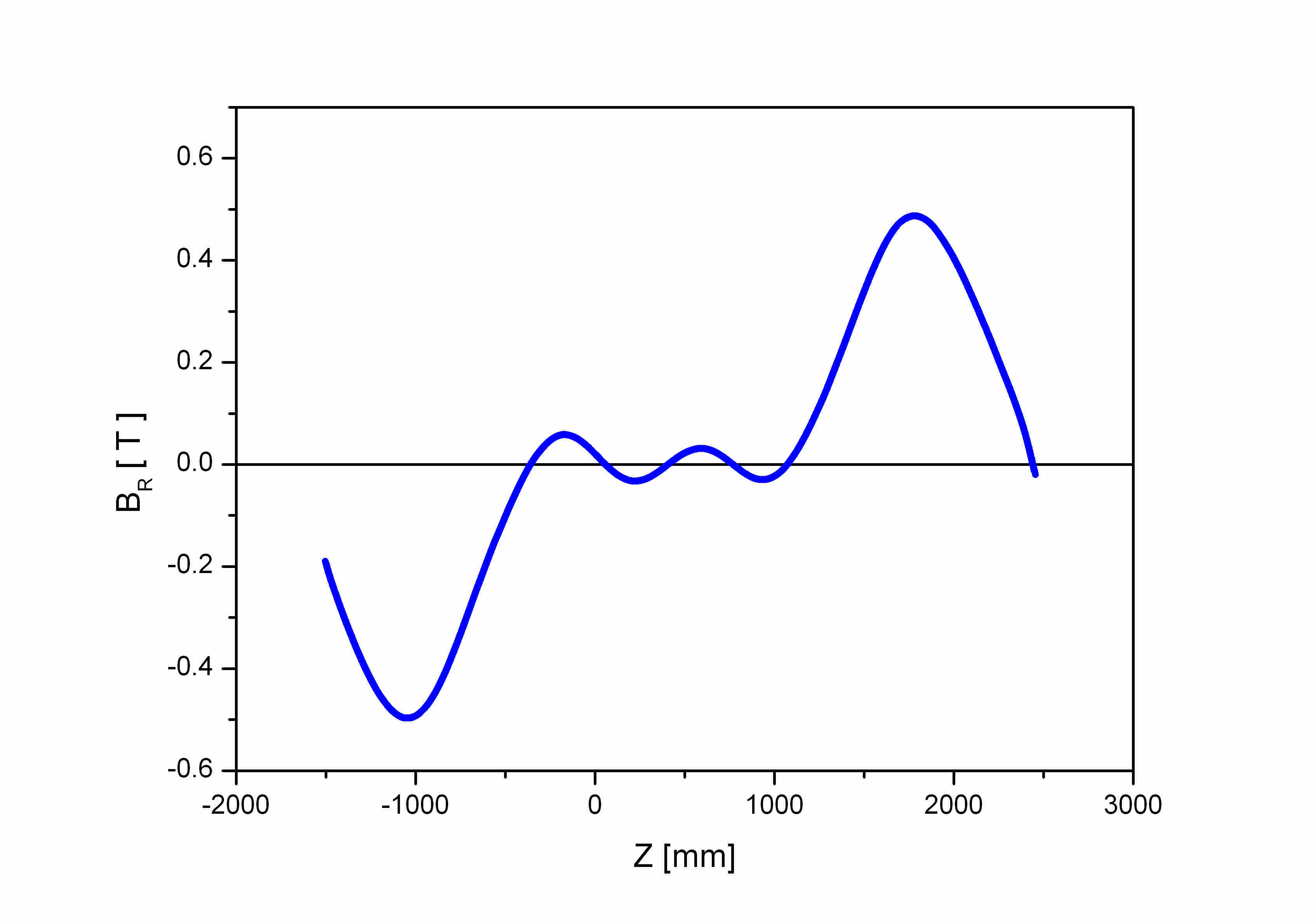}
\caption{Distribution of $B_R$ along the inner surface of the yoke
beam ($R=1490\,mm$)} \label{yoke:Br(z)_r1490}
\end{center}
\end{figure}

Both 2D and 3D finite-element models (FEM) were used to calculate the
deformations and stresses. The 2D yoke model is shown in
Fig.~\ref{yoke:Yoke_model}.  It should be noted here that the
simplifications which went into this model ensure that the results can
be treated as a worst case scenario, and the full system would deform
no more than this model predicts.  The iron yoke plates are loaded by
magnetic pressure, non-uniformly distributed over their surface
according to the radial component of magnetic induction (see
Fig.~\ref{yoke:Br(z)_r1490}).

\begin{figure}[htb]
\begin{center}
\includegraphics[width=\swidth]{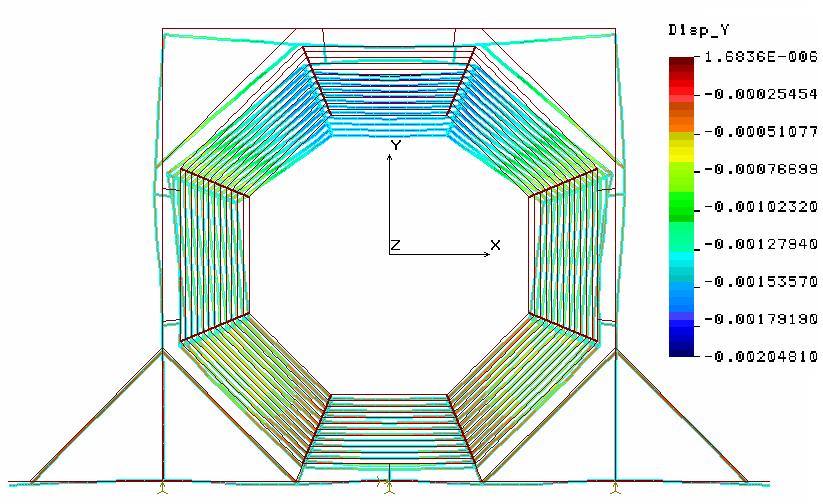}
\caption[Deformation of the yoke due to gravity in the stationary
  positions]{Deformation of the yoke due to gravity in the stationary
  positions, {\it i.e.\ }using 6 support points.}
\label{yoke:Deform_whole2D_3point}
\end{center}
\end{figure}

\paragraph{Deformation due to Gravity}

The yoke frame will be placed on the platform, which will be supported
at 6 points on its footprint in the solenoid assembly and final
positions.  It could be shown that the deformations will be minimal
when placing the support points under the vertical beams and the
centre of yoke at both support frames.

In the stationary position, the yoke gravity load will result in the
maximal vertical height change within the yoke, which is calculated to
be $\Delta y = 1.3\unit{mm}$.  The maximal horizontal change of width
inside the yoke will be $\Delta x = 2.4\unit{mm}$ (see also
Fig.~\ref{yoke:Deform_whole2D_3point}).

\begin{figure}[htb]
\begin{center}
\includegraphics[width=\swidth]{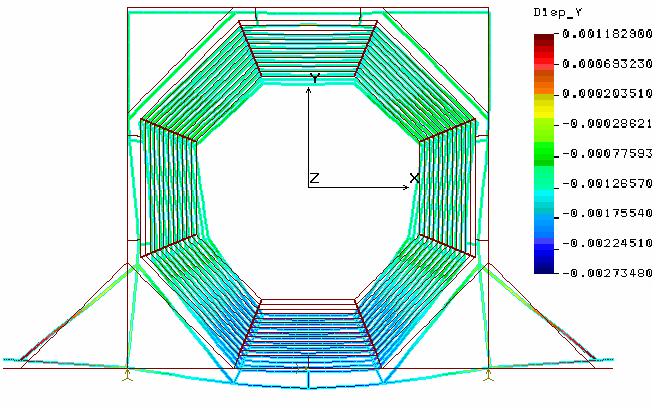}
\caption[Deformation of the yoke due to gravity during the
movement]{Deformation of the yoke due to gravity during the
movement,
  {\it i.e.\ }using 4 support points.}
\label{yoke:Deform_whole2D_2point}
\end{center}
\end{figure}

When the supports will be removed in order to allow the movement of
the system, the platform will remain on the wheels for transportation.
In effect this means that the two central support points will be
released while the outer 4 points will stay under load (see
Fig.~\ref{yoke:Deform_whole2D_2point}).  In this case the situation
will be less ideal than in the stationary position but the
deformations will remain within tolerable limits.  We still
investigate additional reinforcements of the support frame which may
reduce the deformations further.

When the solenoid is being transported, the doors have to be
closed and rigidly attached to the barrel part. After the magnet
is moved to its operation position at the beam line, the platform
is supported on the supports again.

\paragraph{Deformations due to Magnetic and Seismic Loads}

The first (inner) plate will be loaded by a magnetic force of
$105.8\unit{kN}$, the second plate by $20.9\unit{kN}$. The maximal
vertical and horizontal deformations due to gravity load and magnetic
pressure are calculated to be: $\Delta y = 1.9\unit{mm}$, $\Delta x
= \pm 1.1\unit{mm}$.

\begin{figure}[htb]
\begin{center}
\includegraphics[width=\swidth]{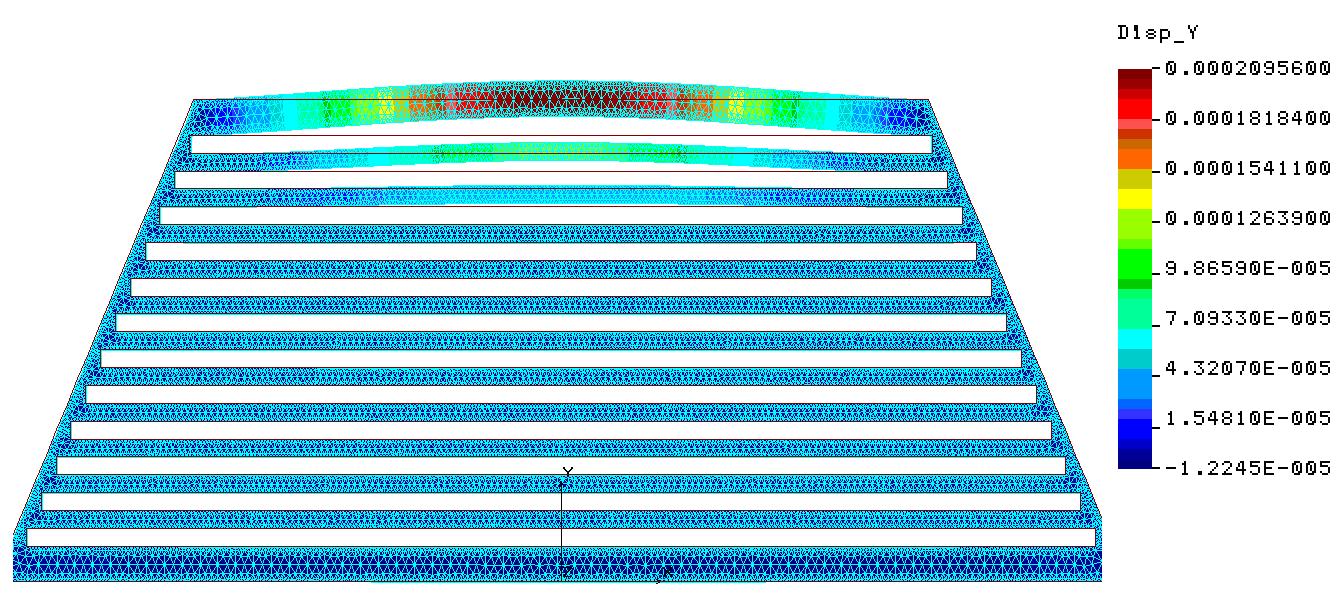}
\caption{Deformation of the yoke beam plates due to magnetic
pressure.} \label{yoke:Beam_deform_mag}
\end{center}
\end{figure}

The estimated deformations of the yoke barrel beam due to magnetic
pressure (additional deformations when the magnet will be switched on)
are shown in Fig.~\ref{yoke:Beam_deform_mag}. The deformations are
calculated at the cross-section near the barrel ends in the area where
the magnetic pressure will be maximal. The deformation will be less
than 0.2\,mm. It will not change the positions of the octagon corners,
but the shape of the inner plates of the beam.

\begin{figure}[htb]
\begin{center}
\includegraphics[width=\swidth]{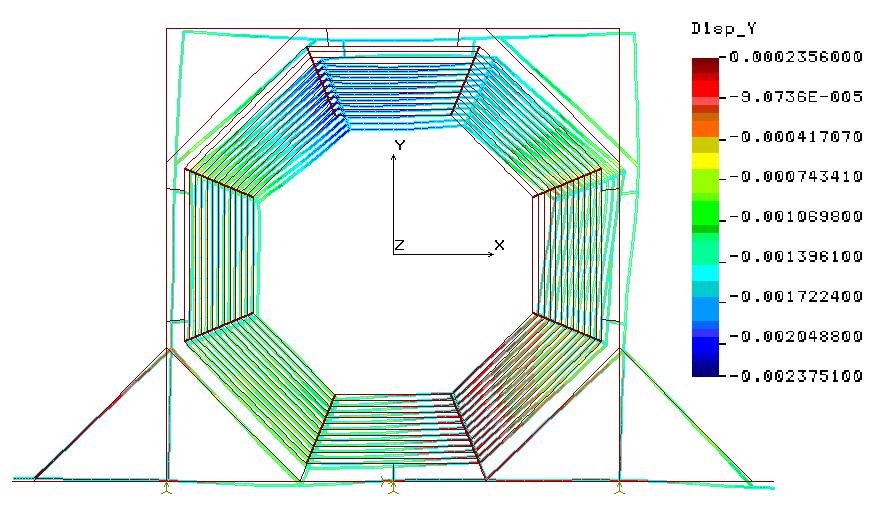}
\caption{Deformation of the yoke due to gravity, magnetic and
seismic loads.} \label{yoke:Weight+mag+seism}
\end{center}
\end{figure}

In case of combined action of gravity, magnetic and seismic loads
the maximal vertical height change within the yoke will be $\Delta y =
1.8\unit{mm}$. The maximal horizontal change of width inside the yoke
will be $\Delta x = 2.7\unit{mm}$ (see Fig.~\ref{yoke:Weight+mag+seism}).

In our design the outer support frame significantly increases the
general rigidity of the yoke.  Not using the frame would lead to a
yoke deformation $\Delta y >7\unit{mm}$.

\begin{figure}[th]
\begin{center}
\includegraphics[width=\swidth]{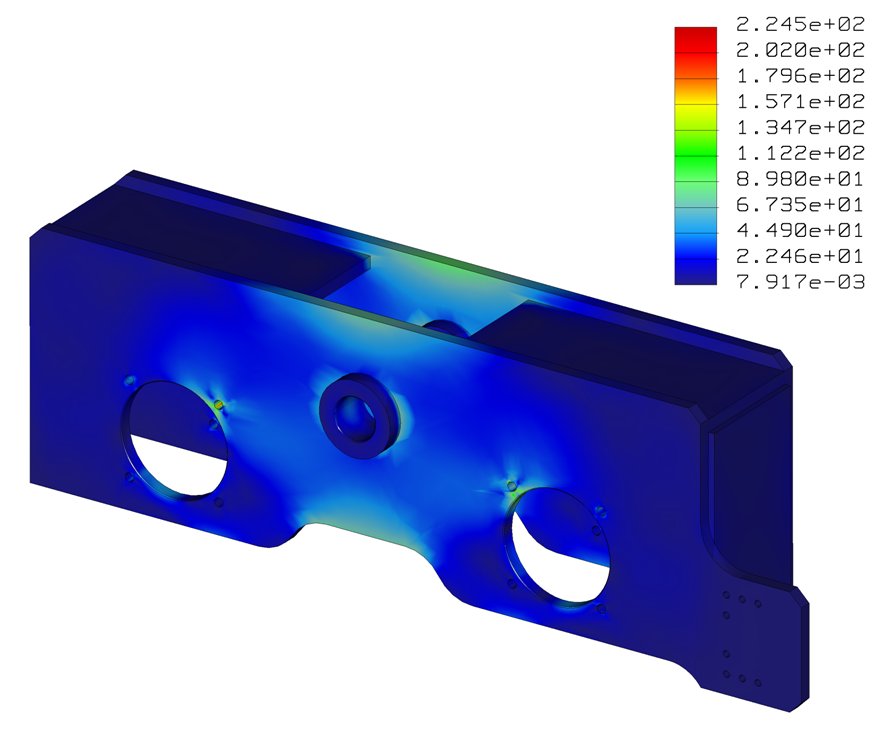}
\caption{Von Mises stress in the body of carriage in MPa}
\label{fig:pos_ts:vms_b_01}
\end{center}
\end{figure}

\begin{figure}[th]
\begin{center}
\includegraphics[width=\swidth]{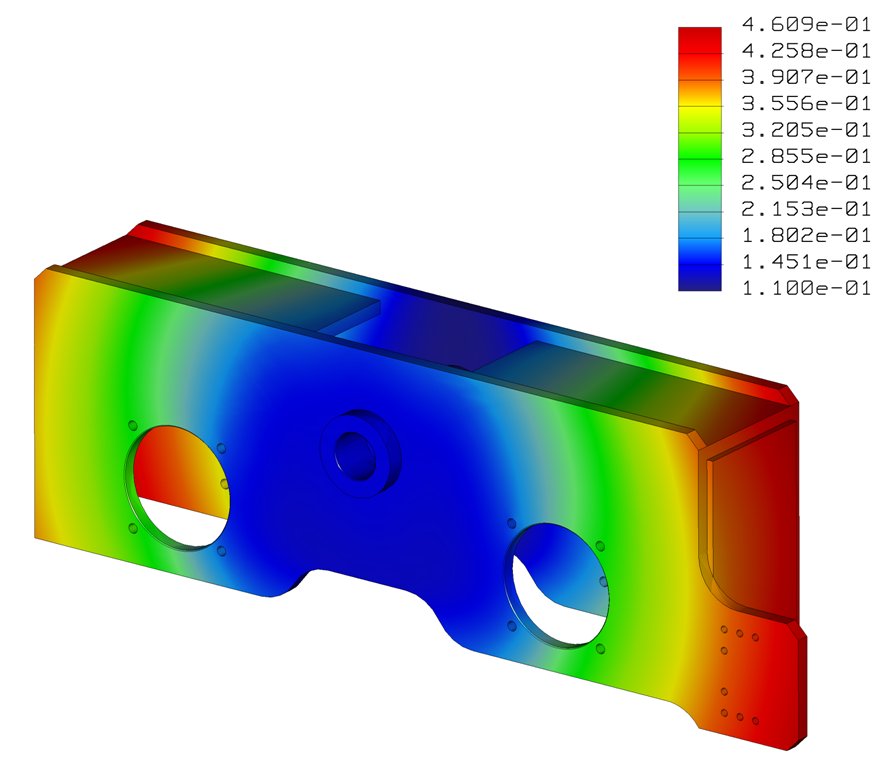}
\caption{Total deformation of the body of the carriage in mm}
\label{fig:pos_ts:td_b_01}
\end{center}
\end{figure}

\begin{figure}[th]
\begin{center}
\includegraphics[width=\swidth]{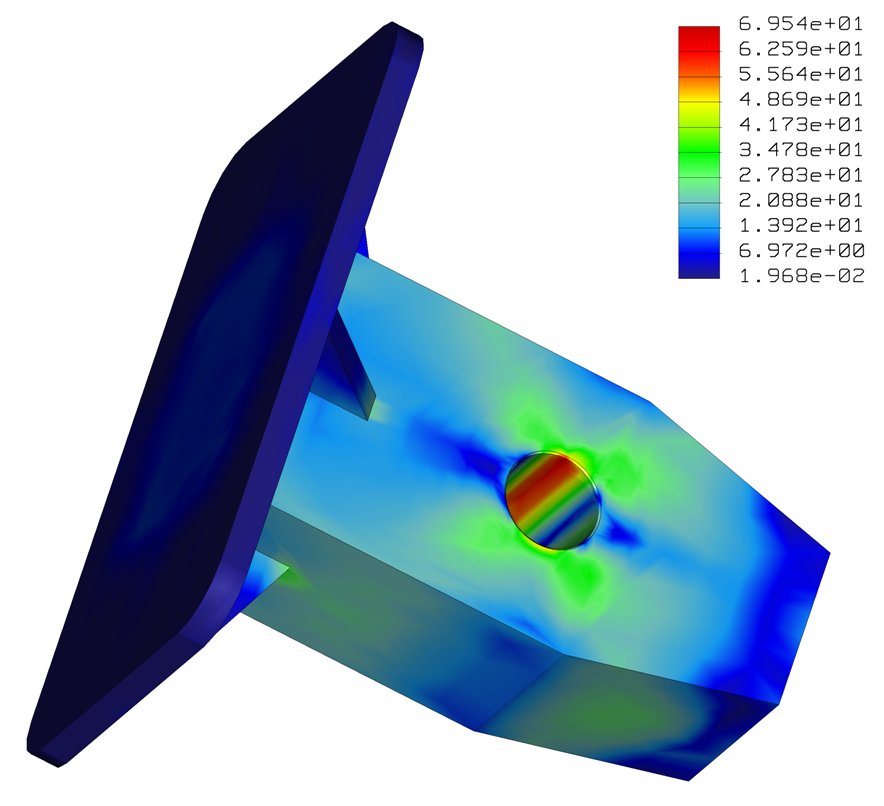}
\caption{Von Mises stress in the central support of carriage in MPa}
\label{fig:pos_ts:vms_support}
\end{center}
\end{figure}

\subsubsection{Rail Carriages and Supports}
\AUTHORS{E.~Lisowski, I.~Lehmann}

Full 3D FEM calculations were performed for
all crucial components of the carriage and support system, {\it
i.e. }the frames of the carriages with their hinge and support
structures, the wheels and their impact on the rail/I-profile, and the
clamps used to suspend the weight during operation and their impact on
the rail/I-profile.

\paragraph{Frame of the Carriage and Central Support}
When the solenoid is to be moved the weight will rest on the 4 carriages.  The
following data were used in the FEM analysis:
\begin{enumerate}
    \item{force acting on support $F=80\unit{tonnes}$;}
    \item{contact connection between pin and support and between
    pin and frame of the carriage;}
  \item{material: steel.}
\end{enumerate}

FEM results for the carriage and the central support are shown in
Figs.~\ref{fig:pos_ts:vms_b_01}, \ref{fig:pos_ts:td_b_01}
and \ref{fig:pos_ts:vms_support}, respectively. Stress values are
below permissible levels, and the maximum displacement does not exceed
0.5\,mm.

\paragraph{Wheel and Rail}
The 8 wheels will transfer the weight to the rails.  The following data
were used in FEM analysis:
\begin{enumerate}
    \item{force acting on support $F=40\unit{tonnes}$;} \item{contact
    connection between rail and wheel and between wheel and its axis
    and support and pin with frame of the carriage;} \item{material:
    steel.}
\end{enumerate}

\begin{figure}[ht]
\begin{center}
\includegraphics[width=\swidth]{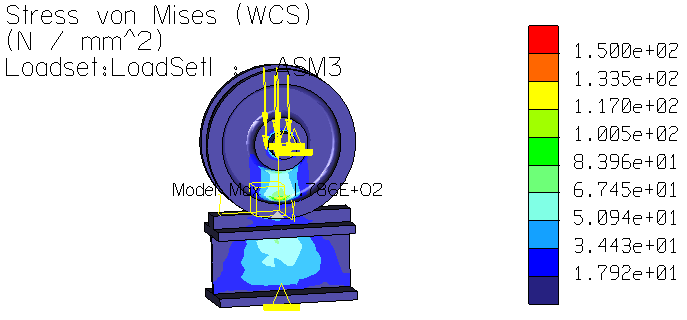}
\caption{Von Mises stress for wheel-rail model in MPa}
\label{fig:pos_ts:vms_r_01}
\end{center}
\end{figure}

\begin{figure}[ht]
\begin{center}
\includegraphics[width=\swidth]{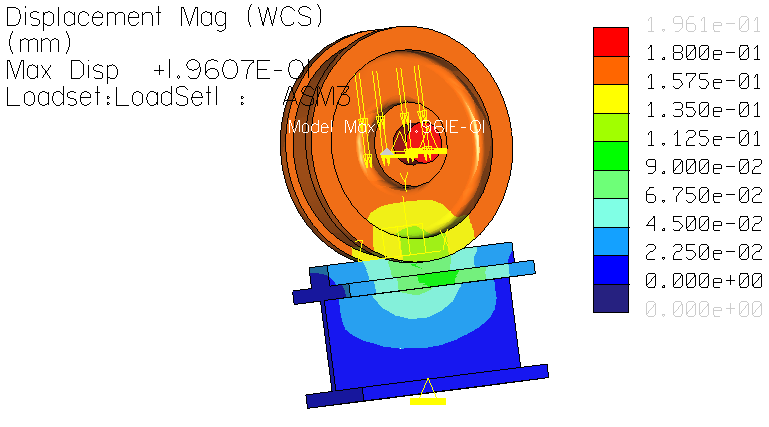}
\caption{Total displacement in mm}
\label{fig:pos_ts:td_r_01}
\end{center}
\end{figure}

FEM results are shown in Fig.~\ref{fig:pos_ts:vms_r_01} and
Fig.~\ref{fig:pos_ts:td_r_01}. Stress values are below the permissible
levels and the maximum displacement is $\sim0.1\unit{mm}$.

\paragraph{Clamp on I Profile}
A clamp which will rest on the I profile below the rail will be used whenever
the system is supported on jacks.  The following data were used in FEM
calculations:
\begin{enumerate}
    \item{force operating acting on support $F=80\unit{tonnes}$;}
    \item{contact connection between clamp and rail;}
  \item{material: steel.}
\end{enumerate}

\begin{figure}[ht]
\begin{center}
\includegraphics[width=\swidth]{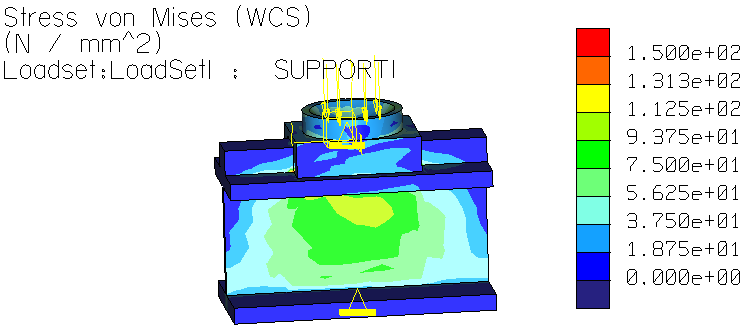}
\caption{Von Mises stress in rail and support}
\label{fig:pos_ts:vms_rs_01}
\end{center}
\end{figure}

\begin{figure}[ht]
\begin{center}
\includegraphics[width=\swidth]{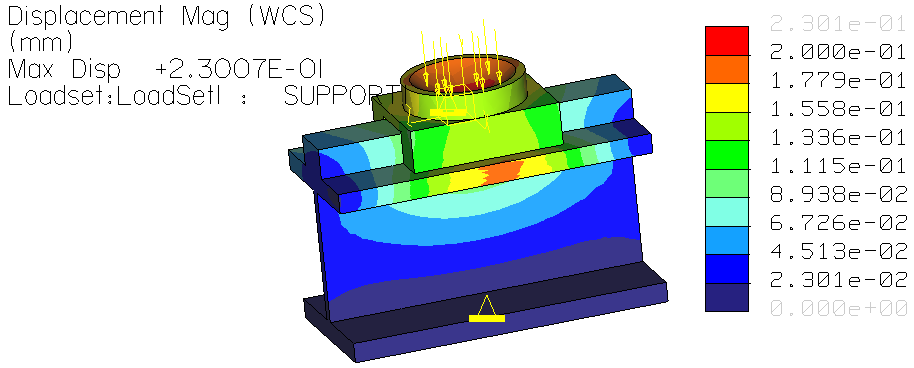}
\caption{Total displacement in rail and support}
\label{fig:pos_ts:td_rs_01}
\end{center}
\end{figure}

FE results are shown in Fig.~\ref{fig:pos_ts:vms_rs_01} and
Fig.~\ref{fig:pos_ts:td_rs_01}. Stress values are below the
permissible stress levels and the maximum displacement is
$\sim0.2\unit{mm}$.

\subsubsection{Deformation of the Upper Yoke Beam}

\begin{figure}[ht]
\begin{center}
\includegraphics[width=\swidth]{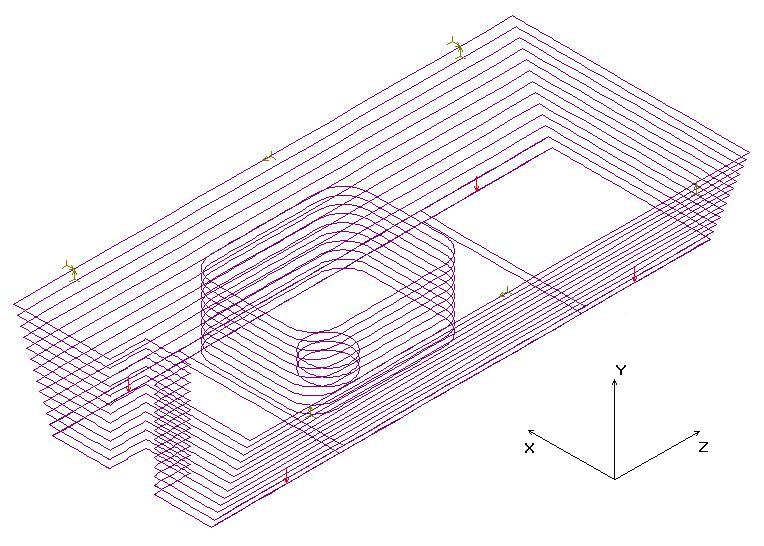}
\caption[3D FE model of the upper yoke beam]{3D FE model of the
upper yoke beam showing the points of application of cryostat and
detectors weight and points of fixation} \label{yoke:Beam_model}
\end{center}
\end{figure}

The 3D model of the upper beam includes 13 steel plates with recesses
for the target and for the chimney (Fig.~\ref{yoke:Beam_model}). In
this model the plates are consolidated by means of
$30\times30\unit{mm}$ steel spacers along the side surfaces of the
beam using the weld seam. The magnetic pressure is applied to the
first two inner plates of the beam. Besides that, the beam is loaded
by the axial force $F_z = 0.23\unit{MN}$ acting at the beam butt-ends
from the yoke end caps.

\begin{figure}[ht]
\begin{center}
\includegraphics[width=\swidth]{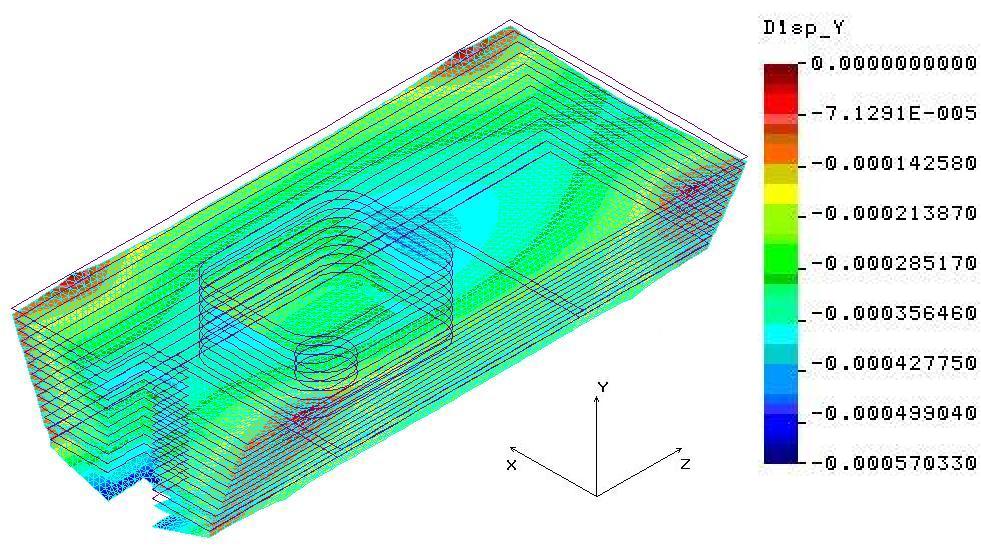}
\caption{Upper beam deformation due to gravity and magnetic
forces}
\label{yoke:Beam_weight+mag}
\end{center}
\end{figure}

The results of computations of deformations due to gravity (without
cryostat) show that the maximal deflection is $0.3\unit{mm}$.  The
deformations due to gravity (with cryostat and detectors) and magnetic
forces are shown in Fig.~\ref{yoke:Beam_weight+mag}.  The resulting
maximal vertical displacement is $0.6\unit{mm}$. That means that the
beams will rigid enough; their sags will be comparable with the
accuracy of manufacturing.

\subsubsection{Stresses in the Yoke and Support Frame}

\begin{figure}[ht]
\begin{center}
\includegraphics[width=\swidth]{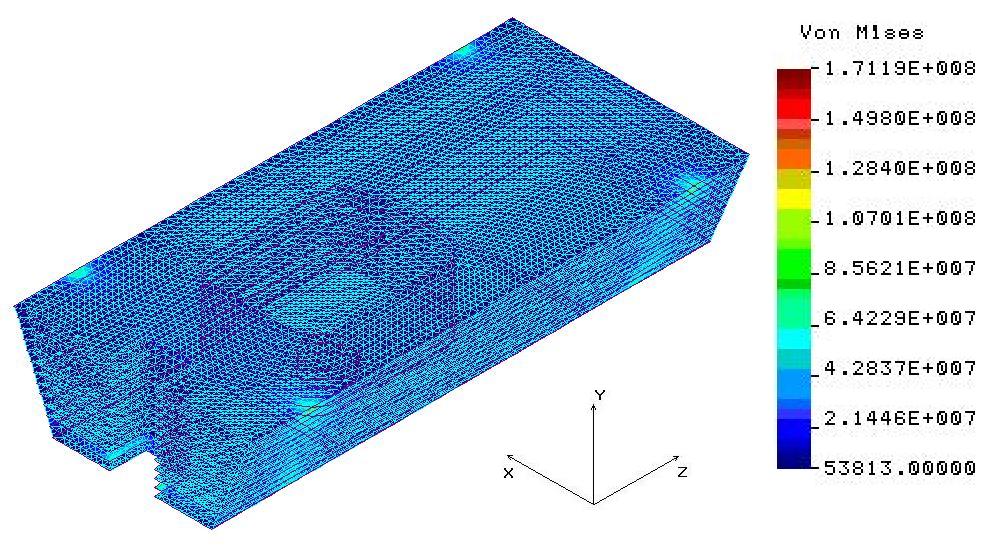}
\caption{Von Mises stress in the upper yoke beam due to gravity
and magnetic forces}
\label{yoke:Yoke_beam_stress}
\end{center}
\end{figure}

According to our calculations the equivalent (Von Mises) stresses in
the yoke beams are within allowable limits $[\sigma] = 140\unit{MPa}$
and the stresses in support frames are $\sigma < 100\unit{MPa}$ both
including and excluding the seismic load.  The distribution of
stresses in the upper yoke beam due to gravity and magnetic forces is
shown in Fig.~\ref{yoke:Yoke_beam_stress}.

\begin{figure}[ht]
\begin{center}
\includegraphics[width=\swidth]{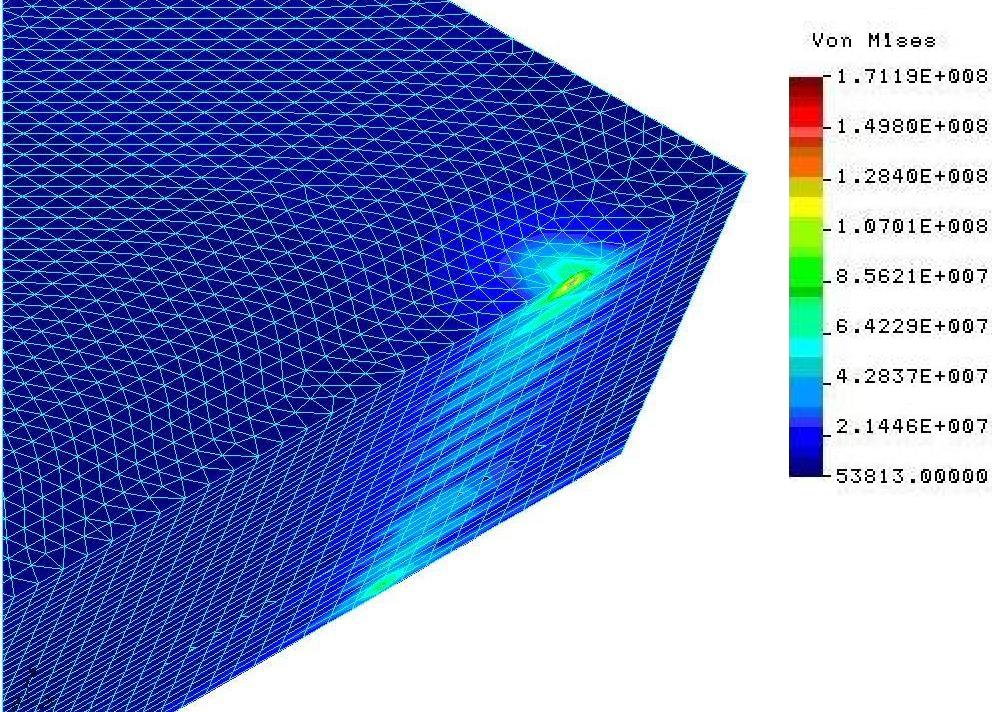}
\caption{Details of the stress distribution in the upper yoke
beam}
\label{yoke:Yoke_beam_stress-2}
\end{center}
\end{figure}

The details of the stress distribution in the upper yoke beam are
shown in Fig.~\ref{yoke:Yoke_beam_stress-2}. The stresses will have
local maxima at the fixation points, where the concentrated force
will be applied.

The material resistibility class of bolts consolidating the yoke beams
is foreseen to be 10.9. This resistibility will be sufficient for
calculated tensile loads $<40\unit{kN}$ per bolt M24.

\subsubsection{Deformations and Stresses in the Downstream End Cap}

\begin{figure}[ht]
\begin{center}
\includegraphics[width=0.6\swidth]{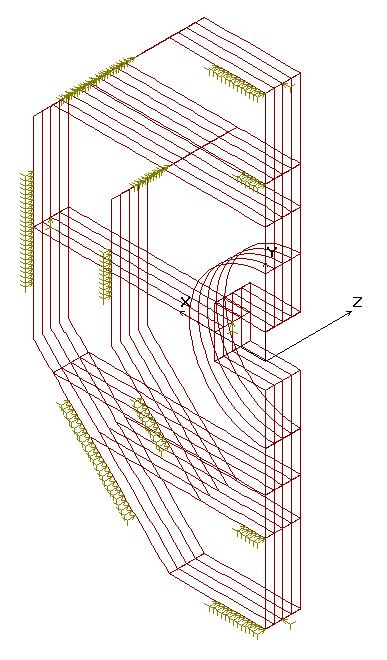}
\caption{3D model of the downstream door}
\label{yoke:Downstream_door_model}
\end{center}
\end{figure}

The 3D model of the downstream door is shown in
Fig.~\ref{yoke:Downstream_door_model}. The connection of the
plates is realised through the welded spacers along the side
surface and through the horizontal welded spacers.

\begin{figure}[ht]
\begin{center}
\includegraphics[width=0.8\swidth]{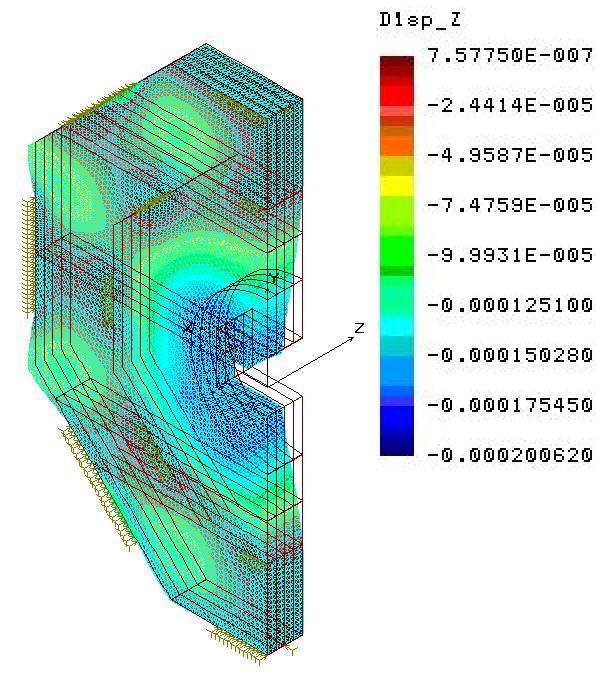}
\caption{Deformation of the downstream door}
\label{yoke:Downstream_door_def}
\end{center}
\end{figure}

\begin{figure}[ht]
\begin{center}
\includegraphics[width=0.8\swidth]{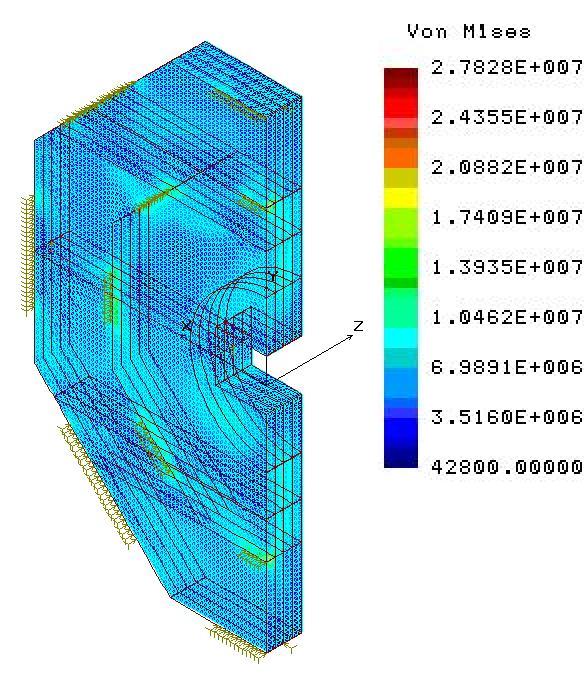}
\caption{Von Mises stresses in the downstream door}
\label{yoke:Downstream_door_stress}
\end{center}
\end{figure}

Axial deformations of the downstream door due to magnetic force are
shown in Fig.~\ref{yoke:Downstream_door_def}. The maximal value of the
displacement is obtained to be $\Delta Z < 0.3\unit{mm}$. The
deformation depends on the number of welded spacers used to connect
the neighbour steel plates. The dependence of deformation on the
number of additional (besides that at the periphery) horizontal
spacers is presented in Table~\ref{yoke:Spacers}. The solution with
three spacers is regarded as optimal. The stresses in the downstream
door are shown in Fig.~\ref{yoke:Downstream_door_stress}. The maximal
value of the stress is anticipated to be $\sigma < 30\, MPa$.

\begin{table}[ht]
\begin{center}
\begin{tabular}{|c|c|}
\hline
  Number of  &  Maximal bending  \\
  welded spacers & deflection [mm] \\
\hline
 No spacers   &  8.1 \\
 1 spacer &  1.1  \\
 3 spacers &  $<0.3$  \\
\hline
\end{tabular}
\end{center}
\caption{Axial bending deflection of the second plate of the
downstream door}
\label{yoke:Spacers}
\end{table}

\subsubsection{Deformations and Stresses in the Upstream End Cap}

\begin{figure}[ht]
\begin{center}
\includegraphics[width=0.7\swidth]{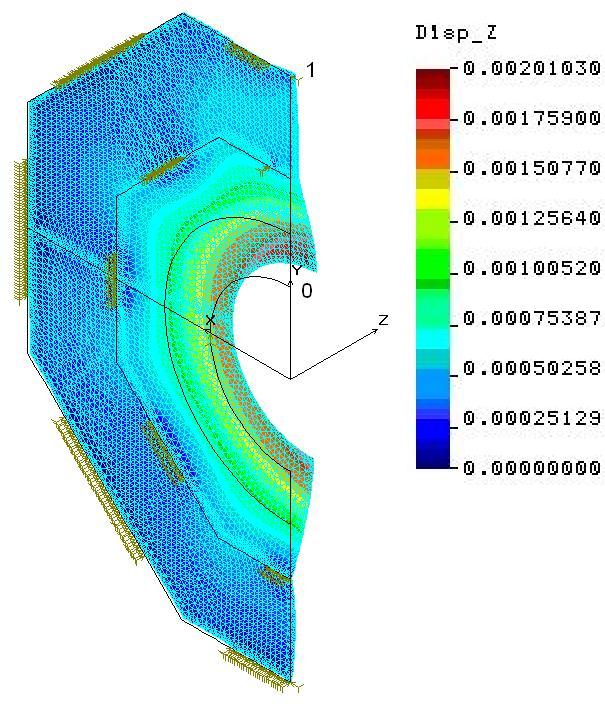}
\caption{Deformation of the upstream door}
\label{yoke:Upstream_door_def}
\end{center}
\end{figure}

\begin{figure}[ht]
\begin{center}
\includegraphics[width=\swidth]{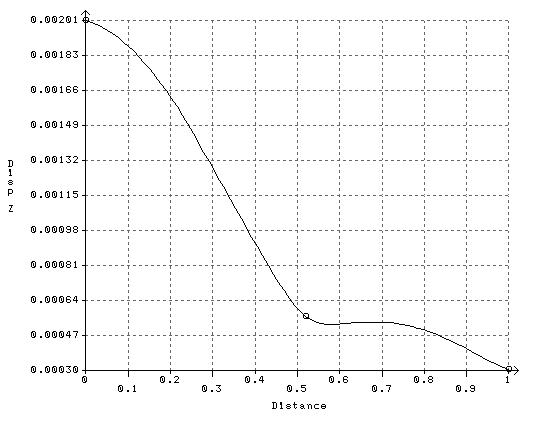}
\caption{Axial displacement [m] of the upstream door along the
line 0-1 (Fig.~\ref{yoke:Upstream_door_def})}
\label{yoke:Upstream_door_plot}
\end{center}
\end{figure}

\begin{figure}[ht]
\begin{center}
\includegraphics[width=0.7\swidth]{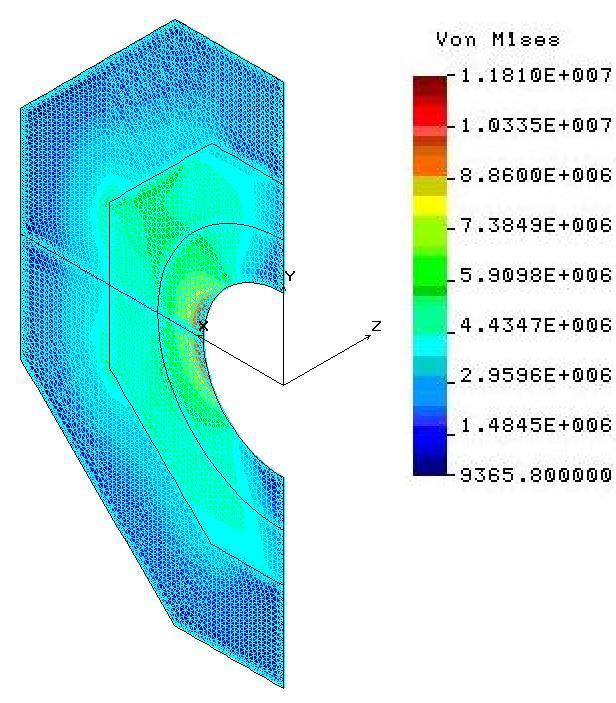}
\caption{Von Mises stresses in the upstream door}
\label{yoke:Upstream_door_stress}
\end{center}
\end{figure}

Axial deformations of the upstream door due to magnetic force are
shown in Fig.~\ref{yoke:Upstream_door_def}. Axial displacements along
the line 0-1 are shown in Fig.~\ref{yoke:Upstream_door_plot}.  The
maximal value of the displacement is calculated to be $\Delta z <
3\unit{mm}$. The stresses in the upstream door are shown in
Fig.~\ref{yoke:Upstream_door_stress}.  The maximal value of the stress
is calculated to be$\sigma < 12\unit{MPa}$.

\subsection{Assembly of the Yoke}
\label{s:sol:yoke:assembly}

The assembly of the flux return yoke in the \Panda hall will be done
in the service area, where sufficient space for detector installation
is foreseen. The structures and the iron segments will be delivered
through an opening in the \Panda hall roof.  The movement of the
single yoke elements, whose weight will be $\sim20\unit{t}$, will be
done using the overhead crane and, if needed, additional mobile cranes
operated from outside the experimental hall.

The process of assembly will start with the installation of the
lower part of the magnet support system. The platform of the
magnet will be mounted on temporary supports to provide stability
of the construction. (The wheels of the platform will be
in a lifted state at this stage.) These supports are the same which
will also be used to support the magnet in its operation position
at the beam line.

The 4 vertical beams of the yoke support frame will be mounted on
the platform. The lower knees supporting the vertical beams  will
be installed.

The lowest beam of the yoke barrel will be placed on
the platform. Then the 2 inclined lower beams will be installed
and attached to the horizontal one by temporary screwing. Then the
template frames that control the form of the internal boundary of
the yoke will be mounted.

The side (vertical) beams of the barrel will be installed and attached
by temporary screwing. The inclined upper barrel beams will be placed
on them. Then the upper knees supporting the outer frame will be
mounted.  The upper barrel beam will then be installed and attached by
temporary screwing. The final fastening of the attachment bolts in the
barrel beams will be performed.

The 2 upper crossbars of the outer frame will be placed on the
vertical bars with the knees. The outer frame will be attached to the
barrel by bolts. The internal template frames will be removed and the
geometric parameters of the yoke will be checked.

The barrel end caps will be installed on their rails at the
solenoid platform. The upper guide frames for the end caps will be
mounted and the door adjustment will be performed. As final test,
the door opening mechanism will be checked.


\svnInfo $Id: perf.tex 714 2009-03-31 17:34:18Z IntiL $

\section{Performance}
\label{s:sol:perf}

\AUTHORS{A.~Bersani, Y.~Lobanov}

\begin{figure*}
 \begin{center}
  \includegraphics[width=\dwidth]{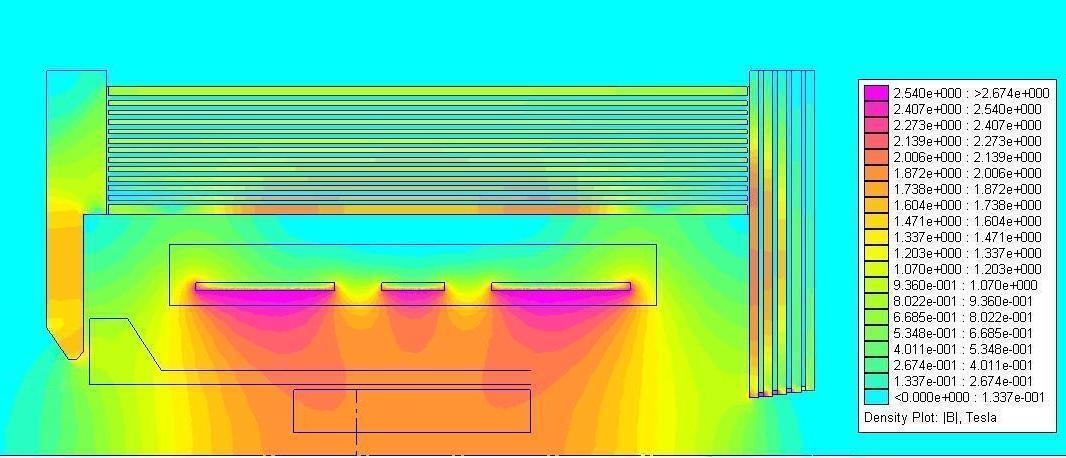}
\caption[Magnetic flux density distribution.]{Magnetic flux density
distribution (cross section for half of the
  horizontal plane where the beam comes in from the left).  The iron
  yoke, coil and cryostat, the barrel DIRC and the outer tracker are
  indicated by their outlines.}
\label{yoke:flux_density}
 \end{center}
\end{figure*}

\subsection{Introduction}

The optimisation of the solenoid magnet design was performed by
independent groups using different techniques and software, both for
finite elements modelling~\cite{Efremov:2008,Bersani:2008}, and
for subsequent analysis~\cite{Koshurnikov:2006}. The final field maps
were calculated using the TOSCA code~\cite{mag:TOSCA_2003} in a
complete 3D model with the dipole magnet operated at different
energies.

As stated above, the coil position and the flux return yoke will not be
symmetric: this will lead to an unbalanced magnetic force directed
parallel to the beam line.  An effort was spent to minimise this
residual axial force, optimising the winding geometry. The final
design features a residual unbalanced axial force of
$2\cdot10^5\unit{N}$ directed from downstream to upstream with respect
to the beam direction.  This force is very sensitive to the
displacement of the coil with respect to its nominal position.  The
force derivative w.r.t.\  displacement will be $4\cdot10^4\unit{N/cm}$,
increasing in the direction of displacement. The support structure and
structural calculation were performed with a safety factor of 4.

The final design is fully compliant with all magnetic requirements,
mechanical constraints and technical guidelines that were described in
the previous sections.  The key aspects of the magnet performance in
terms of fields, operation and beam interference are described in the
following.

\subsection{Field in Tracker Region}

The calculation of magnetic fields and forces has been performed in
the framework of axially-symmetric and 3D FE models, with different
detail level and different symmetry properties in the different phases
of design optimisation. A distribution of the solenoid magnetic flux
density, obtained with an axial-symmetric model and about $10^6$
elements is shown in Fig.~\ref{yoke:flux_density}.

\begin{figure}
\begin{center} \includegraphics[width=\swidth]{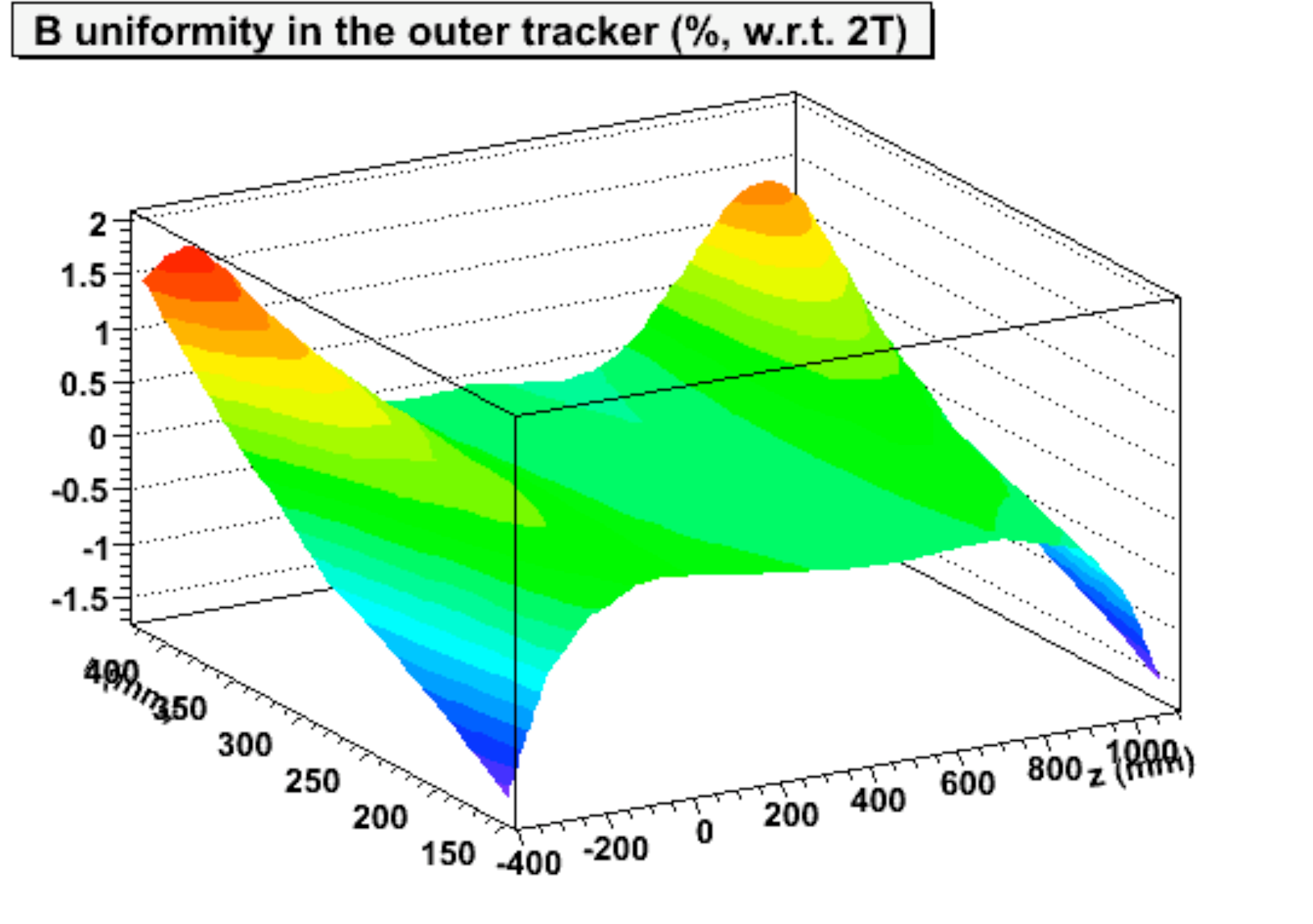}
\caption[Field homogeneity $\Delta B/B$ in the outer tracker
region.]{Field homogeneity $\Delta B/B$ in the outer tracker
region.
  The radius from the target $r$ and the coordinate along the beam
  direction $z$ are given on the lower axes in millimetres.  The
  vertical axis shows the relative difference between the nominal
  field $B_0 = 2$\,T and the actual field in this coordinate in per
  cent.}
\label{f:sol:omo}
\end{center}
\end{figure}

In the optimised solenoid design we have achieved that the field
inhomogeneity, {\it i.e.\ }the deviation $\Delta B/B_0$ from the
nominal field $B_0=2$\,T, stays below $1.6\%$ in any point within the
outer tracker. See also Fig.~\ref{f:sol:omo}.  This undershoots the
request of $\Delta B/B<2\%$ significantly.  The field is even
homogeneous to better than $\pm1\%$ in more than $90\%$ of the tracker
volume.  The resulting calculated distribution is homogeneous to the
per mile level in the central region of the tracker, {\it i.e.\
}$-200\leq z \leq +800$\,mm.  The peaks at the corners cannot be
avoided if keeping to the mechanical constraints from the detectors on
the geometry of the cryostat.

\begin{figure}
\begin{center}
 \includegraphics[width=\swidth]{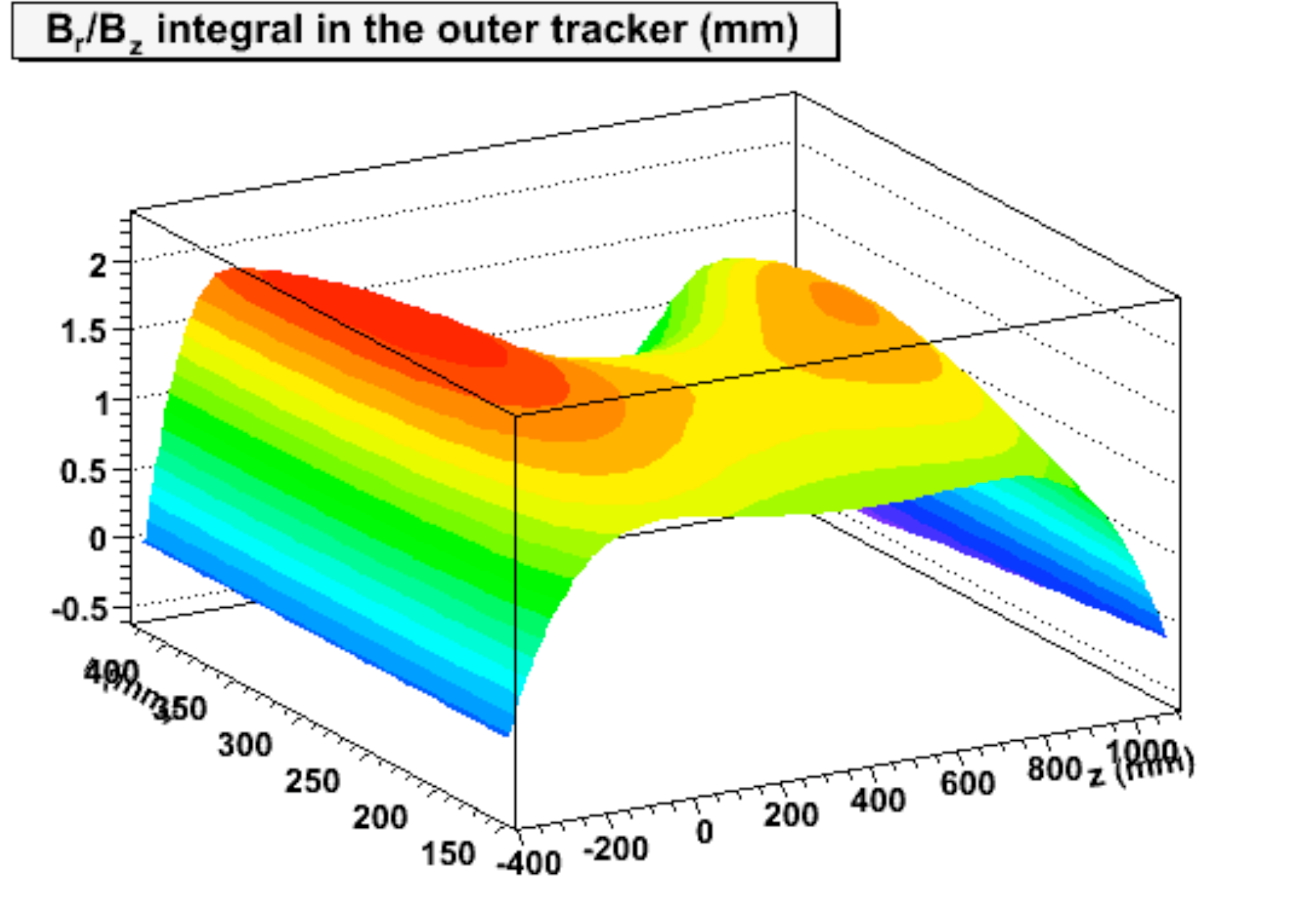}
\caption[Integral of radial field component $I(r,z_0)$ in the
outer tracker region.]{Integral of radial field component
$I(r,z_0)$ in the outer
  tracker region.  The radius from the target $r$ and the coordinate
  along the beam direction $z$ are given on the lower axes in
  millimetres.  The vertical axis shows $I(r,z_0)$ (see text) in
  millimetres.}
\label{f:sol:int}
 \end{center}
\end{figure}

The design also fulfils the requirements concerning the integral of
the radial component of magnetic induction field:
\begin{equation}
I(r,z_0)=\int\limits_{-400}^{z_0}B_r(r,z)/B_z(r,z)\,dz < 2\,mm.
\nonumber
\end{equation}
for $150\unit{mm}\le r\le420\unit{mm}$ and $-400\unit{mm}\le
z_0\le1100\unit{mm}$, {\it i.e.\ }in the outer tracker region. The
optimised design gives $0\unit{mm}\le I(r,z_0)\le2\unit{mm}$ for any
given value of $r$ and $z_0$.  In addition, only positive values of
$I(r,z_0)$ and a smooth behaviour is achieved.  This is especially
crucial for the operation of a Time Projection Chamber (TPC).  A plot
of the $I(r,z_0)$ distribution in the outer tracker region is shown in
Fig.~\ref{f:sol:int}.

\subsection{Fringe Field and Field in Flux Return}

Particular care was taken during the design development on the fringe fields
near the return yoke apertures and in the return yoke itself.

The fringe fields will be critical in several regions:
\begin{enumerate}
\item{DIRC readout regions;}
\item{slits for cables and services in the barrel;}
\item{target recess in the upper octant;}
\item{outside the yoke on the beam axis.}
\end{enumerate}
Problems can arise due to photon detectors operation (point 1),
insufficient iron cross section and local increase of magnetic reluctance
(point 2) or turbo-molecular pumps operation (points 3 and 4). Every
region was separately studied.

The field in the iron flux return yoke and in the doors was studied to
evaluate the development of saturation in the most crucial regions.
This was done in an iterative process in order to minimise the
saturation of the iron.  It is advisable to keep saturation low, as
this enhances the magnet's efficiency, strong local field distortions
may appear close to saturated parts, and nonlinear effects which are
difficult to control appear in forces and fields.

\subsubsection{Magnetic Field in the DIRC Readout Region}

\begin{figure}
\begin{center}
\includegraphics[width=\swidth]{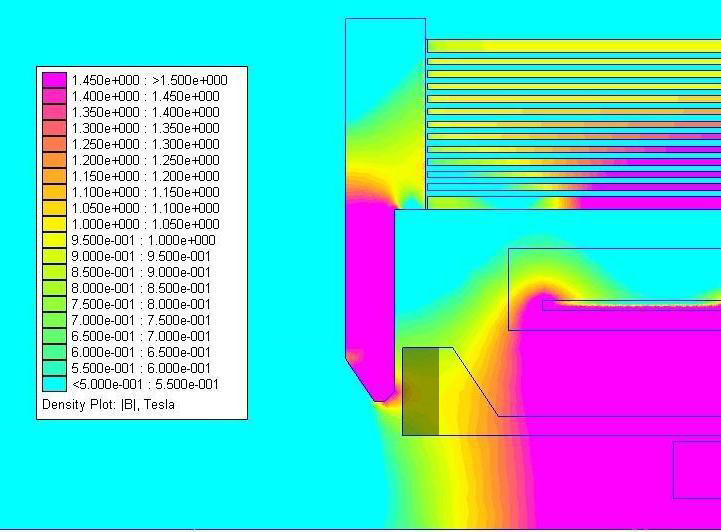}
\caption[Magnetic flux density in the region of the Barrel DIRC
readout.]{Magnetic flux density in the region of the Barrel DIRC
readout.
  The MCPTs will be located on the left within the grey-shaded box
  depicted here.  The field there will remain below 1\,T.}
\label{yoke:DIRC_area}
\end{center}
\end{figure}

The barrel DIRC readout will be installed inside the iron yoke and
will be read out with the recently developed Micro-Channel Plate
Tubes (MCPTs).  These devices are almost insensitive to magnetic field up to
1 or $1.5\unit{T}$, depending on the models and dimensions of the
sensitive area.  The flux density distribution in this area is shown
in Fig.~\ref{yoke:DIRC_area}.

Due to the optimisation of the upstream end cap, the flux density in
the area where the MCPTs will be operated is $|B|<1\unit{T}$.  Since
the technology of MCPTs is relatively new, we foresee a backup
solution where the slabs extend outside the end cap.  Here the readout
would be in a region of much lower fields of few $\unit{mT}$ and could
be even shielded easily further.

\subsubsection{Effect of Cable Routing}

Cut outs in the return yoke barrel are foreseen to allow the
feed-through of most of the cables and service lines for the detectors.
The size of each slit will be $140 \times 420$\,mm$^2$ and there will
be one at each corner of the octagon both at the upstream and
downstream ends of the barrel (see Fig.~\ref{yoke:Barrel}).

The effect of the cut outs on the field parameters has been
extensively studied. Although the magnetic field in the area of the
passages will be disturbed due to redistribution of flux, its
influence on the field in the central region will be insignificant.
This conclusion has been confirmed using complete 3D magnetic field
calculations within the model shown in Fig.~\ref{quarto} that takes
into account the actual shape and dimensions of recesses in the yoke
barrel. The results of calculations of the field inhomogeneity and of
the integral of the radial component of magnetic induction (see
Fig.~\ref{f:sol:omo} and Fig.~\ref{f:sol:int}) show that the
requirements on the field quality are satisfied.

\begin{figure}[th]
 \begin{center}
  \subfigure[$y-z$ plane, $x=0$]
   {\includegraphics[width=\swidth]{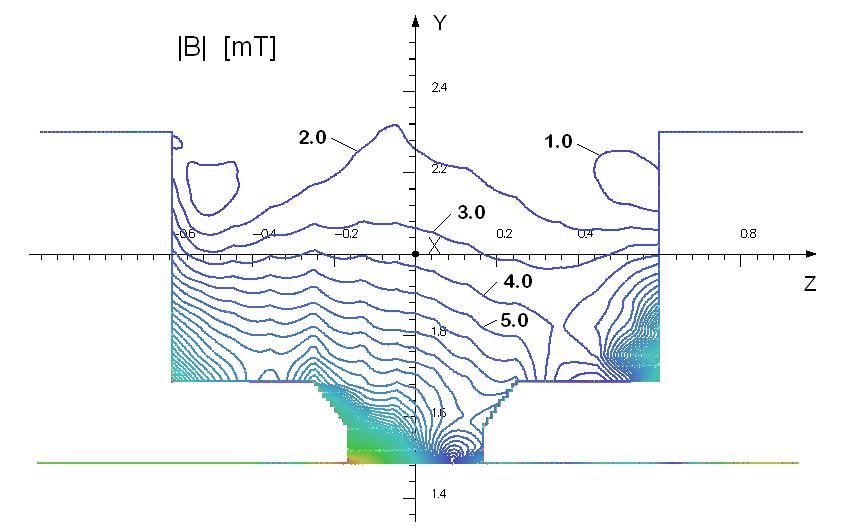}
    \label{yoke:target_recess_flux_YZ}}
  \subfigure[$x-y$ plane, $z=0$]
   {\includegraphics[width=\swidth]{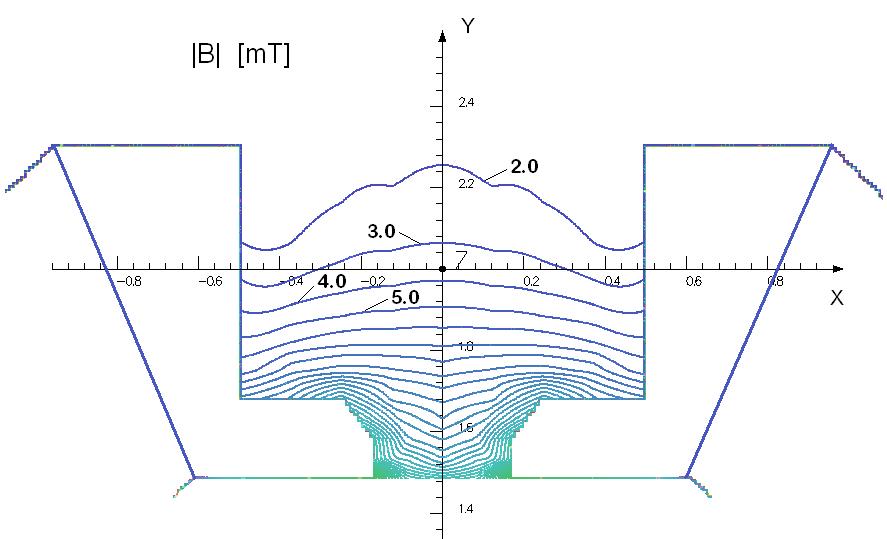}
    \label{yoke:target_recess_flux_XY}}
\caption[Flux density in the target recess.]{Flux density in the
target recess.  The field sensitive
  turbo-molecular pumps will be located at a height of 2.0\,m from the
  beam, {\it i.e.\ }at the level of the $x$ or $z$ axis in the plots.}
\end{center}
\end{figure}

\begin{figure}[th]
 \begin{center}
  \subfigure[Barrel ($Z=1300\,$mm)]
   {\includegraphics[width=\swidth]{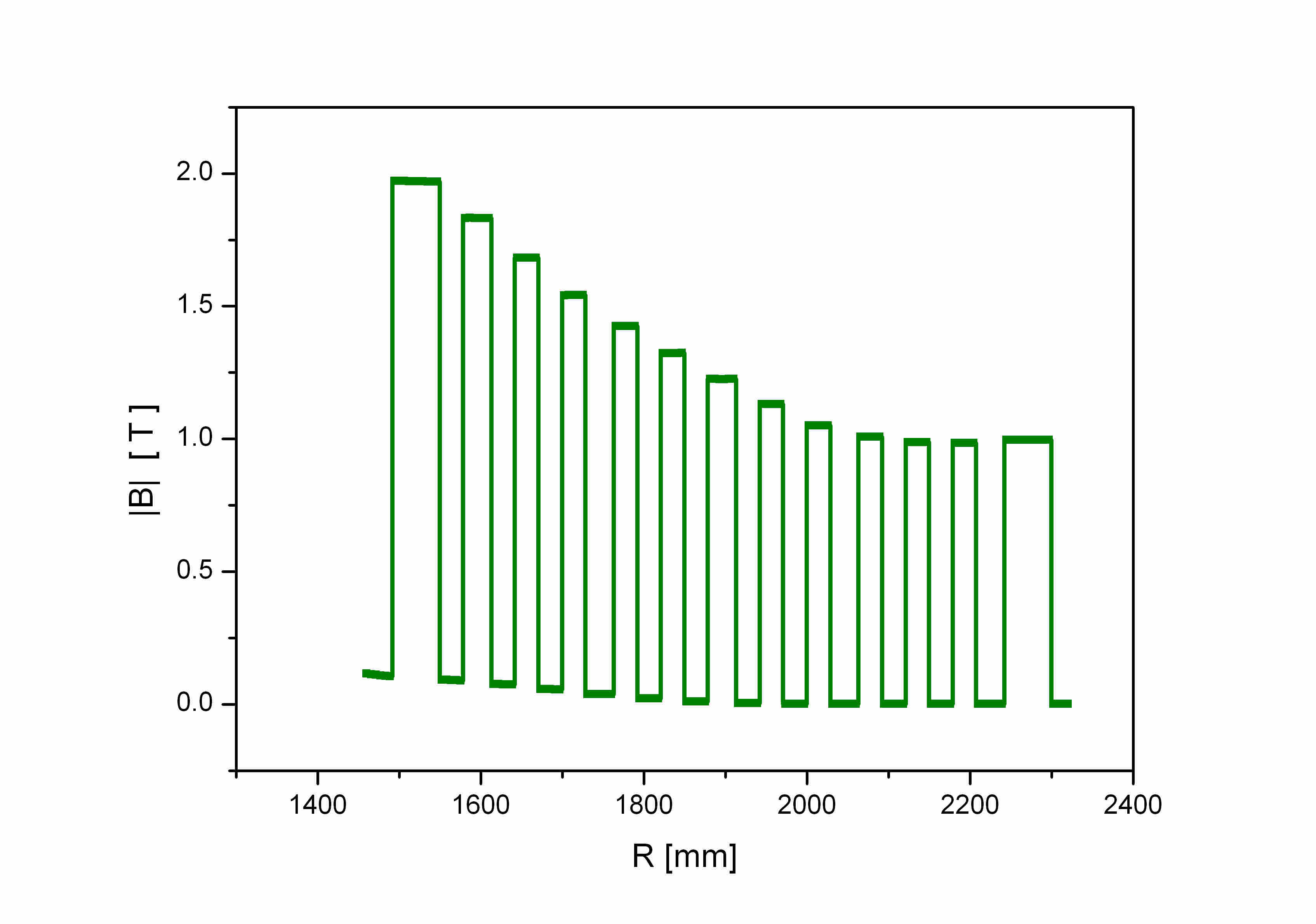}
     \label{yoke:barrel_flux}}
  \subfigure[Downstream end cap ($R=1200\,$mm)]
   {\includegraphics[width=\swidth]{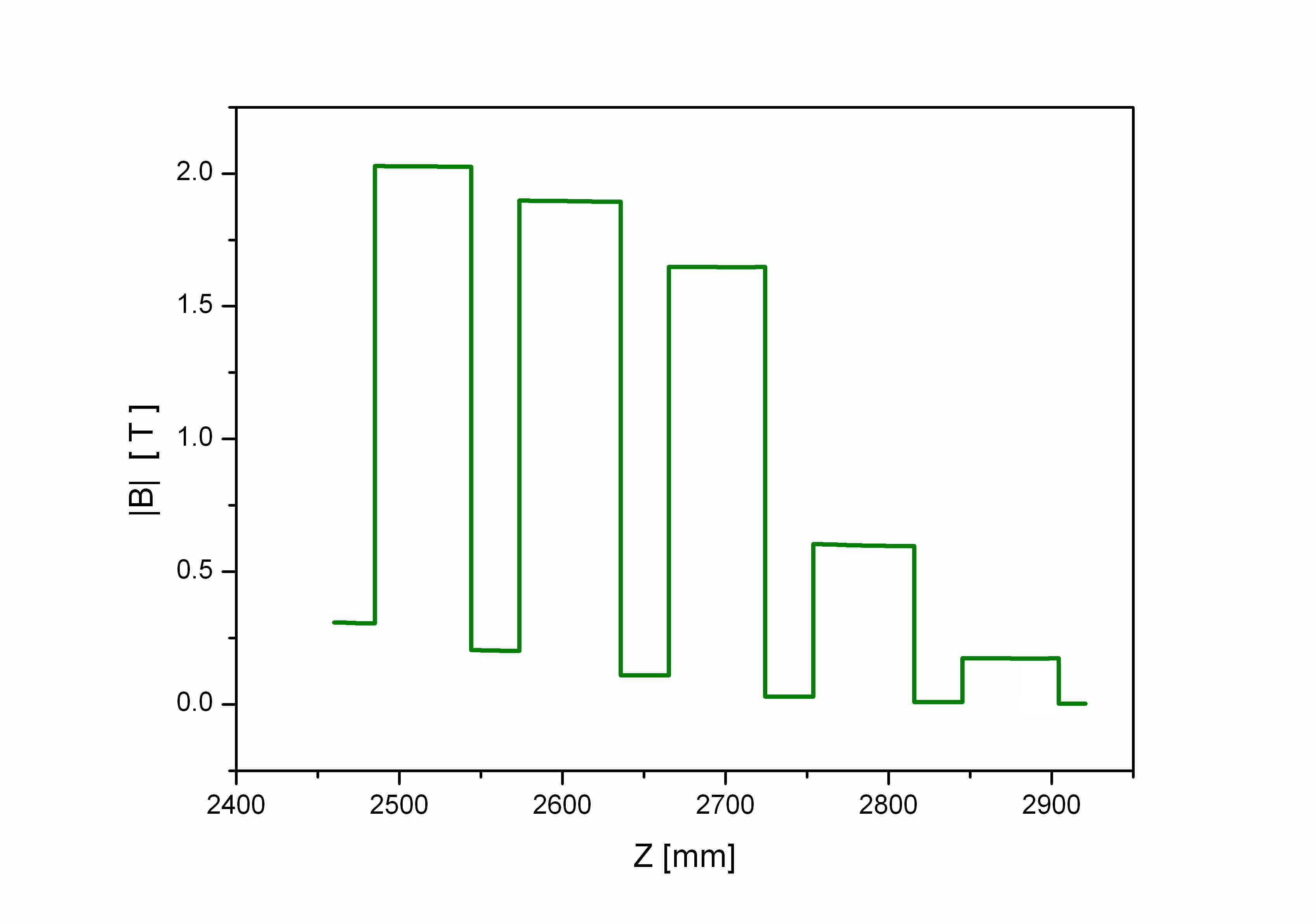}
     \label{yoke:down_endcap_flux}}
\caption{Flux density in different parts of the yoke.}
\end{center}
\end{figure}

\subsubsection{Stray Fields Outside the Yoke}

The magnetic flux distribution in the recess for the target has been
studied, because the generator of the target requires turbo-molecular
pumps, which should be located at a height of 2.0\,m from the beam and
could safely tolerate fields of 5\,mT, but not much higher.  The plots
of this distribution in the vertical $y-z$ plane (containing the beam
axis) and in the $x-y$ plane (perpendicular to the beam axis) are
shown in Fig.~\ref{yoke:target_recess_flux_YZ} and
Fig.~\ref{yoke:target_recess_flux_XY}. The requirement that in the
region of the turbo molecular pumps the magnetic field must not exceed
5 mT is clearly satisfied.

A smaller recess is foreseen in the lower octant, where the
turbo-molecular pumps will be placed further away from the interaction
point.  This recess will have no large effects, neither from the
viewpoint of fringe fields nor from the inducted field distribution.

Downstream from the solenoid the pumps in this region will be placed
more than $1\unit{m}$ off the centre of the beam pipe.  The stray
field outside the flux return yoke will anyhow be well damped in
downstream direction as the orifice in the door will be relatively
small.  Thus no problem from fringe fields arises here.  In contrast
to that in upstream direction, since the opening will be relatively large
and the door will be relatively thin, a slightly higher fringe
field will arise.  This poses no restriction though, as the cable and supply
routing to the inner detectors and the EMC end plug will inhibit the
installation of turbo-molecular pumps closer than $60\unit{cm}$ to the
door anyway.  At this distance also in that region the field will be
tolerable for such pumps.

\subsubsection{Magnetic Induction in the Iron Yoke}

The yoke lamination makes the flux line distribution more complex
than a compact flux return yoke. There is the possibility that high
saturation arises locally, generating deformations of the flux line
distribution near the iron.

Two cases where studied in particular:
\begin{itemize}
\item{the barrel yoke near to the first and last sub-coil;}
\item{the downstream door near the beam pipe hole.}
\end{itemize}
In both cases, small volumes where the flux densities reach
$2\unit{T}$ are found.  These regions, however, will be extremely limited
in size. It was, furthermore, demonstrated that they have no significant
effect.  Two plots of the density profiles in the barrel yoke and in
the downstream door, where the highest values were calculated, are
shown in Figs.~\ref{yoke:barrel_flux} and~\ref{yoke:down_endcap_flux}.

As a result of the optimisation of the upstream door design, the
maximum value of magnetic induction in it is calculated to be
$|B|<1.8\unit{T}$, so there will be no saturation effect in the
upstream door.

\subsection{Solenoid Operation}

The \PANDA solenoid magnet has been designed and optimised to work at
$2\unit{T}$ central field and $5000\unit{A}$ current, but it can be
easily operated at lower currents.  Operation at 60-80\% of the
maximum current is foreseen to allow stable accelerator operation at
low beam momentum.  In this case the solenoid's performance in terms
of homogeneity will be worse, but only slightly.  Usually, the magnet
will be ramped up at the beginning of a measurement period and remain
at its current setting for the whole period, typically for up to half
a year.  Whenever required, it can be ramped up or down within about
half an hour.

\subsubsection{Transient Behaviour}

\paragraph{Normal Power Cycling}

\begin{figure}[ht]
  \begin{center}
    \includegraphics[width=\swidth]{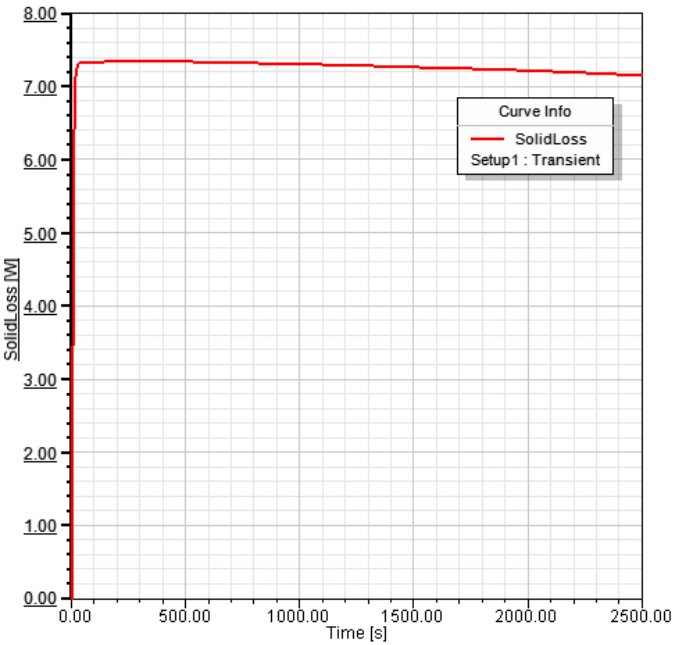}
    \caption{Power loss via eddy currents in the coil former
    during normal power cycling as a function of time.  }
    \label{f:sol:pow_v_t}
  \end{center}
\end{figure}

Power cycling of the solenoid was studied to find the optimal
compromise between energization speed and thermal dissipation. A
calculation was performed using a 3D Finite Element Model (FEM) with the 
Ansoft Maxwell code~\cite{coil:maxwell}.  The result is that the power
dissipation due to eddy currents in the coil former can be kept under
$10\unit{W}$ assuming a ramp up time of $2000\unit{s}$, or, in other
words, a current rise rate of $2.5\unit{A/s}$.  The current ramp and
the power dissipation are plotted in Fig.~\ref{f:sol:pow_v_t}.

\begin{table*}[ht]
\begin{center}
\begin{tabular}{|l|c|c|c|c|}
\hline
Current $[\unit{A}]$) & 5000 & 4000 & 3000 & 2000\\
\hline\hline
Central Induction & $2\unit{T}$ & $1.6\unit{T}$ & $1.2\unit{T}$ & $0.8\unit{T}$ \\
Conductor Peak Field & $2.8\unit{T}$ & $2.4\unit{T}$ & $2\unit{T}$ & $1.5\unit{T}$ \\
$I(r,z_0)$ in the Tracker $[\unit{mm}]$ & $0<I<2$ & $0<I<2$ & $0<I<2.1$ &
$-0.1<I<2.1$\\
Uniformity in the Tracker & $\pm1.6\%$ & $\pm1.8\%$ & $\pm2\%$ & $\pm2.1\%$ \\
Stored Energy & $21\unit{MJ}$ & $14\unit{MJ}$ & $7.7\unit{MJ}$ & $3.4\unit{MJ}$ \\
\hline
\end{tabular}
\caption[Main field parameters of the \PANDA~solenoid at different
  current settings.]{Main field parameters of the \PANDA~solenoid at different
  current settings.  A setting between 3 and 4\,kA is envisaged for
  the operation at low beam momenta.}
\label{t:sol:lowc}
\end{center}
\end{table*}

\begin{figure}[ht]
  \begin{center}
    \includegraphics[width=\swidth]{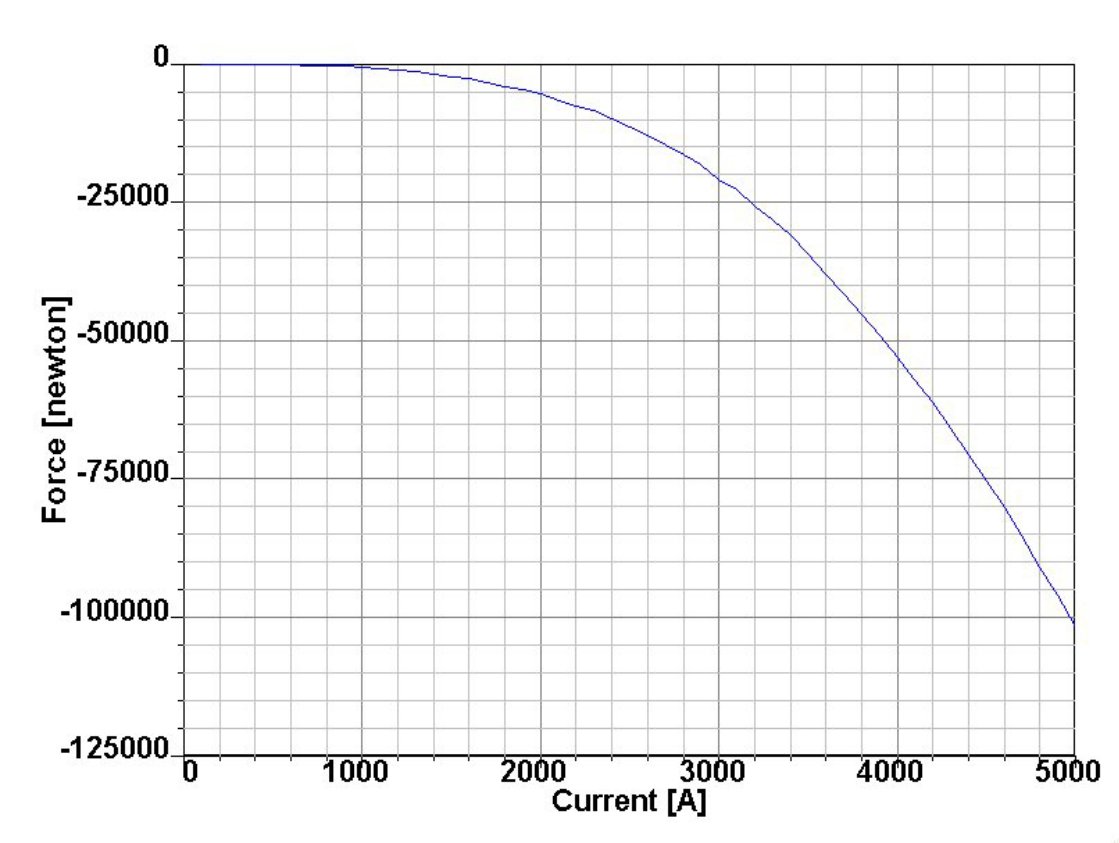}
    \caption[Residual axial force as a function of current during normal
    power cycling.]{Residual axial force as a function of current during normal
    power cycling.  The force will increase steadily and will never change
    direction.}
    \label{f:sol:for_v_t}
  \end{center}
\end{figure}

During power cycling, the force will increase following a quadratic
dependence w.r.t.\ the current.  This is true only if non-linear
effects from the induction will stay negligible.  As discussed above, some
small saturated regions are expected both in the yoke barrel and in
the end doors.  This saturation may induce deviations to the
theoretical behaviour.  However, it could be shown by calculations
that in \PANDA solenoid magnet the deviations from the pure quadratic
law will be negligible (see Fig.~\ref{f:sol:for_v_t}).  The relevant
aspect of this calculation is that the residual axial force is a
steadily increasing function during solenoid energization.  This
feature of the axial force allows a simpler design of the suspension
system of the coil in the cryostat and of the cryostat in the yoke,
compared to a situation in which the force is almost balanced but can
change direction during power cycling.

\paragraph{Emergency Shutdown}

\begin{figure}[ht]
  \begin{center}
    \includegraphics[width=\swidth]{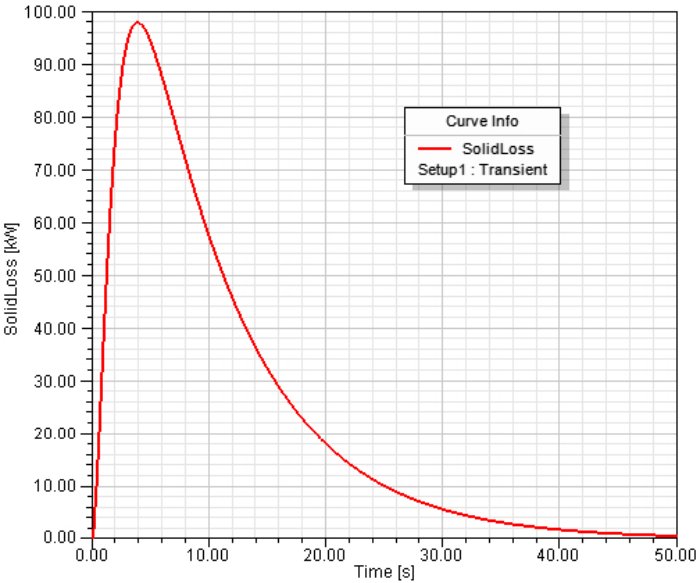}
    \caption{Power loss via eddy currents in the coil former during a quench.}
    \label{f:sol:pow_q}
  \end{center}
\end{figure}

In case of a quench of the magnet, the quench protection system will
open the quench switches and the solenoid will discharge on the
protection resistor, according to the circuit shown in
Fig.~\ref{proto}. Since the inductance of the solenoid will be
$1.7\unit{H}$ and the resistance of the protection resistor will be
$0.1\Ohm$, the discharge time constant will be $17\unit{s}$. This
means that the current will need few of seconds to reach a value in
the order of $100\unit{A}$, at which the solenoid can be considered
switched off. In these $30-60\unit{s}$ approximately $60\%$ of the
stored magnetic energy will be dissipated in the protection resistor
and the remnant will be transformed into heat in the coil. The time
dependence of the power dissipated in the coil former is shown in
Fig.~\ref{f:sol:pow_q}.

The heat which will be generated in the coil former is sufficient to
spread the quench all over the coil. The maximum temperature that will
be reached in the coil will stay below $80\unit{K}$.  This is
comparable to magnets of similar type and poses no difficulties.  It
is anticipated that, after a quench, the whole cold mass will be at
the temperature of the radiation shield or below, {\it i.e.\ }at
$<60\unit{K}$.  This will allow to proceed with a normal cooling
procedure with liquid helium after a quench.

\subsubsection{Reduced Current Operation}

During the operation at beam momenta below 2.25\,GeV/c the solenoid
will be operating at a reduced current between 60 and 80\% of the full
current to reduce tune shifts in HESR.  For this reason, the magnet
performance was evaluated at currents lower than the engineering
current.  A summary of the main parameters for various different currents
is shown in Table~\ref{t:sol:lowc}.

\begin{figure}[ht]
 \begin{center}
  \includegraphics[width=\swidth]{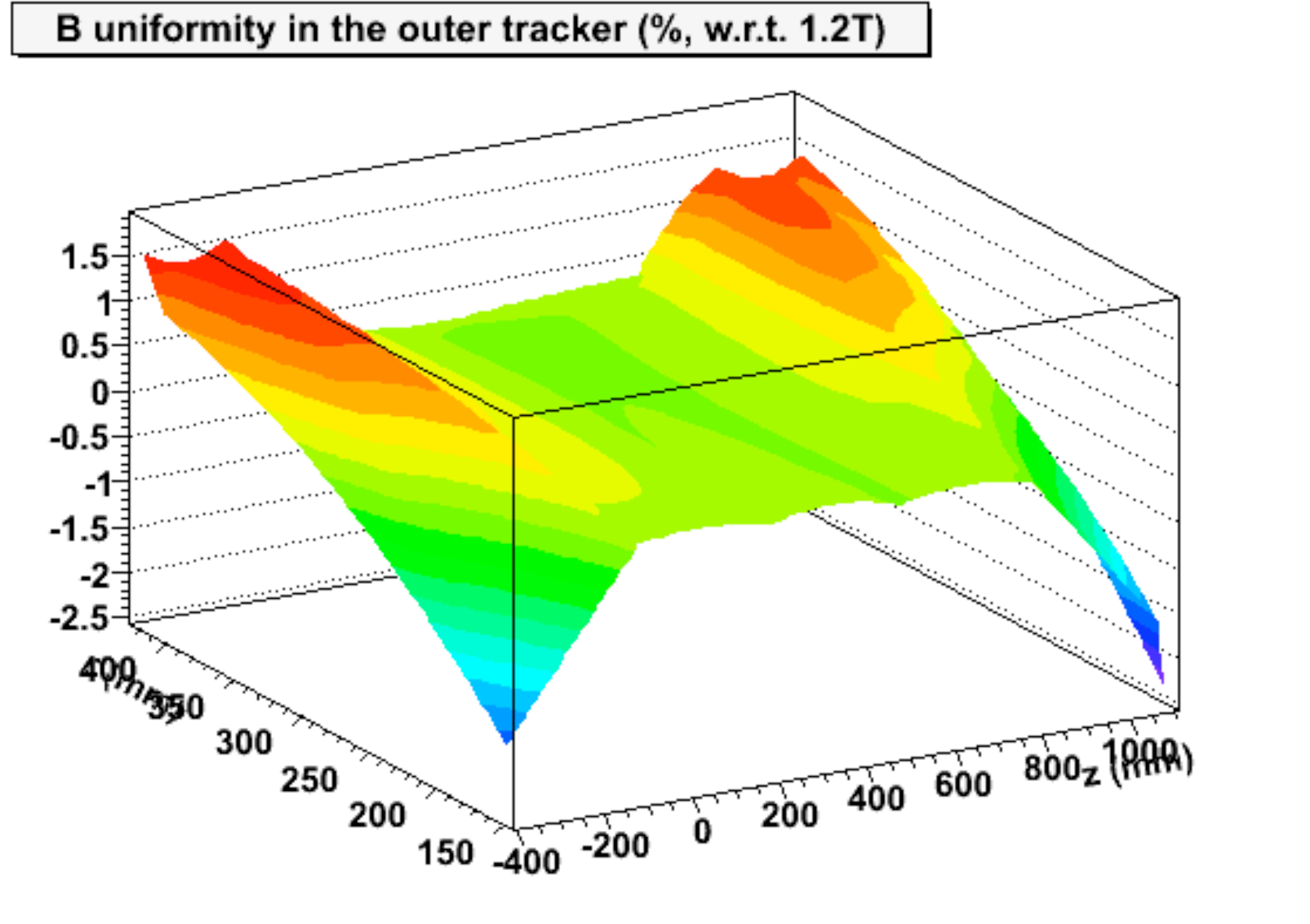}
  \caption{$\Delta B/B$ distribution over the outer tracker region at
  $60\%$ of the nominal current.}
  \label{f:sol:omo06}
 \end{center}
\end{figure}

\begin{figure}[ht]
 \begin{center}
  \includegraphics[width=\swidth]{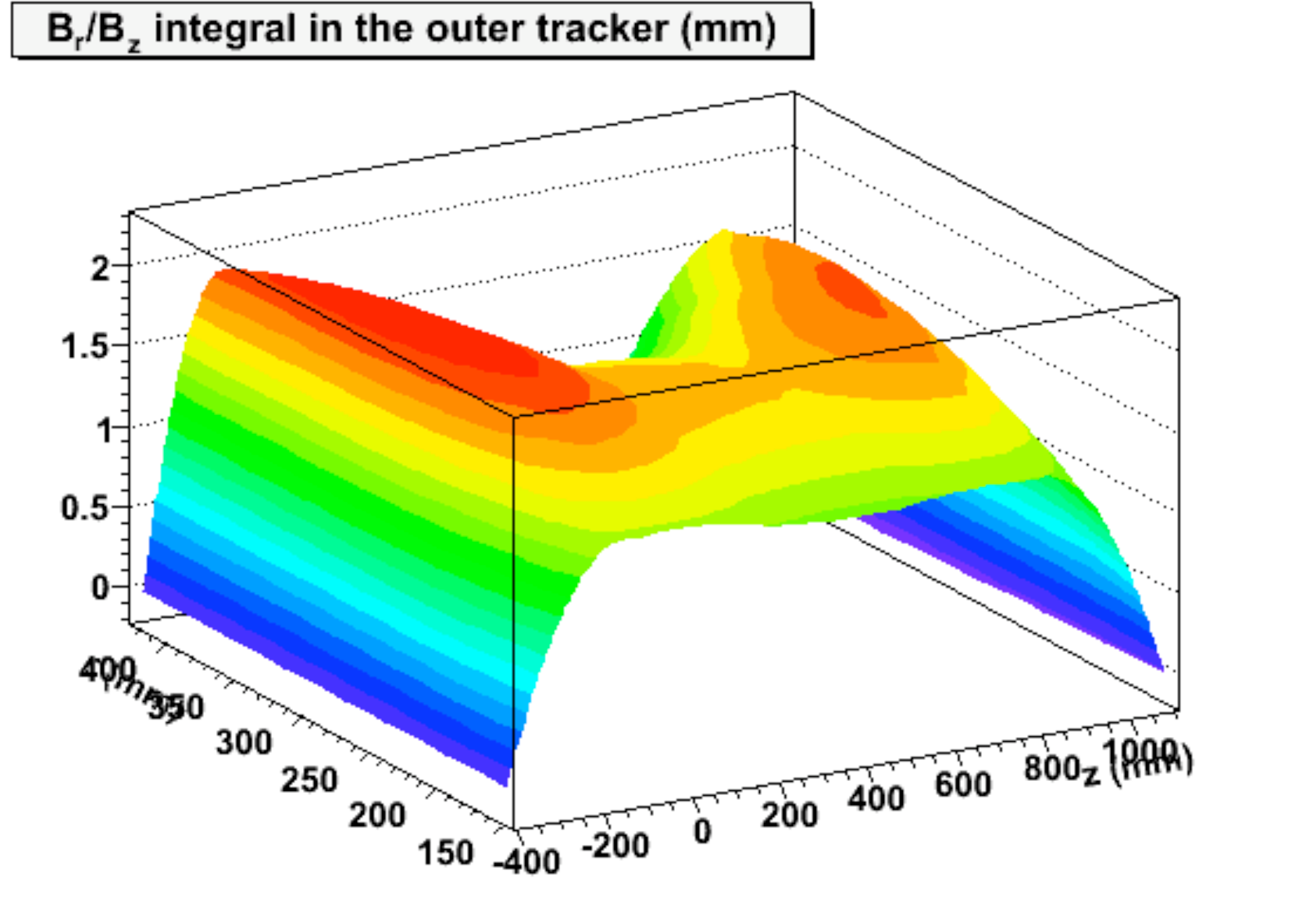}
  \caption{$I(r,z_0)$ distribution over the outer tracker region at
  $60\%$ of the nominal current.}
  \label{f:sol:int06}
 \end{center}
\end{figure}

Since the material properties are not linear the magnetic field will
not simply scale with the current.  Due to that a small increase in
the $I(r,z_0)$ integral needs to be accounted for at lower currents.
The distributions of $\Delta B/B$ and $I(r,z_0)$ in the outer tracker
region are shown in Fig.~\ref{f:sol:omo06} and in
Fig.~\ref{f:sol:int06}, respectively.  This should be compared with
the distribution obtained at the engineering current, which were shown
in Fig.~\ref{f:sol:omo} and Fig.~\ref{f:sol:int}.

\subsubsection{Field Mapping}

Field mapping will be performed using existing equipment at GSI
and with the help of the GSI Magnet Technology Group.  It is envisaged
that one full 3D field map would be determined for two current
settings: at the full field and one lower field setting, the latter
allowing the operation at lowest beam momenta.  This work will be
performed as soon as the magnet is fully commissioned.  Interpolating
between the grid points and maps will allow us to determine the
magnetic field at every point in the spectrometer opening and any
current to sufficiently high accuracy.

\subsection{Influence on the HESR Beam}

The solenoidal field will influence the beam in two ways.  Firstly the
beam optics will be affected even in an ideal configuration.  This
will be dealt with by a counter-solenoid placed in front of the PANDA
solenoid, which will be provided from the accelerator side. Here we
discuss only the effects on the beam slope and mean position due to
possible alignment errors.

To first order, the HESR antiproton beam will pass through the centre of
the \PANDA superconducting solenoid magnet, and the antiproton beam will be
parallel to the solenoid axis. In this ideal scenario there will be
no interaction between the beam and the magnetic field. However, in
reality, the magnet alignment will not be perfect. An imperfect magnet
alignment generates a small magnetic field ($B_p$) perpendicular to
the beam trajectory which bends the beam by a small angle and shifts
the beam trajectory away from its nominal position, i.e.\ generating
an orbit error.  It will not be easy to achieve a magnet alignment better
than a few milli-radians due to its large weight and size.  Another
important reason for the misalignment is that there is always a
difference between the magnetic central line and the mechanical
central line, and even the magnetic central line does not strictly
follow a straight line. Since the magnet misalignment can happen in
any direction, the orbit errors may occur in both horizontal and
vertical directions.

\begin{figure}[ht]
  \begin{center}
  \includegraphics[width=\swidth]{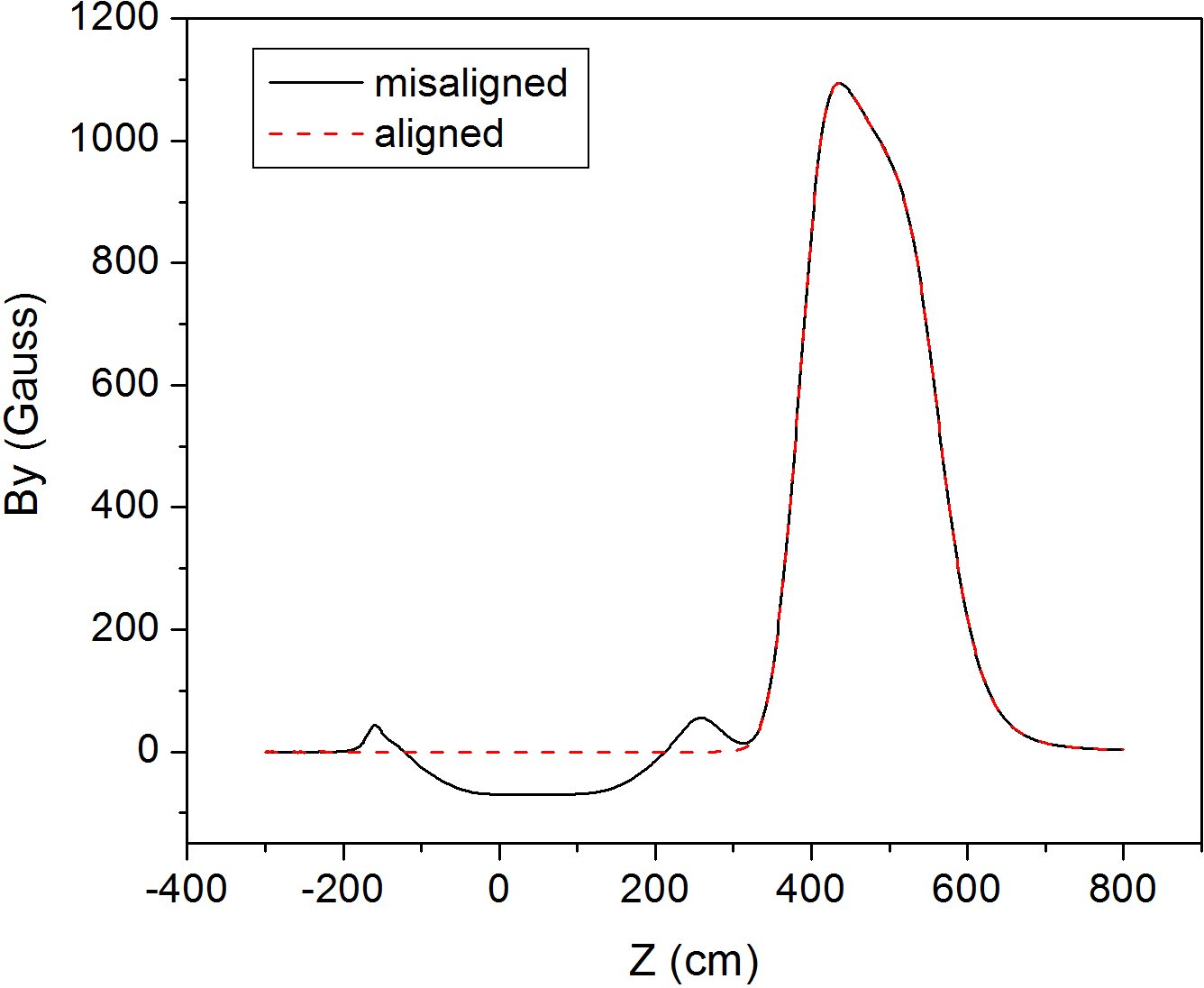}
\caption[Field along the central line, for aligned and misaligned 
  solenoid.]  {Field along the central line for an aligned solenoid
    (red dashed line) and misaligned solenoid by 0.2~degrees (black
    solid line).  In this calculation the dipole has a field integral
    of $\sim$0.2\,Tm and the antiproton momentum is 1.5\,GeV/c.  }
\label{f:sol:field-align1}
  \end{center}
\end{figure}

In order to quantitatively understand the orbit errors caused by the
\PANDA spectrometer, 3D Tosca magnetic field calculations were carried
out. In these calculations, both the solenoid and the Forward
Spectrometer dipole magnet are modelled. A small polar angle
misalignment of $0.2^\circ$ ($\sim$3.5\,mrad) is considered.  For
simplicity, the solenoid is rotated around the x-axis. The coordinate
system used here is the \PANDA coordinate system.  The x-axis is in
the horizontal plane and perpendicular to the antiproton beam, the
y~direction is the vertical direction and the z axis is along the
incoming antiproton beam direction. The magnetic field in the centre
of the solenoid is set to be 2\,T. To make the calculated field as
accurate as possible, around 2~million nodes are used in the Tosca
calculations. Based on the calculated fields, beam trajectories are
calculated for different beam energies. Because of the extension of
the Forward Spectrometer detectors the orbit errors are estimated at
$z=10\,$m.

\begin{figure}[ht]
  \begin{center}
    \subfigure[1.5\,GeV/c beam]{
      \includegraphics[width=\swidth]{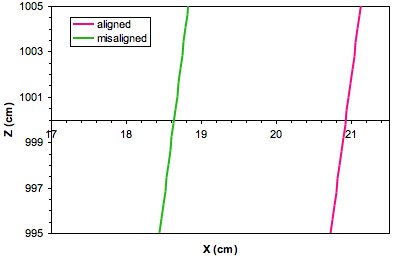}
      \label{f:sol:field-align2a}}
    \subfigure[15\,GeV/c beam]{
      \includegraphics[width=\swidth]{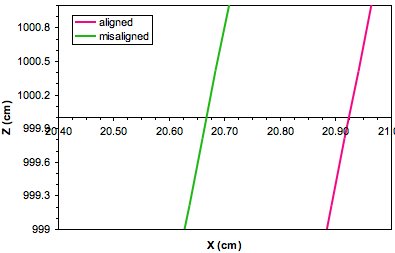}
      \label{f:sol:field-align2b}}
\caption[Antiproton beam trajectories for aligned and misaligned 
         solenoid.]  {Antiproton beam trajectories for aligned (purple
         line) and misaligned solenoid (green line).  }
\end{center}
\end{figure}

In Fig.~\ref{f:sol:field-align1} the calculated magnetic fields $B_y$
are shown along the z-axis at an antiproton momentum of
1.5\,GeV/c. The solid line is for the misaligned solenoid and the
broken line for the properly aligned solenoid. The solenoid is located
in the region between $z=-170$\,cm and $z=290$\,cm. The dipole magnet
is located in the region between $z=350$\,cm and $z=600$\,cm. In the
solenoid region, for a properly aligned solenoid, the field $B_y$ is
calculated to be nearly zero, while for the misaligned solenoid, $B_y$
has a maximum value of -69\,Gauss. The dipole magnet is set to have a
field integral of 0.2\,Tm and it has a peak field of
$\sim$1100\,Gauss. The field integrals $\int{B\,dL}$ from $z=-300$\,cm
to 800\,cm for an aligned and misaligned solenoid are shown in the
bottom of the figure. For the misaligned solenoid, the field integral
is $\sim$7\% smaller than for the properly oriented solenoid.  This
results in a bending angle error of 2.6\,mrad and a beam trajectory
shift at $z=10$\,m of $\sim$23\,mm for a 1.5\,GeV/c antiproton beam,
as shown in Fig.~\ref{f:sol:field-align2a}. The beam bending angle
errors and the trajectory shifts become smaller with increasing beam
energies.  For a 15\,GeV/c beam, the trajectory shift is only
$\sim$2.3\,mm, which is shown in Fig.~\ref{f:sol:field-align2b}.

A small solenoid misalignment of 3.5\,mrad in polar angle can cause a
large orbit error for the \Panda solenoid. For the 1.5\,GeV/c beam, the
trajectory shift could be as large as 23\,mm 10\,m downstream of the
target, while for the 15\,GeV/c beam it is approximately 2.3\,mm.  The
importance of these shifts and correction possibilities within HESR
are currently discussed with the accelerator group.

\svnInfo $Id: int.tex 714 2009-03-31 17:34:18Z IntiL $

\section{Detector and Target Integration}
\label{s:sol:int}

\AUTHORS{I.~Lehmann, A.~Bersani, Y.~Lobanov}

This section describes the details of the integration of the target
and all detectors inside the solenoid.  We only discuss the
installation of the parts which are directly attached to the magnet or
in a way relevant for the design.  Some of the structures are laid out
only conceptually as their detailed design will not affect the overall
construction of the magnet.

\subsection{Overall Assembly Procedure}

The complete magnet will be assembled in the parking position outside
of the beam line. (Please refer to Fig.~\ref{f:over:Panda_Hall} of
the floor plan of the \PANDA hall.)  As a first step the lower frame
of the support structure will be assembled and mounted on the
transport carriage.  This will then be used to subsequently assemble
the barrel part of the flux return yoke (see
Sec.~\ref{s:sol:yoke:assembly}).  Then the additional structures on
top of the barrel will be mounted.  The fully assembled cryostat will
be inserted from the upstream side into the iron yoke.  Only when this
is done the doors of the yoke will be mounted though we have foreseen
the possibility to also install or remove the cryostat with the doors
in place but open.  All individual parts will be aligned during the
assembly but the overall alignment will be performed when the magnet
is fully assembled.  In order to that an optical alignment system will
be used which allows to align the system with respect to the
components of the \HESR ring.

As a next step the cooling lines and control lines will be connected
and the solenoid will be ready for commissioning.  When the
commissioning will be successfully completed and the solenoid will
have been approved the magnet will be powered and the central field
will be mapped precisely at all operational modes.

Field mapping will be performed at this stage. Due to its almost
cylindrical symmetry, the volume inside the solenoid can be measured with
a fine mesh in a half-plane containing the beam axis and in several 
circumferences to verify the actual symmetry. The three components of the
induction field will be measured, with gauges capable of a wide dynamic 
range, to allow a high degree of accuracy up to $2\unit{T}$ over at
least 4 orders of magnitude. The field will be measured in the whole
length of the yoke, up to the cryostat inner radius. Between the 
downward end of the cryostat and the downward door, the field will be 
measured up to the yoke innermost radius. The grid step will be chosen
in the range $2$ to $7\unit{cm}$ according to the required accuracy and
the foreseen field gradient in different regions. The field will be
measured in a second stage on different circumference, around the beam
axis, to measure the deviation from the perfect cylindrical symmetry.
The $z$ and $r$ coordinates where to perform these measurements will be
determined according to the needs of the various sub-detectors. In 
addition, a certain number of fixed gauges will be inserted in the 
muon chamber gaps in the yoke and downward door to check the consistency
of the magnetic calculations.

As a very last step the muon detectors will be introduced into the
gaps of the yoke, all detector systems will be installed with the
foreseen support structures inside the cryostat; and finally the target
system will be inserted from the top and bottom of the yoke.


\begin{figure}[th]
  \begin{center}
    \includegraphics[width=\swidth]{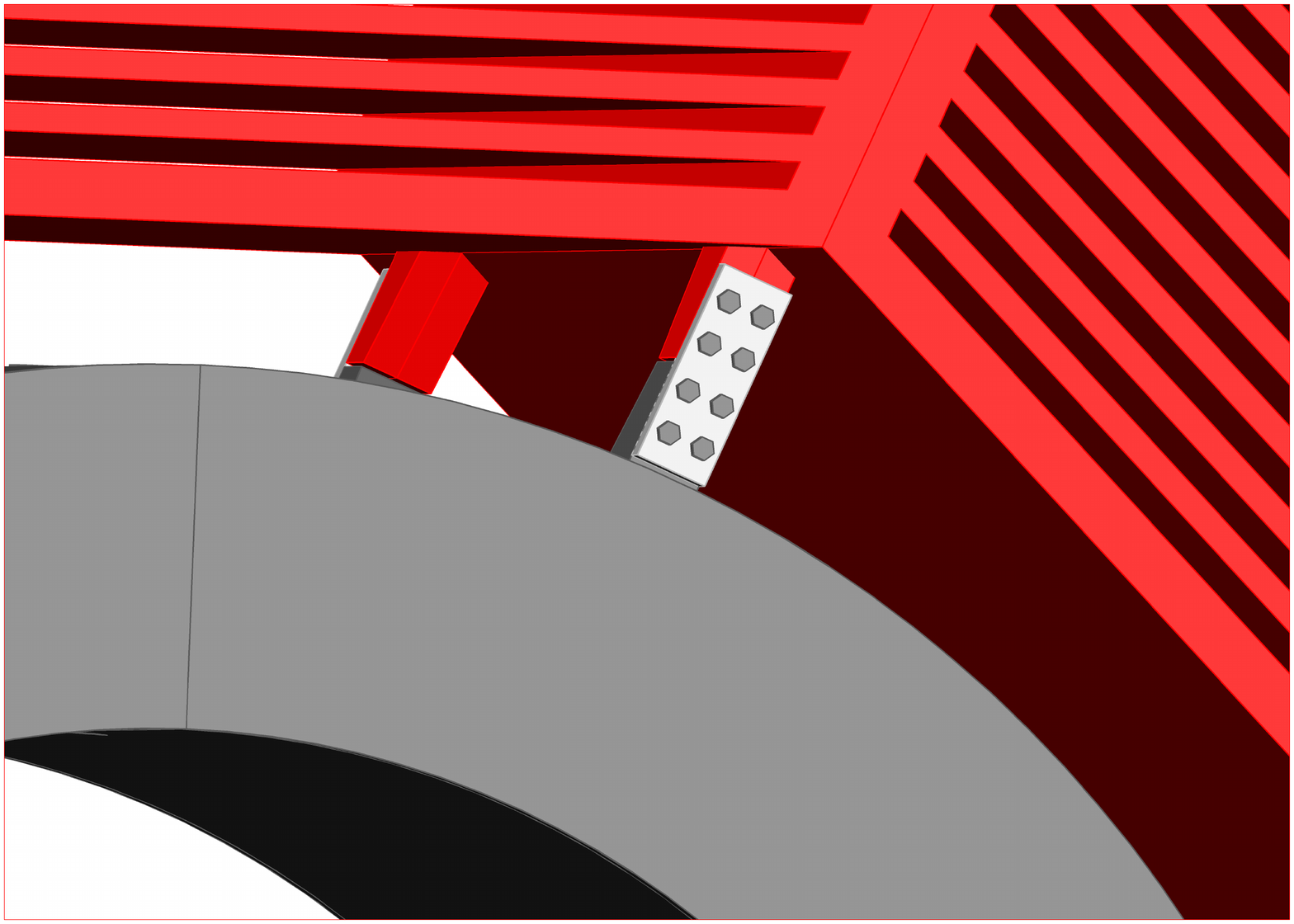}
  \caption[Cryostat attachments to the flux return yoke.]
  {View of 2 of the 16 cryostat attachments to the flux
     return yoke.  The attachment of the mounting points via a plate
     will allow for an easy adjustment of the cryostat within the barrel
     yoke.}
  \label{fig:staffa_prosp}
  \end{center}
\end{figure}

\subsection{Cryostat Mounting in the Yoke}

The coil will be delivered by the manufacturer enclosed inside the
cryostat with the proximity cryogenics already attached to it. Thus
the cryogenic chimney, which will be located at the upstream top end of the
cryostat, will be in place.  The whole system will be handled as a
single object.  A support structure will be produced to place the
whole object into the yoke barrel and allow for its safe mounting.
The cryogenic chimney will be hosted in the dedicated cut out in the
upper octant of the iron yoke.  This structure will, most likely, be a
slightly modified version of the support structure which will have
been manufactured for the handling of the coil and cryostat at the
company before. In principle, the same support beam and movement
system could also be used to install, with minor modifications, the
barrel part of the EMC in the warm bore of the cryostat.

The cryostat will be attached to the instrumented flux return yoke via
16 mounting points, 8 at each end of the cryostat.  The yoke will be
equipped with 16 equivalent mounting points, placed at the
intersections of the beams of the octagonal barrel.  The support
points inside the yoke and on the cryostat will be accurately measured
before the cryostat insertion, to ensure the mechanical tolerances be
satisfied for safe operation. The supports will be bolted together
using 8 M14 bolts each, 4 on the mounting points of the cryostat and 4
on the mounting points from the yoke. The positioning of the support
structures is shown in for one of the 8 identical azimuthal positions
in Fig.~\ref{fig:staffa_prosp}.

The cryostat will be bolted to the yoke and placed, using adequate
spacers, to a precision better than $1\unit{mm}$ with respect to the
nominal position.  Further adjustments of the coil position will be
possible acting on the support rods.  The clearance that allows a fine
position adjustment is visible where the grey and red parts meet in
Fig.~\ref{fig:staffa_prosp}. The cryostat has been designed to behave
as a very rigid body, as specified in the previous sections, so a
deformations of the order of less than $0.2\unit{mm}$ is foreseen after
the installation and the power up w.r.t.\  the nominal position. This 
value is expected to grow no more than an additional $0.1\unit{mm}$
when the detector loads will be mounted to the cryostat.

\begin{figure*}
  \begin{centering}
    \includegraphics[width=\dwidth]{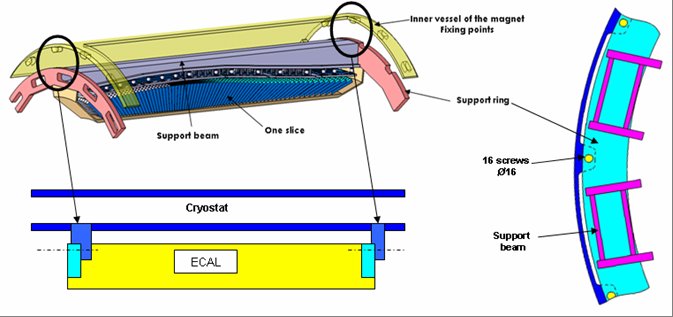}
  \caption[The attachment of the EMC barrel to
    the cryostat.]{Several views detailing the attachment of the EMC barrel to
    the cryostat. Only a small fraction of the barrel is shown.  The
    top left panel shows an exploded view where the brackets on the
    cryostat vessel and the EMC support ring are separated.  The lower
    left and right panels show cross sections of the mounted system in
    the $z-y$ and $x-y$ planes, respectively.}
  \label{fig:EMC-Yoke-Attachment}
  \end{centering}
\end{figure*}

\subsection{Detector Installation}

The detectors to be installed within the solenoid fall into two main
categories: the barrel detectors brought in from the upstream side,
and the forward end cap detectors brought in from the downstream side
of the yoke.  All detectors can be installed and removed with the
magnet fully assembled but will require the doors to be open.  In the 
following, the detectors are listed in the sequence of installation, 
putting emphasis on the
detectors directly attached to magnet components or most crucial for
the design.  The weights of the individual detectors are listed in
Table~\ref{t:over:det-masses}.

\subsubsection{Muon Detectors}

\begin{figure}[ht]
  \begin{center}
    \includegraphics[width=\swidth]{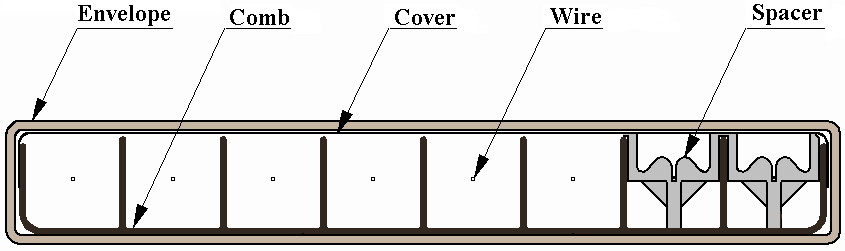}
    \caption{Cross section of a muon MDT module.}
    \label{fig:MDT}
  \end{center}
\end{figure}

The muon detectors can be installed completely independently of the
other detectors. The single detector element will be an array of $8$ Mini 
Drift Tubes, of variable length according to the position where the 
element should be installed. A cross section of an array of MDT is shown 
in Fig. \ref{fig:MDT}. The elements can be inserted individually, 
simply by hands being relatively light ($\sim 1\unit{kg}$ per linear 
metre of array).

The detectors installed in the iron yoke will be slid in the proper
clearances from downstream, with the end doors open. Shorter elements
are foreseen in the upper octant, due to the target recess. The cables 
will be routed through the slits present in the barrel corners, at
downstream end. To ensure a proper cabling scheme, the muon detectors
will be the first system to be install to the yoke.

The detectors installed in the downstream doors will be inserted in
their clearances from the sides. Different lengths are foreseen to 
allow the best coverage of the octagonal surface of the doors.
The cables will be routed from the sides of the end doors.

\subsubsection{Barrel EMC}

\begin{figure}[ht]
  \begin{center} \includegraphics[width=\swidth]
                 {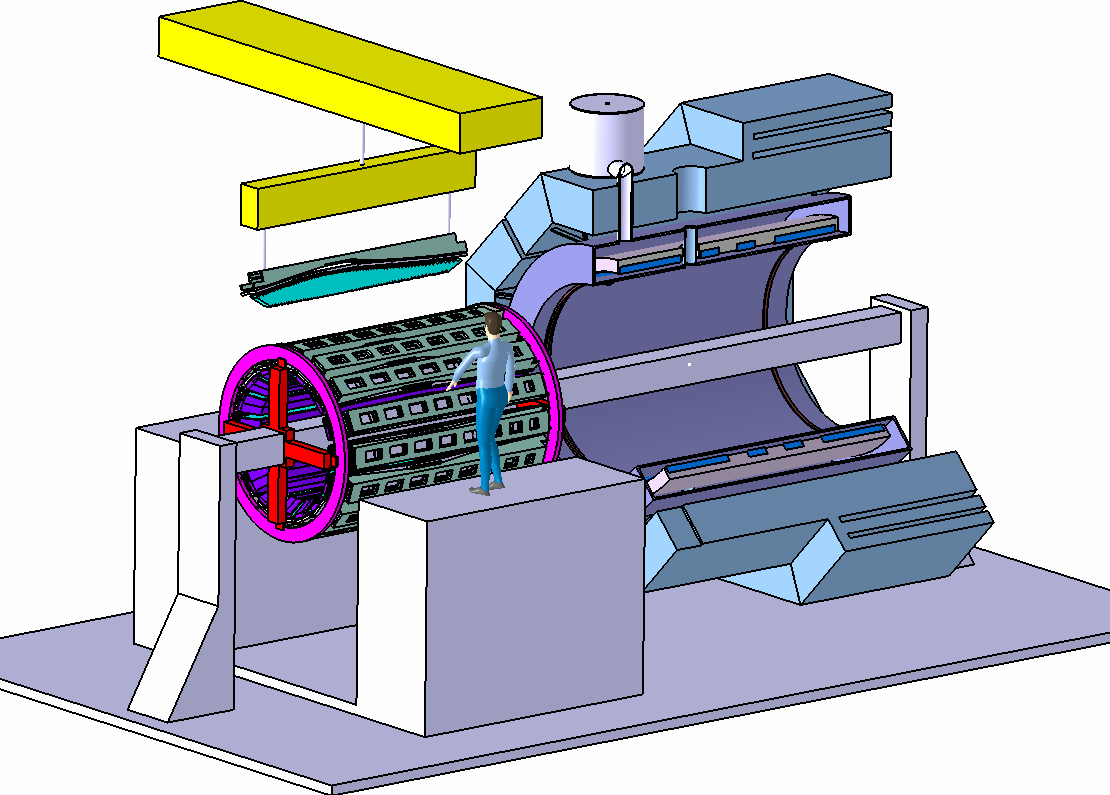}
\caption[Mounting procedure of the Barrel EMC.]{Artistic view showing
    the mounting procedure of the Barrel EMC.  After the assembly of
    all parts in a self-supporting barrel structure outside of the
    solenoid the whole system will be transferred on a central beam
    into the operation position.  After attaching the structure to the
    cryostat the inner beam and support crosses are removed to make
    space for the inner detector installation.}
\label{fig:EMC-mounting}
  \end{center}
\end{figure}

The Barrel Electromagnetic Calorimeter (EMC) will be connected to the
inner surface of the cryostat using 32 dedicated mounting brackets at
the inner rim of the cryostat vessel.  The attachment between the EMC
and the cryostat is shown schematically in
Fig.~\ref{fig:EMC-Yoke-Attachment}.  The design ensures that the fully
assembled barrel of the EMC can be moved in from the upstream side
after the installation of the cryostat inside the flux return yoke
(see Fig.~\ref{fig:EMC-mounting}).  The weight of the Barrel EMC is
estimated to be about $20\unit{t}$, which will be by far the major
contribution to the load that must be sustained by the cryostat from
the detector side.  The cryostat itself has been designed to take into
account this additional load with adequate safety margins.  For
further details about EMC design and installation procedures, please
refer to EMC TDR~\cite{PANDA:TDR:EMC}.

\subsubsection{Barrel DIRC}

The Cherenkov detector based on detection of internally reflected
Cherenkov light covering the large angles (Barrel DIRC) will be
mounted into the solenoid in 2 independent parts.  First the barrel of
quartz slaps will be inserted into the solenoid just inside the Barrel
EMC.  Its weight of 300\,kg will be supported by the common support
system for the inner detectors which is attached to the cryostat.

The ring structure, where read-out boxes with a weight of about
800\,kg are encased, will be mounted separately.  It will be mounted
after the inner detectors (see next section) will be in place. It
is attached to the upstream end of the Barrel yoke by a spider-like
structure supported using the same kind of mount points as in the
downstream end.

\begin{figure}[th]
  \begin{center}
    \includegraphics[width=0.8\swidth]{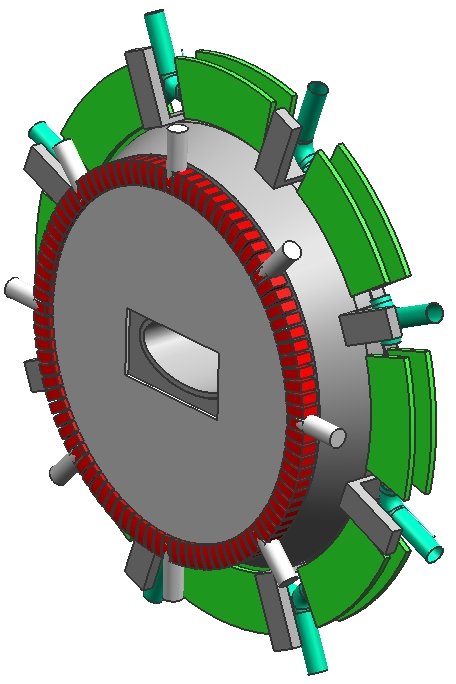}
\caption[End caps of the EMC and Disc DIRC with mounting brackets.]
    {Forward end cap of the EMC and Disc DIRC with mounting brackets
    which will also serve the purpose of guiding the cables and
    supplies to the cut outs in the iron yoke.}
\label{fig:EMC-FwEndCap}
 \end{center}
\end{figure}

\begin{figure}[th]
  \begin{center}
    \includegraphics[width=\swidth]{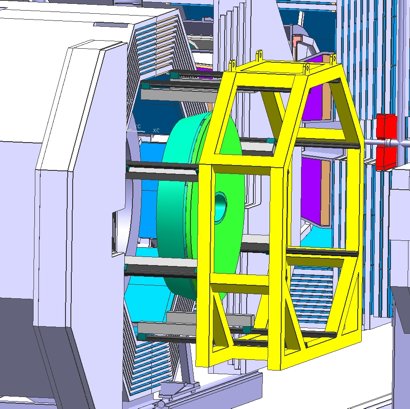}
    \caption[The mounting procedure for the forward
    end cap of the EMC.]{Artistic view of the mounting procedure for the forward
    end cap of the EMC showing the frame which allows the safe
    movement of the system into the flux return yoke.  Some additional
    support structures have been omitted for visibility.}
    \label{fig:EMC-FwEndCap-Mount}
  \end{center}
\end{figure}

\subsubsection{Inner Detectors and Beam Pipe}

The beam pipe and the interaction cross will be installed with the
micro-vertex detector (MVD) and the outer tracker, in a single support
frame.  The pipe with the frame will be prepared outside the
yoke, then the detectors will be mounted on it and it will be placed
in its working position from the upstream side.  It will travel on a
dedicated rail system attached to the lateral surface of the
cryostat.\\

Turbo pumps will be installed to remove the background gas coming from
the target just outside both the upstream and downstream ends of the
solenoid magnet. The locations of the pumps are given in
Table~\ref{t:req:stray-field}.

\begin{figure}[th]
  \centering
    \subfigure[Bird's eye view]{\includegraphics[width=0.8\swidth]
        {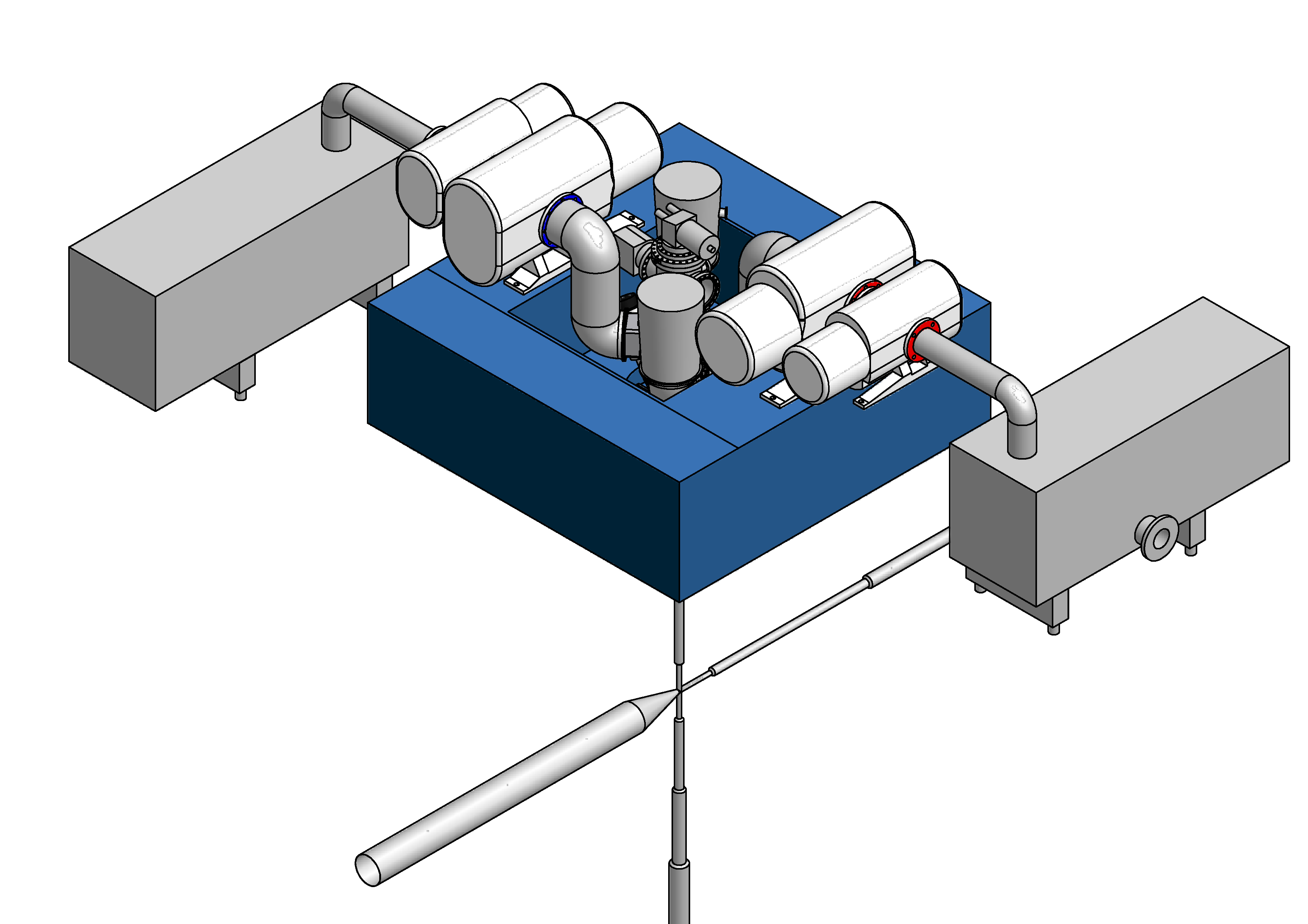}}
    \subfigure[Cross section viewed from upstream]{\includegraphics[width=\swidth]
        {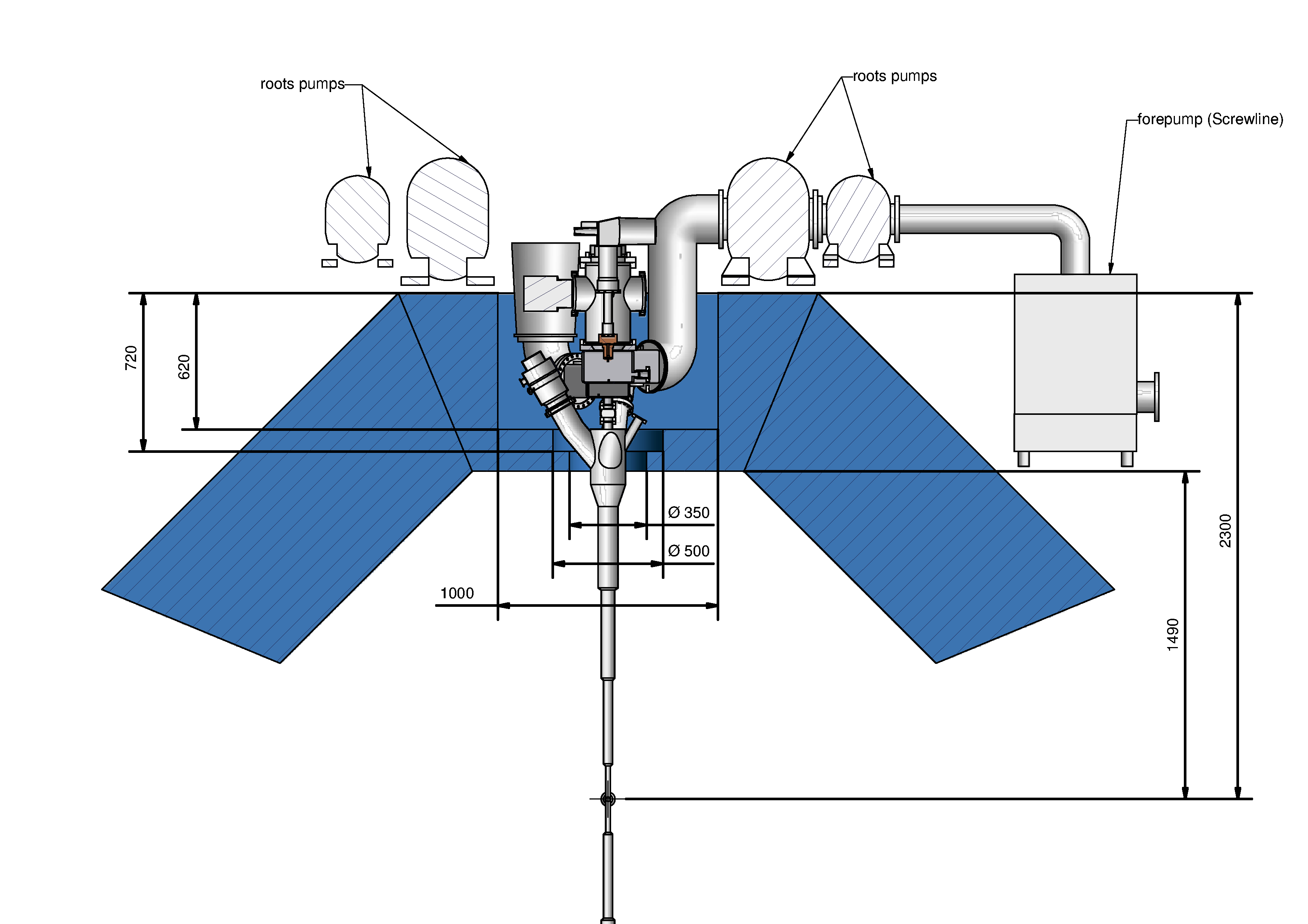}}
  \caption[Integration of the generator of the cluster-jet target.]
  {\label{fig:Target_Generator_CJT} Views of the foreseen
    integration of the generator of the cluster-jet target in the yoke
    of the solenoid.}
\end{figure}

\subsubsection{Forward End Cap Detectors}

The three stations of Gas Electron Multiplier (GEM) detectors will be
mounted as the first item inserted from the downstream end of the
solenoid. The details of their attachment are not yet specified
precisely.  It is envisaged that the comparably small weight of these
detectors will be attached to the same frame which holds the inner
detectors.

The Disc shaped DIRC detector in the forward end cap will be mounted
just in front of the forward end cap of the EMC.  Mechanically it will
be assembled onto the same holding structure and installed together
with the EMC.  The Forward end cap of the EMC is shown in
Fig.~\ref{fig:EMC-FwEndCap}.  Here the DIRC is mounted and covers the
crystals of the EMC.  Once the DIRC is attached the whole system will
be placed on a temporary frame which will allow to safely move the
array to the operational position and mount it to the inner surface of
the yoke (see Fig.~\ref{fig:EMC-FwEndCap-Mount}).

\subsection{Target Integration}

The installation of the target production and recovery stages will
take place after the beam pipe cross with the detectors has been
inserted.  The upper and lower straight sections of the target pipe
will be inserted from the top and bottom of the flux return yoke,
respectively.  The connection inside the solenoid will be done through
fast joints which can be operated from a distance.

A recess in the upper octant of the barrel is needed to host the
turbo-molecular pumps and cryogenics for the production stage, as the
cluster jet target requires to keep a maximum distance of $2\unit{m}$
between the nozzle and the interaction point for safe operation.

The roughing pumps will be installed on the sides of the production
stage, as close as possible to the turbo-molecular pumps, on dedicated
supports.  Views showing the integration of the production stage in
the instrumented flux return yoke are given in
Fig.~\ref{fig:Target_Generator_CJT}.  The recovery stage will be
attached to lower part of the target pipe in a similar way. For that a
smaller recess of 500\,mm depth and a diameter of 1\,m is sufficient.

The target dump will be installed under the yoke barrel. A proper 
clearance under the support structure is foreseen to allow the vacuum
chamber and pumping installation: in addition, a platform suspended
to the yoke carriage is foreseen to allow the hosting of the fore-vacuum
pumps and of the proximity electronics of the dump itself.

\begin{figure}[th]
  \begin{centering}
    \includegraphics[width=\swidth]
            {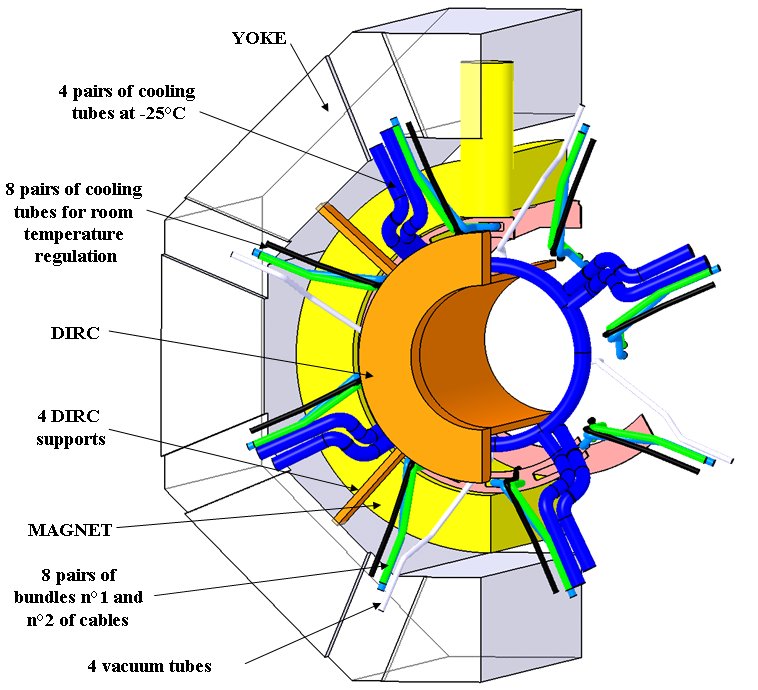}
  \end{centering}
  \caption[Routing of the cooling pipes, 
     supply lines and signal cables of the Barrel 
    EMC.]{Schematic view showing the routing of the cooling pipes 
    (blue) and all other supply lines and signal cables of the Barrel 
    EMC at the upstream end of the solenoid.  For visibility details 
    have been omitted and magnet and detectors are cut in half.}
\label{fig:EMC-supplies}
\end{figure}

\subsection{Cable and Supply Routing}

Cables and pipes coming from the detectors will be guided to the outside
of the yoke in their majority through 16 cut outs in the corners of
the instrumented flux return barrel, 8 at the upstream and 8 at the
downstream end of the barrel, $42\unit{cm}$ wide and $14\unit{cm}$ long. 
(Please also refer to Sec.~\ref{s:sol:yoke:struct} and
Fig.~\ref{yoke:Cut-outs}.)

Together with the supply lines, all signal cables coming from most of
the detectors will be routed through these slits.  It should be
emphasised that all cables will be mounted taking into account
additional magnetic and Laplace forces acting on them.  The upstream
end cap of the EMC and the central inner and outer trackers will have
their supply lines and signal cables guided to the upstream end close
to the beam pipe.  As this will not affect the magnet design this is
detailed elsewhere.  The routing of the Barrel EMC supply lines and
signal cables through the upstream end of the yoke is illustrated
schematically in Fig.~\ref{fig:EMC-supplies}.

\bibliographystyle{tdr_lit}
\bibliography{lit}

\cleardoublepage
\svnInfo $Id: dipole.tex 692 2009-01-30 10:43:35Z IntiL $

\chapter{Forward Spectrometer}
\label{s:dipole}
 
\AUTHORS{I.~Lehmann, J.~L\"uhning, E.~Lisowski}
  
\begin{figure}[htb]
\begin{center}
\includegraphics[width=\swidth]{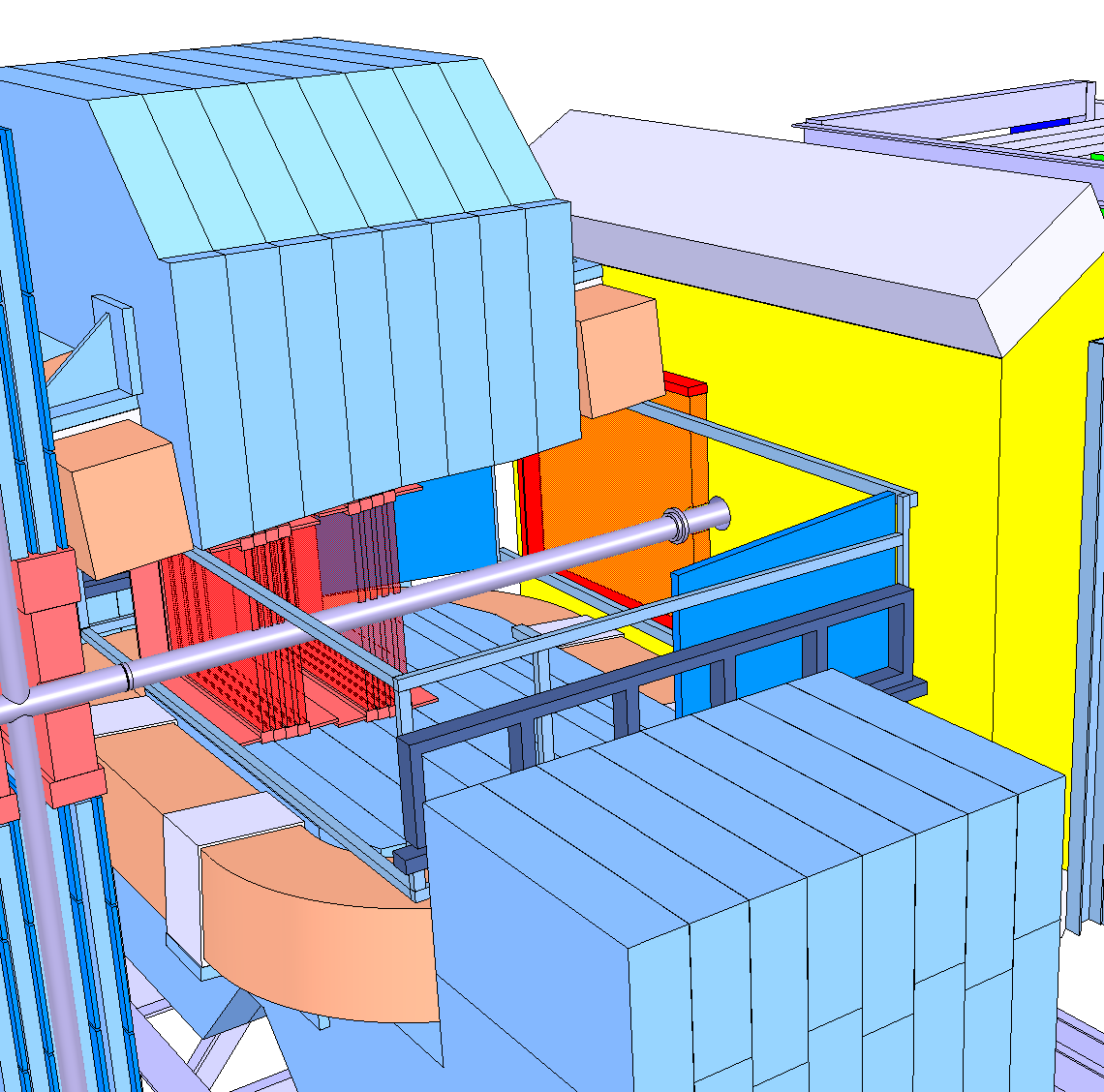}
\caption{View of the dipole magnet cut for visibility and parts of the
forward detectors of PANDA.}
\label{f:dip:zoom}
\end{center}
\end{figure}

The central solenoid will be augmented by a large-aperture dipole with 2\,Tm
bending power.  It will be equipped with drift chambers for particle
tracking and scintillation counters for time-of-flight measurements
inside the aperture, and followed by a full forward detection system.
The \PANDA Forward Spectrometer (FS) will cover angles of up to 5 and 10 degrees
in the vertical and horizontal planes, respectively. A view of part of
the system showing the instrumented dipole is shown in
Fig.~\ref{f:dip:zoom}.
 
The major challenge in the construction of the dipole is the
achievement of a 2\,Tm field integral over a short longitudinal
extent, while maintaining reasonable operational costs.  The length of
the dipole along the beam axis determines the position of the forward
detectors.  However, the area of the forward detectors obviously
increases quadratically with distance to the interaction point.  Thus
it is necessary to minimise the dimensions of the dipole in this
direction.  On the other hand, a shorter dipole with large aperture
requires higher fields which increase the saturation of the iron yoke.
Saturation increases the required currents through the coils
considerably.  The use of a superconducting coil, although
superficially attractive, proved impractical and uneconomic for
several reasons discussed in the following.  Another important feature
of the dipole will be that it will need to be part of the accelerator lattice
of the High Energy Storage Ring (\HESR).  Ramping capabilities
compatible with the accelerator will be ensured by segmenting the yoke
and providing an appropriate power supply.  Extensive optimisation
studies have led to a detailed design, which is described later in
this chapter.

\svnInfo $Id: concept.tex 692 2009-01-30 10:43:35Z IntiL $

\section{Conceptual Design}
\label{s:dipole:concept} 

\begin{figure*}
\begin{center}
\subfigure[Superconducting design]{\includegraphics[height=7cm]
{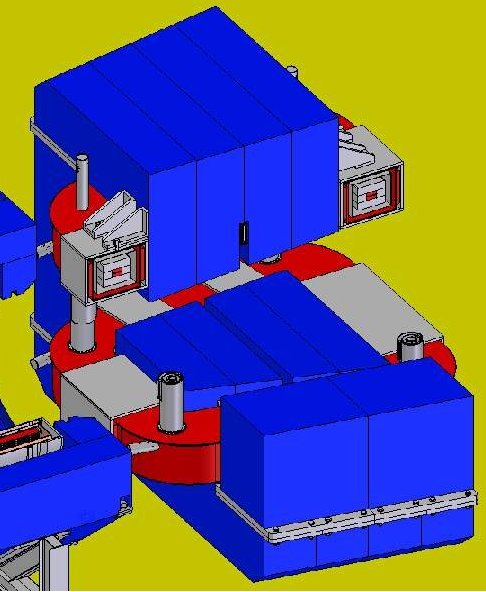}}
\subfigure[Bedstead-type coil design]{\includegraphics[height=7cm]
{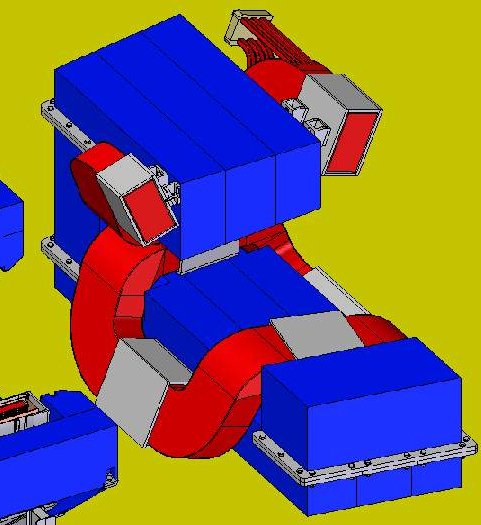}}
\caption[Examples for design options considered for the dipole magnet.]
{Examples for design options considered for the dipole magnet.
(a) Superconducting coils decrease the power consumption in the
cable but impose mechanical constraints and require
refrigeration. (b) A non-planar coil (shown here is the bedstead
type) may allow a rather short design but the field properties are
less convenient than in the planar construction which was finally
chosen.  }
\label{f:dip:discont}
\end{center}
\end{figure*}

The conceptual design of the dipole magnet was detailed by a
collaboration of research groups experienced in the design and
construction of such magnets.  The main contributors were engineers
from JINR, Dubna, physicists and engineers from GSI, Darmstadt, and
physicists from the University of Glasgow.  The design studies
addressed the issues of physics requirements, as laid out above, field
quality, mechanical considerations and economic construction and
operation.  The option to shield the beam from the field by an iron
plate was considered but it proved unacceptable.  The primary reason
not to further consider this option was the required acceptance for zero degree
particles.  Furthermore, large background contributions were expected
if placing large amounts of material close to the beam.  In
Fig.~\ref{f:dip:discont} two options for the coil design are shown
which were considered but finally dismissed.

A design selection has now been made, based on the reasoning and
criteria presented below.  Minor changes may however be introduced
when the details of the construction are evaluated together with the
manufacturer.  A superconducting coil does not provide an attractive
alternative because of the following points.
\begin{itemize}
\item The vertical force on each coil will be higher than 100 tons.  For 
      a superconducting coil this would demand great skill in the
      design of the suspensions.  Apart from the imposed risks and
      costs this would require a considerable amount of cryogenic
      cooling, which would marginalise savings due to reduced power
      consumption.
\item A suspension ensuring low heat losses would require rods in 
      the yoke gap, which would reduce the acceptance of the
      detectors.
\item Our design foresees racetrack shaped coils. The reason for this 
      is a good magnetic stiffness not only on the beam axis but also
      at an horizontal deflection angle of $10^\circ$.  A disadvantage of
      non-circular coils is the fact that they are subject to bending
      moments in the winding plane.  For the large cross sections
      of normal-conducting coils this poses no problem; but
      superconducting coils with their rather small cross sections
      would need additional bending stiffness in the winding plane.
\item The dipole magnet will be part of the \HESR lattice and therefore 
      needs to be ramped during acceleration within 60\,s.  This
      requirement imposes high demands on the conductor and its
      cooling for a superconducting dipole.  Special measures would have 
      to be taken
      to limit coupling and eddy current loss~\cite{Wilson:1983} 
      and to ensure cryogenic stability.
\item The construction cost for a dipole magnet with superconducting 
      coils is significantly higher than using resistive coils.  From
      rough estimates we have deduced factor of 2 higher construction
      costs for a similar superconducting dipole magnet.
\item The operation of a superconducting magnet is more challenging and 
      higher efforts on maintenance and survey are required than in
      the case of resistive coils.  Finally, the risk of downtimes or
      failure of operation is minimal in the latter case.
\end{itemize}
In conclusion, we have taken the decision to pursue the resistive coil
option.  Similar considerations have led to the design choice of other
large aperture dipole magnets (see {\it e.g.\ }the LHCb
magnet~\cite{LHCb}).

The optimum coil shape has been determined based on studies of
current, effective field, stray field and geometrical dimensions. Many
different layouts have been considered, including bedstead and
deck-chair shapes. Finally it was found that racetrack type coils,
i.e.\ laid out in one plane, provide satisfactory performance and lowest
cost simultaneously. 

The effect of field clamping has been studied in detail. It was shown
that, indeed, field clamping is required to separate the fields from
the solenoid and dipole magnets.  It was found that it is advantageous
to extend the upstream field clamp such that it can simultaneously act
as an extension to the range system for the muon identification above
$5^\circ$ and $10^\circ$ in vertical and horizontal angles,
respectively.  This system will replace this field clamp fully, but will
be mechanically independent and will be mounted only after the
installation of all detectors.  Therefore, we treat it as a separate
part in Sec.~\ref{s:dip:det:sol-dip}.  To the downstream side of the
dipole the necessity of field clamping is not yet clear, as the
details of the detectors in that region are not yet defined.  We
foresee the possibility to attach several sheets of iron downstream of
the coils.  Details of their design are not crucial to the design of
the remaining components of the magnet.

\svnInfo $Id: coil_yoke.tex 712 2009-03-31 16:27:19Z IntiL $

\section{Coil and Yoke Design}
\label{s:dipole:coil}
\label{s:dipole:yoke}

\AUTHORS{I.~Lehmann, J.~L\"uhning}

The total length along the beam has been optimised to balance magnet
and detector requirements.  A longer magnet would have lower power
dissipation, however the area the forward detectors have to cover
would increase quadratically with increasing distance from the
interaction point. The solution was to select a magnet length of
2.5\,meters, which achieves the desired field integral without
considerable saturation in the iron yoke.  The overall height and
width are 3.9 and 5.3\,metres, respectively.

\begin{figure}
\begin{center}
\includegraphics[width=\swidth]{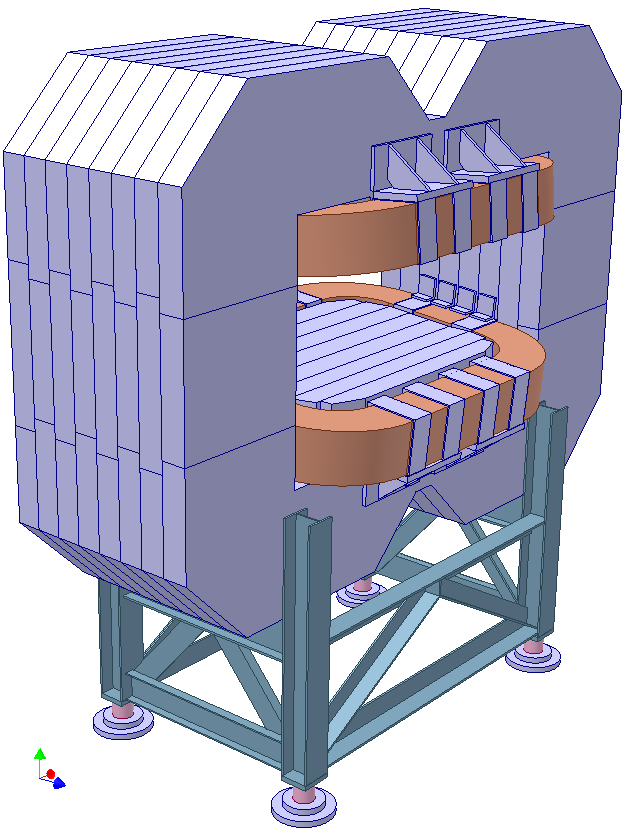}
\caption{View of the large-aperture dipole magnet from the downstream 
  side.}
\label{f:dip:3d_view}
\end{center}
\end{figure}

\begin{figure}[th]
\begin{center}
\includegraphics[width=\swidth]{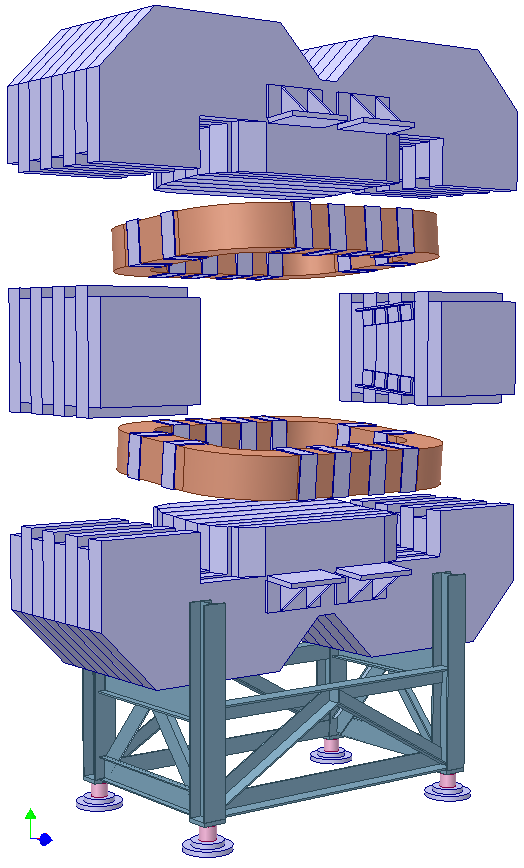}
\caption{Exploded view of the flux return yoke from the downstream 
  side.}
\label{f:dip:expl_view}
\end{center}
\end{figure}

\begin{figure}[ht]
\begin{center}
\includegraphics[width=\swidth]{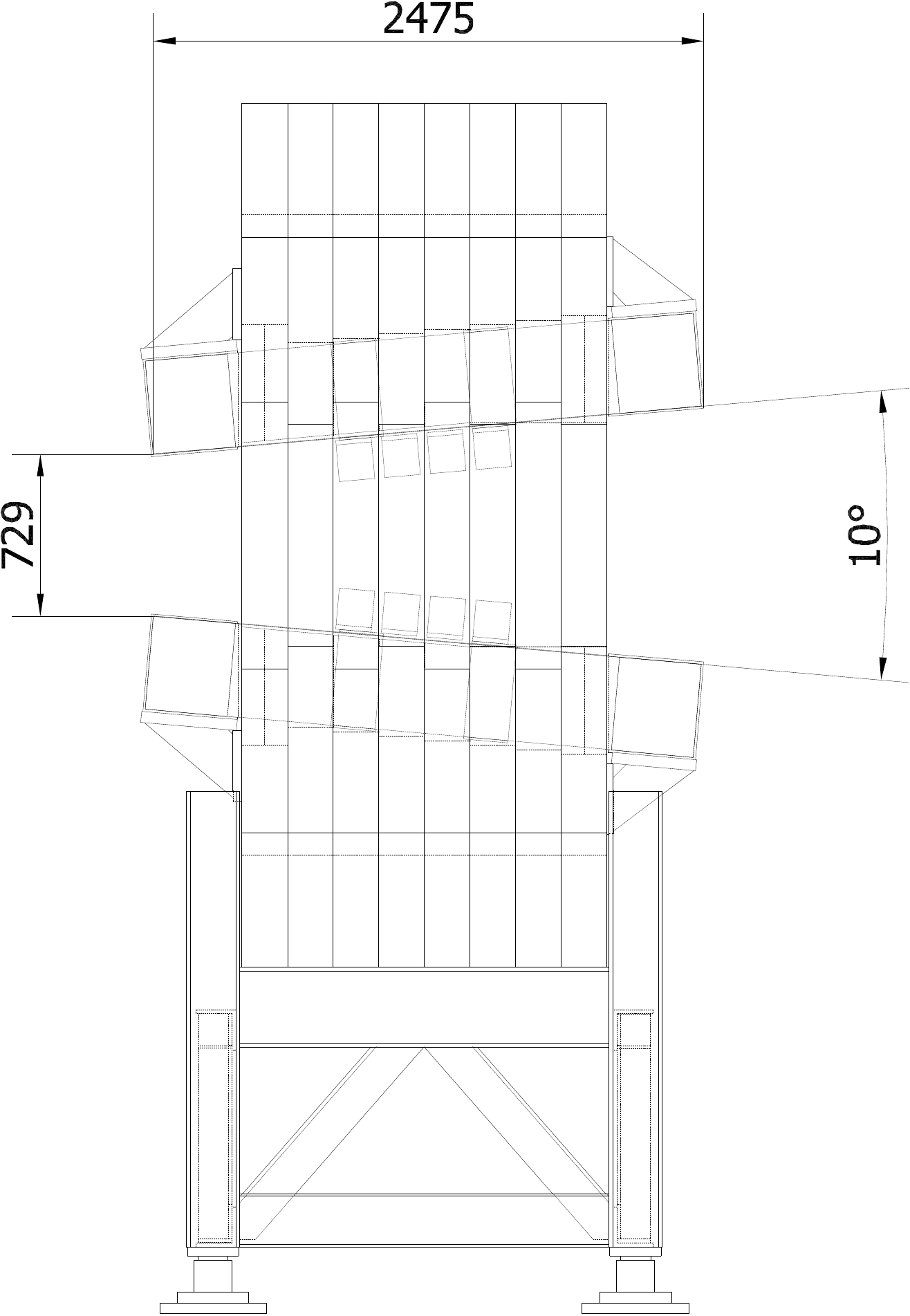}
\caption[Side projection of the dipole magnet.]
{Side projection of the dipole magnet showing the overall
  length and minimal gap in millimetres. The beam comes in from the
  left.  Please note that the opening angle of the coils given here
  does not fall on the $\pm 5^\circ$ line from the interaction point.}
\label{f:dip:proj_vert}
\end{center}
\end{figure}

\begin{figure}[ht]
\begin{center}
\includegraphics[width=\swidth]{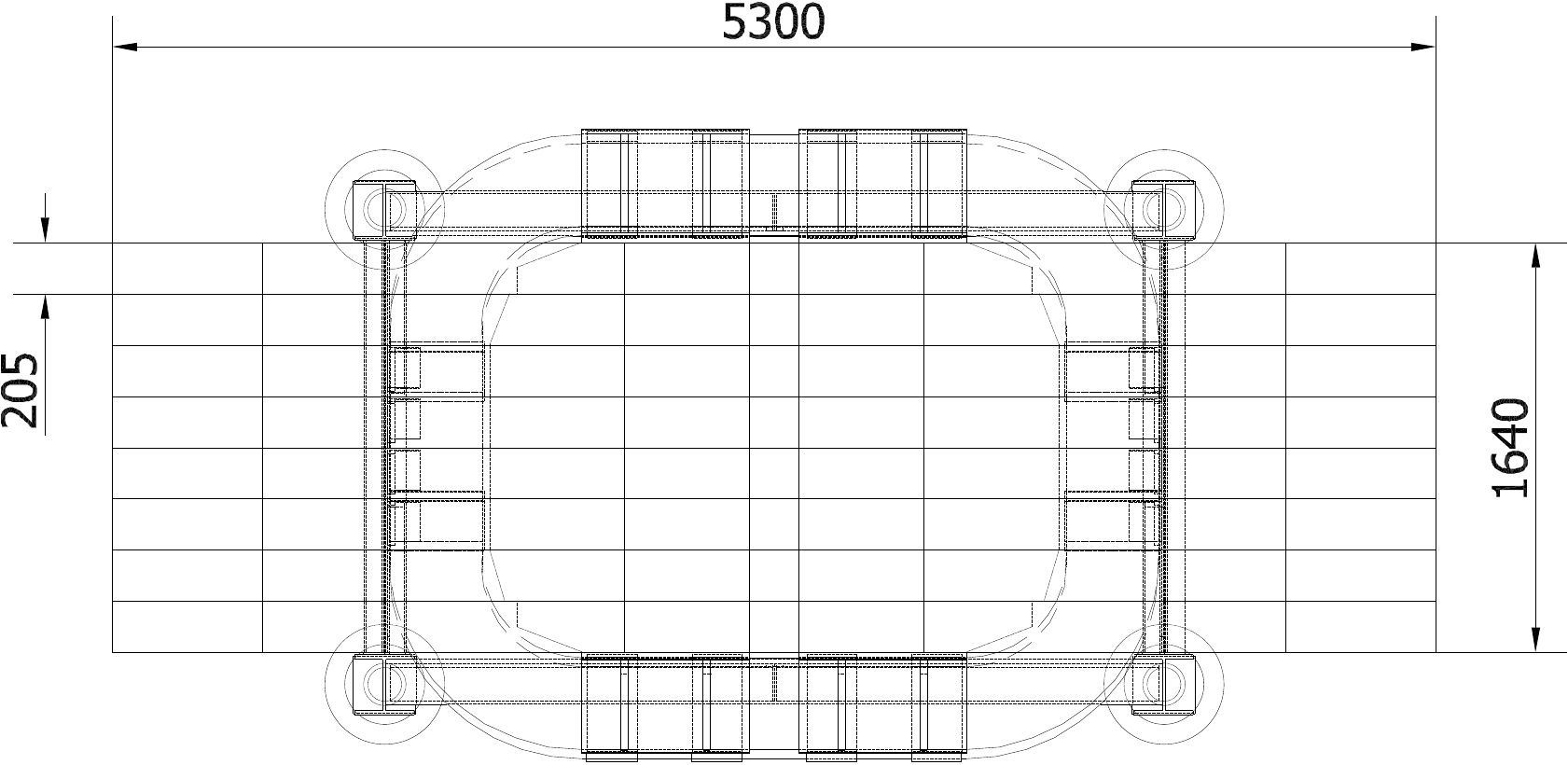}
\caption[Top projection of the dipole magnet.]
{Top projection of the dipole magnet showing the dimensions of 
  the yoke in millimetres.  The beam comes in from the top.}
\label{f:dip:proj_horiz2}
\end{center}
\end{figure}

\begin{figure}[ht]
\begin{center}
\includegraphics[width=\swidth]{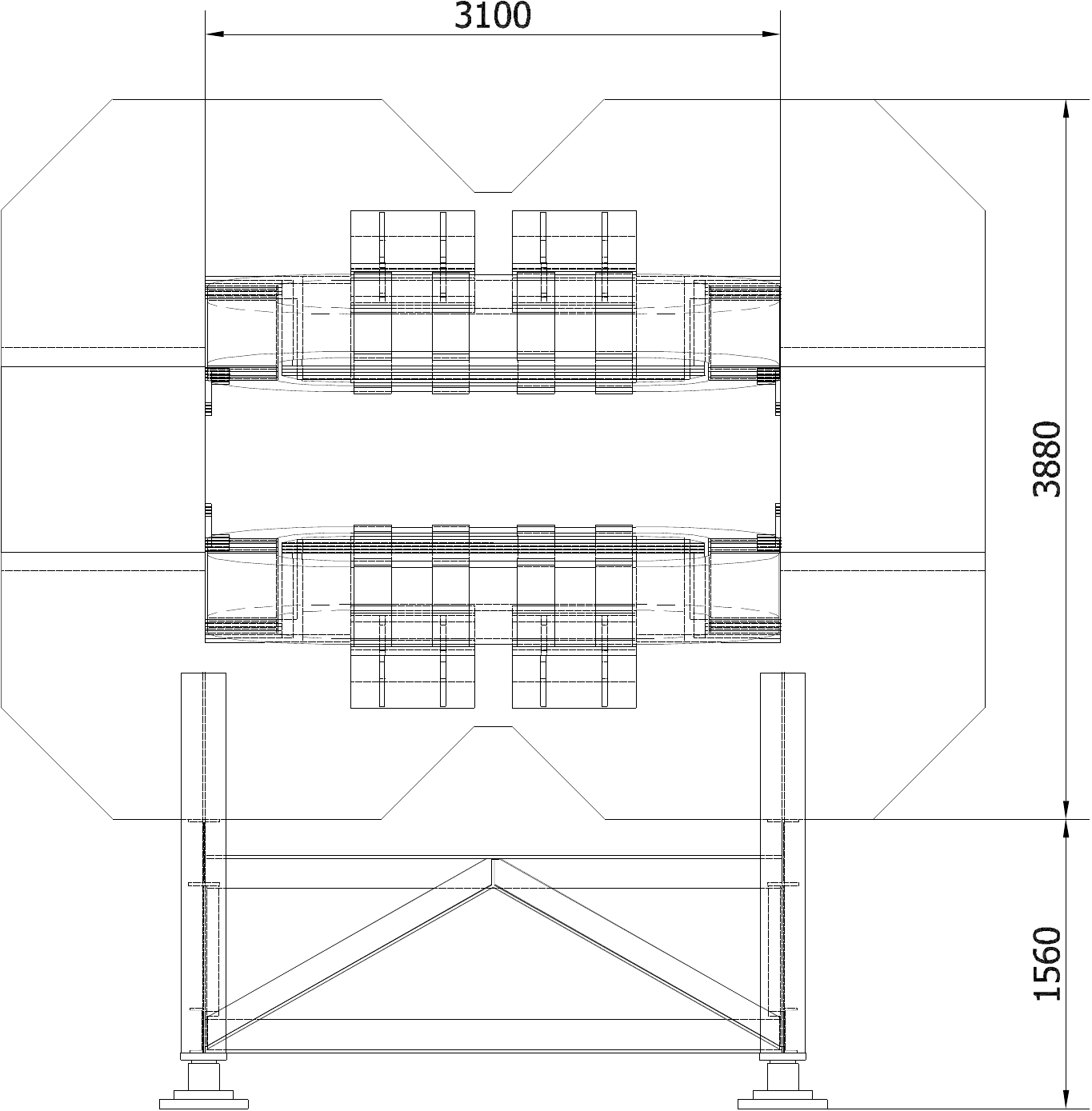}
\caption[Projection from the downstream side of the dipole magnet.]
{Projection from the downstream side of the dipole magnet 
  showing the dimensions of the yoke, the gap opening and the distance
  to the HESR floor in millimetres.}
\label{f:dip:proj_horiz1}
\end{center}
\end{figure}

The yoke is to be built out of plates of low carbon steel. We foresee
AISI 1006 here, which is a high quality, low carbon (0.06\%) magnetic
steel. The flux return yoke will be segmented for two reasons. First of all
a moderate segmentation is mandatory as the magnet needs to be ramped
with the acceleration of particles in HESR.  At the anticipated ramp
time of 60\,s and a plate thickness of 20\,cm the eddy currents will stay
below 5\,A/cm$^2$, the power dissipation below 400\,W, and the delay
between the current and field will be less than two seconds (see 
Sec.~\ref{s:dipole:perf:dynamic}).   Secondly, the
weight of each individual magnet part is below 15\,t and so the crane
in the experimental area can carry each part.  Hence there is no need
for an additional crane for assembly.  The total assembled weight will be
of the order of 220\,t.  An exploded view showing the mounting points
for the coils is shown in Fig.~\ref{f:dip:expl_view}. Three projections
in the three coordinate planes are shown in
Figs.~\ref{f:dip:proj_vert} to~\ref{f:dip:proj_horiz1}, respectively.

\begin{figure}[ht]
\begin{center}
\includegraphics[width=\swidth]{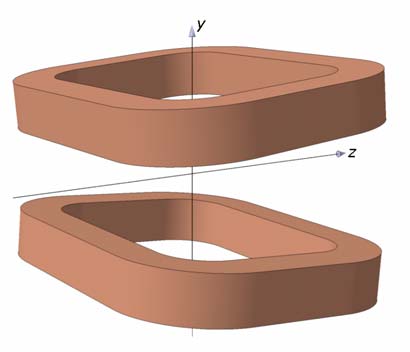}
\caption[Schematic view of the coil arrangement in the dipole magnet. ]
{Schematic view of the coil arrangement in the dipole magnet.  
  The upper and lower coils are inclined in the $z-y$ plane by plus and
  minus $5^\circ$, respectively.  The $z$ axis is defined by the
  antiproton beam trajectory before the entrance into the magnetic
  field of the dipole magnet and the $y$ axis is the vertical
  direction.}
\label{f:dip:coils}
\end{center}
\end{figure}

\begin{figure}[ht]
\begin{center}
\includegraphics[width=\swidth]{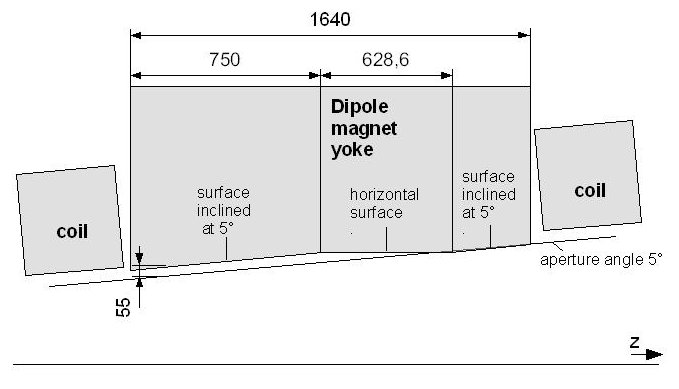}
\caption[Sketch of the pole shoe geometry in the $z-y$ plane.]
{Sketch of the pole shoe geometry in the $z-y$ plane. This 
  shape allows the accommodation of in-gap detectors without
  constraining the vertical acceptance while keeping the field maximal
  and continuous.}
\label{f:dip:pole_shoe}
\end{center}
\end{figure}

The opening of the yoke and its coils are designed such that particles
emitted from the target with vertical angles below $5^\circ$ in the
vertical plane traverse the magnet fully and reach the detectors in
the forward region.  As the mounting structure and the frames of the
in-gap detectors (see Sec.~\ref{s:dipole:int}) use some of the space,
the rectangular upper and lower coils are located 5.5\,cm outside the
upper and lower $5^\circ$ planes (see Fig.~\ref{f:dip:coils}).  The
use of inclined coils rather than horizontal coils reduces the forces
on the muon filter by a factor of 3 and reduces the power consumption.
The pole shoes will be manufactured such that the frame has the
appropriate space while keeping the iron of the flux return closing in
as far as possible to reach a maximum field integral.  In addition the
surfaces will be kept without steps in order to keep field gradients
minimal.  A tentative sketch of the envisaged design in shown in
Fig.~\ref{f:dip:pole_shoe}.  The yoke aperture opens out from 0.80\,m to
1.01\,m in vertical direction while staying constant at 3.1\,m in the
horizontal plane. The constant horizontal aperture width will
facilitate the installation of tracking and time-of-flight detectors
to distinguish slower particles that do not exit the magnet.

Copper is the selected conductor material, despite being slightly more
expensive than aluminium, due to its lower resistivity and better 
ductility.  Only
with copper the required small bending radius can be achieved.  Copper
can cope with higher current densities and reduces operational costs
due to a 37\% reduction in power dissipation. It is calculated that
the increased purchase cost due to the selection of copper rather than
aluminium will be fully compensated by the reduced power consumption costs
in less than one year of operation.  Additionally, the choice of a
copper conductor allows a more compact magnet design, thereby reducing
the cost for the forward detectors. The selected hollow copper
conductor has a $30 \times 24\,$mm$^2$ cross section and a cooling
channel diameter of 10\,mm.  A likely arrangement would be to have 6
double pancakes with $14+14$ windings each in both the upper and lower
coils.  The main parameters of the magnet are listed in
Table~\ref{t:dipole:param}.

\begin{table}[ht]
\begin{center}
\begin{tabular}{|l|c|}
\hline
Item & Value \\
\hline
\multicolumn{2}{c}{Coils}\\
\hline
Type            & Resistive \\
Material        & Copper    \\
Weight          &   $\sim 18$\,t        \\
Arrangement     & Race track, inc.\ $\pm 5^\circ$ \\
Conductor diam. & $30 \times 24\,$mm$^2$ \\
Water cooling   & Channel $\varnothing=10$\,mm \\
Conductor current   & 2.16\,kA \\
Current density & 3.38\,A/mm$^2$ \\ 
Total current   & 727\,kA  \\
Single turn length & 8.68\,m \\
Mean resistivity   & 18.5\,n$\Omega$\,m \\
Total dissipated power  & 360\,kW \\
Inductance (incl.\ yoke)  & 0.87\,H \\
Stored energy             & 2.03\,MJ \\
\hline
\multicolumn{2}{c}{Flux Return Yoke}\\
\hline
Material        & Steel XC06    \\
Lamination      & $\sim 20$\,cm \\
Weight          &  $\sim 200$\,t         \\
Dim.\ ($H \times W \times D$) & $3.88 \times 5.3 \times 1.64$\,m$^3$ \\
Gap opening ($H \times W$) & $0.80-1.01 \times 3.10$\,m$^2$ \\
\hline
\end{tabular}
\caption[Overview of the main parameters of the dipole magnet. ]
{Overview of the main parameters of the dipole magnet. The 
depth $D$ denotes the length along the beam direction $z$ axis.}

\label{t:dipole:param}
\end{center}
\end{table}

We assume that the 24 pancakes will be fed by cooling lines in a
parallel arrangement, where the water inlet and outlet are located on
the inner and on the outer sides of the winding, respectively.  Then
we would require 3.9\,l/s of water and the temperature difference
between inlet and outlet would be 22\,K. The water velocity in the
cooling channels of 10\,mm would reach about 2\,m/s and the pressure
drop would be 8\,bar.

To ensure a proper operation of the dipole magnet, a slow control
system will be built which provides the interface to HESR and
diagnostic systems. The following magnet systems will be monitored
\begin{itemize}
\item \textbf{Interface to \HESR.} This system will guarantee that 
  the control and diagnostic systems of \HESR exchange information and
  are fully synchronised with all systems of the \PANDA dipole. The
  precise layout is not yet defined.
\item \textbf{Current control} will be used to monitor the magnetic 
  field intensity and the stability of the dipole
  magnet.
\item \textbf{Temperature control.} The temperature of the
  copper coils will be monitored by several temperature sensors
  located at different positions on the upper and lower coils.
\item \textbf{Magnetic field control.} Two Nuclear Magnetic Resonance 
  (NMR) probes and 5--10 Hall probes in critical places are foreseen to
  be used to directly measure the magnetic field intensity and
  stability.
\item \textbf{Cooling water flow control.} The flow of cooling water 
  will be controlled by water flow meters at several locations.
\end{itemize}

The chief tasks of the diagnostic system are listed in the following.
\begin{enumerate}
\item Process signals fast to generate alarm and interlock signals for 
  magnet safety in emergency situations.
\item Record control parameters in the normal operation regime and 
  provide data logging.
\item Allow remote control of all parameters, in particular the full 
  integration within the \HESR control systems.  Display the data at
  the operator's console.
\item Process signals to generate control responses to optimise 
  the magnet operation.
\end{enumerate}
The magnet and auxiliary equipment control system will incorporate a
distributed computer control system composed of commercially available
components. Detailed design of the slow control system will be
performed starting from 2010.

The field mapping will be performed using existing equipment at GSI and
with the help of the GSI Magnet Technology Group.  It is envisaged
that for each of 5 different current settings one full 3D field map
would be determined with a grid size of 7, 5 and 2\,cm in $x, y$ and
$z$, respectively.  This work will be performed as soon as the magnet
is fully commissioned and 60 days are estimated for the
measurements.  Interpolating between the grid points and maps would
allow us to determine the magnetic field at every point in the
spectrometer opening and any current to sufficiently high accuracy.

The magnet design has been optimised and is finalised to a high level
of detail.  A few minor modifications to the design might, however, be
imposed by constraints or recommendations from the manufacturer during
the tendering process.

\svnInfo $Id: perf.tex 705 2009-02-04 15:39:35Z IntiL $

\section{Performance}
\label{s:dipole:perf}

\AUTHORS{I.~Lehmann, J.~L\"uhning}

\subsection{Structural Considerations}

The studies of the mechanical stability have concentrated on the coils
and the stand of the dipole magnet.  The bulk mass of the yoke is
stemming from the rigidly connected steel plates which provide a good
stability by themselves.  

\begin{figure}[ht]
\begin{center}
\includegraphics[width=\swidth]{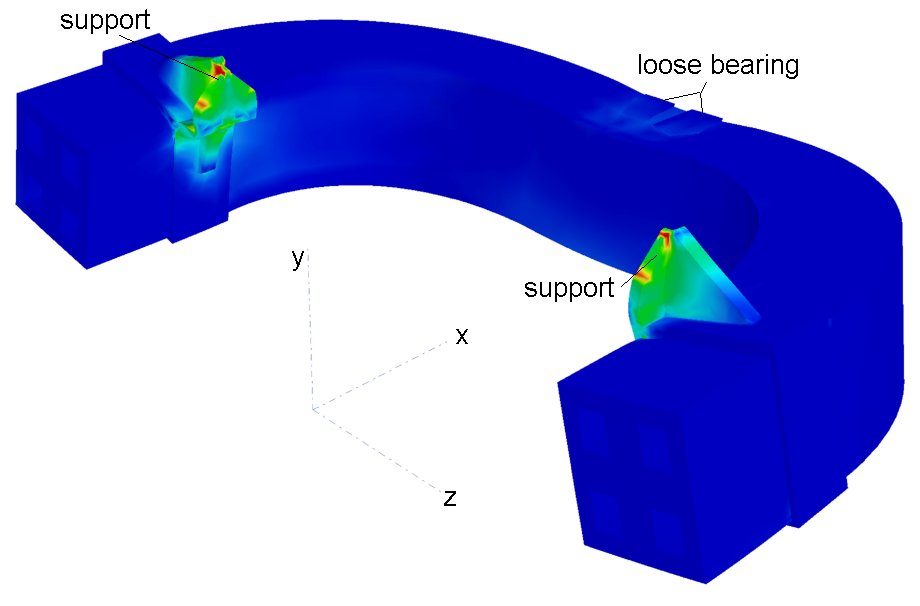}
\caption[Deformation of one half of the upper dipole coil.]
{Deformation of one half of the upper dipole coil, enlarged by 
   1000 times for visibility. The winding is supported at the vertical
   upstream and downstream surfaces of the dipole yoke. The loose
   bearing is in the region between the pole shoe and the flux return
   part of the yoke. The colours show the Von Mises stresses (red
   indicates stresses of about 100\,MPa, light blue 10\,MPa and dark
   blue below 10\,MPa).}
\label{f:dip:CoilDefo}
\end{center}
\end{figure}

Finite Element Model (FEM) calculations were performed with the
Generative Structural Analysis module of CATIA~\cite{mag:CATIA}.  Out
of symmetry reasons it is sufficient to analyse only one half of one
coil.  The result of this calculation is shown in
Fig.~\ref{f:dip:CoilDefo}.  The force input for this model was taken
from a calculation using TOSCA~\cite{mag:TOSCA_2008}.  The total
Lorentz force of one coil will be about 110\,t in vertical direction while
its weight will be only about 10\,t.  It is difficult to represent the
orthotropic properties of the winding precisely.  For our calculation,
however, a coarse approximation of the properties is sufficient.  This
was done by introducing 4 empty channels along the perimeter of the
winding such that the winding is stiffer in longitudinal direction
than in the transverse directions.  The support rings holding the
winding are assumed to consist of aluminium.  Between the support ring
and the winding we assume a 5\,mm thick layer of soft material
(Young's modulus 4\,GPa) in order to avoid high local stresses.  A
soft material was also applied for the loose bearing in the centre of
the magnet.  As visible in Fig.~\ref{f:dip:CoilDefo} those
calculations reveal that the stresses in the windings stay well below
10\,MPa while the stresses in the support stay below 120\,MPa.  These
results show that we have kept a good safety margin in the design of
the coils.

\begin{figure}[ht]
\begin{center}
\includegraphics[width=\swidth]{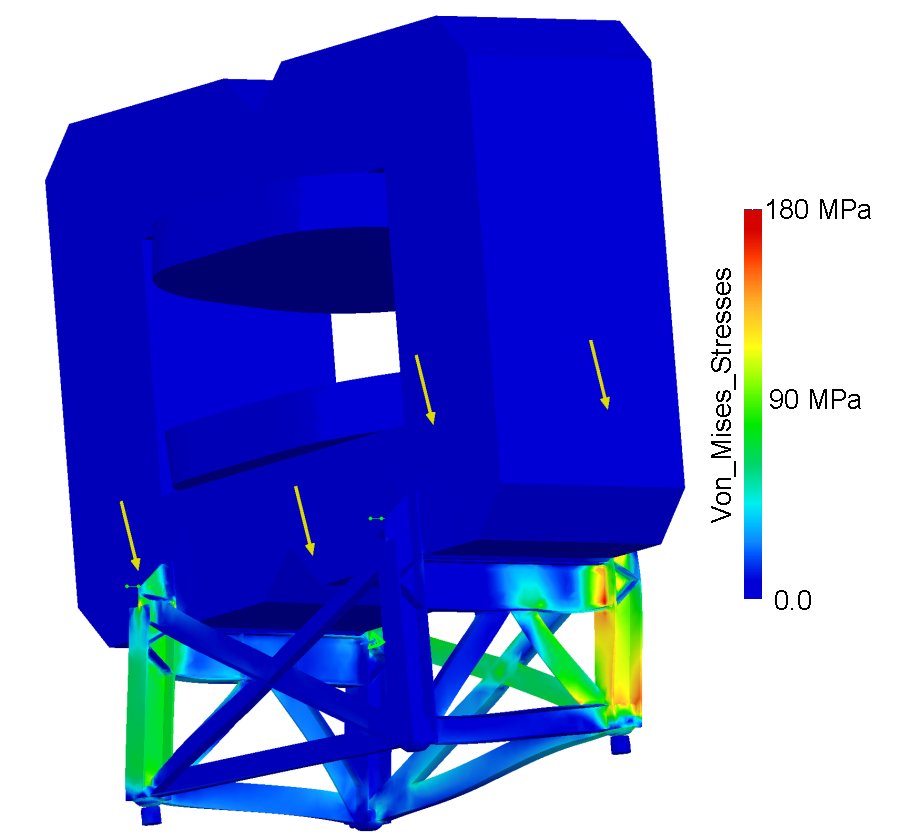}
\caption[Deformation of the dipole support structure in an extreme 
   case]{Deformation of the dipole support structure in the extreme 
   case of a strong seismic load exactly at the very moment where the
   magnet is suspended only by two legs during the alignment procedure
   (see text).}
\label{f:dip:DipSupp}
\end{center}
\end{figure}

The support structure for the large aperture dipole magnet has been
designed to safely support the weight of the magnet and additional
installations even under exceptional situations.  Two extreme cases
have been considered in detail:
\begin{enumerate}
\item a horizontal acceleration of 1.5\,m/s$^2$, \label{i:dip:seis}
\item the whole weight resting on only 2 diagonally opposite legs 
of the support structure. \label{i:dip:2leg}
\end{enumerate}
In the very rare occasion of strong seismic activity case~\ref{i:dip:seis} 
may occur.  Case~\ref{i:dip:2leg} may occur during the
alignment procedure of the dipole, when one leg is pushed up or
released further than the two neighbouring legs ({\it e.g. }with the
help of a hydraulic jack).  The worst case scenario would be if
a seismic activity would lead to a horizontal acceleration in the
direction diagonal to the two supporting legs precisely in the very
moment where the whole load is suspended only by two legs.  Though
this is extremely unlikely, this case has been studied (see
Fig.~\ref{f:dip:DipSupp}).  The maximum stress does not exceed
180\,MPa, {\it i.e. }3/4 of the elastic limit of ordinary structural
steel. Thus, even in this unlikely event the structure would guarantee the
mechanical safety of the dipole magnet.

\subsection{Static Field Properties}

The major objective for the field is to achieve a field integral of
2\,Tm at 15\,\gevc for particles emitted from the interaction area
below horizontal and vertical angles of $10^\circ$ and $5^\circ$,
respectively.  These particles will escape the Target Spectrometer and
reach the \PANDA Forward Spectrometer, where the large-aperture dipole
magnet together with the tracking devices will enable the momentum
reconstruction of charged particles.  Most crucial are the highest
momentum particles, as their bending is smallest.  Particles with
momenta of more than a factor of 15 below the beam momentum will be
deflected so much that they will not transverse the full magnet and
hence experience a smaller field integral.  This poses, however, no 
problem as the momentum is reconstructed very precisely also
with a reduced field integral, because these particles are bend much 
stronger.

\begin{figure}[ht]
\begin{center}
\subfigure[Contour plot of $B_y$ in the $z-y$ plane at $x=0$. 
 The linear scale starts at 0.80\,T (blue) and ends at 1.05\,T
 (lilac).]
{\includegraphics[width=\swidth]{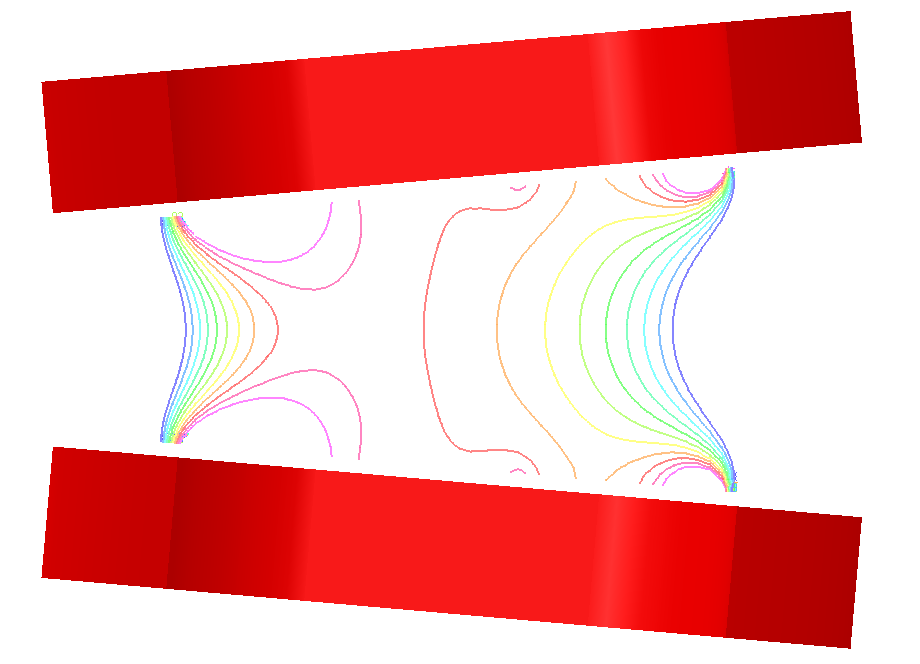}}
\subfigure[View from the downstream side showing the field in a 
  colour map from 0.5\,T (blue) to 1\,T (lilac) in the $y-x$ plane at
  $z=4.75\,$m.]
{\includegraphics[width=\swidth]{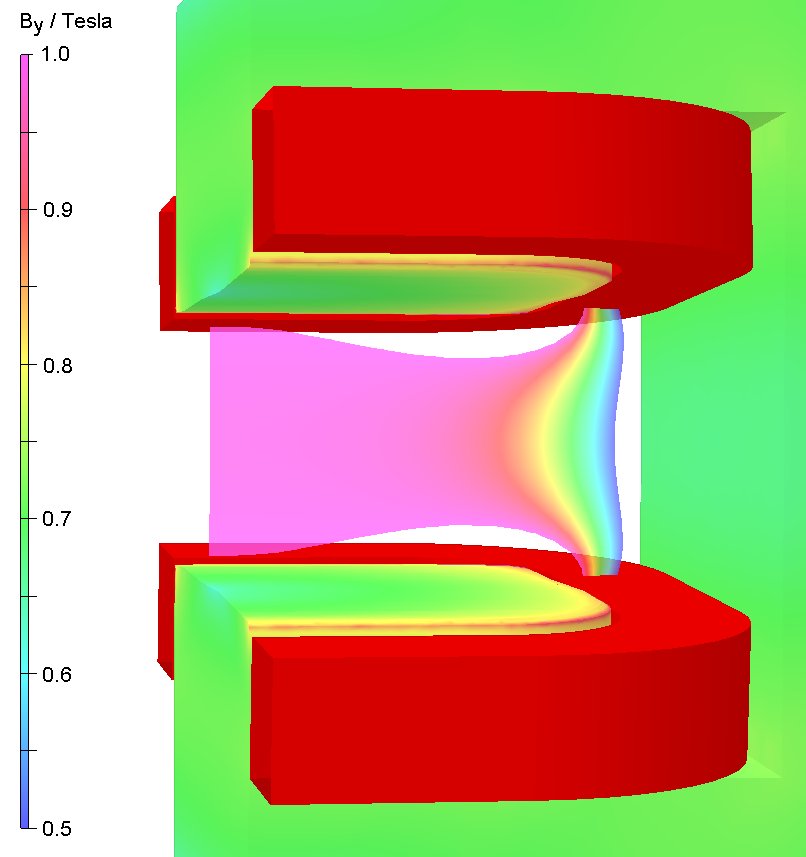}}
\caption[Vertical magnetic field component $B_y$ in the opening of
 the dipole magnet.]{Vertical magnetic field component $B_y$ in the opening of
 the dipole magnet in two different views. }
\label{f:dip:fieldmaps}
\end{center}
\end{figure}

\begin{figure}[th]
\begin{center}
\includegraphics[width=\swidth]{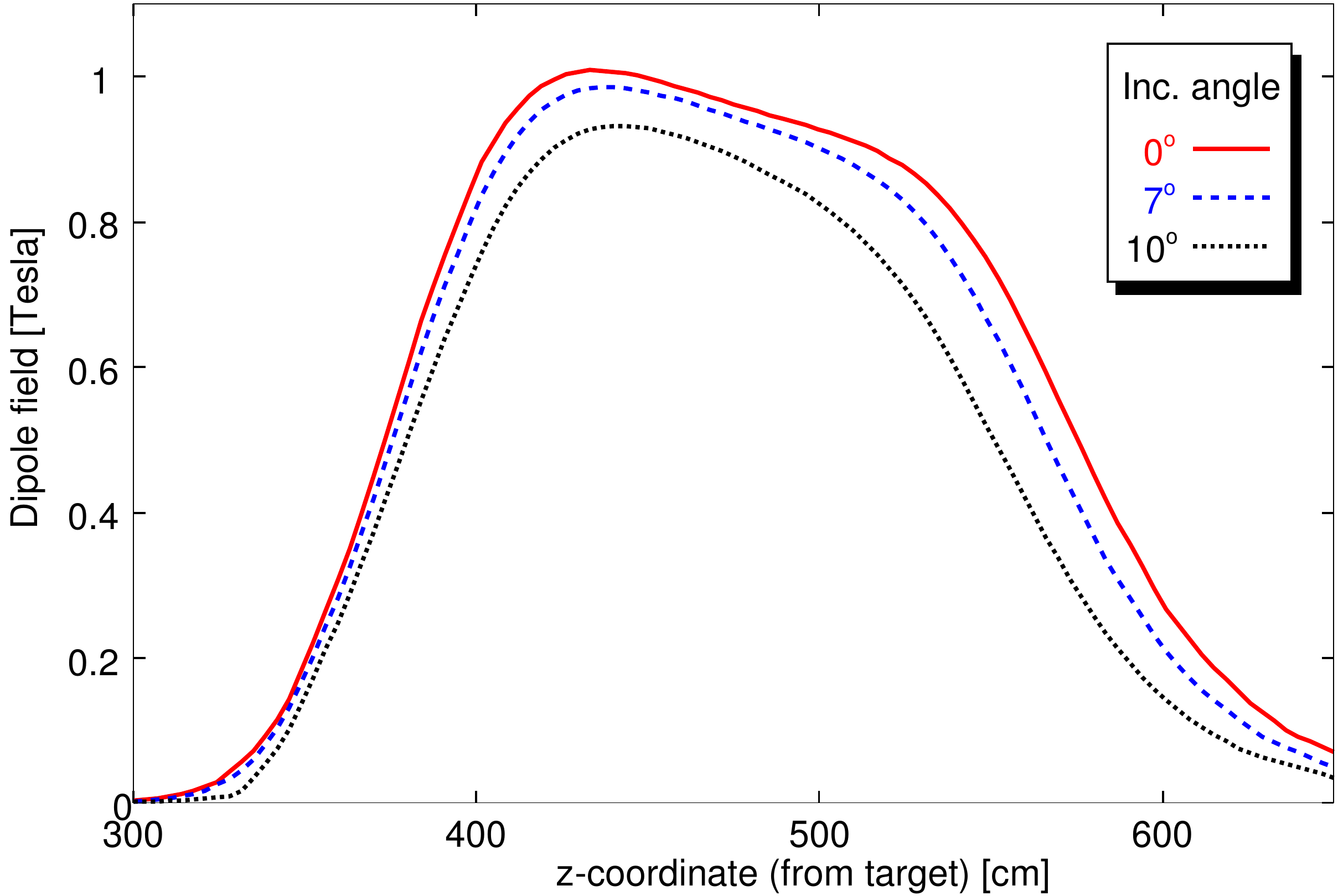}
\caption[Dipole field for several straight trajectories.]
{Dipole field for several straight trajectories in the 
  horizontal plane of the dipole. The tracks originate at the
  interaction point with angles respective to the beam direction of
  $0^\circ$ (red solid curve), $7^\circ$ (blue dashed curve) and
  $10^\circ$ (black dotted curve). }
\label{f:dip:field}
\end{center}
\end{figure}

\begin{figure}[th]
\begin{center}
\includegraphics[width=\swidth]{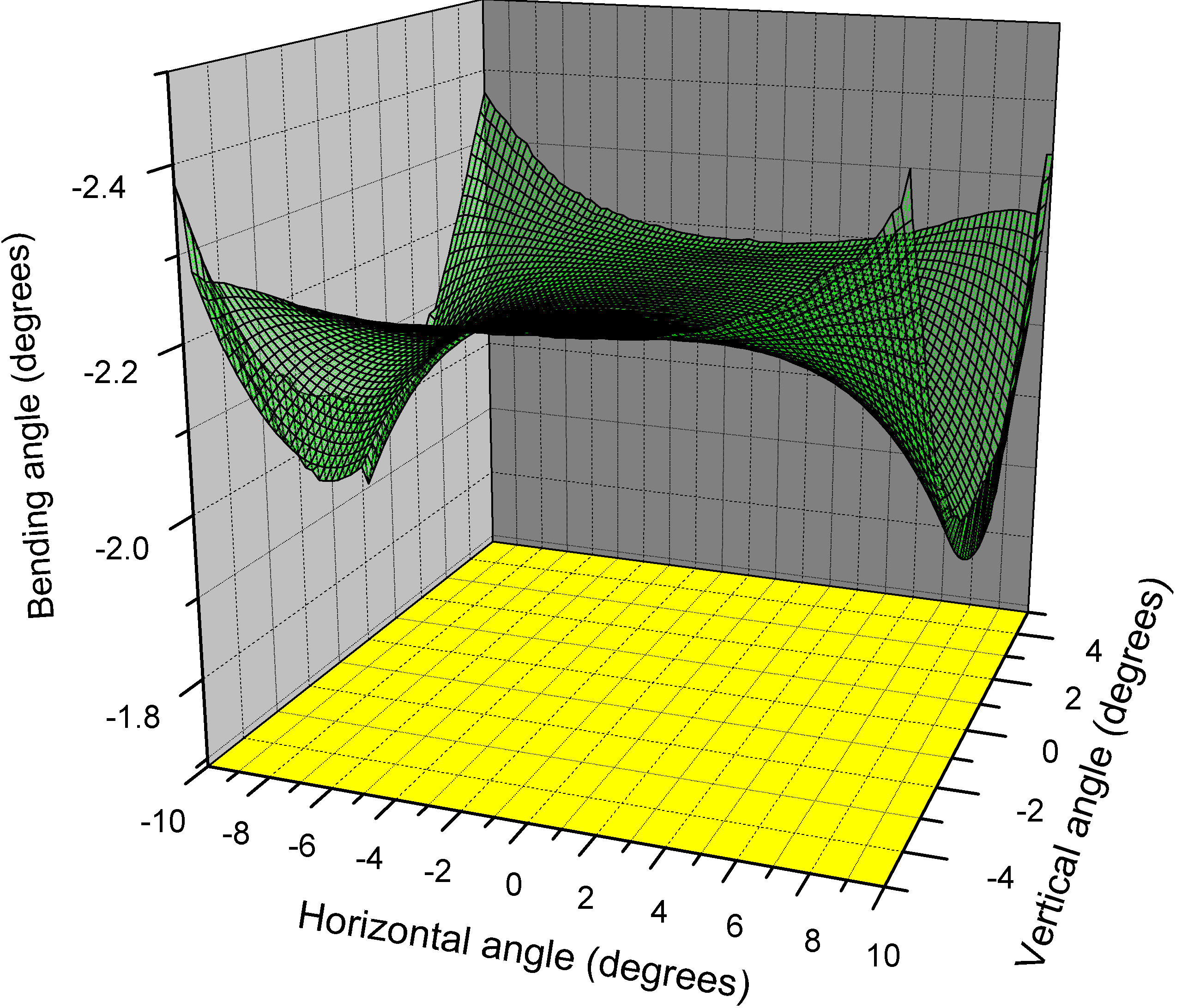}
\caption[Bending angle of tracks with $P=P_{beam}$ through the 
  dipole.]{Bending angle of tracks with $P=P_{beam}$ through the 
  magnetic field of the dipole using ray-trace calculations. The
  bending varies by 10\%, but this can be easily handled by the PANDA
  track reconstruction.}
\label{f:dip:ray-trace}
\end{center}
\end{figure}

The magnet was designed such that the field between the pole shoes
stays as uniform as possible for a magnet with such a large aperture
(about $1 \times 3\,$m).  Field calculations have been performed with
a full 3D model of the magnet using TOSCA~\cite{mag:TOSCA_2008}.
Figs.~\ref{f:dip:fieldmaps} show views of the field in $z-y$ and $y-x$
planes at $x=0$ and $z=4.75\,$m, respectively.  One observes that the
field in the central region is constant within 20\% while it falls off
steeply close to the coil location.  Here the change along the
particle path is unimportant compared to any change within the $x-y$
plane, as the latter influences the field integral experienced by the
particle.  In Fig.~\ref{f:dip:field} the actual field seen by
particles transversing the magnet under three different trajectories
is shown.  The relevant quantity for the momentum reconstruction is,
however, the bending of the tracks due to the passage through the
magnetic field.  Assuming particles with beam momentum the change of
angle induced by the magnetic field has been studied in ray-trace
calculations.  Particles with lower momenta will be bend more than
that and hence can be reconstructed better.  The resulting differences
of a mean bending angle of $2.2^\circ$ are in the order of 10\% at the
very corners of the yoke gap (see Fig.~\ref{f:dip:ray-trace}).

\begin{figure}[th]
\begin{center}
\includegraphics[width=\swidth]{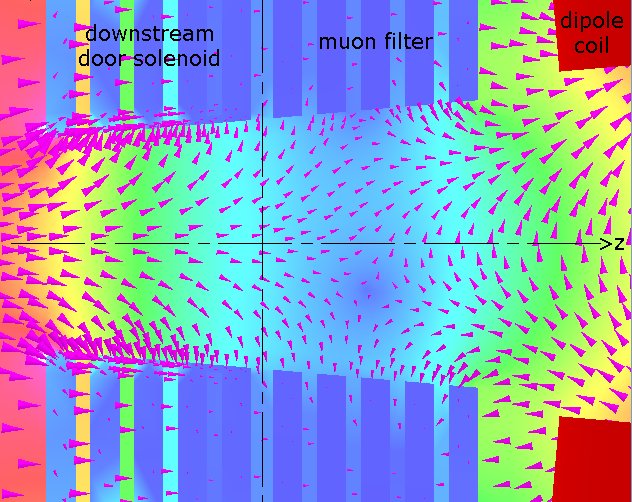}
\caption[The field in the region between the end-cap 
  of the solenoid and the entrance of the dipole.]
  {Detail of the field in the region between the end-cap 
  of the solenoid and the entrance of the dipole magnet.  This view
  corresponds to an $z-y$ slice several centimetres off the axis
  ($x=0$).}
\label{f:dip:fieldmap}
\end{center}
\end{figure}

These differences are easily handled by the track reconstruction
software provided that the actual field is known to the appropriate
detail.  Therefore the magnet is designed such that the field remains
a smooth function in space over the whole volume where particles are
tracked. 

In Fig.~\ref{f:dip:fieldmap} the combined field of the dipole and
solenoid shows how the intermediate muon filter in front of the dipole
will act as an effective field clamp and will help to separate the two
spectrometers magnetically and reduce stray fields.

\begin{figure}[th]
\begin{center}
\includegraphics[width=\swidth]{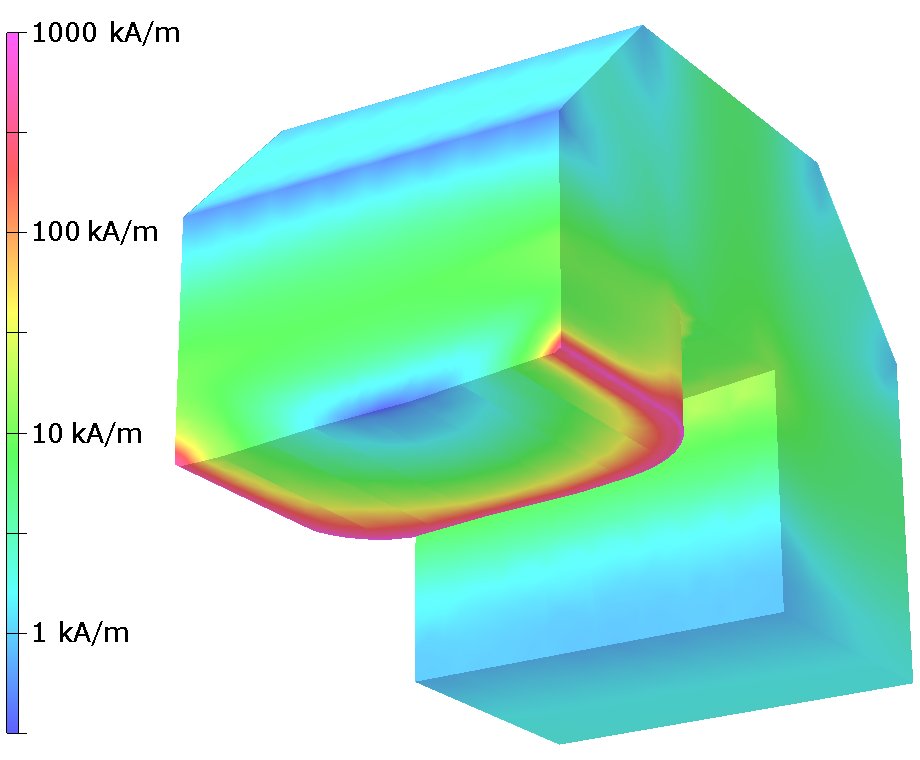}
\caption[Magnetic field strength $H$ in the flux return of the 
 dipole at full current.]
 {Magnetic field strength $H$ in the flux return of the 
 dipole at full current. The colour scale is logarithmic.  Only one
 upper quarter of the magnet is shown as the other quarters show
 identical distributions. The field strength is clearly reaching
 maximal values at the edge of the pole shoes, while it stays two
 orders of magnitude lower in the bulk of the flux return yoke.}
\label{f:dip:SteelSaturation}
\end{center}
\end{figure}

The field strength between the pole shoes of the magnet will not
follow the current in the coils in an absolutely linear way.  One
effect is saturation of the iron which limits the maximum flux and
hence changes the resulting field. This effect is static and will start to
significantly influence the field at about half of the full current.
The saturation will occur mainly at the edges of the pole shoes (see
Fig.~\ref{f:dip:SteelSaturation}).  Thus the actual bending power of
the magnet will be about 2.02\,Tm at full current, while with half the
current 1.05\,Tm will be reached.  This effect of the saturation can be
illustrated by calculating the bending power and inductance normalised by
the fraction of the full current (see also blue lines in 
Figs.~\ref{f:dip:bend-ind}).  The
$\int{B\,dl}$ and the current density in the coils is listed in
Table~\ref{t:dip:currents} for 10 beam momenta.  At full current, the
total energy is 2.03\,MJ, the total inductance is 0.87\,Henry.

\begin{table}[ht]
\begin{center}
\begin{tabular}{|l|c|c|c|}
\hline
$p_\mathrm{beam}$ & $T_\mathrm{beam}$ & $\int{B\,dl}$
 & Total Current \\
$[$GeV/c$]$ & $[$GeV$]$ & $[$Tm$]$ & $[$A$]$ \\
\hline
1.5  &          0.83101 & 0.2 &   69144 \\
3.0  &          2.20503 & 0.4 &  138298 \\
4.5  &          3.65850 & 0.6 &  207480 \\
6.0  &          5.13465 & 0.8 &  276821 \\
7.5  &          6.62019 & 1.0 &  346557 \\
9.0  &          8.11050 & 1.2 &  416917 \\
10.5 &          9.60357 & 1.4 &  488486 \\
12.0 &         11.09835 & 1.6 &  562660 \\
13.5 &         12.59429 & 1.8 &  641346 \\
15.0 &         14.09104 & 2.0 &  726962 \\
\hline
\end{tabular}
\caption[Dipole parameters for 10 antiproton beam momenta.]
{Dipole parameters for 10 antiproton beam momenta 
  $p_\mathrm{beam}$ ranging between 1.5 and 15\,GeV/c.  Listed are the
  corresponding kinetic energy $T_\mathrm{beam}$, the integral of the
  field along the $z$ axis $\int{B\,dl}$ and the total current in
  the coils.}
\label{t:dip:currents}
\end{center}
\end{table}

\begin{figure}[th]
 \begin{center} 
\includegraphics[width=\swidth]{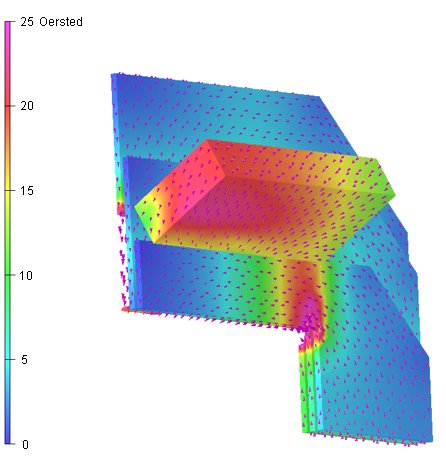}        
\caption[Stray fields downstream of the dipole magnet.]
{Stray fields downstream of the dipole magnet. Shown is 
 half of the upper downstream field clamp of the dipole and a
 representative volume where the read out of a potential RICH detector
 would be located.  The maximum field in this volume would be
 25\,Gauss.}
\label{f:dip:StrayRICH}
\end{center}
\end{figure}

\subsection{Stray Fields}

The detectors downstream of the dipole will be affected by stray fields,
in particular some detectors which have read-out systems sensitive to
magnetic fields.  As their design is not yet defined in detail,
definite limits for the field strength are not finally set.  We have,
however, enough flexibility to control the fields in that region by
adding field clamps downstream of the coils.  As an exemplary scenario
we have calculated the field for the most crucial system. The closest
detector with possibly field sensitive read-out is the RICH detector.
We have studied the field at the location of its likely read-out,
though this detector is not yet fully designed.  The outcome is
illustrated in Fig.~\ref{f:dip:StrayRICH}.  The use of a field clamp
with 3 layers of iron with 4\,cm thickness each will result in stray
fields below 25\,Gauss at the location of a possible RICH readout.
This calculation does not include possible additional shieldings at
the location of the read out.  All other detection systems are either
located much further away or foresee field insensitive read-out
systems.

\subsection{Dynamic Properties}
\label{s:dipole:perf:dynamic}

The dipole will form part of the accelerator lattice deflecting the
beam and hence needs to be ramped for every cycle in a synchronised
fashion with the other bending magnets in the \HESR.  This dynamical
change of currents in the coils and subsequent change of field
strength induces eddy currents in the flux return yoke.  These
counteract the magnetisation and demagnetisation at ramp-up and
ramp-down, respectively.  Thus the field changes slower than the
current does.  The effect will be decreased by the lamination but still
needs to be compensated in order to guarantee beam stability over the
full ramp.

\begin{figure}[ht]
 \begin{center}
 \subfigure[Bending power]
  {\includegraphics[width=\swidth]{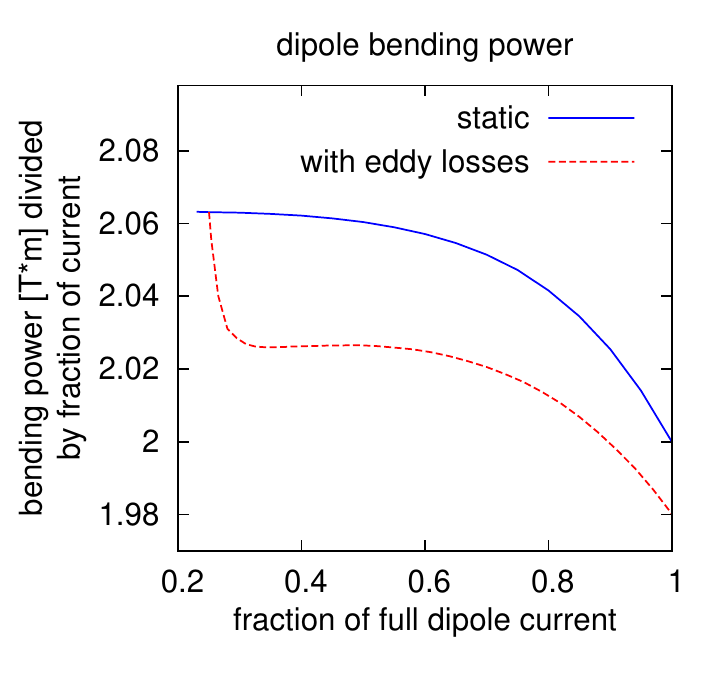}
  \label{f:dip:bend}}
 \subfigure[Inductance]
  {\includegraphics[width=\swidth]{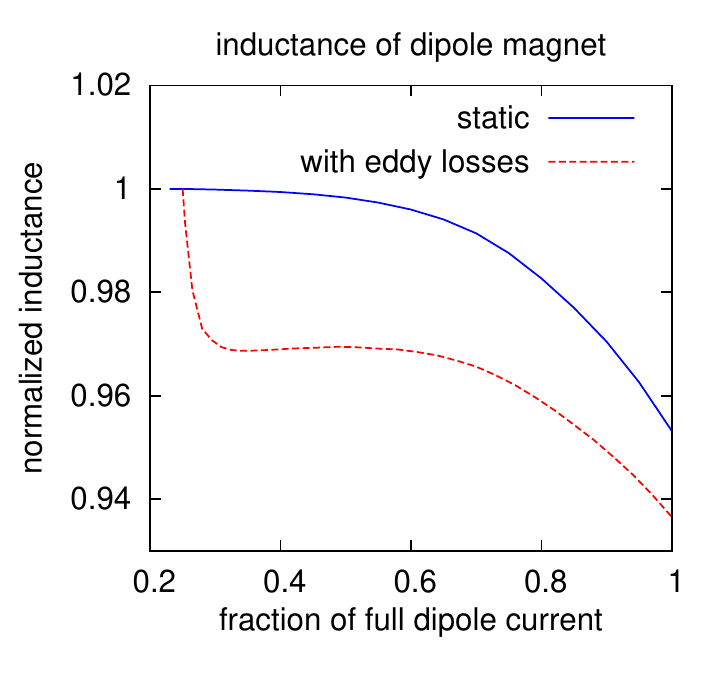}
  \label{f:dip:ind}}
 \caption[Normalised bending power and inductance {\it versus}
    current in the coils of the dipole magnet.]
    {Normalised bending power (a) and inductance (b) {\it versus}
    current in the coils of the dipole magnet.  Shown are two cases:
    in blue the magnet is ramped very slowly such that eddy losses can
    be neglected, while the red curves indicate the situation in the
    fast ramping foreseen at \HESR, {\it i.e.\ }in 1 minute from 25\%
    to full nominal current.}
 \label{f:dip:bend-ind}
 \end{center}
\end{figure}

\begin{figure}[th]
\begin{center}
\includegraphics[width=0.9\swidth]{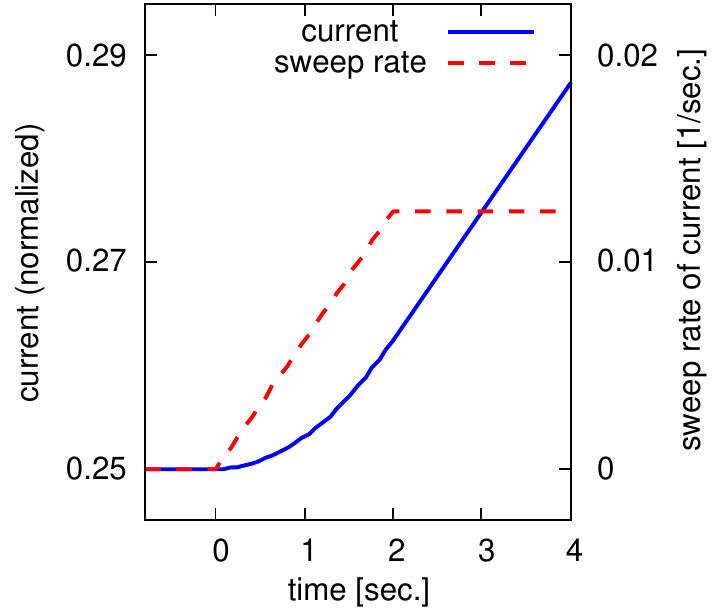}
\caption[Envisaged start of the ramp procedure.]
{Envisaged start of the ramp procedure.  A smooth transition 
is reached by increasing the sweep rate linearly from 0 to 1.25\% over a
period of 2\,s.}
\label{f:dip:ramp-start}
\end{center}
\end{figure}

\begin{figure}[th]
 \begin{center}
  \includegraphics[width=\swidth]{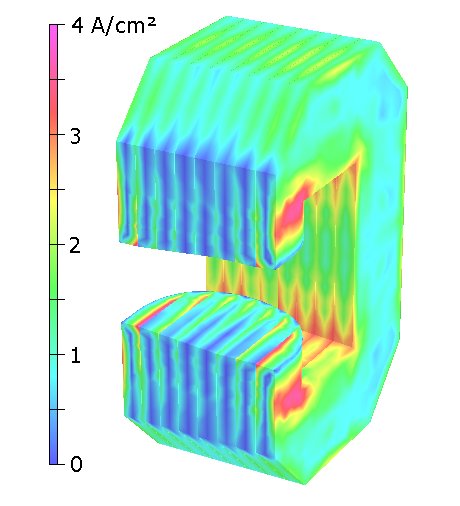}

  \caption[Density
    of the eddy currents 12 seconds after the start of the ramp.]
    {View of half of the dipole magnet indicating the density
    of the eddy currents 12 seconds after the start of the ramp --
    colour coding from 0 (dark blue) to the maximum value of
    4\,A/cm$^2$ (pink).}

\label{f:dip:eddy}
\end{center}
\end{figure}

The time foreseen to accelerate antiprotons from injection momentum
(3.8\,GeV/c) to full momentum (15\,GeV/c) in the \HESR is 60 seconds.
Thus it is required to change the current in the coils with a rate of
1.25\% of the nominal dipole current per second. The ramp-up procedure
is envisaged as following.  Before injection the magnet is ramped to
25\% of the nominal current $I_\mathrm{nom}$.  At this plateau the
current is kept constant for a short while, which allows the
(re)injection of antiprotons.  Meanwhile the eddy currents will drop.
For our calculations we assume now the following sequence.  The
ramp-up is started by a parabolic current increase over 2\,s.  This
avoids discontinuities in $dI/dt$ which may harm the power supplies.
Hence, the actual current
$I = I_\mathrm{nom} \left[ \frac{I_\mathrm{inj}}{I_\mathrm{nom}} 
+ p_0 \, t^2 \right] \, , $
where $t$ is the time in seconds, and the current at injection
$I_\mathrm{inj} = 0.25\, I_\mathrm{nom}$.  Setting the parameter of
the polynomial $p_0$ to $2.6 \times 10^{-3}\,\mathrm{s}^{-2}$ guarantees a
moderate delay, such that the system reaches a steady ramping speed of
$dI/dt = 0.0125\, I_\mathrm{nom}/ \mathrm{s}$ after 2\,s.
The power supply would need to provide additional 25\,V to compensate
the resistance due to induction if considering a nominal current of 2.1\,kA in
the coils. Please also refer to Fig.~\ref{f:dip:ramp-start}. 

In this scenario with the moderate lamination of 20\,cm the eddy
currents induced in the dipole yoke reduce the inductance and the
bending power only by less than 2\% (see Fig.~\ref{f:dip:bend-ind}).
The total power losses due to the eddy currents will stay below 400\,W
during the whole ramp time.  In Fig.~\ref{f:dip:eddy} one half of
the yoke is shown indicating the density of the eddy currents.

\subsection{Influence on HESR Beam}

In order to quantitatively understand the orbit errors caused by the
\PANDA spectrometer, 3D Tosca magnetic field calculations are carried
out.  For the dipole the major source of uncertainty is the non
uniform saturation in the yoke iron as it is ramped with the beam
momentum.  10 different excitation currents corresponding to 10
different antiproton beam energies are used for the dipole, which are
shown in Table~\ref{t:dip:TOSCA}. The excitation currents are
selected such that the different energy antiproton beams have the same
bending angle.

\begin{table}[ht]
\begin{center}
\begin{tabular}{|l|c|c|c|}
\hline
$p_\mathrm{beam}$ & \multicolumn{2}{c|}{$\int{B\,dl}$} 
 & Relative \\
      & nom.  & model       & deviation \\
$[$GeV/c$]$ & $[$Tm$]$ & $[$Tm$]$ &  \\
\hline
 1.5 &    0.2 &   0.200000 &    $+4.2850 \times 10^{-7}$ \\
 3.0 &    0.4 &   0.400000 &    $+1.9000 \times 10^{-7}$ \\
 4.5 &    0.6 &   0.600001 &    $+1.5333 \times 10^{-6}$ \\
 6.0 &    0.8 &   0.800006 &    $+7.8725 \times 10^{-6}$ \\
 7.5 &    1.0 &   1.000019 &    $+1.9074 \times 10^{-5}$ \\
 9.0 &    1.2 &   1.200045 &    $+3.7556 \times 10^{-5}$ \\
 10.5 &   1.4 &   1.399972 &    $-2.0050 \times 10^{-5}$ \\
 12.0 &   1.6 &   1.600052 &    $+3.2750 \times 10^{-5}$ \\
 13.5 &   1.8 &   1.800047 &    $+2.6321 \times 10^{-5}$ \\
 15.0 &   2.0 &   1.999948 &    $-2.6110 \times 10^{-5}$ \\
\hline
\end{tabular}
\caption[Dipole parameters for 10 antiproton beam momenta.]
{Dipole parameters for 10 antiproton beam momenta 
  $p_\mathrm{beam}$ ranging between 1.5 and 15\,GeV/c as in
  Table~\ref{t:dip:currents}.  Here the nominal integral of the field
  along the $z$ axis $\int{B\,dl}$ is compared to calculations using a
  realistic model in TOSCA (nom.\ and model, respectively).  The
  relative deviations stay at a level of $10^{-5}$ or below.}
\label{t:dip:TOSCA}
\end{center}
\end{table}

\begin{figure}[ht]
  \begin{center}
    \includegraphics[width=\swidth]{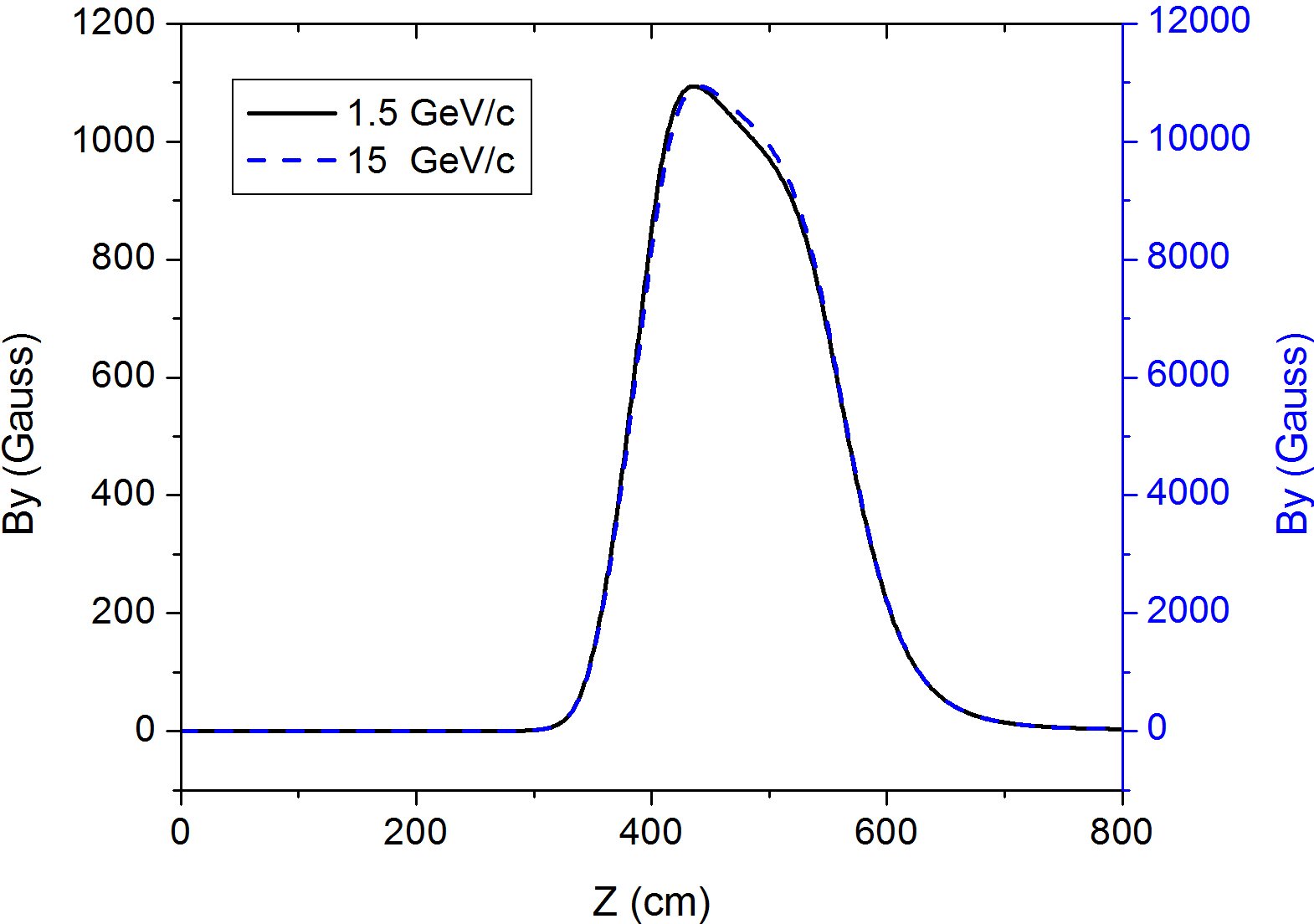}
    \caption[Field distribution along the central line for different
      dipole settings.]{Field distribution along the central line for different
      dipole settings for a 1.5\,GeV/c (solid black line and left scale) 
      and a 15 GeV/c beam (dashed blue line and right scale).  }
    \label{f:dip:beam1}
  \end{center}
\end{figure}

In Fig.~\ref{f:dip:beam1}, the dipole magnetic fields along the z axis
are shown. The dashed line is for the 15\,GeV/c antiproton beam, and
the solid line is for the 1.5\,GeV/c antiproton beam. To compare the
differences in the field distributions, the scale for the 1.5\,GeV/c
antiproton beam is a factor of 10 larger than the one for the
15\,GeV/c beam.  It is observed that the field changes slightly its
shape when scaling between those extremes.

As mentioned in the previous paragraph, the field distribution
differences arise from a consideration of non uniform saturation in the iron. The dipole
will have a field integral of 2\,Tm. The maximum peak field in the dipole
mid plane is around 1.1\,T, while the maximum magnetic field in the
iron pole shoe is as high as 2.9\,T, which appears at one of the
corners. The magnetic field distribution through the pole shoes will be not
uniform and in addition the iron saturation will be non uniform. The
saturation itself will not affect the field distribution too much, but
the influence of the non uniform saturation will be more serious. To reduce
the non uniform saturation effect, a chamfered or rounded pole shoe
could be used. 

\begin{figure}[ht]
  \begin{center}
    \includegraphics[width=\swidth]{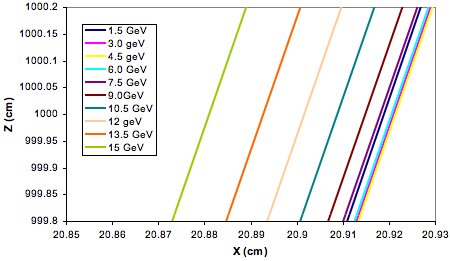}
    \caption[Beam trajectories of antiproton beam of momentum from 15
      to 1.5\,GeV/c.]{Beam trajectories of antiproton beam of momentum from 15
      to 1.5\,GeV/c.  the leftmost side on is for a 15\,GeV/c beam,
      while on the rightmost site is that for a 3\,GeV/c beam. The blue
      line is the trajectory for the 1.5\,GeV/c beam.}
    \label{f:dip:beam3}
  \end{center}
\end{figure}

In Fig.~\ref{f:dip:beam3}, beam trajectories of antiproton beams in a
range of momenta from 1.5 to 15\,GeV/c are shown.  These calculations
assume a perfectly aligned solenoid.  On the leftmost side is the
trajectory for a 15\,GeV/c beam, while on the rightmost site is that
for a 3\,GeV/c beam. The distance between these two trajectories at
$z=10$\,m is around $0.4$\,mm. The other trajectories are located in
between these two extremes, and except for the 1.5\,GeV/c beam, a beam
trajectory is normally located on the left side of a beam trajectory
having a smaller energy. The blue line is the trajectory for the
1.5\,GeV/c beam. The distance between the 3\,GeV/c beam trajectory and
the 1.5\,GeV/c beam trajectory at $z=10$\,m is around 0.03\,mm. It
should be pointed out that the field errors from the TOSCA simulation
(see Table~\ref{t:dip:TOSCA}) also contribute to the orbit shift shown
in Fig.~\ref{f:dip:beam3}, and has the largest effect on the beam
trajectory of the 1.5\,GeV/c beam. Other beams will also suffer from the
same effect, but the orbit variations are much smaller.

In summary the non uniform saturation effect in the dipole magnet will also
introduces a small trajectory shift which can be as large as
0.4\,mm. This could be reduced by using chamfered or rounded pole
shoes. But since this is a minor contribution, it is not clear yet if
this will be necessary.

\svnInfo $Id: platforms.tex 714 2009-03-31 17:34:18Z IntiL $

\section{Detector Integration}
\label{s:dipole:int}
\label{s:dipole:platforms}

\AUTHORS{E.~Lisowski, J.~Smyrski}

The construction of supports for the Forward Spectrometer
detectors is closely connected with the design of the \PANDA dipole
magnet, since the supports must be integrated in the yoke, to avoid
any reduction of the aperture of the magnet,
leading to limitations of the angular and momentum
acceptance of the system.  Besides, the detector supports will be
mounted directly on the magnet yoke in order to reach high
reproducibility of the detectors position with respect to the magnetic
field. This is of special importance for the drift chambers, which 
reconstruct the momentum of charged particles on the basis of
deflection of their trajectories in the dipole magnetic field. 

The Forward Spectrometer will be equipped with three pairs of tracking
drift detectors, hence six independently operating detectors, each
containing four double-layers of straw tubes.  The first pair of drift
detectors (DC1, DC2) will be mounted in front of the dipole magnet,
the third (DC5, DC6) further downstream, behind the magnet.  The
second pair (DC3, DC4) will be installed inside the magnet gap in
order to track low momentum particles hitting the magnet yoke inside
the gap. The drift detectors in each pair will be mounted on a common
support frame.  This solution allows to reach a high accuracy of
relative positioning of the detectors -- better than 0.1\,mm.  It also
eases the installation of the detectors in the experimental position.

High mechanical precision of the mounting
elements, used to fix the support frames on the magnet yoke, will
allow to reach a high reproducibility of the detector position
with respect to the magnet yoke, when dismounting the detector themselves
for servicing and mounting them again on the magnet.  The frames will 
be rigid enough to hold the weight of the detectors, which is evaluated 
to be about 50\,kg for the smallest one -- DC1 -- and about 400\,kg for 
the largest one -- DC6.  The frame construction will also guarantee 
high stability of the drift detector positions with respect to the magnet 
-- on the level of 0.05\,mm -- during periods of data taking.

The detector components of the Forward Spectrometer
foreseen for the particle identification (the RICH detector,
the TOF wall, the forward muon detectors) 
and for the calorimetry (the electromagnetic calorimeter) 
as well as the luminosity monitor 
will be mounted on a common platform. 
The platform will be used to support the detectors on a
proper height with respect to the beam line. The total mass of the
detectors is about 60 tonnes.  The platform will be equipped with a
driving system allowing to move the detectors between the experimental
area and the parking position.  The driving system should guarantee a
reproducibility of the detector positions with respect to the magnet
yoke better than 1\,mm.  

In the four following subsections we describe
respectively: 
\begin{itemize}
\item{the construction of supports for mounting detectors between 
the solenoid and dipole magnet;}
\item{the construction of support frame for detectors
inside the magnet gap;} 
\item{the design of  supports for mounting
the drift detectors at the exit of the dipole magnet;}
\item{the design of the movable platform for the Forward Spectrometer detectors.}
\end{itemize}  
The presented design of the detector supports was worked
out by the groups from the Cracow University of Technology (CUT) and
from the Jagiellonian University (JU), in close cooperation with the
whole \PANDA collaboration.

\subsection{Detectors Between Solenoid and Dipole}
\label{s:dip:det:sol-dip}

\subsubsection{Drift Chambers DC1 and DC2}

\begin{figure}[ht]
\begin{center}
\includegraphics[width=\swidth]{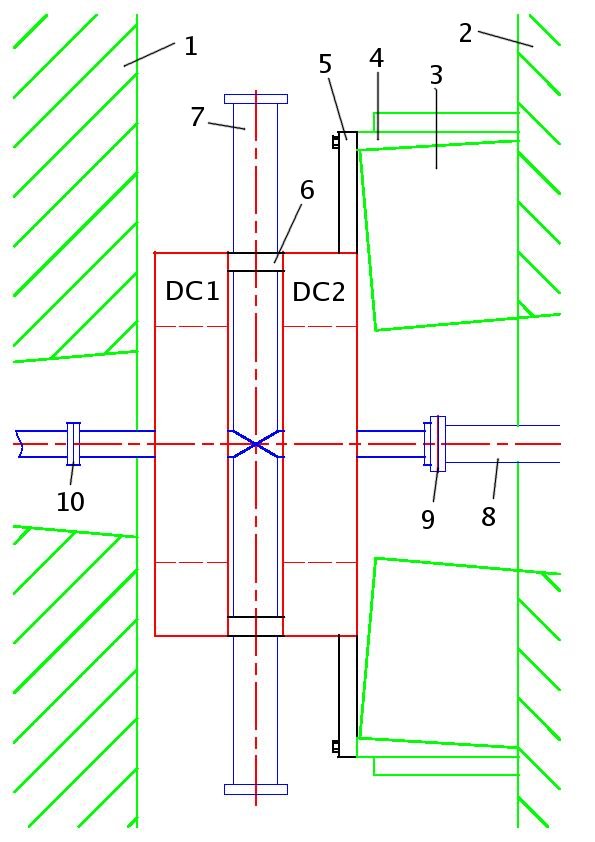}
\caption[Pair of drift detectors DC1 and DC2 mounted on the yoke of 
the dipole magnet.]{Pair of drift detectors DC1 and DC2 mounted on the yoke of 
the dipole magnet:
1~--~TS forward door, 2~--~yoke of the dipole magnet, 
3~--~coil, 4~--~coil holder,
5~--~steel profile for hanging the detectors, 6~--~connecting element,
7~--~vacuum pipe, 8~--~beam pipe, 9,10~--~beam pipe connections.}
\label{fig:sdc:DC1-2_sv}
\end{center}
\end{figure}

The first pair of drift detectors (DC1, DC2) will be installed in
the space between the Target Spectrometer  downstream door and the
yoke of the dipole magnet. The chambers will be mounted on the cross
formed by the beam pipe and a vertical pipe used for pumping (see
Fig.~\ref{fig:sdc:DC1-2_sv}). After that the detectors will be connected
together by means of steel joints and the vacuum pipe will be attached
to the detector frames by means of two clamps. The connection between the
detectors will guarantee the required high reproducibility of the
relative positioning.

During the installation of the drift detectors, the TS downstream door will
be opened.  The detectors will be mounted on the dipole magnet yoke by
means of vertical steel profiles attached to the horizontal coil
holders (see Fig.~\ref{fig:sdc:DC1-2_sv}).  After that the segment of
the beam pipe passing through the DC1-DC2 pair will be connected to the
TS beam pipe with proper flange connection. This connection is located 
inside the TS downstream door opening. The following step will be the
insertion of the frame which will hold the detectors inside
the dipole gap (see next section). This will be rolled in from downstream 
direction and the beam pipe will be connected with a flange foreseen at 
the $z$-position of the dipole magnet coils.  The access to this
connection will be limited by the coils to about 70\,cm in the vertical
direction and to 60\,cm in the horizontal direction by the remaining
free space between the chambers DC2 and DC3.  This space will be sufficient
for humans to access the beam pipe and connect the flanges safely.

\subsubsection{Muon Filter}

Surrounding the vacuum pipe and the tracking detectors DC1 and DC2,
a removable muon filter will be placed which serves as a continuation of
the muon detectors in the downstream door for the high momentum part
of the muon spectrum. 

The filter will consist of five layers of 6cm thick iron interleaved with
the same drift tube detectors used within the laminated solenoid yoke.
It is segmented in four individual blocks (upper-right, upper-left,
lower-right, lower-left). The four segments will have cut-outs to respect
the enclosed detectors and equipments and can be removed easily as blocks
by crane to give access to the drift chambers and the beam and vacuum
pipes.

The iron of the muon filter, in addition, will decouple the magnetic
fields of the solenoid and the dipole and reduces the stray field of
the dipole in the region between the two magnets.

\subsection{Yoke Gap Fittings}

\subsubsection{Support Frame}

\begin{figure}[htb]
\begin{center}
\includegraphics[width=\swidth]{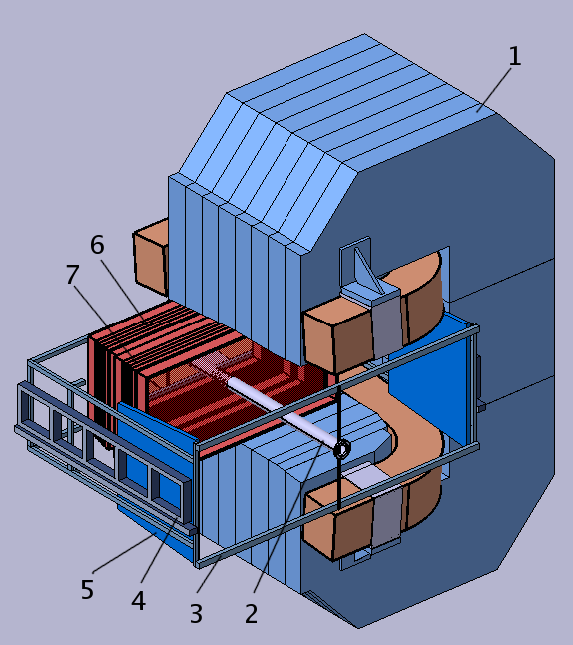}
\caption[Support frame for detectors inside the dipole magnet gap.]
{Support frame for detectors inside the dipole magnet gap: 
1~--~magnet yoke, 
2~--~beam pipe, 3~--~movable frame, 4~--~rails mounted on the wall of magnet,
5~--~time of flight detector, 6~--~DC3, 7~--~DC4.}
\label{fig:ygf:mag}
\end{center}
\end{figure}

The design of the frame foreseen for supporting detectors between the
poles of the dipole magnet is shown in Fig.~\ref{fig:ygf:mag}.  The
frame will be made of standard closed stainless-steel profiles.  It will have a
trapezoidal shape fitting the aperture of the magnet and will be supported
by two rails mounted on the side walls of the magnet gap.  The rails
will be used for sliding the frame in the gap during the installation.
The frame is foreseen to support the drift detectors DC3 and DC4 as
well as two TOF side walls.  It will be also used for supporting a
3\,m long part of the beam pipe crossing the magnet gap.

\subsubsection{Detector Installation}

The installation of detectors in the frame will be done with the frame
rolled out of the magnet gap.  The positions of the drift detectors with respect
to the frame will be fixed with the precision better than 0.1\,mm by
means of reference pins.  The beam pipe will be inserted in the
central openings in the detectors and subsequently it will be hanged
on the frame by means of two vertical bars.  Afterwards, the detectors
will be cabled and connected to supply lines.  The installation of the
whole assembly inside the gap of the dipole magnet will be
accomplished just by rolling it inside the gap.  Excesses foreseen in
the guidance of the detector cables and the supply lines should allow
for a free movement of the structure during the installation.  
The movement of
the frame inside the magnet gap will be limited by mechanical blockade.
Positioning pins mounted on the blockade system are used to fix
precisely the position of the support frame with respect to the magnet
yoke.

\subsection{Drift Chambers at the Exit of the Dipole}

\begin{figure}
\begin{center}
\includegraphics[width=\swidth]{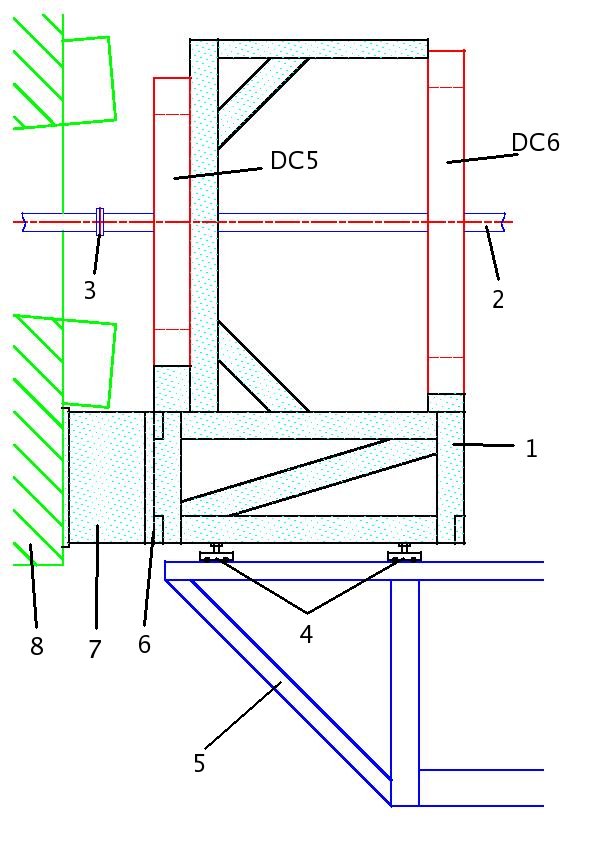}
\caption[Side view of the support frame for the drift detectors DC5 and DC6.]
{Side view of the support frame for the drift detectors DC5 and DC6:
1~--~support frame, 2~--~beam pipe, 
3~--~beam pipe connection, 4~--~adjustable feet, 
5~--~forward platform, 6~--~connection between support frame
and cantilever, 7~--~steel cantilever, 8~--~dipole magnet yoke.}
\label{fig:sdc:support1_sv}
\end{center}
\end{figure}

\begin{figure}[ht]
\begin{center}
\subfigure[]{
\includegraphics[width=0.9\swidth]{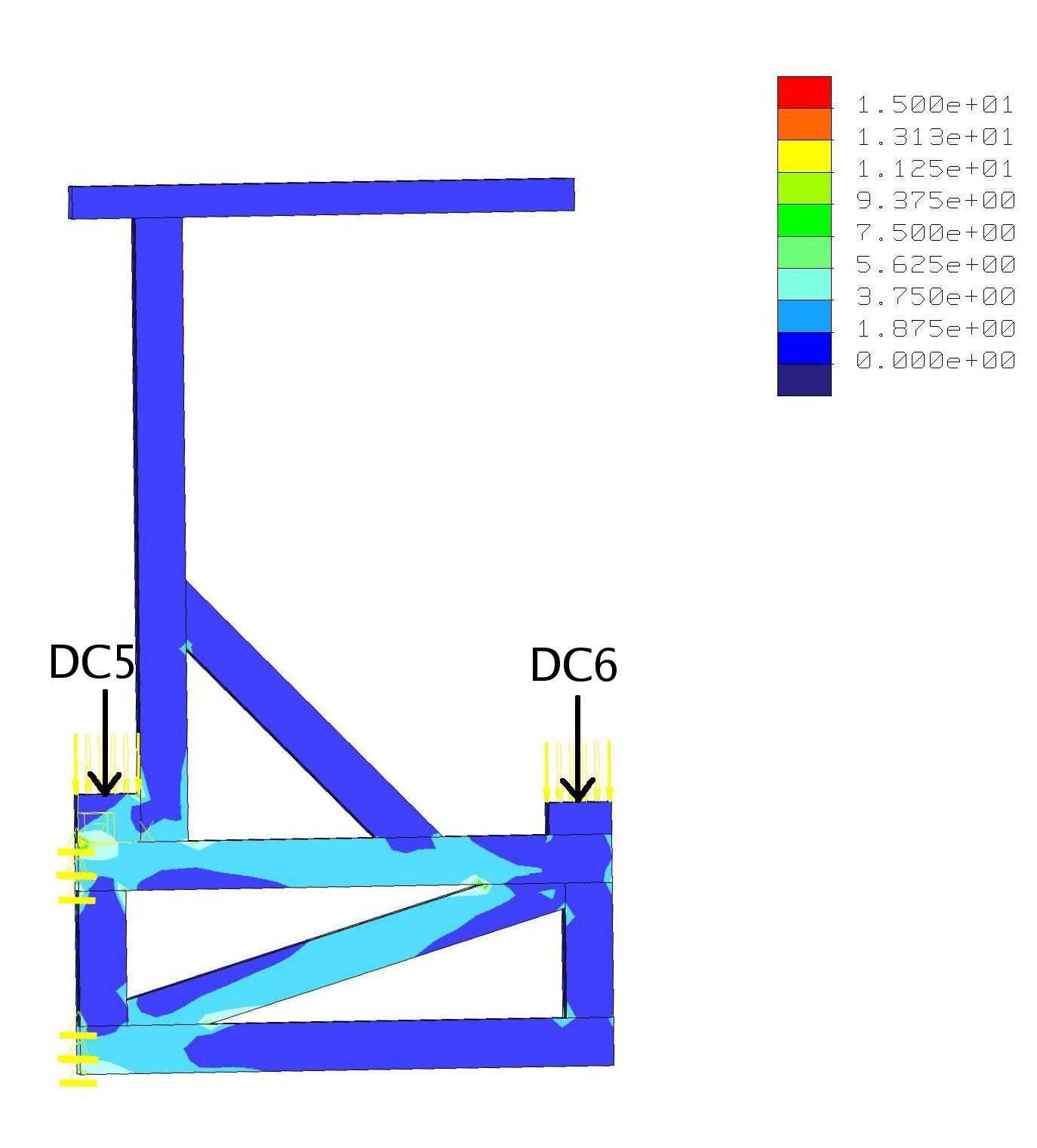}
\label{fig:sdc:Stress_5-6}}
\subfigure[]{
\includegraphics[width=0.9\swidth]{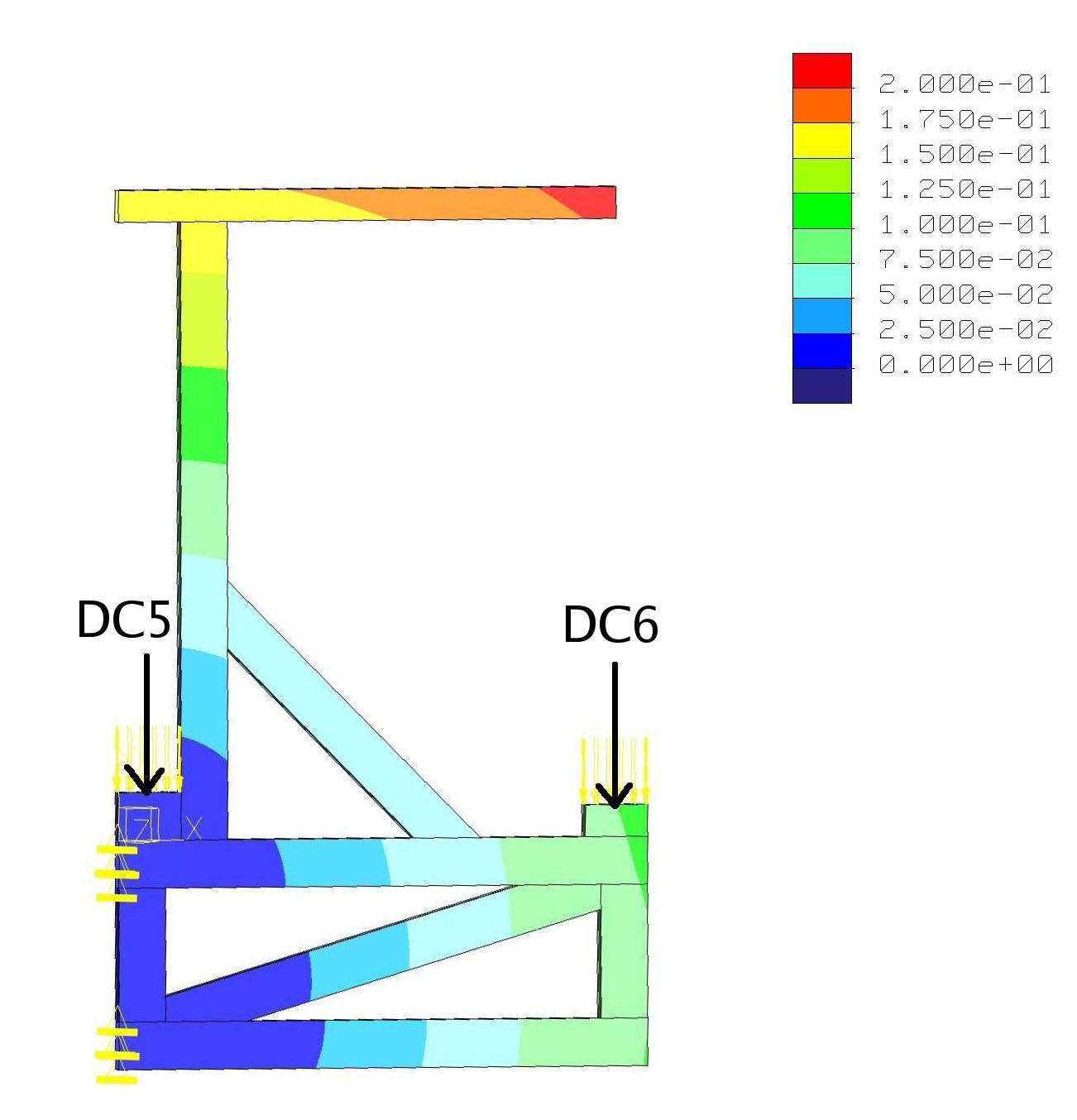}
\label{fig:sdc:Disp_5-6}}
\caption[Von Mises stresses in MPa and deformation 
of the support frame.]{a) Von Mises stresses in MPa and b) deformation in mm 
of the support frame 
due to weight of DC5 (300\,kg) and DC6 (400\,kg).}
\end{center}
\end{figure}

The support for the third pair of drift chambers, DC5 and DC6, will be built as a
frame with a rectangular box shape with attached bracket to hold
them in proper orientation and position (see Fig.~\ref{fig:sdc:support1_sv}).
The support will be constructed using standard stainless steel profiles.  For the
total weight of the chambers pair, about 700\,kg, a profile with
transverse dimensions of 50\,mm~x~150\,mm and wall thickness of 4\,mm
will guarantee sufficient stiffness of the support.  A side view of
the support frame is shown in Fig.~\ref{fig:sdc:support1_sv}.  In the
service area, the frame will stand on four feet on the platform for the
forward detectors.  After installation of the drift detectors, the
frame will be transferred on the platform to the experimental area and
it will be hanged on two steel cantilevers that will be permanently
fixed to the dipole magnet yoke.  To assure reproducibility of the
frame position with respect to the yoke, precisely machined junctions
between the frame and the cantilevers are foreseen.  After mounting
the frame on the cantilevers, the frame feet can be released, since
the load is taken over by the cantilevers.

The stresses in the frame under the expected load of DC5 (300\,kg) and
DC6 (400\,kg) lie in a safe range below 7\,MPa (see
Fig.~\ref{fig:sdc:Stress_5-6}).  The maximum deformation of the frame
is equal to 0.2\,mm (see Fig.~\ref{fig:sdc:Disp_5-6}) and is
acceptable from the point of view of the required positioning
reproducibility.

\subsection{Platform for Forward Detectors}

\subsubsection{Construction}

\begin{figure*}[htb]
\begin{center}
\includegraphics[width=1.6\swidth]{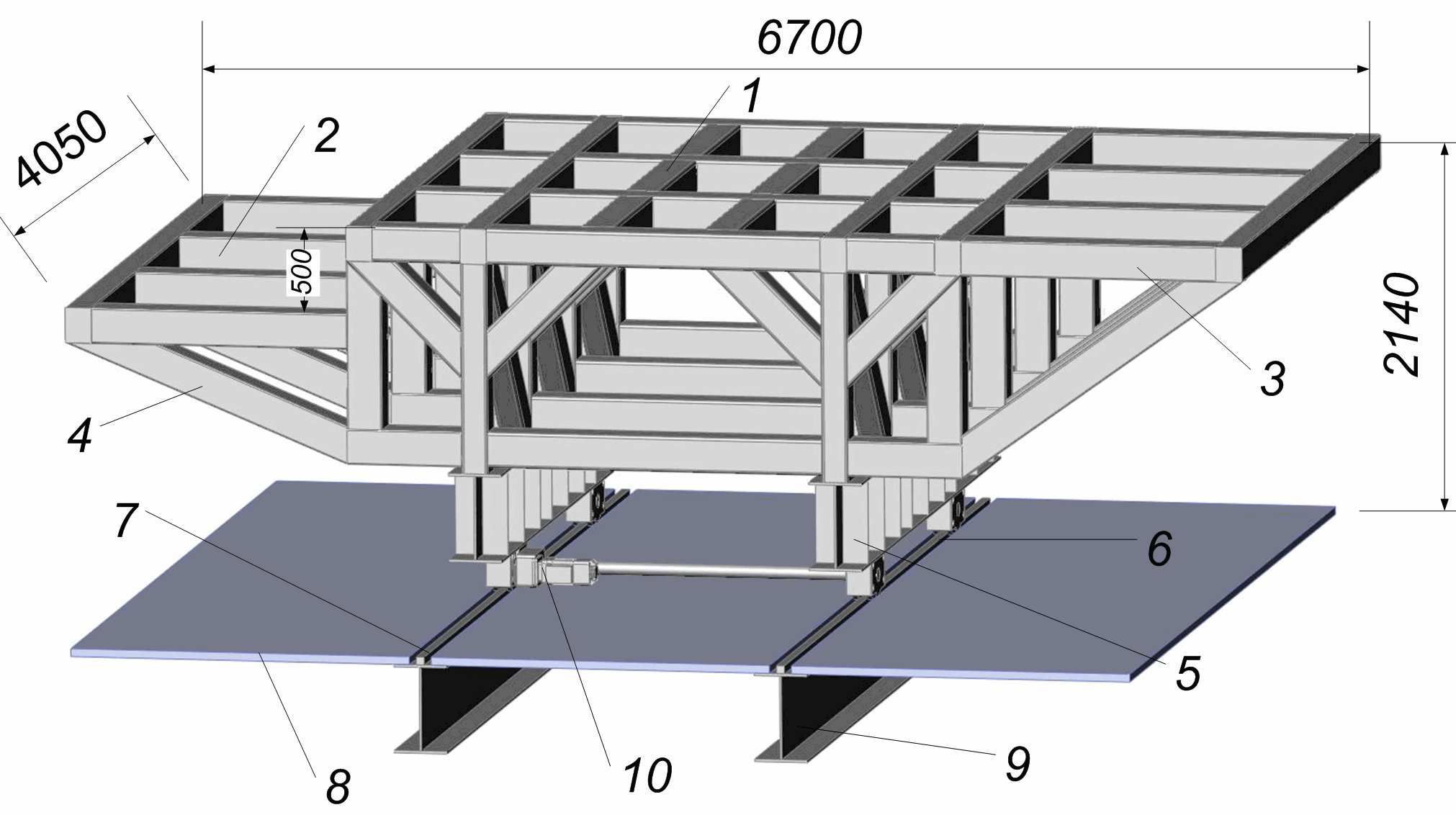}
\caption[Model of the platform for the forward detectors.]
{Model of the platform for the forward detectors:
1~--~main platform, 2~--~lowered part of platform, 3,4~--~framework made of closed profiles,
5~--~carrying beams, 6~--~wheel sets, 7~--~wheel tracks, 
8~--~plate determining the floor level, 9~--~reinforcing beams,
10~--~driving unit.}
\label{fig:pfd:model}
\end{center}
\end{figure*}

The components of the Forward Spectrometer detection system
including the RICH detector, the TOF-wall, the electromagnetic 
calorimeter and the forward muon detectors, as well as the 
luminosity monitor, will be placed on a common platform 
which will be used to support them on a proper height with respect 
to the beam line. It will also allow to move the detectors from 
the experimental area to the service position. A view of the proposed 
platform is shown in Fig.~\ref{fig:pfd:model}.

The area of the platform will be 6.70\,m~x~4.05\,m and the height will be 2.14\,m. 
The heaviest detectors, the electromagnetic calorimeter (8\,tonnes), 
and the forward muon detectors (49\,tonnes),
will be set up in the central part of the platform.
The RICH detector and the forward TOF-wall, with a total mass of about 
2\,tonnes, will be placed near the dipole magnet on a lowered part of 
the platform. The platform will be $\sim0.5\unit{m}$ lower in the upstream
segment to host the RICH detector. The lowered part of the platform will 
be also used to support the drift detectors DC5 and DC6 during the 
transfer between the service position and the experimental area.  
The far end of the platform in downstream direction 
will be used for supporting the luminosity monitor.

The platform will be built as a framework based on two carrying beams
with attached wheel sets.  In the footing, below the floor level,
reinforcing beams with attached wheel tracks will be placed.

To build the carrying beams we will use a construction profile
commonly used in the industry according to the standard EN~10024.  The
profile will be additionally strengthened by vertical ribs.  The same
type of profile will be also used to build the reinforcing beams.  The
platform framework will be built of rectangular steel profiles
(material: S355J2, according to EN-10025-2).  The total mass of the
platform is 5.5\,tonnes.

\subsubsection{Driving system}
For the driving system of the detector platform, we plan to use wheels
with two rims moving on rectangular bars welded to rails.  The chosen
wheel diameter is 250\,mm and the spacing of wheel rims is by
$2s=1$\,mm larger than the width of the rail bars equal to
$k=110$\,mm.  Wheels will be mounted in blocks which will be fastened
under the carrying beams (Fig.~\ref{fig:pfd:model}).  To move the
platform, with its total foreseen load of 60\,tonnes, four wheel blocks
will be sufficient.  The platform will be driven by a single
electric motor, with transmission shaft transferring the torque to two
wheels.  Movement of the platform with maximum velocity of 0.05\,m/s
can be realised by an electric motor with the power of 0.5\,kW.  The
electric motor will be equipped with a gear box and a brake and will
be steered using an electronic system of movement control.

\subsubsection{Stress calculations}

\begin{figure}[ht]
\begin{center}
\subfigure[]{
\includegraphics[width=\swidth]{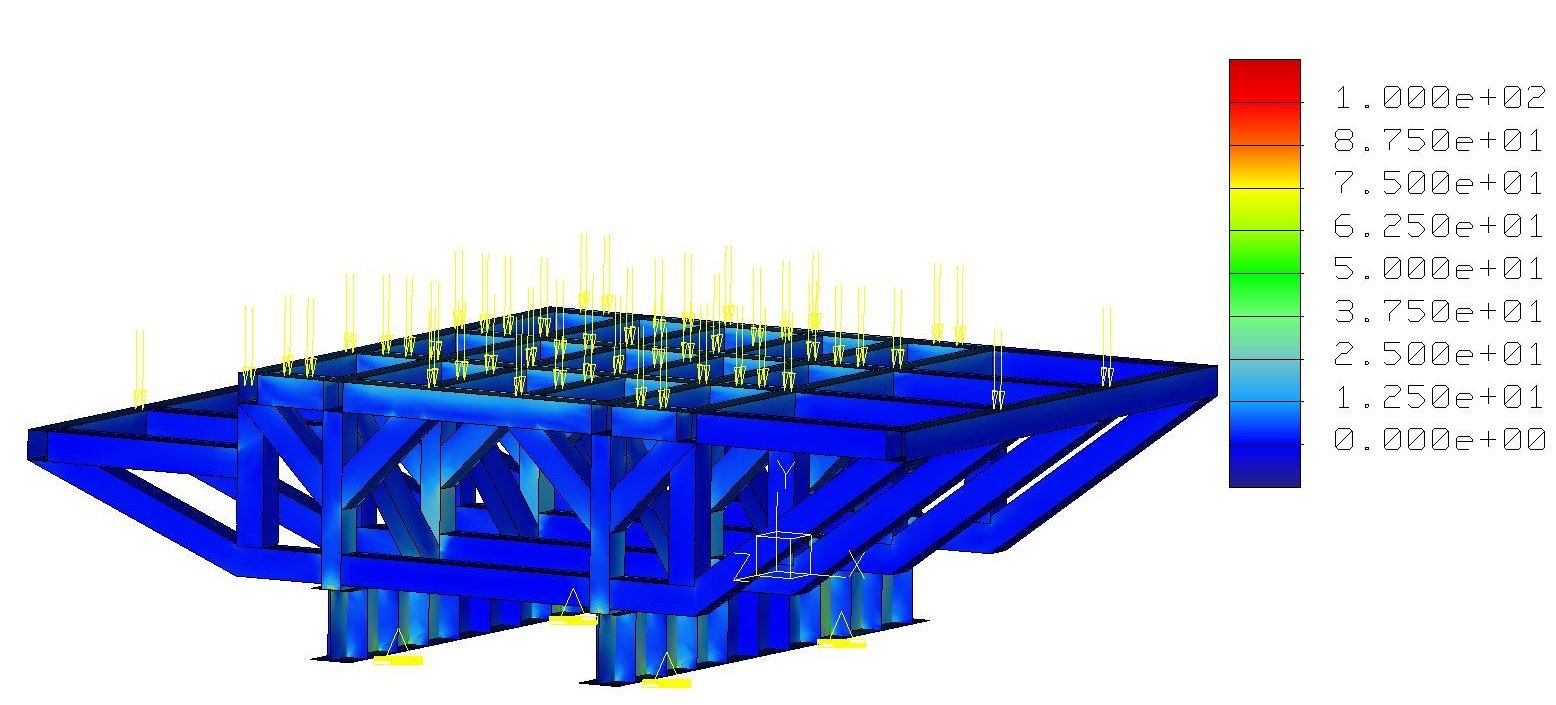}
\label{fig:pfd:stress_60t}}
\subfigure[]{
\includegraphics[width=\swidth]{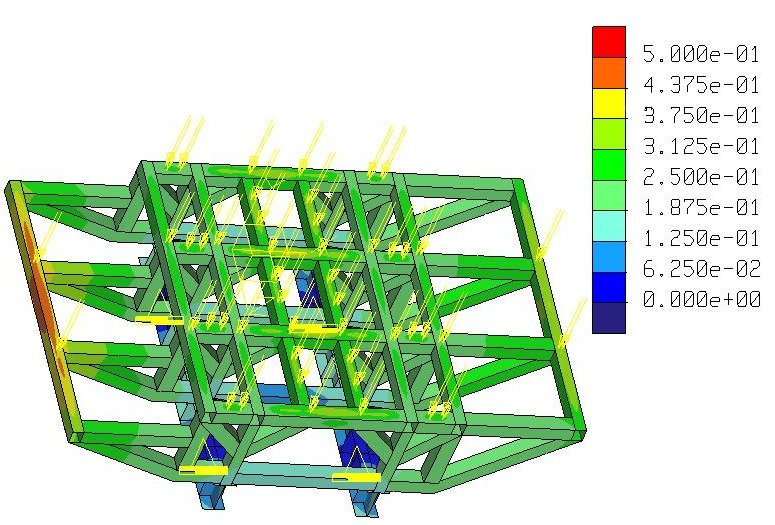}
\label{fig:pfd:disp_60t}}
\caption[Von Mises stresses in MPa and deformation of the platform.]
{ a) Von Mises stresses in MPa and b) deformation in mm of the platform due
to load of 60\,tonnes.}
\end{center}
\end{figure}

Calculations of mechanical stress in the platform were carried out
using the application Pro/Engineer - Pro/Mechanica.  Under the
expected load of 60\,tonnes, the maximum stress will be lower than 75\,MPa
(see Fig.~\ref{fig:pfd:stress_60t}) and is acceptable for the used
material.  Fig.~\ref{fig:pfd:disp_60t} presents corresponding
displacement contours.  The maximum displacement will not exceed
0.5\,mm and thus is acceptable from the point of view of the required
positioning accuracy of the detectors.

\subsubsection{Transport to FAIR}

For transfer from manufacturer site to FAIR, the platform 
it will be divided into
three subunits: central platform and two side platforms. The
dimensions of these subunits allow for delivery of each of them by
conventional means of transport: trucks or railway.  After transport
to FAIR - to the PANDA experimental hall - the three subunits will be
assembled by means of screwing and welding.

\bibliographystyle{tdr_lit}
\bibliography{lit}


\cleardoublepage
\svnInfo $Id: orga.tex 709 2009-02-10 09:03:42Z IntiL $ 

\chapter{Organisation}
\label{s:manage}
\label{s:orga}

\AUTHOR{I.~Lehmann}

The two large spectrometer magnets will be built by seven groups from
universities and research institutes in Germany, Italy, Russia, Poland
and the UK. The institutions and their abbreviations are listed in
Table~\ref{t:orga:inst}.

\begin{table}[htb]
\begin{center}
\begin{tabular}{|l l|}
\hline
Abr.\ & Institution \\
\hline
\hline
CUT   & Cracow University of Technology, \\ 
      & Krak\'ow, Poland \\
FZJ   & Forschungszentrum J\"ulich, \\
      & J\"ulich, Germany \\
Gla   & University of Glasgow, \\
      & Glasgow, United Kingdom \\
GSI   & GSI Helmholtzzentrum f\"ur \\
      & Schwerionenforschung GmbH, \\
      & Darmstadt, Germany \\
INFN  & INFN, Sezione di Genova, \\
      & Genova, Italy \\
JINR  & Joint Institute for Nuclear Research, \\ 
      & Dubna, Russia \\
UJ    & Jagiellonian University, \\ 
      & Krak\'ow, Poland \\
\hline
\end{tabular}
\end{center}
\caption{Table of abbreviations and institutions responsible for the 
  design and construction of the magnets at \PANDA.}
\label{t:orga:inst}
\end{table}

\section{Work Packages and Responsibilities}

Five distinct primary work packages have been identified. These comprise
the design and construction of the following items.
\begin{enumerate}
\item The coil and cryostat of the Target Spectrometer. The overall
  responsibility is taken by INFN, Genoa.
\item The flux return yoke of the Target Spectrometer, which will serve
  simultaneously as multi-layer absorber for a large-angle muon
  detection system. The overall responsibility is taken by JINR,
  Dubna.
\item The large-aperture dipole magnet for the Forward
  Spectrometer. The overall responsibility is taken by the University
  of Glasgow.
\item The support structures and detector mountings for all of the 
  Forward Spectrometer. The overall responsibility is taken jointly by
  the Jagiellonian University, Krak\'ow and the Cracow University
  of Technology.
\item The railing systems and movement of the whole Target 
  Spectrometer and the platform with the Forward Spectrometer
  detectors. The overall responsibility is taken jointly by 
  GSI, Darmstadt and Cracow University of Technology.
\end{enumerate}
The Forschungszentrum J\"ulich takes care of the \PANDA
spectrometers' integration into the HESR, which is particularly
important for the dipole, since it will be part of the
accelerator/storage ring lattice. 

The responsibility for the detector mountings inside the Target
Spectrometer will remain with the individual detector groups, while the
magnet group will provide only mounting points.  A detailed list of
work packages has been worked out, and responsibilities have been
identified, which have been approved by all groups.  These are listed
in Table~\ref{t:orga:wp}, where the responsible institutions are
listed by their abbreviations specified in Table~\ref{t:orga:inst}.
Where responsibilities are shared the main responsible is given first
and the secondary is indicated in brackets.  The main responsible
institution will always supervise the work package and take
responsibility for a timely and full completion.

\begin{table*}[htb]
\begin{center}
\begin{tabular}{|l l |c|}
\hline
\multicolumn{2}{|l|}{Task}                                                & Responsible    \\
\hline
\hline
\multicolumn{3}{c}{\bf Instrumented Flux Return} \\
\hline
& Design, documentation, tender, supervision of construction              & JINR   \\
& Material procurement, manufacturing, assembly and transport preparation & JINR   \\
& Assembly and tests at company                                           & JINR   \\
& Transport to FAIR                                                       & JINR   \\
& Interface for muon system                                               & JINR   \\
\hline
\multicolumn{3}{c}{\bf Coil \& Cryostat}  \\
\hline
& Cold mass cooling circuit design                                        & INFN   \\
& Selection of cable                                                      & INFN   \\
& Mandrel and winding (spacers, connectors...) design                     & INFN   \\
& Mechanical suspension of the cold mass in the cryostat                  & INFN   \\
& Design of intermediate temperature shields        				        & INFN           \\
& Feed lines and turret                                                   & INFN(FZJ)      \\
& Coil Protection system                                                  & INFN           \\
& Coil diagnostic and DAQ                                                 & INFN(GSI)      \\
& Tender and procurement                                                  & INFN           \\
& Follow-up of the cable procurement \& tests                             & INFN           \\
& Follow-up of the coil construction                                      & INFN           \\
& Follow-up of the cryostat construction                                  & INFN           \\
& Follow-up of the cryogenic turret feed lines                            & INFN           \\
& Coil, cryostat, turret final assembly at manufacturer site              & INFN           \\
& Cryogenic and electric tests at company                                 & INFN           \\
& Transport to FAIR                                                       & INFN           \\
\hline
\multicolumn{3}{c}{\bf Dipole Magnet}  \\
\hline
& Final dipole design                                                     & Gla      \\
& Procurement and quality assurance (no assembly)                         & Gla      \\
& Dipole slow control                                                     & Gla      \\
\hline
\multicolumn{3}{c}{\bf Detector Support Structures}  \\
\hline
& Platform for the FS detectors                                           & CUT      \\
& Supports in the dipole magnet gap                                       & CUT      \\
& Wire chamber supports of the FS                                         & UJ       \\
& Absorber system for muon filtering between TS and FS                    & CUT      \\
\hline
\multicolumn{3}{c}{\bf Magnet Support Structures}  \\
\hline
& Rail system and moving of solenoid                                      & GSI(CUT) \\
& Solenoid support structure                                       & JINR     \\
& Dipole support structure                                                & Gla      \\
\hline
\multicolumn{3}{c}{\bf Assembly and Commissioning at FAIR}  \\
\hline
& Assembly of yoke and support structures                          & JINR   \\
& Installation of cryostat and supply lines                        & INFN (JINR) \\
& Assembly of dipole magnet and power supply                       & Gla\\
& Alignment of magnets                                                    & GSI (FZJ)      \\
& Commissioning of solenoid                                        & INFN (JINR)\\
& Commissioning of dipole                                          & Gla\\
& Field mapping for both magnets                                          & GSI     \\
& Assembly of detector supports in the FS                          & CUT(UJ) \\
\hline
\end{tabular}
\end{center}
\caption[Table of work packages listing the individual work packages 
and the responsible institutions.]
{Table of work packages listing the individual work packages 
and the responsible institutions by the abbreviations as listed in
Table~\ref{t:orga:inst}.  The institutions listed in brackets are
responsible for a specific part of the WP but the overall
responsibility is taken by the leading institution.}
\label{t:orga:wp}
\end{table*}

\section{Timelines}
\label{s:time}

The timelines are driven by the envisaged start of commissioning of
HESR in 2014.  In particular, the
Forward Spectrometer needs to be in place in order to allow any beam
operation, as the dipole forms part of the accelerator lattice. In
order to allow for field mapping, detector installation and
commissioning the magnets and support structures must be in place and
commissioned by 2013.

\section{General Safety Aspects}

The design details and construction of the magnets including the
infrastructure for operation will be done according to the safety
requirements of FAIR and the European and German safety regulations.

All electrical equipment will comply to the legally required
safety code and concur to standards for large scientific installations
following guidelines worked out at CERN to ensure
the protection of all personnel working at or close to the components
of the \PANDA experimental facility.  Power supplies will be mounted
safely and independently from large mechanical loads.  Hazardous
voltage supplies and lines will be marked visibly and protected from
damage by any equipment which may cause forces to act on them.  All
supplies will be protected against overcurrent and overvoltage and
have appropriate safety circuits and fuses against short cuts.  All
cabling and connections will use non-flammable halogen-free materials
according to up-to-date standards and will be dimensioned with proper
safety margins to prevent overheating.  A safe ground scheme will be
employed throughout all electrical installations of the experiment.
Smoke detectors will be mounted in all appropriate locations. The more
specific safety considerations are discussed in the respective
sections throughout the document.

\begin{landscape}

\begin{figure}
   \includegraphics[width=24cm]{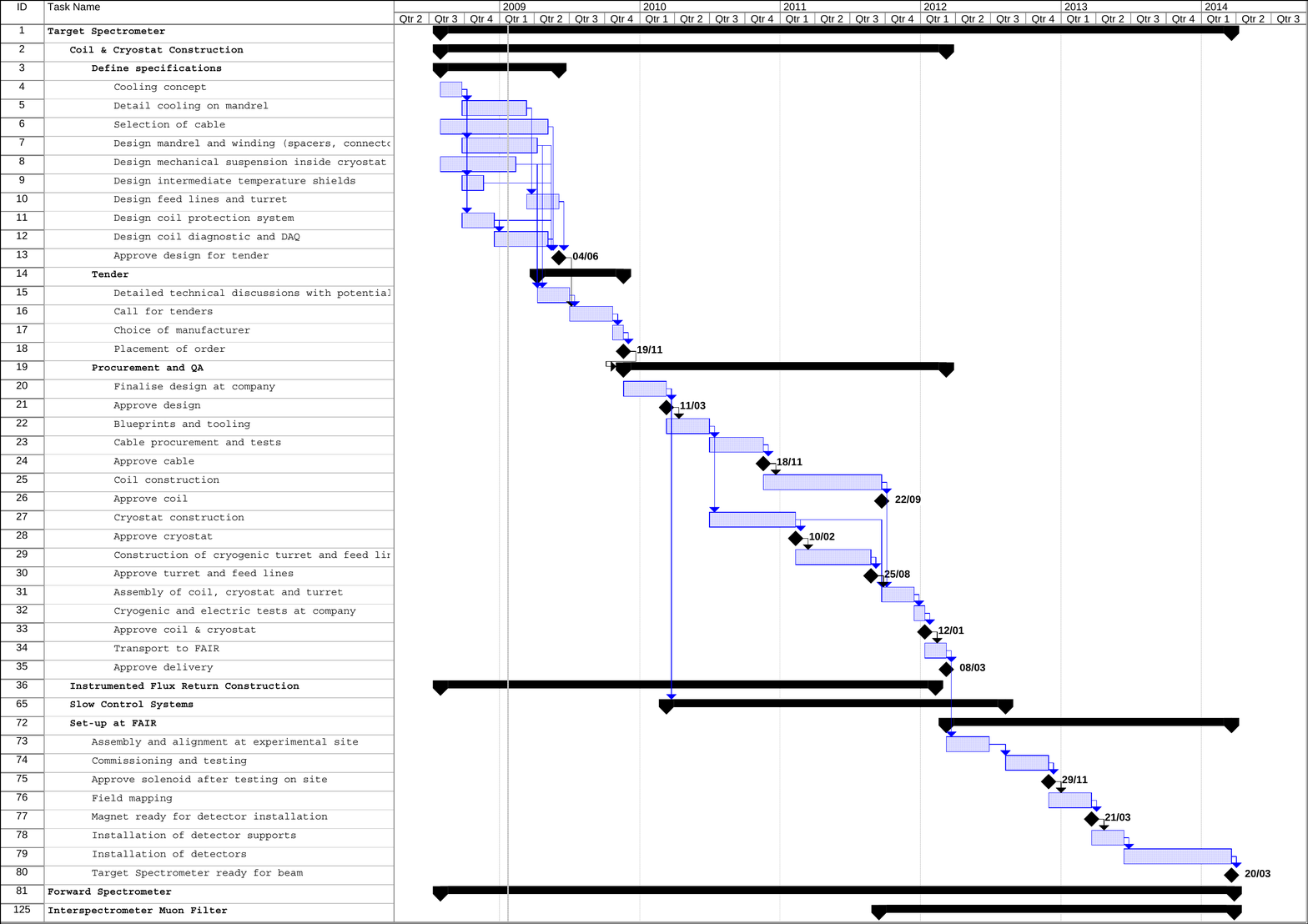}
   \caption{Timelines highlighting the coil and cryostat of the Target
   Spectrometer.}
\end{figure}

\end{landscape}
\begin{landscape}


\begin{figure}
   \vspace*{-1cm}
   \includegraphics[width=24cm]{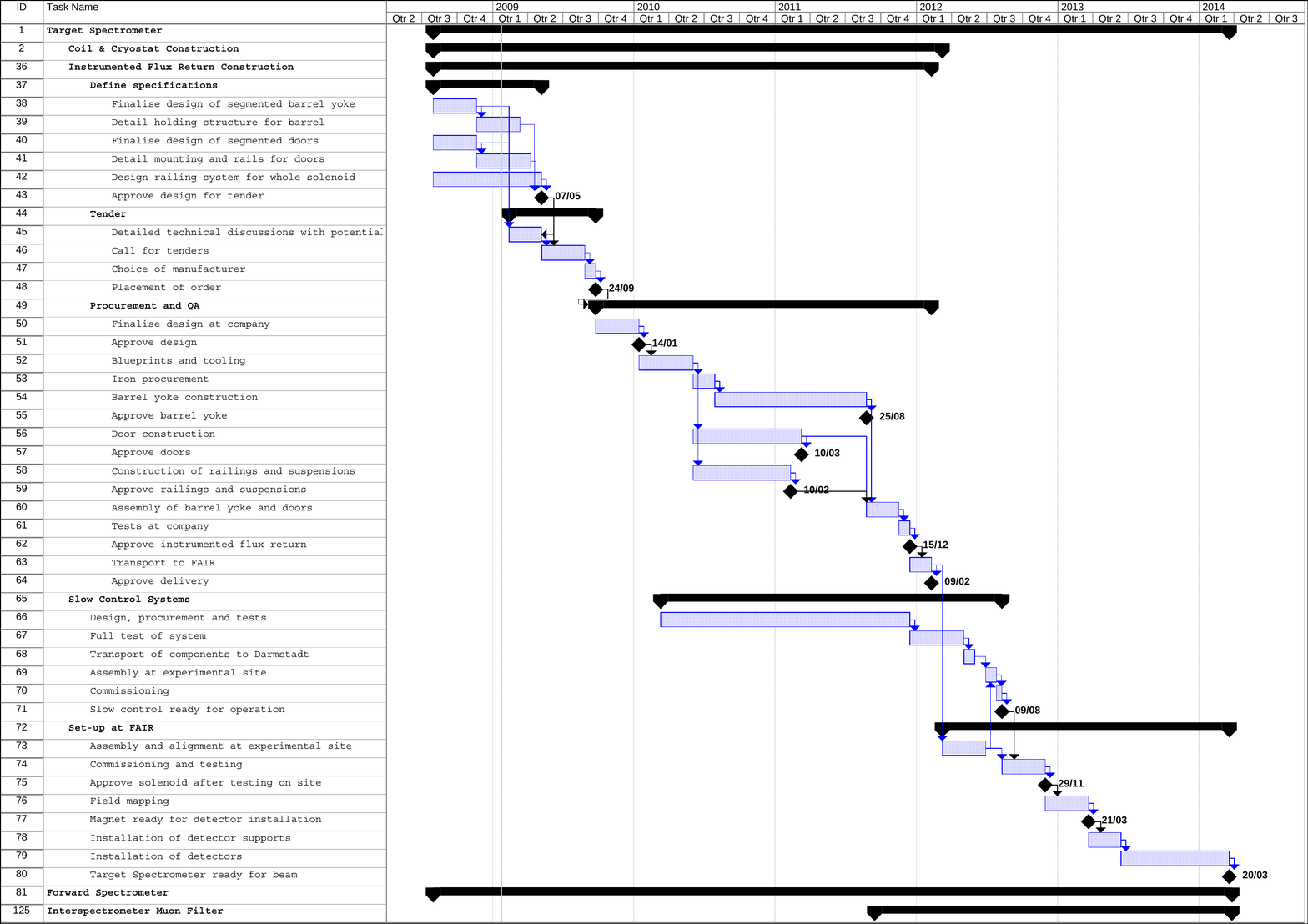}
   \caption{Timelines highlighting the instrumented flux return and 
     the slow control systems of the Target Spectrometer.}
\end{figure}

\end{landscape}
\begin{landscape}
\begin{figure}
   \includegraphics[width=24cm]{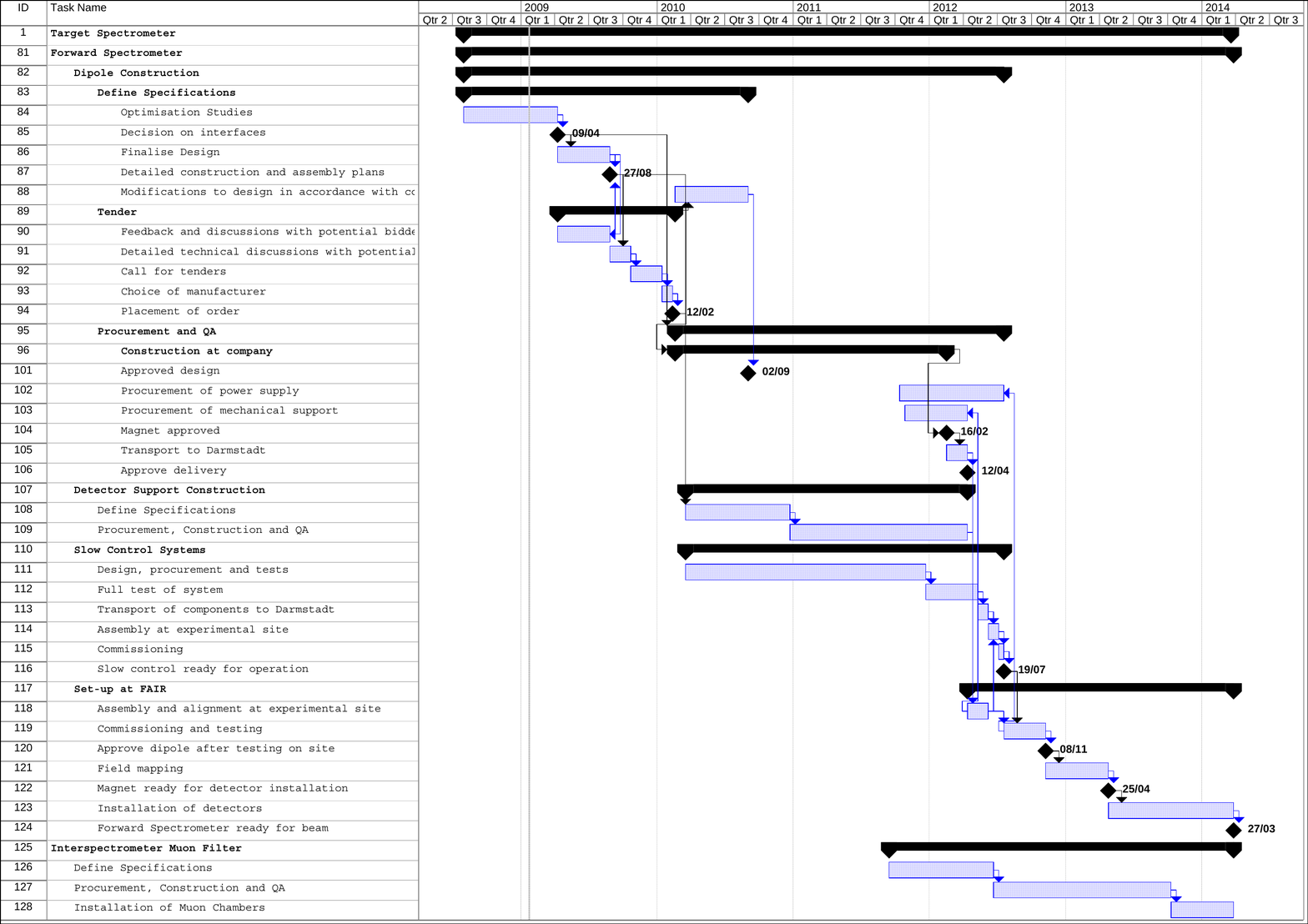}
   \caption{Timelines highlighting the Forward Spectrometer and intermediate
   muon filter.}
\end{figure}

\end{landscape}



\appendix
\refstepcounter{chapter}
\onecolumn
%
\svnInfo $Id: ackno.tex 711 2009-02-10 21:37:26Z IntiL $

\begin{center}
\vspace*{2cm}
{\Large\bf Acknowledgments}
\addcontentsline{toc}{chapter}{Acknowledgements}
\vskip 2cm
\begin{minipage}[t]{8cm}
\sloppy\large
We acknowledge financial support from:
the Bundesministerium f\"ur Bildung und Forschung (bmbf),
the Deutsche Forschungsgemeinschaft (DFG),
the University of Groningen, Netherlands,
the GSI Helmholtz\-zentrum f\"ur Schwer\-ionen\-forschung GmbH, Darmstadt,
the Helmholtz-Gemeinschaft Deutscher Forschungszentren (HGF),
the Schweizerischer Nationalfonds zur F\"orderung der wissenschaftlichen
Forschung (SNF),
the Russian funding agency ``State Corporation for Atomic Energy Rosatom'',
the CNRS/IN2P3 and the Universit\'e Paris-sud,
the British funding agency ``Science and Technology Facilities
Council'' (STFC),
the Instituto Nazionale di Fisica Nucleare (INFN),
the Swedish Research Council,
the Polish Ministry of Science and Higher Education,
the European Community FP6 FAIR Design Study: DIRACsecondary-Beams,
contract number 515873,
the European Community FP6 Integrated Infra\-structure Initiative:
HadronPhysics, contract number RII3-CT-2004-506078,
the INTAS, 
and the Deutscher Akademischer Austauschdienst (DAAD).
\end{minipage}
\end{center}
\vfill
%
%

%
\cleardoublepage
\refstepcounter{chapter}
\onecolumn
%
\svnInfo $Id: acronyms.tex 710 2009-02-10 12:13:38Z IntiL $

\onecolumn

\chapter*{List of Acronyms}
\addcontentsline{toc}{chapter}{List of Acronyms}
\begin{acronym}
\acro{CAD}{Computer Aided Design} 
\acro{CUT}{Cracow University of Technology, Krak\'ow, Poland}
\acro{DAQ}{Data Acquisition}
\acro{DIRC}{Detector for Internally Reflected Cherenkov Light}
\acro{EMC}{Electromagnetic Calorimeter} 
\acro{FAIR}{Facility for Antiproton and Ion Research}
\acro{FEM}{Finite Element Model}
\acro{FS}{\PANDA Forward Spectrometer}
\acro{FZJ}{Forschungszentrum J\"ulich} 
\acro{GEM}{Gas Electron Multiplier}
\acro{GSI}{GSI Helmholtzzentrum f\"ur Schwerionenforschnung GmbH, 
      Darmstadt, Germany} 
\acro{HEP}{High Energy Physics}
\acro{HESR}{High Energy Storage Ring}
\acro{HL}{High Luminosity (mode of the \HESR)}
\acro{HR}{High Resolution (mode of the \HESR)}
\acro{INFN}{Istituto Nazionale di Fisica Nucleare, Sezione di Genova,
      Genova, Italy}
\acro{IP}{Interaction Point (beam-target crossing at \PANDA)}
\acro{JINR}{Joint Institute for Nuclear Research, Dubna, Russia}
\acro{LoI}{Letter of Intent (for \PANDA)} 
\acro{MDT}{Micro Drift-Tube (used for the muon detectors)}
\acro{MVD}{Micro Vertex Detector} 
\acro{NMR}{Nuclear Magnetic Resonance}
\acro{PANDA}{Antiproton Annihilations at Darmstadt}
\acro{RICH}{Ring Imaging Cherenkov}
\acro{RMS}{Root Mean Square}
\acro{STT}{Straw Tube Tracker} 
\acro{TDR}{Technical Design Report} 
\acro{TPC}{Time Projection Chamber}
\acro{TPR}{Technical Progress Report (for \PANDA)} 
\acro{TS}{\PANDA Target Spectrometer} 
\acro{UJ}{Jagiellonian University, Krak\'ow, Poland}
\end{acronym}
\vfill
%
%

%
\cleardoublepage
\refstepcounter{chapter}
\onecolumn
\listoffigures
\addcontentsline{toc}{chapter}{List of Figures}

%
\cleardoublepage
\refstepcounter{chapter}
\onecolumn
\listoftables
\addcontentsline{toc}{chapter}{List of Tables}

\end{document}